\documentclass[10 pt]{article}
\usepackage[latin1]{inputenc}
\usepackage[english]{babel}
\usepackage{latexsym}
\usepackage{amsmath}
\usepackage{amsfonts}
\usepackage{amssymb}
\usepackage{amsthm}
\ifx\macrosloaded\relax\endinput\else\let\macrosloaded\relax\fi
\newtheorem{thm}{Theorem}[section]

\newtheorem{cor}{Corollary}[section]

\catcode`\@=11
\newcommand{\eqinsec}{\relax\@addtoreset{equation}{section}}
\renewcommand{\theequation}{\ifx\showlabels\iftrue\the\id\else\thesection.\arabic{equation}\fi}
\catcode`\@=13
\newcounter{supeq}
\newenvironment{subeq}
\stepcounter{equation} 
\def\theequation{\ifx\showlabels\iftrue\the\id\else\thesection.\arabic{equation}\fi}
\newtoks\id
\newcommand{\eqlabel}[1]{\label{#1}\global\id={(#1)}}
\newcommand{\medn}{\medskip\noindent}
\newcommand{\tr}{\mbox{tr}}
\newcommand{\be}{\begin{equation}}
\newcommand{\eeq}{\end{equation}}
\newcommand{\bea}{\begin{eqnarray}}
\newcommand{\eea}{\end{eqnarray}}
\newcommand{\beaa}{\begin{eqnarray*}}
\newcommand{\eeaa}{\end{eqnarray*}}
\newcommand{\bseq}{\begin{subeq}}
\newcommand{\eseq}{\end{subeq}}
\newcommand{\ba}{\begin{array}}
\newcommand{\ea}{\end{array}}
\newcommand{\eql}{\eqlabel}

\def \rectangle#1#2{\hbox{\vrule\vbox to #2
{\hrule\hbox to
#1{\hfil}\vfil\hrule}\vrule}}
\newcommand{\edd}{\end{document}}

\renewcommand{\c}{\cdot}
\newcommand{\NI}{\noindent}

\newcommand{\Lb}{\underline{L}}

\newcommand{\Si}{\Sigma}
\newcommand{\ga}{\gamma}
\newcommand{\Ga}{\Gamma}

\newcommand{\dd}{\mbox{${\bf D}$}}

\newcommand{\lap}{\mbox{$\bigtriangleup$}}
\newcommand{\lapp}{\mbox{$\bigtriangleup  \mkern-13mu /$\,}}
\newcommand{\nab}{\mbox{$\nabla$}}
\newcommand{\nabb}{\mbox{$\nabla \mkern-13mu /$\,}}
\newcommand{\prb}{\mbox{$\pr \mkern-9mu /$\,}}

\newcommand{\Us}{\mbox{$U\mkern-13mu /$\,}}

\newcommand{\Fs}{\mbox{$F\mkern-13mu /$\,}}

\newcommand{\ddb}{\mbox{$\dd \mkern-13mu /$\,}}
\newcommand{\BBb}{\mbox{$B \mkern-11mu /$\,}}
\newcommand{\pr}{\partial}

\newcommand{\hot}{\widehat{\otimes}}

\newtheorem{Le}{Lemma}[section]

\newcommand{\Lie}{\mbox{$\cal L$}}

\newcommand{\lie}{\hat{\Lie}}

\newcommand{\nn}{\nonumber}
\newcommand{\Sch}{{Schwarzschild}}

\newcommand{\chib}{\underline{\chi}}

\newcommand{\de}{\delta}
\newcommand{\De}{\Delta}
\newcommand{\e}{\epsilon}
\newcommand{\xib}{\underline{\xi}}
\newcommand{\chih}{\hat{\chi}}

\newcommand{\chibh}{\underline{\hat{\chi}}}

\newcommand{\mub}{\underline{\mu}}

\newcommand{\und}[1]{\underline{#1}}

\newcommand{\Nb}{\und{N}}

\newcommand{\Cb}{\und{C}}

\newcommand{\ub}{{\und{u}}}
\renewcommand{\c}{\cdot}

\newcommand{\M}{{\cal M}}

\renewcommand{\aa}{\underline{\alpha}}
\newcommand{\bb}{\underline{\beta}}
\renewcommand{\a}{\alpha}
\renewcommand{\b}{\beta}
\newcommand{\dual}{\mbox{}^{\star}\!}

\newcommand{\si}{\sigma}

\newcommand{\ro}{\rho}

\newcommand{\divv}{\mbox{div}\mkern-19mu /\,\,\,\,}

\newcommand{\curll}{\mbox{curl}\mkern-19mu /\,\,\,\,}

\newcommand{\curl}{\mbox{curl}}

\newcommand{\om}{\omega}
\newcommand{\oom}{\Omega}

\newcommand{\omb}{\underline{\omega}}

\newcommand{\etab}{\underline{\eta}}
\newcommand{\la}{\lambda}

\newcommand{\ep}{\epsilon}

\newcommand{\dddd}{{\bf D} \mkern-13mu /\,}

\newcommand{\QQ}{{\widetilde{\cal Q}}}
\newcommand{\QQb}{\underline{\widetilde{{\cal Q}}}}

\newcommand{\Liee}{{\cal L}}    
\newcommand{\Shift}{X}  
\newcommand{\trr}{{\mbox{\rm tr}}}  

\newcommand{\lu}[2]{^{(#2)}\mkern -1.5mu #1}

\newcommand{\acc}{\bar{K}}

\newcommand{\Th}{\Theta}

\newcommand{\li}[1]{{\lu{{\bf i}}{#1}}}
\newcommand{\lj}[1]{{\lu{{\bf j}}{#1}}}
\newcommand{\lm}[1]{{\lu{{\bf m}}{#1}}}
\newcommand{\lmm}[1]{{\lu{\und{{\bf m}}}{#1}}}

\newcommand{\ML}{\!\!\!\!\!\!\!\!\!}
\newcommand{\un}{\underline}

\newcommand{\tW}{\tilde{R}}
\newcommand{\rb}{r_{\!b}}
\def\th{\theta}
\def\ze{\zeta}
\begin{document}
\author{Giulio CACIOTTA , Francesco NICOL\`O \footnote{Dipartimento di Matematica,
Universit\`{a} degli Studi di Roma ``Tor Vergata", Via della Ricerca Scientifica, 00133-Roma, Italy}
\\
\\Universit\`{a} degli Studi di Roma ``TorVergata"}
\title{\LARGE{The non linear perturbation of the Kerr spacetime in an external region.}}
\date{\today}
\maketitle
\begin{abstract}
{\NI We prove, outside the influence region of a ball of radius $R_0$ centered in the origin of the initial data hypersurface, $\Si_0$, the existence of global solutions near to Kerr spacetime, provided that the initial data are sufficiently near to those of Kerr; This external region is the ``far" part of the outer region of the perturbed Kerr spacetime. Moreover if we assume that the corrections to the Kerr metric decay sufficiently fast, $o(r^{-3})$, we prove that the various null components of the Riemann tensor decay in agreement with the ``Peeling theorem". }
\end{abstract}
\newpage
\tableofcontents
\newpage
\section{Introduction}\label{S1}

\subsection{The problem and the results}

\NI The problem of the global stability for the Kerr spacetime is a very difficult and open problem. The more difficult issue is that of proving the existence of solutions of the vacuum Einstein equations with initial data ``near to Kerr" in the whole outer region up to the event horizon, which is also an unknown of the problem.

\NI What is known up to now relative to the whole outer region are some relevant uniform boundedness results for solutions to the wave equation in the Kerr spacetime used as a background spacetime, see Dafermos-Rodnianski, \cite{Daf-Rod:Kerr} and references therein.\footnote{See also for the $J=0$ case, \cite{Blue} and references therein.} 
\smallskip

\NI If we consider the existence problem in a external region sufficiently far from the Kerr event horizon for a slow rotating Kerr spacetime the result is included,
in the version of Minkowski stability result proved by S.Klainerman and one of the present author, F.N., see \cite{Kl-Ni:book} and also \cite{Ch-Kl:book}. In this case one of the present authors, F.N., recently proved, \cite{Ni;Peel}, that the asymptotic behaviour of the Riemann components is in agreement with the ``Peeling theorem" if the corrections to the Kerr initial data decay sufficiently fast.
\smallskip

\NI In this paper we face and solve a more difficult problem, namely we prove in an external region which will be defined in detail later on, see the remark after the statement of Theorem \ref{final version} in Section \ref{Final result section}, but basically is the region where\footnote{$r$ is a radial coordinate which will be defined later on.} $r\geq R_0$ and $M/R_0\leq \la$ with $\la$ sufficiently small,
the non linear stability of the Kerr spacetime for any $J\leq M^2$ for an appropriate class of initial data near to Kerr. 

\NI Moreover if we restrict the class of initial data, once subtracted the initial data of the Kerr part, to those which decay toward the spacelike infinity faster than $r^{-3}$, we show again that the null asymptotic decay of the Riemann tensor is in agreement with the ``Peeling theorem".
We can now state our main result in a somewhat preliminary version, the full version of the theorem can be found in Section \ref{Final result section}.
\begin{thm}\label{preliminary version}
Assume that initial data  are given on $\Si_0$ such that, outside of a ball centered in the origin of radius $R_0$, they are different from the ``Kerr initial data of a Kerr spacetime with mass $M$ satisfying 
\[\frac{M}{R_0}<<1\ \ ,\ \ J\leq M^2\]
for some metric corrections decaying faster than $r^{-3}$ toward spacelike infinity together with its derivatives up to an order $q\geq 4$
, namely \footnote{The components of the metric tensor written in dimensional coordinates. $f=o_q(r^{-a})$ means that $f$ asymptotically
behaves as $o(r^{-a})$ and its partial derivatives $\partial^kf$, up to order $q$ behave as $o(r^{-a-k})$.}
\bea
g_{ij}=g^{(Kerr)}_{ij}+o_{q+1}(r^{-(3+\frac{\ga}{2})})\ \ ,\ \ k_{ij}=k^{(Kerr)}_{ij}+o_{q}(r^{-(4+\frac{\ga}{2})})
\eea
where $\ga>0$. Let us assume that the metric correction $\de g_{ij}$, the second fundamental form correction $\de k_{ij}$ are sufficiently small, namely a function made by $L^2$ norms on $\Si_0$ of these quantities is small,
\bea
{\cal J}(\Si_0,R_0; \de{^{(3)}\!}{\bf g}, \de{\bf k})\leq \varepsilon\ ,\eql{smallnesscond}
\eea
then this initial data set has a unique development, ${\widetilde{\M}}$, defined outside the domain of influence of $B_{R_0}$. Moreover ${\widetilde{\M}}$
can be foliated by a canonical double null foliation $\{C(u),\Cb(\ub)\}$  whose outgoing leaves $C(u)$ are complete\ \!\footnote{By this we mean that the null geodesics generating $C(u)$ can be indefinitely extended toward the future.} and the various null components of the Riemann tensor relative to a null frame associated to this foliation decay as expected from the Peeling theorem.
\end{thm}

\NI The proof of this result depends on many previous results, the result on the stability for the Minkowski spacetime in the external region proved by S. Klainerman and  F.N., in \cite{Kl-Ni:book}, a result which, at its turn, is based on the seminal work by D.Christodoulou and S.Klainerman, \cite{Ch-Kl:book} and, concerning the proof of the peeling decay, important ideas of the proof come from the previous work by S.Klainerman and F.N. , \cite{Kl-Ni:peeling}, and the recent \cite{Ni;Peel}.

\NI Before presenting the main new ideas used in this paper we observe that the two results proved in this work, the global stability in the external region and the asymptotic Riemann decay in agreement with the peeling, are basically independent; relaxing the decay conditions on the initial data we can prove the stability with a worst null asymptotic decay. In this paper we prove the two results together, but it should be easy to realize how to enlarge the class of initial data to prove only the first one, proceeding as in \cite{Kl-Ni:peeling} where it has been shown in detail, for the perturbed Minkowski spacetime, how the spacelike decays of the initial data are connected to the null decay of the Riemann components.

\NI Finally, as we said, many steps to prove Theorem \ref{preliminary version} have been discussed in previous works, here we prove in every detail the new part of this result, namely Theorem \ref{Theorem to prove} which is the core result to obtain, via a bootstrap mechanism, the global existence and the decay satisfying the ``Peeling theorem".

\NI In the remaining part of the introduction we examine the difficulties one encounters to prove this result and how they have been  overcomed.

\subsection{The global existence in an external region around the Kerr spacetime}

\NI To understand the problems arising perturbing around the Kerr spacetime solution of the Einstein equations it is appropriate first to remember  how the problem of perturbing around the Minkowski spacetime solution has been solved. The general strategy is usually called  ``bootstrap mechanism": one proves with a local existence result that there is a finite region $V$, whose metric satisfies the Einstein equations, endowed with some specific properties, mainly that some norms associated to the metric components and its derivatives are bounded by a (small) constant, then assumes that the largest possible region, $V_*$, where these bounds, we call ``bootstrap bounds", hold is finite and finally proves that, if the initial data are sufficiently small, this region can be extended showing that the previous bounds can be improved. Therefore, to avoid a contradiction, this region should coincide with the whole spacetime.

\NI To be a little less schematic we assume that $V_*$ is endowed with a foliation made by outgoing and incoming null cones, $\{C(\la)\}$ and $\{\Cb(\nu)\}$, and that the norms we assume bounded are those relative to the connection coefficients and to the components of the Riemann tensor. 

\NI The central part of the proof is, therefore, to show that these norms can have better bounds. To do it we use in the manifold $V_*$ the structure equations and the Bianchi equations. The structure equations take the form of transport equations for the connection coefficients along the incoming and outgoing cones and of elliptic Hodge systems on the two dimensional surfaces intersections of the incoming and an outgoing cones, $S=C\cap\Cb$ . These equations can be seen as inhomogeneous equations whose inhomogeneous part depends on the Riemann components. The Bianchi equations, at their turn, can be written as transport equations for the Riemann components along the cones whose inhomogeneous part is made by products of the Riemann components and the connection coefficients.

\NI To use these equations we make a sort of linearization, namely we consider the Riemann components as external sources satisfying the ``bootstrap bounds" and show, using the equations for connection coefficients, that these bounds (of the connection coefficients) can be improved. Then we control the Riemann components; to do it we use in a crucial way the fact that it is possible to define for the Riemann components some quantities which play the role of the energy norms, in association to the Bianchi equations. These norms are basically weighted $L^2$ integrals, along the outgoing and incoming cones, of the Bel-Robinson tensor. Assuming the connection coefficients satisfying the ``bootstrap bounds" we can prove that these energy norms are bounded by the corresponding initial data norms and using also the Bianchi equations obtain, finally, better norms for the Riemann components.

\NI This is a very schematic description of some of the main ideas developed in \cite{Kl-Ni:book} to prove the Mikowski stability result for the external region.
 
\NI If we ask ourselves how to transport this strategy to perturb non linearly around the Kerr spacetime solution instead than around the Minkowski solution, we realize immediately that the main difference is, in broad terms,  that now we are perturbing around a solution different from zero, while the Minkowski spacetime can be considered
a ``zero solution". With that we mean that in the Minkowski spacetime all the connection coefficients are identically zero with the exception of the second null fundamental forms $\chi$ and $\chib$ which, due to the Minkowski spherical simmetry, reduce to the two scalar functions $\tr\chi=2r^{-1}$ and $\tr\chib=-2r^{-1}$. Moreover all the Riemann components are identically zero. Therefore to prove that the norms are bounded and small we do not have to subtract the connection coefficients or the Riemann components associated to the Minkowski solution as they are (with the exception of $\tr\chi$ and $\tr\chib$) identically zero. 

\NI As the Kerr spacetime is not a ``zero solution" some kind of subtraction has to be done.
This ``subtraction" mechanism is delicate as we are not looking for a linearly perturbed solution\footnote{Observe that also the linear perturbation around Kerr has some problems due again to the fact that the Riemann tensor in Kerr spacetime is different from zero, see \cite{Blue} and \cite{Ca-Ni-Ra}.} and 
it is realized through four different steps:
\smallskip

i) The first step consists in defining the ``bootstrap assumptions" for $V_*$ not for the connection coefficients and the Riemann tensor norms, but for their corrections, that is for the connection coefficients and the Riemann tensor components to which we have subtracted their Kerr parts.\footnote{More precisely the Kerr part ``projected" on the $V_*$ foliation, see the details in subsection \ref{SSdeO}} Simbolically,
\bea
\de O=O-O^{(Kerr)}\ \ ,\ \ \de R=R-R^{(Kerr)}
\eea
and in some detail, see \cite{Kl-Ni:book}, Chapter 3, for all the definitions,
\beaa
&&\de\chi=\chi-\chi^{(Kerr)}\ ,\ \de\chib=\chib-\chib^{(Kerr)}\ ,\ \de\ze=\ze-\ze^{(Kerr)}\nn\\
&&\de\om=\om-\om^{(Kerr)}\ ,\ \de\omb=\omb-\omb^{(Kerr)}....
\eeaa
 \beaa
 \de\a=\a-\a^{(Kerr)}\ \ ,\ \ \de\b=\b-\b^{(Kerr)}\ \ ,\ .....
 \eeaa
\smallskip

ii) The second  step consists in writing the structure equation (in the $V_*$ region) instead that for the connection coefficients , $\chi,\chib,\ze,\om,\omb....$ for their corrections.\footnote{The technical details for this ``Kerr decoupling" for the connection coefficients are discussed later on.}

\NI As we said before, recalling the proof for the Minkowski stability, these equations have inhomogeneous terms which depend on the Riemann tensor and, in this case, on the correction to the Riemann tensor 
\[\de R=R-R^{(Kerr)}\ .\]
Once we have these modified structure equations we can use them to obtain better estimates for the norms of these connection coefficients corrections, $\de O$, provided we have a control for the norms of the Riemann components corrections, $\de R$. 
\smallskip

iii) The third step consists in obtaining estimates for the Riemann components corrections. This requires to subtract the Kerr part of the Riemann tensor. This cannot be done in a direct way as the basic step to control the Riemann norms is to prove the boundedness of the energy type norms, generically denoted by $\cal Q$. These norms are weighted $L^2$ integrals of the Bel-Robinson tensor which is a quadratic expression on the Riemann tensor.

\NI We would need analogous norms for the $\de R$ corrections and we also would like a well definite positivity for the integrand, see the explicite expression of the $\cal Q$ norms in Minkowski case in \cite{Kl-Ni:book}, Chapter 3, to get from them estimates for the (correction of the) null Riemann components norms. To obtain it we have to proceed in a different way based, intuitively, on the fact that the Kerr spacetime is stationary and $\frac{\partial}{\partial t}$ is a Killing vector field. Therefore if, instead of considering the Riemann components, we consider their time derivatives, they do not depend anymore on the Kerr part of the Riemann tensor, their initial data can have a better decay and if we could control their $\cal Q$ norms we could then obtain a good control of the $\de R$ norms in $V_*$ and also a good asymptotic decay along the null directions. 

\NI This argument as presented is not rigorous, in the perturbed Kerr spacetime $\frac{\partial}{\partial t}$ is not anymore a Killing vector field,  but it turns out that the basic idea can be implemented in the following sense:

\NI Let us consider instead of the time derivative of the Riemann tensor its (modified) Lie derivative $\lie_{T_0}R$,\footnote{See later for its precise definition.} where ${T_0}$, whose precise definition will be given later on, equal to $\frac{\pr}{\pr t}$ in the Kerr spacetime, is not anymore a Killing vector field, but only ``nearly Killing",\footnote{With ``nearly Killing" we mean that its deformation tensor is small with respect to some Sobolev norms.} then we define some $\tilde{\cal Q}$ norms relative to $\lie_{T_0}R$ with appropriate weights and prove that they are bounded in terms of the corresponding quantities written in terms of the initial data; from it we can prove, after quite a few steps, that the $\de R$ norms are small and satisfy appropriate decays.
\smallskip

\NI The steps just sketched require, to be fully understood, some more details; as we said the region $V_*$ is the largest possible region where the so called ``bootstrap assumptions" hold, in this region there is a foliation made by null incoming and outgoing cones $\{\Cb(\nu), C(\la)\}$, which are the level surfaces  $u=\la$, $\ub=\nu$ of the functions $u(p),\ub(p)$ solutions of the eikonal equations, see later and \cite{Kl-Ni:book}, Chapter 3, for a detailed discussion.
The foliation suggests to use the affine parameters of the null geodesics generating the null cones,  $\la, \nu$, as coodinates,  hereafter denoted again $u,\ub$, and some angular variables $\om^1,\om^2$  to define an arbitrary point on the two dimensional surface $S(u,\ub)=C(u)\cap\Cb(\ub)$.

\NI In $V_*$, thought as a subset of $R^4$ where $(u,\ub,\om^1,\om^2)$ is a generic point,
we can define two metrics the first being the Kerr metric, the other being the perturbed Kerr metric written again in the same coordinates, see subsubsection \ref{ss211}: 
\bea
&&\ML{\bf g}_{(Kerr)}\nn\\
&&\ML=-4\oom_{(Kerr)}^2dud\ub+\ga^{(Kerr)}_{ab}\!\left(d\om^a-X_{(Kerr)}^a(du+d\ub)\right)\!\!\left(d\om^b-X_{(Kerr)}^b(du+d\ub)\right)\nn\\
&&\nn\\
&&\ML{\bf g}_{(pert.Kerr)}\nn\\
&&\ML=-4\oom^2dud\ub+\ga_{ab}\!\left(d\om^a-({X^a_{(Kerr)}}du+X^ad\ub)\right)\!\!\left(d\om^b-({X^b_{(Kerr)}}du+X^bd\ub)\right)\ .\ \ \ \ \ \ \ \ \eql{newmeta}
\eea
where the Kerr metric written in these coordinates  is the one used in \cite{Is-Pr}.

\NI This means that $\{V_*,{\bf g}_{(Kerr)}\}$ can be thought as a region of the (outer) Kerr spacetime, while $\{V_*,{\bf g}_{(pert.Kerr)}\}$ a region of the perturbed external Kerr spacetime whose existence we want to prove. The choice of the same coordinates will turn out important to get, subtracting the Kerr part, the transport equations for the connection coefficients corrections $\de O$.

\NI Moreover in $V_*$ we can build a null frame adapted to this foliation $\{e_3,e_4,e_1,e_2\}$ where $e_1,e_2$ are vector fields orthonormal and tangent to the two dimensional surfaces $S=C(\la)\cap\Cb(\nu)$ while $e_3,e_4$ are proportional to the null geodesics generating the cones and in the previous coordinates have the following expressions, as discussed in detail later on,
\bea
e_4=\frac{1}{\oom}\!\left(\frac{\partial}{\partial\ub}+X\right)\ \ ;\ \ \e_3=\frac{1}{\oom}\!\left(\frac{\partial}{\partial u}+X_{(Kerr)}\right)\ ,\eql{newmet1a}
\eea
while if we were considering the Kerr metric we should have, with \[X_{(Kerr)}=\om_B\frac{\pr}{\pr\om^2}=\om_B\frac{\pr}{\pr\phi}\ ,\]
where $\om_B$ is defined in eq. \ref{ombdef},
\bea
e^{(Kerr)}_4=\frac{1}{\oom_{(Kerr)}}\!\left(\frac{\partial}{\partial\ub}+X_{(Kerr)}\right)\ \ ;\ \ \e^{(Kerr)}_3=\frac{1}{\oom_{(Kerr)}}\!\left(\frac{\partial}{\partial u}+X_{(Kerr)}\right) .\ \eql{newmet1aa}
\eea
From this  we define
\bea
T_0=\frac{\oom}{2}(e_3+e_4)-\frac{X+X_{(Kerr)}}{2}=\frac{\pr}{\pr u}+\frac{\pr}{\pr\ub}\ .
\eea

\NI The last important thing to point out  is that the integrands of the $\tilde{\cal Q}$ norms have to  be a sum of non negative terms, see for instance \cite{Kl-Ni:book}, Chapter 3 equations (3.5.1),...,(3.5.3), which requires that the Bel-Robinson tensor have to be saturate by appropriate vector fields,
linear combinations of $e_3$, $e_4$ with positive weights. Therefore as in \cite{Kl-Ni:book} we saturate the Bel-Robinson tensor with the following vector fields 
\bea
&&T=\frac{1}{2}(e_4+e_3)\nn\\
&&S=\frac{1}{2}(\tau_+e_4+\tau_-e_3)\nn\\
&&K=\frac{1}{2}(\tau_+^2e_4+\tau_-^2e_3)
\eea
where
\[\tau_+=\sqrt{1+\ub^2}\ \ ,\ \ \tau_-=\sqrt{1+u^2}\ ,\]
These vector fields, when we perturb Minkowski spacetime, are nearly Killing, while here they are not Killing vectors even in the Kerr spacetime.\footnote{
Perturbing Minkowski spacetime  we can choose $T=T_0$ as in Minkowski spacetime $T=\frac{\pr}{\pr t}$.} The relevance of the $T,S,K$ vector fields in the present case is connected to the fact that they are non spacelike fields in the region outside the ergosphere, made by $e_3$ and $e_4$, the null vectors of the frame adapted to the foliation which have the property that the fields $N=\oom e_4, \Nb=\oom e_3$ are equivariant vector fields.

\subsection{The decay of the Riemann components, the Peeling}
Beside the proof of the Kerr stability in a region with $r\geq R_0>>M$, we prove that the null Riemann components have a null asymptotic decay consistent with the ``Peeling theorem". This result has already been obtained by one of the authors, F.N. , \cite{Ni;Peel}, if we restrict ourselves to the perturbation of a very slow rotating Kerr spacetime or to a ``very external region" which was defined through the condition
\bea
M\leq {\tilde\la}R_0^{\frac{1}{2}}
\eea
where  ${\tilde\la}$ is a small number depending on the smallness of the initial data. As discussed in \cite{Ni;Peel}, the advantage of restricting to this ``much far" region is that, in this case, we do not have to prove again a global existence result as, in that case, the Kerr part of the metric  can be considered itself a perturbation of the Minkowski spacetime satisfying the conditions of \cite{Kl-Ni:book}.
In the present paper, once we have proved a global existence result, the way to prove again the ``peeling decay" is basically the same as the one discussed in  \cite{Ni;Peel}, 
we sketch now the main ideas involved and we refer to  \cite{Ni;Peel} and to the next sections for a more detailed discussion.

\NI In \cite{Ch-Kl:book} and in \cite{Kl-Ni:book}, the null asymptotic behaviour of some of the null components of the Riemann tensor, specifically the $\a$ and the $\b$ components, see later for their definitions, is different from the one expected from the ``Peeling Theorem", \cite{Wald}, as the decay proved there is slower. 
More precisely the components $\a$ and $\b$ \footnote{Components relative to a null frame adapted to the null outgoing and incoming cones which foliate the ``external region".} do not follow the ``Peeling theorem"as they decay as $r^{-\frac{7}{2}}$ while we expect $r^{-5}$ and $r^{-4}$ respectively.\footnote{In principle some $\log$ powers can be present, see J.A.V.Kroon, \cite {Kr:log2}.}
In a subsequent paper, \cite{Kl-Ni:peeling}, S.Klainerman and F.N. proved that the decay suggested from the ``Peeling theorem" could be obtained assuming a stronger spacelike decay for the initial data. Unfortunately that result requires an initial data decay too strong for proving the ``peeling decay" in spacetimes near to Kerr. 
To show how this result has to be improved let us first recall it in some more detail. In \cite{Kl-Ni:peeling}  the following result was proved:
\smallskip

\NI{\bf Theorem:}
{\em Let assume that on $\Si_0/B$ the metric and the second fundamental form have the following asymptotic behaviour
\footnote{Here $f=O_q(r^{-a})$ means that $f$ asymptotically
behaves as $O(r^{-a})$ and its partial derivatives $\partial^kf$, up to order $q$ behave as $O(r^{-a-k})$. Here with $g_{ij}$ we mean the components written in Cartesian coordinates.}
\bea
&&g_{ij}={g_S}_{ij}+O_{q+1}(r^{-(3+\ep)})\nn\\
&&{k}_{ij}=O_{q}(r^{-(4+\ep)})\eql{1.1b}
\eea
where ${g_S}$ denotes the restriction of the
Schwarzschild metric on the initial hypersurface:
\[g_S=(1-\frac{2M}{r})^{-1}dr^2+r^2(d\theta^2+\sin\theta^2d\phi^2)\ .\] 
Let us assume that a smallness condition for the initial data is satisfied.\footnote{The details of the smallness condition are in \cite{Kl-Ni:peeling}. }
Then along the outgoing null hypersurfaces $C(u)$ (of the external region) the following limits hold, with $\ep'<\ep$ and $\ub$ and $u$ the generalization of the Finkelstein variables $u=t-r_*\ \,\ \ u=t+r_*$ in the {\Sch} spacetime:\footnote{$\aa,\bb,...$ are the null components of the Riemann tensor defined with respect to a null frame adapted to the double null foliation, see \cite{Kl-Ni:book} Chapter 3, and later on.}
\bea
&&\lim_{C(u);\ub\rightarrow\infty}r(1+|u|)^{(4+\ep)}\aa=C_0\nn\\
&&\lim_{C(u);\ub\rightarrow\infty}r^2(1+|u|)^{(3+\ep)}\bb=C_0\nn\\
&&\lim_{C(u);\ub\rightarrow\infty}r^3\ro=C_0\nn\\
&&\lim_{C(u);\ub\rightarrow\infty}r^3\si=C_0\eql{peel1}\\
&&\lim_{C(u);\ub\rightarrow\infty}r^4(1+|u|)^{(1+\ep)}\b=C_0\nn\\
&&\sup_{(u,\ub)\in {\cal K}}r^{5}(1+|u|)^{\ep'}|\a|\leq C_0\ .\nn
\eea}

\NI This result was obtained, basically, in two steps. 
The first one consisted in proving that a family of energy-type norms, $\tilde{\cal Q}$, made with the Bel-Robinson tensor associated to the Riemann tensor $R$, of the same type as those used to prove the global existence near Minkowski in \cite{Kl-Ni:book}, but with a different weight in the integrand, were bounded in terms of the same norms relative to the initial data. The new weights are obtained multiplying the previous weights by a function $|u|^{\ga}$ with appropriate $\ga>0$ and the central point is that the extra terms appearing in the ``Error" we have to estimate to control the boundedness of the $\tilde{\cal Q}$ norms have a definite sign and can be discarded. This allowed to prove, from the boundedness of the $\tilde{\cal Q}$ norms,  that the various null components of the Riemann tensor, beside the decay in $r$, have a decay factor in the $|u|$ variable. In the second step  it was proved that, integrating along the incoming cones, the extra decay in the $|u|$ variable can be transformed in an extra decay in the $r$ variable proving the final result.
\smallskip

\NI As we said this result cannot be immediately translated to the present case as the required decay for the initial data does not admit initial data near to Kerr. The way out, which was already used in \cite{Ni;Peel}, is the one already used to prove global stability: we have to subtract the Kerr part, namely instead of looking directly to the decay of the Riemann components we look to the decay of their time derivatives, more precisely the (modified) Lie derivative with respect to the nearly Killing vector field $T_0$. This  basically subtracts the Kerr part, allows to prove the boundedness of the modified $\tilde{\cal Q}$ norms and finally proves that,with a ``time" integration, first, $\de R$ satisfies bounds which allow to extend the region $V_*$ and prove globale existence and, second, shows that the null components of the Riemann tensor satisfy the peeling decay. More details are given in the introduction to \cite{Ni;Peel} and the complete proof is given in the following sections.
\bigskip

\NI {\bf Acknowledgments:} {\em The initial part of this work has been done during a visit of one of the authors, F.N. , to the Institut Mittag-Leffler (Djursholm, Sweden), where he was invited for the General Relativity semester and where he enjoyed many scientific discussions. Moreover the same author  is deeply indebted to S. Klainerman for pointing to him the importance  of considering the Lie derivative, with respect to  the ``time" vector field $T_0$, of the Riemann tensor to obtain more detailed estimates for the various components of the Riemann tensor. Beside F.N. is also indebted for many illuminating discussions he had with him about this subject and many related ones. We also want to state clearly that the present result is deeply based on the previous works \cite{Kl-Ni:book}, \cite{Kl-Ni:peeling} and on the original fundamental work by D.Christodoulou and S.Klainerman, \cite{Ch-Kl:book}. Therefore nothing has been ``gracefully\footnote{An adverb sometimes very improperly used.}  acknowledged", but all the due credits have been explicitely given at the best of our knowledge.}

\section{The Bootstrap assumptions}\label{S.2}
 

\NI The main difference with \cite{Kl-Ni:book} and also \cite{Kl-Ni:peeling} are that in this case the initial data we are considering are a perturbation of Kerr initial data (in the external region of $\Si_0$) and that, therefore, we do not assume the ADM mass small; moreover due to the Kerr metric the decay of the initial data metric does not satisfy the assumptions required in \cite{Kl-Ni:peeling}. 

\NI We denote with $\cal O$ the connection coefficient norms defined as in \cite{Kl-Ni:book} and we make specific assumptions on them. 
We denote by $\cal R$ the norms associated to the various null Riemann components, where $\ro-\overline{\ro}$ and $\si-\overline{\si}$ are in place of $\ro$ and $\si$.

The choice of norms as $|\c|_{\infty}$, $|\c|_{p,S}$ or $|\c|_{L^2}$ norms follows the scheme of \cite{Kl-Ni:book} and will be discussed later on, here we are interested to their weight factors and to their smallness, therefore we denote all the norms with $|\c|$ and which kind of norms they represent follows exactly the pattern of \cite{Kl-Ni:book}.
\smallskip

\subsection{The null canonical foliation and the metric in $V_*$}

\NI As we said in the introduction differently from \cite{Ni;Peel} 
one has to prove again the existence of a global (external) spacetime, non linear perturbation of the Kerr one. This is done with a strategy similar to the one used in \cite{Kl-Ni:book} with some important differences. First the $\cal Q$ norms for the Riemann tensor defined in \cite{Kl-Ni:book} are substituted by some $\tilde{\cal Q}$ norms associated to the ``time derivative" of the Riemann tensor, $\lie_{T_0}R$, 
second, to rebuild from the knowledge of $\lie_{T_0}R$ the complete Riemann tensor one has to control the connection coefficients in a very detailed way, namely to have norm estimates for the difference $\de O$ between the connection coefficients and their Kerr part. 
The control of the $\de O$ terms, as the control of the complete connection coefficients in \cite{Kl-Ni:book}, is made writing the transport equations for these terms which are obtained starting from the structure equations for the whole connection coefficients and subtracting their Kerr parts. 

\NI More precisely it is clear that the Kerr connection coefficients satisfy transport equations similar to those in \cite{Kl-Ni:book} with respect to a double null cone foliation of the Kerr spacetime. On the other side the connection coefficients of the perturbed Kerr spacetime satisfy the transport equations written with respect to a double null cone foliation of the perturbed Kerr spacetime. This implies that to subtract the Kerr part (to these trasport equations) we have also to control the difference between the double null cone foliation associated to the Kerr spacetime and the one associated to the perturbed Kerr spacetime and prove that this is ``small". This also has to be part of the bootstrap mechanism and requires that the bootstrap assumptions on the connection coefficients terms imply analogous estimates at the level of the corrections to the metric components, we denote hereafter globally $\de{\cal O}^{(0)}$. 


\NI The subtraction of the Kerr part to the transport equations for the connection coefficients 
is a crucial step of the proof. The transport equations on the other side require $V_*$ endowed with a double null cone canonical  foliation which is the first of the ``Bootstrap assumptions" we make.
\smallskip

\NI {\bf The double null cone foliation:} 
Let us assume, for a moment, that a, possibly finite, region $V_*$ exists whose boundary is made by the union of $V_*\cap\Si_0$, $\pr{V_*}_1$ and 
$\pr{V_*}_2$, the first part being a spacelike hypersurface the second and the third two null hypersurfaces, the first incoming and the second outgoing, let us also assume that in this region we have a metric whose components, $\{g_{\mu\nu}\}$,  satisfy the Einstein vacuum equations. We can solve (at least locally) the eikonal equation 
\bea
g^{\mu\nu}\pr_{\mu}w\pr_{\nu}w=0
\eea
choosing as initial data a function\footnote{$\ub_0$ defines an appropriate radial foliation of $\Si_0/B_{R_0}$, see subsection \ref{SS3.6}.} $\ub_0$ on $V_*\cap\Si_0$ or a function $u_0$ on  $\pr{V_*}_1$. Let us call $\ub(p)$ and $u(p)$ these solutions respectively. Let assume that the level hypersurfaces $\ub(p)=\ub , u(p)=u$ define two family of null hypersurfaces we call null cones in analogy with the Minkowski case and denote $\Cb(u)$, $C(\ub)$. These two families form what we call a ``double null cone foliation". 

\NI The null vector fields of the geodesics generating the $C$ and $\Cb$ null ``cones" are
\bea
L^{\mu}=-g^{\mu\nu}\pr_{\nu}u\frac{\pr}{\pr x^{\nu}}\ \ ;\ \ \Lb^{\mu}=-g^{\mu\nu}\pr_{\nu}\ub\frac{\pr}{\pr x^{\nu}}\eql{LLbexpr}
\eea
and the ``lapse function" $\oom$ is defined through the relation
\[{\bf g}(L,\Lb)=-(2\oom^2)^{-1}\ .\]
Associated to the double null cone foliation we can define now two null fields $\{e_3,e_4\}$ in the following way:
\bea
e_4=2\oom L\ \ ,\ \ e_3=2\oom\Lb \eql{nulle3e4}
\eea
such that 
\[{\bf g}(e_3,e_4)=0\ .\]

\NI Given the double null foliation we can define $S(u,\ub)=C(u)\cap\Cb(\ub)$, each $S(u,\ub)$ being a codimension 2 surface. On each $S(u,\ub)$ we define two vector fields $\{e_a\}$, $a\in\{1,2\}$ orthonormal to $e_3,e_4$. We have therefore at each point $p\in V_*$ a null orthonormal frame.
Observe that the foliation made by the two dimensional surfaces $\{S(u,\ub)\}$ is null outgoing and null incoming integrable. This means that the distributions $\lap$ and $\underline{\lap}$ made by $\{e_4,e_1,e_2\}$ and by $\{e_3,e_1,e_2\}$ respectively are integrable.\footnote{This follows immediately by the definition of the $S(u,\ub)$ surfaces, anyway to prove it explicitely let us assume that (locally) the hypersurface ${\cal K}:=\{p\in V_*|w(p)=const\}$ be such that for any $q\in {\cal K}$ we have $T{\cal K}_q=\lap$. This implies that the normal vector  to $T{\cal K}_q$, $N=g^{\mu\nu}\pr_{\nu}\frac{\pr}{\pr x^{\mu}}$, has to satisfy ${\bf g}(N,e_4)={\bf g}(N,e_a)=0$. Therefore $N$ has to be proportional to $e_4$ which implies that it must be such that
\[g^{\mu\nu}\pr_{\mu}w\pr_{\nu}w=0\ .\]
Therefore $w$ satisfies the eikonal equation and as $u(p)$ satisfies the eikonal equation we can conclude that $C(u)$ for an arbitrary fixed $u$ is the ``integral" of the distribution $\lap$. Exactly the same argument works for the distribution $\underline\lap$.}

\NI Recall also that the integrability property of the $S$-foliation implies that the connection coefficients $\xi$ and $\xib$ are identically zero
and that the second null fundamental forms are symmetric. This follows immediately from the Frobenius condition that $[e_4,e_a]$, and $[e_b,e_a]$ cannot have a contribution from $e_3$ which implies that
\[{\bf g}([e_4,e_a],e_4)={\bf g}([e_b,e_a],e_4)=0\ .\]

\NI The frame $\{e_4,e_3,e_1,e_2\}$ is called the ``adapted (to the double null foliation) frame". 
Of course one can have different double null foliations choosing different ``initial data", and in the future among the initial data we are going to choose some specific ones we will discuss later on, see subsections \ref{lastlslice}, \ref{SS3.6} and also \cite{Kl-Ni:book}, Chapter 3, and the foliation associated with these data will be called ``the double null canonical foliation".
\smallskip

\NI In conclusion the first of the ``bootstrap assumptions" we are stating is the following one:
\bigskip

\NI {\bf (I) : $V_*$ is endowed with a double null canonical foliation.}
\medskip

\subsubsection{The ``adapted" coordinates }\label{ss211}
\NI Once we have assumed that in $V_*$ a double null canonical foliation exists, in ``adapted" coordinates the metric has the following form:
\begin{thm}\label{metricform}
let us assume $V_*$ be endowed with a double null cone foliation, then, in the appropriate coordinates the metric tensor has the following form:
\bea
{\bf g}(\c,\c)=-4\oom^2dud\ub+\ga_{ab}\big(d\om^a-({X_{(Kerr)}}^adu+X^ad\ub)\big)\!\big(d\om^b-({X_{(Kerr)}}^bdu+X^bd\ub)\big)\ ,\ \ \ \ \ \ 
\eea
where
\bea
\oom=\sqrt{-\frac{{\bf g}(L,\Lb)}{2}}\ \ ,\ \ X=X^a\frac{\pr}{\pr\om^a}\ \ \ ,\ \ \ X_{(Kerr)}=\om_B\frac{\pr}{\pr\om^2}
\eea
and the coordinates $\{u,\ub,\om^1,\om^2\}$ are defined in the course of the proof.
\end{thm} 
\NI{\bf Proof:} The null vector fields of the geodesics generating the $C$ and the $\Cb$ null ``cones" are
\bea
L^{\mu}=-g^{\mu\nu}\pr_{\nu}u\frac{\pr}{\pr x^{\nu}}\ \ ;\ \ \Lb^{\mu}=-g^{\mu\nu}\pr_{\nu}\ub\frac{\pr}{\pr x^{\nu}}\eql{LLbexpr}
\eea
and the ``lapse function" $\oom$ is defined through the relation
\[{\bf g}(L,\Lb)=-(2\oom^2)^{-1}\ .\]
Let us choose $u$ and $\ub$ as coordinates. As $u(p)=u$ and $\ub(p)=\ub$ satisfy the eikonal equation it follows that
\bea
g^{uu}=g^{\ub\ub}=0\ .
\eea
Therefore choosing as coordinates $\{u,\ub,x^1,x^2\}$ , where the coordinates $\{x^a\}$ are still generic ones, we have from \ref{LLbexpr},
\bea
L=-g^{u\ub}\left(\frac{\pr}{\pr\ub}+\frac{g^{au}}{g^{u\ub}}\frac{\pr}{\pr x^a}\right)\ \ ;\ \ \Lb=-g^{u\ub}\left(\frac{\pr}{\pr u}+\frac{g^{a\ub}}{g^{u\ub}}\frac{\pr}{\pr x^a}\right)\ .
\eea
It is easy to prove, see \cite{Kl-Ni:book} Chapter 3, that the vector fields
\bea
N=\left(\frac{\pr}{\pr\ub}+\frac{g^{au}}{g^{u\ub}}\frac{\pr}{\pr x^a}\right)\ \ ;\ \ \Nb=\left(\frac{\pr}{\pr u}+\frac{g^{a\ub}}{g^{u\ub}}\frac{\pr}{\pr x^a}\right)
\eea
are equivariant. This means that the diffeomorphism generated by them sends a surface $S(u,\ub)=C(u)\cap\Cb(\ub)$ to another surface $S$ on the same outgoing or incoming cone respectively.

\NI To specify the choice of the ``angular" coordinates $\{x^1,x^2\}$, we proceed in two steps. We consider the diffeomorphism generated by $\Nb$, we denote it $\underline{\Phi}_{\la}$, which sends $S(u,\ub)$ to $S(u+\la,\ub)$. Let $p\in S(u,\ub)$ there exists a point $p_0\in\Cb\cap\Si_0$ such that $p=\underline{\Phi}(u; p_0)$. Let us denote the ``angular" coordinates on $\Si_0$ , $\om_0^1,\om_0^2$ and make the following choice for the angular coordinates of $p$
\[x^1(p)=\om^1_0(p_0)\ \ ,\ \ x^2(p)=\om^2_0(p_0)\ .\]
Therefore the integral curve of the vector field $\Nb$ , $\underline{\Phi}(\la; p_0)$, in these coordinates is
\bea
&&\underline{\Phi}^u(\la; \{\ub,\om^b_0\})=\la\nn\\
&&\underline{\Phi}^{\ub}(\la; \{\ub,\om^b_0\})=\ub\\
&&\underline{\Phi}^a(\la; \{\ub,\om^b_0\})=\om^a_0\nn
\eea
and
\[\Nb=\frac{\pr}{\pr u}\ .\]
To have an expression for $N$ and $\Nb$ as similar as possible to the one in Kerr spacetime for the corresponding quantities in the Pretorius-Israel coordinates, see \cite{Is-Pr}, we perform  a change of coordinates from \[\{x^{\mu}\}\equiv\{u,\ub,x^1,x^2\}\ \ \mbox{to}\ \ \{u,\ub,\om^1,\om^2\}\equiv\{y^{\mu}\}\] where
\bea
\om^a=x^a+f^a(\la,\ub,\{\om^b\})\ .\eql{changeofc}
\eea
and from it
\bea
\Nb=\Nb^{\mu}\frac{\pr}{\pr x^{\mu}}=\left(\Nb^{\mu}\frac{\pr y^{\nu}}{\pr x^{\mu}}\right)\frac{\pr}{\pr y^{\nu}}=\frac{\pr}{\pr u}+\frac{\pr f^a}{\pr\la}\frac{\pr}{\pr \om^a}\ .
\eea
We choose $f^a(\la,\ub,\{\om^b\})$ such that
\bea
\frac{\pr f^a}{\pr\la}=\de^a_2\ \!\om_B(\la,\ub,\om^1)
\eea
where the explicit expression of the function $\om_B(\la,\ub,\om^1)$ will be given later on. Therefore with this change of coordinates we have
\bea
\Nb=\frac{\pr}{\pr u}+\om_B\frac{\pr}{\pr \om^2}\equiv\frac{\pr}{\pr u}+X_{(Kerr)}.
\eea
Once we have defined $\Nb$ with this coordinates choice there is no more freedom in the expression of the equivariant vector field $N$; in fact  as $V_*$ is not flat it follows that $[N,\Nb]\neq 0$ and an explicit calculation, see \cite{Kl-Ni:book} Chapter 3, gives
\bea
[N,\Nb]=-4\oom^2\zeta(e_a)e_a\ ,\eql{comm}
\eea
where $\ze$ is the connection coefficient called ``torsion" which in the adapted frame we are going to define, has the following expression
\bea
\ze(e_a)=\frac{1}{2}{\bf g}(\dd_{e_a}e_4,e_3)\ .
\eea
Therefore we can write
\bea
N=\frac{\pr}{\pr\ub}+X
\eea
where
\bea
X={\om_B}\frac{\pr}{\pr\phi}+\de X=X_{(Kerr)}+\de X\eql{newX}
\eea
and $\de X$ satisfies the equation, which will be needed later on,
\bea
\pr_{N}X_{(Kerr)}-\pr_{\Nb}(X_{(Kerr)}+\de X)=-4\oom^2\zeta(e_a)e_a\ .\eql{eqfordeX}
\eea
Associated to the double null cone foliation we can define an adapted null orthonormal frame in the following way:
\bea
&&\ML e_4=2\oom L=\frac{1}{\oom}N\ \ ,\ \ e_3=2\oom\Lb=\frac{1}{\oom}\Nb\nn\\
&&\ML e_1=e_1^1\frac{\pr}{\pr\om^1}\ \ ,\ \ e_2=e_2^2\frac{\pr}{\pr\om^2}\ ,\eql{ortnullpertframe}
\eea
where $e_1,e_2$ are $S$ tangent vector fields orthonormal and orthogonal to $e_3,e_4$. Moreover with this definition it follow immediately that
\[{\bf g}(e_3,e_4)=-2\ .\]
Once we have defined to adapted null frame in $V_*$ we can write at the generic point of $V_*$ indentified by the coordinates $\{u,\ub,\th,\phi\}$ the  inverse metric as
\bea
g^{\mu\nu}=-2\left(e_4^{\mu}e_3^{\nu}+e_3^{\mu}e_4^{\nu}\right)+\sum_{a=1}^2e_a^{\mu}e_a^{\nu}
=-\frac{2}{\oom^2}\left(N^{\mu}\Nb^{\nu}+\Nb^{\mu}N^{\nu}\right)+\sum_{a=1}^2e_a^{\mu}e_a^{\nu}\ .\ \ \ \ \ \eql{invmet}
\eea
With this choice of $N$ and $\Nb$ we have
\bea
&&\ML g^{\mu\nu}=-\frac{1}{2\oom^2}\left(N^{\mu}\Nb^{\nu}+\Nb^{\mu}N^{\nu}\right)+\sum_{a=1}^2e_a^{\mu}e_a^{\nu}\nn\\
&&\ML =-\frac{1}{2\oom^2}\left(\big(\de^{\mu}_{\ub}+X^a\de^{\mu}_{a}\big)\big(\de^{\nu}_{u}+{X_{(Kerr)}}^b\de^{\nu}_{b}\big)+\big(\de^{\mu}_{u}+{X_{(Kerr)}}^b\de^{\mu}_{b}\big)\big(\de^{\nu}_{\ub}+X^a\de^{\nu}_{a}\big)\right)+\ga^{\mu\nu}\nn\\
&&\ML=-\frac{1}{2\oom^2}\left[\left(\de^{\mu}_{\ub}\de^{\nu}_{u}+\de^{\mu}_{u}\de^{\nu}_{\ub}\right)
+X^c\left(\de^{\mu}_{c}\de^{\nu}_{u}+\de^{\nu}_{c}\de^{\mu}_{u}\right)+{X_{(Kerr)}}^d\!\left(\de^{\mu}_{d}\de^{\nu}_{\ub}+\de^{\nu}_{d}\de^{\mu}_{\ub}\right)\right]+\ga^{ab}\de^{\mu}_a\de^{\nu}_b\ .\nn
\eea
and the inverse matrix, that is the metric tensor, is:
\bea
{\bf g}(\c,\c)=-4\oom^2dud\ub+\ga_{ab}\big(d\om^a-({X_{(Kerr)}}^adu+X^ad\ub)\big)\!\big(d\om^b-({X_{(Kerr)}}^bdu+X^bd\ub)\big)\ \ \ \ \ \ \ \eql{pertKerr1}
\eea
completing the proof of the theorem.
\smallskip

\NI The previous result has still a certain arbitrariness as $e_3,e_4$ depend on the foliation which at its turn depends on the choice of the initial conditions on $V_*\cap\Si_0$ and on $\pr{V_*}_1$. The natural choice is that of a foliation ``near" to the one we use in the Kerr spacetime, namely the Pretorius-Israel one. To discuss this point let us recall some aspects of this foliation, \cite{Is-Pr}. 

\NI In \cite{Is-Pr} we have
\bea
\ub=\frac{t+r_*}{2}\ \ ;\ \ u=\frac{t-r_*}{2}\ ,
\eea
where $r_*$ is the solution of the eikonal equation described in \cite{Is-Pr}.\footnote{To specify it uniquely we still have to give the initial data, we discuss it later on.} the following definitions hold
\bea
\om_B=\frac{2mar_b}{\Si R^2}\ \ ,\ \ \oom=\sqrt{\frac{\Delta}{R^2}}\ ,\  \la=\sin^2\th_*.\eql{ombdef}
\eea
where
\bea
&&\De=r_b^2+a^2-2Mr\nn\\
&&\Si=r_b^2+a^2\cos^2\theta\nn\\
&&\Si R^2={(r_b^2+a^2)^2-\Delta a^2\sin^2\theta}
\eea
and $r_b$ is the Boyer-Lindquist radial coordinate.
With these definitions the Kerr null frame ``adapted" to the P-I double null foliation is \footnote{The explicit definition of $\cal L$ is given in \cite{Is-Pr}.}
 \bea
&&e^{(Kerr)}_3=2{\oom^{(Kerr)}}\Lb={R\over\sqrt{\De}}(\pr_t-\pr_{r_*}+\om_B\pr_\phi)=\frac{1}{{\oom^{(Kerr)}}}\!\left(\frac{\pr}{\pr u}+{\om_B}\frac{\pr}{\pr\phi}\right)\nn\\
&&e^{(Kerr)}_4=2{\oom^{(Kerr)}}L={R\over\sqrt{\De}}(\pr_t+\pr_{r_*}+\om_B\pr_\phi)=\frac{1}{{\oom^{(Kerr)}}}\!\left(\frac{\pr}{\pr\ub}+{\om_B}\frac{\pr}{\pr\phi}\right)\nn\\
&&e^{(Kerr)}_{1}=\frac{R}{{\mathcal L}{\sin2\th_*}}\frac{\partial}{\partial\th_*}=\frac{R}{{\mathcal L}}\frac{\partial}{\partial\la}\eql{2.37}\\ 
&&e^{(Kerr)}_{2}=\frac{1}{R\sin\theta}\frac{\partial}{\partial\phi}\ ,\nn
\eea
where $\cal L$ and $\la$ are defined in \cite{Is-Pr}, see also equations \ref{PQLdef}.
Let us denote the $\{\th_*,\phi\}$ coordinates \footnote{Observe that in the Pretorius-Israel frame the variable which stays near to the spherical coordinate $\th$ of Minkowski spacetime is $\th_*$ and not $\th$.}  again $\{\om^1,\om^2\}$, a point of the Kerr spacetime is specified  assigning $\{u,\ub,\om^1,\om^2\}$, moreover in the Kerr spacetime the null hypersurfaces $u=const$, $\ub=const$ define the double null foliation. We can, therefore, reobtain as done before, starting from the null orthonormal frame the Kerr metric written in the $\{u,\ub,\om^1,\om^2\}$ coordinates and  the result is
\bea
&&\ML{\bf g}_{(Kerr)}(\c,\c)\eql{Kerrm1}\\
&&\ML=-4\oom_{(Kerr)}^2dud\ub+\ga^{(Kerr)}_{ab}\!\!\left(d\om^a-X_{(Kerr)}^a(du+d\ub)\right)\!\!\left(d\om^b-X_{(Kerr)}^b(du+d\ub)\right)\ .\nn
\eea
where 
\bea
&&X_{(Kerr)}=\om_B\frac{\pr}{\pr \phi}\ \ ,\ \ \oom_{(Kerr)}=\sqrt{\frac{\Delta}{R^2}}\nn\\
&&\ga_{11}=\frac{{\cal L}^2{(\sin2\th_*)}^2}{R^2}\ ,\ \ga_{12}=0\ ,\ \ga_{22}=R^2\sin^2\th\ .
\eea
which is exactly the metric written in \cite{Is-Pr}. Proceeding as before we obtain the equivariant vector fields in the Kerr spacetime
\bea
N^{(Kerr)}=\left(\frac{\pr}{\pr\ub}+{\om_B}\frac{\pr}{\pr\phi}\right)\ \ \ ,\ \ \ \Nb^{(Kerr)}=\left(\frac{\pr}{\pr u}+{\om_B}\frac{\pr}{\pr\phi}\right)
\eea
which satisfies \ref{comm} 
\beaa
[N,\Nb]=-4\oom^2\zeta(e_a)e_a
\eeaa
with 
\bea
\zeta^{(Kerr)}(e_a)e_a=\left(-\frac{QR\sin\th}{2\Si}{\pr_{r_b}\om_B}\right)e_{\phi}\ .
\eea
It is now possible to compare the foliation associated to the Kerr spacetime and to the perturbed Kerr spacetime. As said in the introduction we consider the region $V_*$ as a subset of $R^4$ where a generic element (point) is the 4-ple $(u,\ub,\om^1,\om^2)$. If we endow $V_*$ with the metric \ref{Kerrm1}, $(V_*,{\bf g}_{(Kerr)})$ is a submanifold of the Kerr spacetime, if the metric associated to $V_*$ is \ref{pertKerr1} then $(V_*,{\bf g})$ is a submanifold of the global perturbed Kerr spacetime whose existence we are proving. Comparing the two metrics \ref{Kerrm1} and \ref{pertKerr1}
\beaa
&&\ML{\bf g}_{(Kerr)}(\c,\c)\eql{Kerrmet}\\
&&\ML=-4\oom_{(Kerr)}^2dud\ub+\ga^{(Kerr)}_{ab}\left(d\om^a-X_{(Kerr)}^a(du+d\ub)\right)\left(d\om^b-X_{(Kerr)}^b(du+d\ub)\right)\ .\nn
\eeaa
and\footnote{In the perturbed Kerr spacetime we defined $\om_B$ as in the Kerr spacetime; this requires some care. In fact we define it as a function  in the $\{u,\ub,\om^1,\om^2\}$  coordinates while in Kerr spacetime $\om_B=\om_B(r_b,\th,\phi)$  and $r_b,\th,\phi$ are Boyer Lindquist coordinates. Therefore using the Pretorius-Israel coordinates we can rewrite it as a function  $\om_B=\om_B(r_*,\th_*,\phi)$ and finally following \cite{Is-Pr} we express $r_*$ as a function of $u$ and $\ub$. This is the expression we use in $(V_*,{\bf g})$.}
\beaa
{\bf g}(\c,\c)=-4\oom^2dud\ub+\ga_{ab}\big(d\om^a-({X_{(Kerr)}}^adu+X^ad\ub)\big)\!\big(d\om^b-({X_{(Kerr)}}^bdu+X^bd\ub)\big)\ \ \ \ \ \ \ 
\eeaa
we see that at the metric level the components which are modified due to the initial data perturbation of the ``Kerr initial data" are
\bea
\de\oom=\oom-\oom_{(Kerr)}\ \ ,\ \ \de X^a=X^a-X_{(Kerr)}^a\ \ ,\ \ \de\ga_{ab}=\ga_{ab}-\ga^{(Kerr)}_{ab}\ .
\eea
We will have to prove that these corrections are, in some appropriate norms, small  once the connection coefficients satisfy the ``Bootstrap assumptions". Therefore we define the following norms, we denote globally $\de{\cal O}^{(0)}$, which will be estimated later on,
\bea
&&|r^2|u|^{1+\de}\de\oom|_{\infty}\equiv \sup_{V_*}|r|u|^{2+\de}\de\oom|\eql{deOmet}\\
&&|r^2|u|^{2+\de}\de X|_{\infty}\equiv \sup_{V_*}\sum_a|r^2|u|^{2+\de}\de X^a|\ \ ,\ \ 
||u|^{1+\de}\de \ga|_{\infty}\equiv\sup_{V_*}\sum_{ab}||u|^{1+\de}\de \ga_{ab}|\ \ .\nn
\eea

\subsection {The connection coefficients and Riemann tensor bootstrap assumptions:}\label{DcanfolBass}
Once 
that $V_*$ is endowed with a double null canonical foliation we have to specify the remaining conditions we assume satisfied in $V_*$. They are relative to the connection coefficients and to the Riemann components, more precisely to their difference from the analogous quantities in the Kerr spacetime and are stated in the following.

\subsubsection {The $\cal O$ connection coefficient norms:}\label{SSdeO}

\NI 

\NI To prove our result, we write any connection coefficient as the sum of the Kerr connection coefficient part plus a correction and make assumptions for the estimates of the correction parts. There is nevertheless an important technical aspect to remember. The $O$ connection coefficients are tensor fields tangent to the $S$ two dimensional surfaces associated to the double null cone foliation of the perturbed Kerr spacetime, while the $O^{(Kerr)}$ connection coefficients are tensor fields tangent to the $S_{(Kerr)}$ two dimensional surfaces. Therefore their difference would not be an $S$-tangent tensor. To avoid this problem we recall that each $O^{(Kerr)}$ can be written as, assuming it, for instance, a $(0,2)$ covariant tensor,
\bea
O^{(Kerr)}_{\mu\nu}={\Pi^{(Kerr)}}^{\ro}_{\mu}{\Pi^{(Kerr)}}^{\si}_{\nu}H^{(Kerr)}_{\ro\si}\eql{Otg02}\ ,
\eea
where $H^{(Kerr)}$ is not an $S$-tangent tensor and $\Pi^{(Kerr)}$ projects on the $TS_{(Kerr)}$ tangent space. Therefore we substitute $O^{(Kerr)}$ with ${\hat O}$ 
\bea
{\hat O}_{\mu\nu}=\Pi^{\ro}_{\mu}\Pi^{\si}_{\nu}H^{(Kerr)}_{\ro\si}\ .\eql{Otg01}
\eea
where $\Pi$ projects on the $TS$ tangent space. It will then be needed to control that the differences $({\hat O}-O^{(Kerr)})$ are small, this will be proved later on, see subsection \ref{SS2.5}.

\NI Therefore in our bootstrap assumptions we write each connection coefficient as
\[O={\hat O}+\de O\]
and we make bootstrap assumptions on the norms of the $\de O$ parts, we denote globally as $\cal{\de O}$, while for the Kerr part of the connection coefficients we control their $sup$ norms.
These (assumed) norm estimates  control  the smallness and  the decay along the outgoing or incoming cones of these corrections. The decay is expressed in terms of a ``radial variable" denoted by $r$ which is defined, at a generic point $p\in V_*$, as 
\bea
r=r(u,\ub)=\frac{1}{\sqrt{4\pi}}|S(u,\ub)|^{\frac{1}{2}}\ .\eql{rdefa}
\eea
where $S(u,\ub)$ is the surface to which the point $p$ belongs.
It is important to observe that this variable, although not far from, is different from the radial variable $r_*=\ub-u$ used by Israel and Pretorius, \cite{Is-Pr}, in the Kerr spacetime and also with respect to the quantity $\ub-u$ of the perturbed Kerr spacetime. All the decay will be expressed in this variable and in the $|u|$ variable and we show, in subsection \ref{rcoord}, that all the various radial coordinates appearing in the Kerr spacetime and in the perturbed Kerr spacetime stay near to $r(u,\ub)$, (of course in this external region). Therefore hereafter with $r$ we always mean the quantity defined in \ref{rdefa}, a well defined function of $u$ and $\ub$. 
\smallskip

\NI Finally we have to specify which are the norms we are using for the connection coefficients. They are, as in \cite{Kl-Ni:book}, or  $\sup$ norms, $|\c|_\infty$, or $(p,S)$-norms, $|\c|_{p,S}$, with $p\in[2,4]$, where
\bea
|f|_{p,S}=\left(\int_S|f|^pd\mu_S\right)^{\frac{1}{p}}\ .\eql{psnormdef}
\eea
If $f$ denotes a generic connection coefficient the pointwise norm of the integrand is made with the restriction of the metric on $S$, $\ga_{ab}$. Even if we are considering the norm of the Kerr part we use the $\ga$ metric associated to the perturbed metric ${\bf g}$, equation \ref{pertKerr1}, which does not change, apart from a constant, the estimates for the Kerr part. In fact from the bootstrap assumptions it follows, as we will show in detail, that the metric $\ga_{ab}$ stays near to $\ga^{(Kerr)}_{ab}$, see {\bf step IV} of subsection \ref{S.S.stepsfortheproof}.
\subsubsection{Decay of Kerr connection coefficients}
We start recalling the general definitions of the connection coefficients, see \cite{Kl-Ni:book} Chapter 3, for more details,
\bea
&&\xi_{a}=\frac{1}{2}{\bf g}(\dd_{e_4}e_4,e_{a})\ \ \ ,\ \ 
\xib_{a}=\frac{1}{2}{\bf g}(\dd_{e_3}e_3,e_{a})\nn\\
&&\eta_{a}=-\frac{1}{2}{\bf g}(\dd_{e_3}e_{a},e_4)\ ,\ 
\etab_{a}=-\frac{1}{2}{\bf g}(\dd_{e_4}e_{a},e_3)\eql{riccicoeffdef}\\
&&\om=-\frac{1}{4}{\bf g}(\dd_{e_4}e_3,e_4)\ \ ,\ \ 
\omb=-\frac{1}{4}{\bf g}(\dd_{e_3}e_4,e_3)\nn\\
&&\ze_a=\frac{1}{2}g(\dd_{e_a}{e_4},{e_3})\ .
\eea
\NI The decay of the Kerr part of the connection coefficients, those for the tangential derivatives follows in the standard way their weight getting an extra $r$ for each tangential derivative, is the following one:
\bea
&&\tr\chi^{(Kerr)}=\frac{O(1)}{r}+\frac{O(M)}{r^2}+\frac{O(M^2)}{r^3}+....\nn\\
&&\tr\chib^{(Kerr)}=\frac{O(1)}{r}+\frac{O(M)}{r^2}+\frac{O(M^2)}{r^3}+....\nn\\
&&|r\tr\chi^{(Kerr)}|\leq \kappa\ \ ;\ \ |r\tr\chib^{(Kerr)}|\leq \kappa\eql{trchi-b}
\eea
\bea
&&|r^2\om^{(Kerr)}|\leq \kappa M\ \ ;\ \ |r^2|\om^{(Kerr)}|\leq \kappa M\eql{bass1a}
\eea
\bea
&&|r^3\chih^{(Kerr)}|\leq  \kappa{aM}\leq\kappa{M^2}\ \ ;\ \ |r^3\chibh^{(Kerr)}|\leq  
\kappa{aM}\leq\kappa{M^2}\nn\\
&&|r^4\nabb\tr\chi^{(Kerr)}|\leq  \kappa{aM}\leq\kappa{M^2}\ \ ;\ \ |r^4\nabb\tr\chib^{(Kerr)}|\leq\kappa{aM}
\leq \kappa{M^2}\ \ \ \ \ \eql{bass1b}\\
&&|r^3\ze^{(Kerr)}|\leq  \kappa{aM}\leq\kappa{M^2}\nn\\
&&|r^4\nabb\om^{(Kerr)}|\leq \kappa{aM}\leq \kappa{M^2}\ \ ,\ \ |r^4\nabb\omb^{(Kerr)}|
\leq \kappa{aM}\leq\kappa{M^2}\nn
\eea
where $\kappa>1$ is a definite adimensional constant.
\smallskip

\NI Observe that the norm estimates in  \ref{trchi-b} refer to those connection coefficients which are different from zero in Minkowski, the norms estimates in  \ref{bass1a} refer to those connection coefficients which are different from zero in {\Sch}, but zero in Minkowski and the norm estimates in \ref{bass1b}  refer to those connection coefficients or derivatives of connection coefficients which are different from zero in Kerr, but zero in {\Sch} and in Minkowski.

\NI These ``Kerr decays" can be easily obtained by dimensional arguments recalling that the Kerr metric, written in the Boyer-Lindquist coordinates, in the limit $M\rightarrow 0$, $a$ kept fixed and different from zero, reduces to the Minkowski metric written in the  ``oblate coordinates" and as we know that in Minkowski spacetime all connection coefficients, with the exception of $\tr\chi$ and $\tr\chib$, are zero, it follows that performing an $\frac{1}{r}$ expansion of the connection coefficients which are zero in Minkowski, all the coefficients of the terms in $\frac{1}{r^k}$ which depend only on $a$ and not on $M$ must be identically zero. 

\NI The dimensional argument goes as following: we assume the metric written in ``cartesian" coordinates, namely coordinates with the dimension of a length: $L^1$. Then as the metric tensor has dimension $L^2$ its components $g_{\mu\nu}$ has dimension $L^0$. Proceeding in the same way it follows that the connection coefficient components have dimension $L^{-1}$ and the same happens for their norms.

Therefore as, for instance, $[\chi]=L^{-1}$ it follows that in Kerr it must be
\[\chih^{(Kerr)}=\frac{O(aM)}{r^3}\ .\]
Moreover as in Kerr spacetime $a=J/M$ and we consider those spacetimes where $J\leq M^2$, it follows that $a\leq M$ and the Kerr coefficient satisfy the condition
\[|r^3\chih^{(Kerr)}|\leq \kappa M^2\ .\]

\NI The bounds for $\om^{(Kerr)}$ and $\omb^{(Kerr)}$ and for $\tr\chi^{(Kerr)}$ and $\tr\chib^{(Kerr)}$ are different as they reflect the fact that $\om^{(Kerr)}$, $\omb^{(Kerr)}$ are different from zero also in {\Sch} and $\tr\chi^{(Kerr)},\tr\chib^{(Kerr)}$ even in Minkowski.

\subsection{The $\de{\cal O}$ connection coefficient norms}\label{SSconcoef}

\NI The assumptions 
\bea
\de{\cal O}\leq \ep_0\eql{Oboosestwithdec}
\eea
 summarizes the conditions on the connection coefficients and their tangential derivatives, $\nabb^l$, with $0\leq l\leq 4$,\footnote{The bounds foe $\eta$ and $\etab$ follow from these ones in the adapted frame, see for details \cite{Kl-Ni:book} Chapter 3.}
\bea
&&|r^{2+l}|u|^{2+\de}\nabb^l\de\tr\chi|\leq  \ep_0\ \ ;\ \ |r^{1+l}|u|^{3+\de}\nabb^l\de\tr\chib|\leq  \ep_0\nn\\
&&|r^{2+l}|u|^{2+\de}\nabb^l\de\chih|\leq  \ep_0\ \ ;\ \ |r^{1+l}|u|^{3+\de}\nabb^l\de\chibh|\leq  \ep_0\nn\\
&&|r^{2+l}|u|^{2+\de}\nabb^l\de\ze|\leq  \ep_0\eql{deOcond}\\
&&|r^{2+l}|u|^{2+\de}\nabb^l\de\om|\leq \ep_0\ \ ,\ \ |r^{1+l}|u|^{3+\de}\nabb^l\de\omb| \leq \ep_0\ .\nn
\eea

\NI{\bf Remark:} {\em The choice of the $\de$ coefficient is connected to the decay of the Riemann tensor. $\de>0$ is required to prove the decay in agreement with the peeling. $\de\leq 0$ could also be chosen, as discussed in the introduction, but in this case a weaker decay for (some components of) the Riemann tensor follows.
The factor $|u|^{\de}$ could also describe, symbolically, a log factor, for instance 
$|r^2|u|^{2+\de}\de\tr\chi|\leq  \ep_0$ could mean 
\[|\de\tr\chi|\leq c\ep_0\frac{1}{(\log{|u|/M)^\de}r^2|u|^2}\ .\]
This last definition is in agreement with the statement we do later that all the constants denoted generically with, $c,c',c'',c''',\tilde{c},....$ are adimensional. In the general case $|u|^{\de}$ has to be replaced by $(|u|/R_0)^\de$.}
\smallskip

\NI In conclusion the second of the ``bootstrap assumptions" we are stating is the following one:
\bigskip

\NI {\bf (II) : In $V_*$ the connection coefficient norms satisfy the following bounds: $\de{\cal O}\leq \ep_0$ .}

\subsection{the $\de\cal R$ null Riemann component norms:}
We start recalling the general definitions, \cite{Kl-Ni:book} Chapter 3, of the null components of the Riemann tensor with respect to the adapted null frame, where $W$ is a generic Weyl tensor and $X,Y$ are $S$-tangent vector fields,
\bea
&&\a({W})(X,Y)={W}(X,e_4,Y,e_4)\ \ ,\ \ \b({W})(X)=\frac{1}{2}{W}(X,e_4,e_3,e_4)\nn\\
&&\ro({W})=\frac{1}{4}{W}(e_3,e_4,e_3,e_4)\ \ ,\ \ \si({W})=\frac{1}{4}{^\star{W}}(e_3,e_4,e_3,e_4)\eql{3.1.19za}\\
&&\bb({W})(X)=\frac{1}{2}{W}(X,e_3,e_3,e_4)\ \ ,\ \ \aa({W})(X,Y)={W}(X,e_3,Y,e_3)\ .\nn
\eea
As done for the connection coefficients, to prove our result we must ``subtract" the Kerr part of the Riemann tensor.
This implies that writing all the various components of the Riemann tensor as
\[R=R^{(Kerr)}+\de R\]
we have to make the bootstrap assumptions on the norms relative to the ``correction part", we denote the set of these norms $\de{\cal R}$. The norms we choose bounded in the ``bootstrap assumptions" are the sup norms of the various null components of $\de R$ and their tangential derivatives.\footnote{Once we control the tangential derivatives, via the Bianchi equations we control also the derivatives along $e_3$ and $e_4$, see \cite{Kl-Ni:book} for the more delicate control of $\dd_{e_4}\a$ and $\dd_{e_3}\aa$.} The Bootstrap assumption can then be written  in a compact way as
\bea
\de{\cal R}\leq \ep_0\ \ .\eql{bass3}
\eea
which written in detail are, with $l\leq q-1$: 
\begin{eqnarray}
&&\ML\sup_{{V_*}}r^{5+l}|u|^{\frac{\ep}{2}}|\nabb^l\a(\de R)| \leq \ep_0\ \ ,\quad\sup_{V_*}r^{4+l}|u|^{1+{\frac{\ep}{2}}} |\nabb^l\b(\de R)|\leq \ep_0\nn\\
&&\ML\sup_{V_*}r^{3+l}|u|^{2+{\frac{\ep}{2}}}|\nabb^l(\ro(\de R)-\overline{\ro(\de R)})| \leq \ep_0\ \ ,\quad \sup r^{3+l}|u|^{2+{\frac{\ep}{2}}}|\nabb^l(\si(\de R)-\overline{\si(\de R)})|
\leq \ep_0\nn\\
&&\ML\sup_{V_*}r^{2+l}|u|^{3+{\frac{\ep}{2}}}|\nabb^l\bb(\de R)|\leq \ep_0\ \ ,\quad\sup_{V_*}r^{1+l}|u|^{4+{\frac{\ep}{2}}}|\nabb^l\aa(\de R)|\leq \ep_0\ \ .\ \eql{Riemest1}
\end{eqnarray} 
\NI Finally we have to add a bootstrap assumption for  for $\overline{\ro(\de R)}$,
\bea
\sup_{V_*}|r^3|u|^{2+{\frac{\ep}{2}}}{\overline{\ro(\de R)}}|\leq  \ep_0\ ,\eql{rodeRass}
\eea
where $\overline{\ro}$ denotes the average over the $S$ surface.
\medskip

\NI{\bf Remarks:}

\NI i) The exponent $\frac{\ep}{2}$ appearing here in the decay factors has to be assumed equal to $\de$ to complete the proof of the main theorem, Theorem \ref{Theorem to prove}\ .
\smallskip

\NI ii) The estimate for $\a(\de R)$, the same argument holds for the other components, is not yet the estimate for $\a(R)$ as we have to add $\a(R^{(Kerr)})$, the same, of course, for all the remaining components. In fact all the Riemann components contain also a Kerr part, even the components different from $\ro$ and $\si$. This is due to the fact that we are not using the principal null direction frame, see \cite{Ch} and later on for its precise definition. Their precise estimates are given later on in subsection \ref{SSrieminKerr}.
\smallskip

\NI In conclusion the third of the ``bootstrap assumptions" we are stating is the following one:
\bigskip

\NI {\bf (III) : In $V_*$ the norms of the components of the correction to the Kerr Riemann tensor, $\de R$, satisfy the following bounds: $\de{\cal R}\leq \ep_0$ .}
\medskip
\section{The results}
\subsection{The main theorem}
\begin{thm}\label{Theorem to prove}
If the bootstrap assumptions, \ref{Oboosestwithdec}, \ref{bass3} hold with $\de=\frac{\ep}{2}$, the previous smallness constants satisfy
\bea
\frac{\ep_0^2}{R_0^3}<\varepsilon<\ep_0\ \ , \frac{M}{R_0}<<1\ ,\eql{assneeded1}
\eea
and the smallness of the initial data is controlled by a function $\cal J$ explicitely defined in subsection \ref{SS3.6}, equation \ref{3.412},
\bea
{\cal J}(\Si_0,R_0; \de{^{(3)}\!}{\bf g}, \de{\bf k})\leq \varepsilon\ ,\eql{small2}
\eea 
then in the $V_*$ region the following bounds are satisfied:
\bea
&&\de{\cal O}\leq\frac{\ep_0}{2}\ \ ,\ \  {\de\cal R}\leq \frac{\ep_0}{2} \eql{exp1}\\
&&\nn\\
&& \sup_{V_*}|r^3|u|^{2+{\frac{\ep}{2}}}{\overline{\ro(\de R)}}|\leq  \frac{\ep_0}{2} \ \ .\eql{exp1b}
\eea
\end{thm}
\NI If we prove this result then we have basically proved the global existence and the peeling properties in the external region defined by the condition
\[ |u|\geq R_0\ ,\]
see also subsection \ref{extension}.

\subsection{The structure of the proof of Theorem \ref{Theorem to prove}}\label{S.S.stepsfortheproof}
The proof of the Main Theorem \ref{Theorem to prove} is long and is divided in many steps we list in the folowing,
\medskip

\NI{\bf 0 step:} {\bf Dimensional discussion }
\medskip

\NI{\bf I step:} {\bf The definition of the $\tilde{\cal Q}$ norms}
\medskip

\NI We denote $\tilde{\cal Q}$ the $\cal Q$ norms associated to ${\tilde R}=\lie_{T_0}R$, their explicit expression will be given in the sequel.
\medskip

\NI{\bf II step:} {\bf the estimate of $\overline{\ro({\tilde R})}$ }
\medskip

\NI The first theorem to prove says that under the bootstrap assumptions \ref{Oboosestwithdec}, \ref{bass3} and initial data such that on $\Si_0$:
\[{\tilde{\cal Q}}_{\Si_0}\leq \varepsilon^2\ \ ;\ \ \sup_{\Si_0/B_{R_0}}|r^{6+}\overline{\ro({\tilde R})}|\leq c\varepsilon\]
then 
\bea
\tilde{\cal Q}\leq c\varepsilon^2+\frac{M}{R_0}\ep_0^2\ \ ;\ \ \sup_{V_*}|r^{3}|u|^{3+\frac{\ep}{2}}\overline{\ro({\tilde R})}|\leq \frac{\ep_0}{2}\ .
\eea
\medskip

\NI{\bf III step:}

\NI In this step, proceeding basically as in \cite{Ni;Peel}, one proves that from the results in the first two steps all the norms $\de\cal R$ are bounded by $\frac{\ep_0}{2}$. This proof is very long and is made by many consecutive lemmas; it requires also the control, based on the $\de{\cal O}$ bootstrap assumptions, of the (components of the) deformation tensor ${^{(T_0)}\!\pi}$.
\medskip

\NI{\bf IV step:}

\NI This step completes the proof of Theorem \ref{Theorem to prove} showing that under the results of the previous steps we have
\bea
\de{\cal O}\leq\frac{\ep_0}{2}\ \ .\eql{Oresult}
\eea
This last step is done in three parts to do in a precise order, namely:
\smallskip

\NI a) First using the bootstrap assumptions for the $\de{\cal O}$ norms, $\de{\cal O}\leq \ep_0$, we prove that, under the initial data assumptions, the metric component corrections norms satisfy, 
\bea
\de{\cal O}^{(0)}\leq c{\ep_0} . \eql{O0est}
\eea
\smallskip

\NI b) Using this result plus the initial data assumptions and the bootstrap assumptions, we obtain that the connection coefficient norms satisfy
\bea
\de{\cal O}\leq  \frac{\ep_0}{2}\ . 
\eea

\NI c) Finally we prove that under this result also the metric components corrections have better estimates, namely
\bea
\de{\cal O}^{(0)}\leq \frac{\ep_0}{2}\  . 
\eea
This last result will be needed with all the other ones to perform step VI.
\medskip

\NI{\bf V step:} 
\NI In this step we collect all the initial data conditions required in the previous steps and show how they can be satisfied imposing a global decay condition for the initial data and the smallness condition \ref{small2}\ .
\medskip

\NI{\bf VI step:}
In this step we show how, by a basically local existence theorem, in view of the previous results we can extend the region $V_*$ showing that it has to coincide with the whole spacetime.
\smallskip

\subsection{Proof of the various steps}
\subsubsection{0 step: Dimensional discussion:}

\NI $\varepsilon$ and $\ep_0$ have dimension $L^3$, all the constants $c$'s are adimensional, 
$\kappa,\de$ are also adimensional. The correction to the various power decay are considered as adimensional ( see the remark in subsection \ref{SSconcoef}). We call a constant ``independent" if it does not depend on $a,M,\ep_0,\varepsilon$. 

\subsubsection{I step: The definition of the $\tilde{\cal Q}$ norms.}
These norms, we denote ${\tilde Q}$, are the analogous of the $Q$ terms defined in \cite{Kl-Ni:book}, Chapter 3,
\begin{eqnarray}\label{norme1}
&& \QQ(u,\ub) = \QQ_1(u,\ub)+\QQ_2(u,\ub)+\sum_{i=2}^{q-1}{\QQ}_{(q)}(u,\ub)\nn\\
&& \QQb(u,\ub) = \QQb_1(u,\ub)+\QQb_2(u,\ub)+\sum_{i=2}^{q-1}{\QQb}_{(q)}(u,\ub)\	,
\end{eqnarray}
where, with $S(u,\ub)\subset V_*$, we denote
\[V(u,\ub)=J^{(-)}(S(u,\ub))\cap{V_*}\ ,\]
\begin{eqnarray}\label{norms1}
\QQ_1(u,\ub) &\equiv & \int_{C(u)\cap V(u,\ub)}|u|^{5+\ga}Q(\lie_T
{\tW})(\bar{K},\bar{K},\bar{K},e_4)\nonumber\\
& &+\int_{C(u)\cap{V(u,\ub)}}|u|^{5+\ga}Q(\lie_O\tW)(\bar{K},\bar{K},T,e_4)\nonumber\\
\QQ_2(u,\ub) &\equiv &\int_{C(u)\cap{V(u,\ub)}}|u|^{5+\ga}Q(\lie_O\lie_T
\tW)(\bar{K},\bar{K},\bar{K},e_4)\nonumber\\
& &+\int_{C(u)\cap{V(u,\ub)}}|u|^{5+\ga}Q(\lie^2_O\tW)(\bar{K},\bar{K},T,e_4)\label{Q_2}\\
& & +\int_{C(u)\cap{V(u,\ub)}}|u|^{5+\ga}Q(\lie_S\lie_T\tW)(\bar{K},\bar{K},\bar{K},e_4)\nn
\end{eqnarray}
\bea
\widetilde{\cal Q}_{(q)}(u,\ub)&\equiv&\sum_{i=2}^{q-1}\left\{\int_{C(u)\cap
V(u,\ub)}|u|^{5+\ga}Q(\lie_{T}\lie_{O}^{i}\tW)(\acc,\acc,\acc,e_4)\right.\nn\\
&&\left.+\int_{C(u)\cap V(u,\ub)}|u|^{5+\ga}Q(\lie_{O}^{i+1}\tW)(\acc,\acc,T,e_4)\right.\eql{2.2}\\
&&\left.+\int_{C(u)\cap V(u,\ub)}|u|^{5+\ga}Q(\lie_{S}\lie_{T}\lie_O^{i-1}\tW)(\acc,\acc,\acc,e_4)\right\}\nn
\eea
\begin{eqnarray}\label{norms2}
\QQb_1(u,\ub)& \equiv &\int_{\un{C}(\ub)\cap{V(u,\ub)}}|u|^{5+\ga}Q(\lie_T{\tilde R})(\bar{K},\bar{K},\bar{K},e_3)\nonumber\\
& &+\int_{\un{C}(\ub)\cap{V(u,\ub)}}|u|^{5+\ga}Q(\lie_O{\tilde R})(\bar{K},\bar{K},T,e_3)\ \ ,\nonumber
\end{eqnarray}
\begin{eqnarray}
\QQb_2(u,\ub) &\equiv &\int_{\un{C}(\ub)\cap{V(u,\ub)}}|u|^{5+\ga}Q(\lie_O\lie_T{\tilde R})(\bar{K},\bar{K},\bar{K},e_3)\nonumber\\
& &+\int_{\un{C}(u)\cap{V(u,\ub)}}|u|^{5+\ga}Q(\lie^2_O{\tilde R})(\bar{K},\bar{K},T,e_3)\label{Qb_2}\\
& & +\int_{\un{C}(\ub)\cap{V(u,\ub)}}|u|^{5+\ga}Q(\lie_S\lie_T{\tilde R})(\bar{K},\bar{K},\bar{K},e_3)\nonumber
\end{eqnarray}
\bea
{\QQb}_{(q)}(u,\ub)&\equiv&\sum_{i=2}^{q-1}\left\{\int_{C(\la)\cap
V(\la,\nu)}|u|^{5+\ga}Q(\lie_{T}\lie_{O}^{i}\tW)(\acc,\acc,\acc,e_3)\right.\nn\\
&&\left.+\int_{C(u)\cap V(u,\ub)}|u|^{5+\ga}Q(\lie_{O}^{i+1}\tW)(\acc,\acc,T_0,e_3)\right.\eql{2.4}\\
&&\left.+\int_{C(u)\cap V(u,\ub)}|u|^{5+\ga}Q(\lie_{S}\lie_{T_0}\lie_O^{i-1}\tW)(\acc,\acc,\acc,e_3)\right\}\ .\nn
\eea
and
\begin{eqnarray}
\QQ_1({\Si_0}) &\equiv &
\int_{\Si_0\cap{V_*}}|u|^{5+\ga}Q(\lie_T {\tilde R})(\bar{K},\bar{K},\bar{K},T)\nonumber\\
& & +\int_{\Si_0\cap{V_*}}|u|^{5+\ga}Q(\lie_O{\tilde R})(\bar{K},\bar{K},T,T)\\
\QQ_2({\Si_0}) &\equiv & \int_{\Si_0\cap{V_*}}|u|^{5+\ga}Q(\lie_O\lie_T {\tilde R})(\bar{K},\bar{K},\bar{K},T)\nonumber\\
& & +\int_{\Si_0\cap{V_*}}|u|^{5+\ga}Q(\lie_O^2{\tilde R})(\bar{K},\bar{K},T,T)\nonumber\\
& &+ \int_{\Si_0\cap{V_*}}|u|^{5+\ga}Q(\lie_S\lie_T{\tilde R})(\bar{K},\bar{K},\bar{K},T)\ \ .
\end{eqnarray}
\bea
\sum_{i=2}^{q-1}\widetilde{\cal Q}_{(q)}({\Si_0})&\equiv&
\sum_{i=2}^{q-1}\left\{\int_{\Si_0\cap{V_*}}|u|^{5+\ga}Q(\lie_{T}\lie_{O}^{i}\tW)(\acc,\acc,\acc,T)\right.\nn\\
&&\left.+\int_{\Si_0\cap{V_*}}|\la|^{5+\ga}Q(\lie_{O}^{i+1}\tW)(\acc,\acc,T,T)\right.\eql{2.2a}\\
&&\left.+\int_{\Si_0\cap{V_*}}|\la|^{5+\ga}Q(\lie_{S}\lie_{T}\lie_O^{i-1}\tW)(\acc,\acc,\acc,T)\right\}\nn
\eea
We also introduce the following quantity: 
\begin{equation}
\QQ_{{V_*}}\equiv\sup_{\{u,\ub|S(u,\ub)\subseteq{V_*}\}}\{\QQ(u,\ub)+\QQb(u,\ub)\}\ .
\end{equation}
The initial conditions have to be be chosen in such a way that the $\QQ, \QQb$ norms are bounded on $\Si_0$,
\bea
\QQ_1({\Si_0})+\QQ_2({\Si_0})+\sum_{i=2}^{q-1}\widetilde{\cal Q}_{(q)}({\Si_0})\leq \varepsilon^2\ .\eql{Initcond1}
\eea

\NI Let us discuss the main differences between these norms and the similar norms defined in \cite{Kl-Ni:book}, Chapter 3:
\smallskip

i) First of all in these $\tilde{\cal Q}$ norms a term associated to $\overline{\ro({\tilde R})}$ is not present, differently from the definitions in \cite{Kl-Ni:book}.
\smallskip

ii) The second fundamental difference is that we are considering these norms not associated to the Riemann tensor $R$, but to the Weyl tensor ${\tilde R}=\lie_{T_0}R$ which can be interpreted as the ``time derivative" of $R$; this is one of the steps, as discussed in the introduction, required to separate the Kerr part of the Riemann tensor from the whole Riemann tensor.
\smallskip

iii) The third difference is that the weights of these $\tilde{\cal Q}$ norms have an extra decay factor, $|u|^{5+\ga}$, with $\ga>0$. 
This factor has the effect, in a complicated way shown in the following lemmas, of giving a better decay for the various null components of $\lie_{T_0}R$ . 
This better decay is required for two reasons, the first one is  to guarantee that the decay of  Riemann tensor be in agreement with  the ``peeling decay", this requires the exponent $5+\ga$, and the second one that it has to allow  to recover the decay of the Riemann tensor by ``time integration".\footnote{To perform the ``time integration" the exponent could be less the $5+\ga$ if we do not require the peeling decay.}

\NI The central point in the definition of these $\tilde{\cal Q}$ norms is that the exponent $5+\ga$ is chosen in such a way that, once we prove that these norms are bounded in $V_*$, which depends only on the assumption that $M/R_0$ be sufficiently small, then the null components of $\lie_{T_0}R$, $\a,\b,...$ decay with an exponent $5+\frac{\ep}{2}, 4+\frac{\ep}{2},3+\frac{\ep}{2},....$ depending on the null components we are considering where
\bea
\ep=\ga\ .
\eea
\NI\ Let us also observe that assigned a value for $\ga$ in the $\tilde{\cal Q}$ norms if the $\tilde{\cal Q}|_{\Si_0}$ norms are bounded then the Riemann components on $\Si_0$ have to decay with an $\hat{\ep}>{\ga}$ which implies that the correction of the metric components have to decay on $\Si_0$ toward spacelike infinity as $O(r^{-(3+\frac{\hat{\ep}}{2})})$.

\NI On the other side starting from the $\tilde{\cal Q}$ norm with a weight $|u|^{5+\ga}$ it follows that the sup norm estimates for the Riemann components have the same factor $\ga$ implying that there is a loss of decay going from the initial data decay to the decay of the Riemann components in the whole $V_*$ and , therefore, on the whole spacetime. Finally the connection coefficients satisfy the structure equations which depend on the Riemann tensor so that the exponent factor $\de$ must be equal to $\ga/2$, see later on, and $\de$ must coincide with the exponent $\ep/2$ introduced in lemmas \ref{Lemma2}, 
\ref{Lemma3}, \ref{Lemma4}. In conclusion in the bootstrap assumptions for the $\de{\cal O}$ norms we choose $\ga=2\de=\ep \in (0,{\hat{\ep}})$. 
\smallskip

iv) The last important observation  is that in the definition
of $\tilde R=\lie_{T_0}R$ the vector field $T_0$ is 
\bea
T_0=\frac{\oom}{2}(e_3+e_4)-\frac{X_{(Kerr)}+X}{2}\ .
\eea
$T_0$ is a ``nearly" Killing vector field if the corrections to the Kerr metric are small and equal to $\frac{\pr}{\pr t}$ in Kerr spacetime. Viceversa in the $\tilde{\cal Q}$ norms definition 
\bea
T=\frac{\oom}{2}(e_3+e_4)
\eea
is one of  the vector fields saturating the Bel-Robinson tensor and also appearing in the various Lie derivatives of $\tilde R$.
The reason for the use of both $T$ and $T_0$ is that to go back from $\tilde R$ to the ``time derivatives" of the components of $R$ controlling both the norm smallness and the "peeling decay" we need  $T_0$ which is nearly a Killing vector field. Viceversa to obtain from the control of the $\tilde{\cal Q}$ norms the control of the null components norms of $\tilde R$ it is crucial, to have their integrands made by positive terms which requires to use the vector field $T=\frac{\oom}{2}(e_3+e_4)$.
\medskip

\subsubsection{II step: the estimate of $\overline{\ro({\tilde R})}$ and of $\tilde{\cal Q}$ in $V_*$.}
We prove the following theorem
\begin{thm}\label{TH2.2}
Assume in $V_*$ the bootstrap assumptions \ref{Oboosestwithdec}, \ref{bass3}, assume that on $\Si_0$ the initial data are such that 
\bea
\sup_{\Si_0/B_{R_0}}|r^{6+\frac{\ep}{2}}\overline{\ro({\tilde R})}|\leq \varepsilon;\ \ {\tilde{\cal Q}}_{\Si_0}\leq\varepsilon^2
\eea
then in $V_*$ the following estimates hold,
\bea
\ML\sup_{V_*}|r^{3}|u|^{3+\frac{\ep}{2}}\overline{\ro({\tilde R})}|\leq c_1^{\frac{1}{2}}\bigg(\varepsilon+\left(\frac{M}{R_0}\right)^{\!\!\frac{1}{2}}\!\!\ep_0\!\bigg)\ \ ;\ \ \tilde{\cal Q}\leq c_1\!\!\left(\varepsilon^2+\frac{M}{R_0}\ep_0^2\right)
\eea
where $c$ is an adimensional constant independent from $M$.
\end{thm}

\NI{\bf Proof:} The proof follows the lines of the analogous proof in \cite{Ni;Peel}, therefore it uses a bootstrap inside the region $V_*$. The rationale for it is the following: to prove the estimate for $\overline{\ro({\tilde R})}$ we need to control the other null components of $\tilde R$. These are controlled in terms of the $\tilde Q$ norms which at their turn require, to be controlled, an estimate for $\overline{\ro({\tilde R})}$.\footnote{The bootstrap we use here is different from the bootstrap we need to extend the region $V_*$, in fact the main bootstrap assumptions refer to the null components of $\de R=R-R^{(Kerr)}$ and not to those of ${\tilde R}=\lie_{T_0}R$\ .}

\NI Let $V_*=J^{(-)}(S(u(R_0),\ub_*))$, let us define $\tilde{V}\subset V_*=J^{(-)}(S(u(R_0),\tilde{\ub}))$ and $\tilde{\ub}\leq \ub_*$ the largest value of the affine variable $\ub$ such that in $\tilde{V}$ the following estimates hold:\footnote{$u(R_0)$ is the value of $u|_{\Si_0}=r_*$ associated to $r=R_0$.}
\bea
\sup_{\tilde{V}}|r^{3}|u|^{3+\frac{\ep}{2}}\overline{\ro({\tilde R})}|\leq c_1^{\frac{1}{2}}\bigg(\varepsilon+\left(\frac{M}{R_0}\right)^{\!\!\frac{1}{2}}\!\!\ep_0\!\bigg)\ \ ;\ \ \tilde{\cal Q}\leq c_1\!\!\left(\varepsilon^2+\frac{M}{R_0}\ep_0^2\right)\eql{subboostass}
\eea
with $c_1>1$. 
\NI Writing the transport equation for $r^3\overline{\ro}({\tilde R})$ we obtain,\footnote{The derivation of this equation is in \cite{Kl-Ni:book}, Chapter 5.} (all the Riemann null components refer to ${\tilde R}=\lie_{T_0}R$), 
\beaa
&&\frac{d}{d\la}{(r^3\overline{\ro})}=\frac{r^3}{|S(\la,\nu)|}\int_{S(\la,\nu)}\frac{1}{2}(\overline{\oom tr\chib}-\oom\tr\chib)(\ro-{\overline{\ro}})\nn\\
&&\ \ \ \ \ \ \ \ \ \ \ \ +\frac{r^3}{|S(\la,\nu)|}\int_{S(\la,\nu)}\oom\left(-\divv\bb-2\eta\cdot\bb-\frac{1}{2}\chih\cdot\aa+\ze\cdot\bb\right)\nn\\
&&=\frac{1}{|S(\la,\nu)|}\int_{S(\la,\nu)}\left[\frac{(\overline{\oom tr\chib}-\oom\tr\chib)}{2}r^3(\ro-{\overline{\ro}})
+\oom r^3\left(\divv\oom\c\bb-2\eta\cdot\bb-\frac{1}{2}\chih\cdot\aa+\ze\cdot\bb\right)\right]\nn
\eeaa
Therefore
\bea
&&\frac{d}{d\la}{(r^3\overline{\ro})}=\frac{1}{|S(\la,\nu)|}\int_{S(\la,\nu)}G\nn
\eea
where
\bea
G\equiv\left[\frac{(\overline{\oom tr\chib}-\oom\tr\chib)}{2}r^3(\ro-{\overline{\ro}})
+\oom r^3\left(\divv\oom\c\bb-2\eta\cdot\bb-\frac{1}{2}\chih\cdot\aa+\ze\cdot\bb\right)\right]\ \ \ \ \ \ 
\eea
Integrating along the incoming cones we obtain
\bea
|r^{3}\overline{\ro}|(u,\ub)\leq |r^{3}\overline{\ro}|(u_0,\ub)+\int_{u_0}^u|r^{3}G|(u',\ub)
\eea
Using the sup norms estimates for the connection coefficients, easily derived from \ref{trchi-b}, \ref{bass1a}, \ref{bass1b} and the bootstrap assumptions \ref{Oboosestwithdec}, we have, recalling that in the external region $|u|\leq r$,
\bea
&&|r^{3}G|\leq c\left(\frac{M^2}{r^2|u|}|r ^{3}(\ro-{\overline{\ro}})|+\frac{M^2}{r^2|u|}|r^{2}|u|\bb|
+\frac{M^2}{r^2|u|}|r|u|^2\aa|_{p,S}\right)\eql{Gest}\\
&&\leq c\left(\frac{M^2}{r^2|u|^{4+\frac{\ep}{2}}}|r ^{3}|u|^{3+\frac{\ep}{2}}(\ro-{\overline{\ro}})|+\frac{M^2}{r^2|u|^{4+\frac{\ep}{2}}}|r^{2}|u|^{4+\frac{\ep}{2}}\bb|
+\frac{M^2}{r^2|u|^{4+\frac{\ep}{2}}}|r|u|^{5+\frac{\ep}{2}}\aa|\right)\nn
\eea
where the norms of the $\tilde R$ null Riemann components in \ref{Gest} can all be bounded by the ${\tilde{\cal Q}}^{\frac{1}{2}}$ quantity,\footnote{Recall that as we said in the previous step the exponent $|\ga|$ and the initial data have been chosen exactly in the way to have these $\tilde R$ null Riemann components bounded by ${\tilde{\cal Q}}^{\frac{1}{2}}$ quantities.}
which, at its turn, is bounded by $c_1^{\frac{1}{2}}\!\!\left(\varepsilon+\left(\frac{M}{R_0}\right)^{\!\frac{1}{2}}\!\ep_0\right)$. Therefore we have
\bea
&&\ML|r^{3}|u|^{3+\frac{\ep}{2}}\overline{\ro}|(u,\ub)\!\leq\! \left(|r^{6+\frac{\ep}{2}}\overline{\ro}|(u_0,\ub)+c\frac{M^2}{R_0^{2}}\tilde{\cal Q}^{\frac{1}{2}}\right)\!
\leq\! \left(\varepsilon+c\frac{M^2}{R_0^{2}}c_1^{\frac{1}{2}}\!\!\left(\varepsilon+\left(\frac{M}{R_0}\right)^{\!\frac{1}{2}}\!\ep_0\right)\right)\nn\\
&&\leq c\!\left(1+\frac{M^2}{R_0^{2}}c_1^{\frac{1}{2}}\right)\varepsilon+cc_1^{\frac{1}{2}}\!\left(\frac{M}{R_0}\right)^{\!\frac{5}{2}}\ep_0
< c_1^{\frac{1}{2}}\bigg(\varepsilon+\left(\frac{M}{R_0}\right)^{\!\!\frac{1}{2}}\!\!\ep_0\!\bigg)
\eea
provided $\varepsilon$ and $M/R_0$ have been chosen sufficiently small and $c_1>c^2$, so that
\bea
\left(1+c\frac{M^2}{R_0^{2}}c_1^{\frac{1}{2}}\right)< c_1^{\frac{1}{2}}\ \ , c\!\left(\frac{M}{R_0}\right)^{\!2}< 1\ .\eql{extregcond1}
\eea

\NI{\bf The control of the $\tilde{\cal Q}$ norms in $V_*$\ .}
We are left to prove that in $\tilde V$ the second condition of \ref{subboostass}
\[\tilde{\cal Q}\leq c_1\!\!\left(\varepsilon^2+\frac{M}{R_0}\ep_0^2\right)\ ,\] can be improved. This requires the estimate of the error ${\tilde {\cal E}}=\tilde{\cal Q}-\tilde{\cal Q}_{\Si_0}$. The estimate of the error proceeds basically as in \cite{Kl-Ni:book}, the main differences being that we are considering $\tilde{R}$ instead of $R$ and that in the definition of the $\tilde{\cal Q}$ norms there are extra weight factors. Let us examine some terms. 

\NI Let the ${\tilde Q}$ terms be the analogous of the $Q$ terms defined in \cite{Kl-Ni:book}, Chapter 3, with the extra decay factor, $|u|^{5+\ga}$.
\smallskip

\NI Let $V\subset{\tilde V}$ and consider one of the terms in ${\int_{V_{({u},\ub)}}|Div{\tilde Q}(\lie_{O}\tilde{R})_{\b\ga\de}(K^{\b}K^{\ga}T^{\de})|}$ namely the first term of
\bea
\int_{V_{(u,\ub)}}\tau_{+}^4|u|^{5+\ga}D(O,\tilde{R})_{444}&=&\frac{1}{2}\int_{V_{(u,\ub)}}\tau_{+}^4|u|^{5+\ga}
\a(\lie_{O}\tilde{R})\cdot\Theta(O,\tilde{R})\nn\\
&-&\int_{V_{(u,\ub)}}\tau_{+}^4|u|^{5+\ga}\b(\lie_{O}\tilde{R})\cdot\Xi(O,\tilde{R})
\eea
We have 
\bea
&&\bigg|\int_{V_{(u,\ub)}}\tau_{+}^{4}|u|^{5+\ga}\a(\lie_{O}\tilde{R})\cdot\Th(O,\tilde{R})\bigg|\nn\\
&&\leq\left(\sup_V\int_{C(u';[\ub_0,\ub])}\ub'^{4}|u|^{5+\ga}|\a(\lie_{O}\tilde{R})|^2\right)^{\frac{1}{2}}\!
\int_{u_0}^{u}du'\left(\int_{C(u';[\ub_0,\ub])}\ub'^{4}|u|^{5+\ga}|\Th(O,\tilde{R})|^2\right)^{\frac{1}{2}}\nn\\
&&\leq c\QQ_V^{\frac{1}{2}}\!\!\int_{u_0}^{u}du'\sum_{i=0}^3
\left(\int_{C(u';[\ub_0,\ub])}\ub'^{4}|u|^{5+\ga}|\Th^{(i)}(O,\tilde{R})|^2\right)^{\frac{1}{2}}
\eql{5.4.4}
\eea
The part with $i=0$ describes a term which is not present in the error estimate in \cite{Kl-Ni:book}, the reason being that in that case the Riemann tensor $R_{\mu\nu\ro\si}$ satisfied the Bianchi equation $D^{\mu}R_{\mu\nu\ro\si}=0$, while here
\[D^{\mu}\Lie_{T_0}R_{\mu\nu\ro\si}\neq 0\ .\]  
We treat this term separately later on. Let us consider now the part with $i=1$,
\bea
&&|\tilde{\cal E}|\leq c\QQ_V^{\frac{1}{2}}\!\int_{u_0}^{u}du'\left(\int_{C(u';[\ub_0,\ub])}\ub'^{4}|u|^{5+\ga}|\Th^{(1)}(O,\tilde{R})|^2\right)^{\frac{1}{2}}\nn\\
&&\leq c\QQ_V^{\frac{1}{2}}\!\left\{\int_{u_0}^{u}du'\left(\int_{C(u';[\ub_0,\ub])}\ub'^{4}|u|^{5+\ga}|{^{(O)}\!\pi}|^2|\nabb\a(\tilde{R})|^2\right)^{\frac{1}{2}}
+\c\c\c\c \right\}\nn\\
&&\leq c\QQ_V^{\frac{1}{2}}\int_{u_0}^{u}du'\left(\int_{C(u';[\ub_0,\ub])}\ub'^{4}|u|^{5+\ga}|{^{(O)}\!\pi}|^2|\nabb\a(\tilde{R})|^2\right)^{\frac{1}{2}}+\c\c\c\nn\\
&&\leq c\QQ_V^{\frac{1}{2}}\int_{u_0}^{u}du'\left(\int_{C(u';[\ub_0,\ub])}\ub'^{2}|u|^{5+\ga}|{^{(O)}\!\pi}|^2|\a(\lie_O\tilde{R})|^2\right)^{\frac{1}{2}}\nn\\
&&\leq c\QQ_V\int_{u_0}^{u}du'\sup_V\frac{1}{r}|{^{(O)}\!\pi}|\leq c\QQ_V\int_{u_0}^{u}du'\sup_V\frac{1}{r^2|u|}|r|u|{^{(O)}\!\pi}|\nn\\
&&\leq c\QQ_V\frac{1}{R_0^{2-}}\left(\int_{u_0}^{u}du'\frac{1}{|u'|^{1+}}\right)\sup_V|r|u|{^{(O)}\!\pi}|\nn
\eea
Observe that (by dimensional reasons) in Kerr we have \footnote{There are no terms proportional to $\frac{O(M)}{r}$ as the  \Sch\ spacetime is spherical symmetric.}
\[|r^2{^{(O)}\!\pi^{(Kerr)}}|\leq O(aM)\leq O(M^2)\ ,\]
and the bootstrap assumptions imply analogous estimates in $\tilde V$,
\bea
|r|u|{^{(O)}\!\pi}|\leq O(aM)\leq O(M^2)\ .
\eea
Substituting in the previous expression we have
\bea
|\tilde{\cal E}|\leq c\QQ_V\frac{1}{R_0^{2}}M^2\leq c\frac{M^2}{R_0^2}\QQ_V\ .
\eea
Assuming for a moment that this is the only error contribution it would follow that, if 
\bea
\frac{M^2}{R_0^2}<<1\eql{extregcond2}
\eea
then
\bea
\QQ_V\leq \frac{c}{1-c\frac{M^2}{R_0^2}}\QQ_{\Si_0}\leq c\QQ_{\Si_0}\leq c\varepsilon^2\ .
\eea
\smallskip

\NI We look now at the term with $i=0$, absent in  \cite{Kl-Ni:book}, that is to the error contribution proportional to
\bea
\QQ_V^{\frac{1}{2}}\!\!\int_{u_0}^{u}du'
\left(\int_{C(u';[\ub_0,\ub])}\ub'^{4}|u|^{5+\ga}|\Th^{(0)}(O,\tilde{R})|^2\right)^{\frac{1}{2}}\ .
\eql{5.4.4a}
\eea
This term arises from the first term of
\bea
D^{\a}(\lie_O{\tilde R})_{\b\ga\de}=D^{\a}(\lie_O\lie_{T_0}R)_{\b\ga\de}=\lie_OD^{\a}(\lie_{T_0}R)_{\b\ga\de}+\sum_{i=1}^3J^i(O;{\tilde R})_{\b\ga\de}\ .\ \ 
\eea
As, see \cite{Kl-Ni:book}, Chapter 6, eqs.(6.1.6)
\bea
\lie_OD^{\a}(\lie_{T_0}R)=\lie_OJ(T_0,R)=\lie_O(J^1(T_0,R)+J^2(T_0,R)+J^3(T_0,R))\ \ \ 
\eea
the term we have to estimate is
\bea
\QQ_V^{\frac{1}{2}}\!\!\int_{u_0}^{u}du'
\left(\int_{C(u';[\ub_0,\ub])}\ub'^{4}|u|^{5+\ga}|\lie_OJ(T_0,R)|^2\right)^{\frac{1}{2}}\ .
\eql{5.4.4ba}
\eea
This term has to be estimated in a different way from the previous ones as it cannot be bounded in terms of the $\tilde{\cal Q}$, in fact it depends on $R$ instead than on $\tilde R$.
It is easy to recognize that all the terms in which we can decompose $J(T_0,R)$ can be estimated in the same way, apart from $J^0$, therefore we consider only $J^1(T_0,R)$ and again all the various terms which compose it can be estimated in the same way, see \cite{Kl-Ni:book} pages 245-247. They all have the structure
\bea
{^{(T_0)}\!\pi}DR+{^{(T_0)}\!\pi}\frac{R}{r}
\eea 
and recalling that $T_0$ is nearly Killing it follows that the norm of the generic term of $\lie_OJ^1(T_0,R)$ can be estimated in the following way, neglecting for the moment which norm we are considering and, for simplicity considering all these norms being $\sup$ norms. Therefore, with $\si>0$,
\bea
|\lie_OJ^1(T_0,R)|\leq |{^{(T_0)}\!\pi}||DR|\leq \frac{\ep_0}{r^2|u|^{2+\de}}\frac{M}{r^4}\leq c\frac{\ep_0}{r^5|u|^{2+\de+\si}}\frac{M}{R_0^{1-\si}}\ ,
\eea
remembering that $\lie_0$ does not improve the decay. Substituting this estimate in \ref{5.4.4ba} we obtain
\bea
&&\QQ_V^{\frac{1}{2}}\!\!\int_{u_0}^{u}du'
\left(\int_{C(u';[\ub_0,\ub])}\ub'^{4}|u|^{5+\ga}|\lie_OJ(T_0,R)|^2\right)^{\frac{1}{2}}\nn\\
&&\leq c\QQ_V^{\frac{1}{2}}\ep_0\frac{M}{R_0^{1-\si}}\!\!\int_{u_0}^{u}du'
\left(\int_{C(u';[\ub_0,\ub])}\ub'^{4}|u|^{5+\ga}\frac{1}{r^{10+2\si}|u|^{4+2\de+2\si}}\right)^{\frac{1}{2}}\nn\\
&&\leq c\QQ_V^{\frac{1}{2}}\ep_0\frac{M}{R_0^{1-\si}}\!\!\int_{u_0}^{u}du'
\!\left(\int_{|u|'}^{\infty}dr r^6|u'|^{5+\ga}\frac{1}{r^{10+2\si}|u'|^{4+2\de+2\si}}\right)^{\frac{1}{2}}\nn\\
&&\leq c\QQ_V^{\frac{1}{2}}\ep_0\frac{M}{R_0^{1-\si}}\!\!\int_{u_0}^{u}du'|u'|^{\frac{1}{2}+\frac{\ga}{2}-\de}
\!\left(\int_{|u|'}^{\infty}dr \frac{1}{r^{4+2\si}}\right)^{\frac{1}{2}}
\eql{5.4.4c}
\eea
Let us choose now $\ga$ and $\de$ such that\footnote{To complete the bootstrap procedure we have to require the equal sign, see subsubsection \ref{ss342} .}
\bea
\de\geq \frac{\ga}{2}\eql{gaderel}
\eea
then  the last integral can be estimated in the following way:
\bea
&&\QQ_V^{\frac{1}{2}}\!\!\int_{u_0}^{u}du'
\left(\int_{C(u';[\ub_0,\ub])}\ub'^{4}|u|^{5+\ga}|\lie_OJ(T_0,R)|^2\right)^{\frac{1}{2}}\nn\\
&&\leq c\QQ_V^{\frac{1}{2}}\ep_0\frac{M}{R_0^{1-\si}}\!\!\int_{u_0}^{u}du'|u'|^{\frac{1}{2}}\frac{1}{|u'|^{\frac{3}{2}+\si}}
\leq c\QQ_V^{\frac{1}{2}}\ep_0\frac{M}{R_0}\nn\\
&&\leq c\left(\QQ_V\frac{M}{R_0}+\ep_0^2\frac{M}{R_0}\right)\ .
\eea
If this were the only error term we would write
\bea
|\tilde{\cal E}|\leq c\left(\QQ_V\frac{M}{R_0}+\ep_0^2\frac{M}{R_0}\right)
\eea
and
\bea
\QQ_V\leq \QQ_{\Si_0}+c\frac{M}{R_0}\QQ_V+c\frac{M}{R_0}\ep_0^2
\eea
implying
\bea
&&\QQ_V\leq\frac{1}{1-c\frac{M}{R_0}}\QQ_{\Si_0}+\frac{c}{\left(1-c\frac{M}{R_0}\right)}\frac{M}{R_0}\ep_0^2<\frac{c_1}{2}\!\!\left(\varepsilon^2+\frac{M}{R_0}\ep_0^2\right)
\eea
choosing $c_1$ appropriately. 
To complete the proof we consider some other relevant terms of the error. More specifically we estimate the part of the error:
\[{\int_{V_{({u},\ub)}}|Div{\tilde Q}(\lie_{T}\tilde{R})_{\b\ga\de}(K^{\b}K^{\ga}K^{\de})|}\ .\]

\NI Let us consider one of the term associated to ${\int_{V_{({u},\ub)}}|Div{\tilde Q}(\lie_{T}\tilde{R})_{\b\ga\de}(K^{\b}K^{\ga}K^{\de})|}$ namely the first term of
\bea
\int_{V_{(u,\ub)}}\tau_{+}^6|u|^{5+\ga}D(T,\tilde{R})_{444}&=&\frac{1}{2}\int_{V_{(u,\ub)}}\tau_{+}^6|u|^{5+\ga}\a(\lie_{T}\tilde{R})\cdot\Theta(T,\tilde{R})\nn\\
&-&\int_{V_{(u,\ub)}}\tau_{+}^6|u|^{5+\ga}\b(\lie_{T}\tilde{R})\cdot\Xi(T,\tilde{R})
\eea 
Looking at the first term in the r.h.s. we have
\bea
&&\bigg|\int_{V_{(u,\ub)}}\tau_{+}^{6}|u|^{5+\ga}\a(\lie_{T}\tilde{R})\cdot\Th(T,\tilde{R})\bigg|\nn\\
&&\leq\left(\sup_V\int_{C(u';[\ub_0,\ub])}\ub'^{6}|u|^{5+\ga}|\a(\lie_{T}\tilde{R})|^2\right)^{\frac{1}{2}}\!
\int_{u_0}^{u}du'\left(\int_{C(u';[\ub_0,\ub])}\ub'^{6}|u|^{5+\ga}|\Th(T,\tilde{R})|^2\right)^{\frac{1}{2}}\nn\\
&&\leq c\QQ_V^{\frac{1}{2}}\!\!\int_{u_0}^{u}du'\sum_{i=1}^3
\left(\int_{C(u';[\ub_0,\ub])}\ub'^{6}|u|^{5+\ga}|\Th^{(i)}(T,\tilde{R})|^2\right)^{\frac{1}{2}}\ .
\eql{5.4.4b}
\eea
Let us consider the term with $i=1$,
\bea
&&|\tilde{\cal E}|\leq c\QQ_V^{\frac{1}{2}}\!\int_{u_0}^{u}du'\left(\int_{C(u';[\ub_0,\ub])}\ub'^{6}|u|^{5+\ga}|\Th^{(1)}(T,\tilde{R})|^2\right)^{\frac{1}{2}}\nn\\
&&\leq c\QQ_V^{\frac{1}{2}}\!\left\{\int_{u_0}^{u}du'\left(\int_{C(u';[\ub_0,\ub])}\ub'^{6}|u|^{5+\ga}|{^{(T)}\!\pi}|^2|\nabb\a(\tilde{R})|^2\right)^{\frac{1}{2}}
+\c\c\c\c \right\}\nn\\
&&\leq c\QQ_V^{\frac{1}{2}}\int_{u_0}^{u}du'\left(\int_{C(u';[\ub_0,\ub])}\ub'^{6}|u|^{5+\ga}|{^{(T)}\!\pi}|^2|\nabb\a(\tilde{R})|^2\right)^{\frac{1}{2}}+\c\c\c\nn\\
&&\leq c\QQ_V^{\frac{1}{2}}\int_{u_0}^{u}du'\left(\int_{C(u';[\ub_0,\ub])}\ub'^{4}|u|^{5+\ga}|{^{(T)}\!\pi}|^2|\a(\lie_O\tilde{R})|^2\right)^{\frac{1}{2}}\nn\\
&&\leq c\QQ_V\int_{u_0}^{u}du'\sup_V{^{(T)}\!\pi}|\leq c\QQ_V\int_{u_0}^{u}du'\sup_V\frac{1}{r^2}|r^2{^{(T)}\!\pi}|\nn\\
&&\leq c\QQ_V\left(\int_{u_0}^{u}du'\frac{1}{|u'|^{3}}\right)\sup_V|r^2|u|{^{(T)}\!\pi}|\leq c\QQ_V\frac{\sup_V|r^2|u|{^{(T)}\!\pi}|}{R^2_0}\nn
\eea
Recalling that ${^{(T)}\!\pi}$ is not zero in Kerr spacetime, but, see Lemma \ref{pi(T)exp}, satisfies the inequality 
\[\sup_V|r^3{^{(T)}\!\pi}|\leq O(M^2)\ ,\]
we have the following ``dimensional estimate" in $V_*$
\[\sup_V|r^2|u|{^{(T)}\!\pi}|\leq O(M^2)\ ,\]
which substituted in the previous expression gives
\bea
|\tilde{\cal E}|\leq c\frac{M^2}{R^2_0}\QQ_V
\eea
which again will require, to prove Theorem \ref{TH2.2}, that
\[c\frac{M^2}{R^2_0}<1\ .\]
\smallskip

\NI Let us look now to some other terms of the error involving the estimate of $\ro({\tilde R})$ which have to be treated differently. We look again at the error term
\[\int_{V_{(u,\ub)}}\tau_{+}^6|u|^{5+}\a(\lie_{T}\tilde{R})\cdot\Theta(T,\tilde{R})\]
and from $\Theta(T,\tilde{R})$ we pick the term $\tr\chi\ \!{^{(T)}\!j}\ro({\tilde R})$, therefore the term we are going to estimate is
\bea
&&\int_{V_{(u,\ub)}}\tau_{+}^6|u|^{5+\ga}\a(\lie_{T}\tilde{R})\tr\chi{^{(T)}\!j}\ro({\tilde R})\leq 
c{\tilde{\cal Q}}_V^{\frac{1}{2}}\int_{u_0}^udu'\left(\int_{C}\tau_{+}^6|u|^{5+\ga}\tr\chi^2|{^{(T)}\!i}|^2\ro({\tilde R})^2\right)^{\frac{1}{2}}\nn\\
&&\leq c{\tilde{\cal Q}}^{\frac{1}{2}}\int_{u_0}^udu'\left(\int_{R_0}^{\infty}dr r^8|u|^{5+\ga}\frac{1}{r^2}\left(\frac{M^4}{r^4|u|^2}\right)\!
\left(\frac{M}{R_0}\right)^{\!2}\frac{\bigg(\varepsilon+\left(\frac{M}{R_0}\right)^{\!\!\frac{1}{2}}\!\!\ep_0\!\bigg)^2}{r^6|u'|^{6}}\right)^{\frac{1}{2}}\nn\\
&&\leq c{\tilde{\cal Q}}_V^{\frac{1}{2}}\frac{M^3}{R_0}\bigg(\varepsilon+\left(\frac{M}{R_0}\right)^{\!\!\frac{1}{2}}\!\!\ep_0\!\bigg)
\int_{u_0}^udu'\left(\int_{R_0}^{\infty}dr\frac{r^8|u|^{5+\ga}}{r^{12}|u'|^{8}}\right)^{\frac{1}{2}}\nn\\
&&\leq c{\tilde{\cal Q}}_V^{\frac{1}{2}}\frac{M^3}{R_0}\bigg(\varepsilon+\left(\frac{M}{R_0}\right)^{\!\!\frac{1}{2}}\!\!\ep_0\!\bigg)\int_{u_0}^udu'\frac{1}{|u'|^{\frac{3}{2}}}\left(\int_{R_0}^{\infty}dr\frac{1}{r^{4}}\right)^{\frac{1}{2}}\leq c{\tilde{\cal Q}}_V^{\frac{1}{2}}\frac{M^3}{R_0^{3}}\bigg(\varepsilon+\left(\frac{M}{R_0}\right)^{\!\!\frac{1}{2}}\!\!\ep_0\!\bigg)\nn\\
&&\leq c\frac{M^6}{R_0^{6}}{\tilde{\cal Q}}_V+c\varepsilon^2
\eea
in agreement with the result we want to obtain if
\bea
c\frac{M^6}{R_0^{6}}<1\ .
\eea
\NI{\bf Remark:} {\em An important remark is needed here: the estimates discussed here for some of the error terms we have to control to prove the boundedness of the $\tilde{\cal Q}$ norms, are far from giving a complete proof of the result; in fact the error term is made by a very large number of integrals, of the order of one hundred, and their complete estimates would take a large number of pages. On the other side these estimates have been done in a very complete way for the $\cal Q$ norms in \cite{Kl-Ni:book}, Chapter 6, and what we want to show here is that, apart some differences explicitely examinated, the way of controlling the present error terms follows exactly the same pattern and therefore its proof is just a consequence of the proof given there.  }

\subsubsection{III step: the control of the $\de \cal R$ norms}
In the previous step we have proved that in $V_*$ the following estimates hold
\bea
\ML\sup_{V_*}|r^{3}|u|^{3+\frac{\ep}{2}}\overline{\ro({\tilde R})}|\leq c_1^{\frac{1}{2}}\bigg(\varepsilon+\left(\frac{M}{R_0}\right)^{\!\!\frac{1}{2}}\!\!\ep_0\!\bigg)
\ \ ;\ \ \tilde{\cal Q}\leq c_1\!\!\left(\varepsilon^2+\frac{M}{R_0}\ep_0^2\right)\ .
\eea
\medskip

\NI  In this third step using the estimates proved in Theorem \ref{TH2.2} we  prove estimates \ref{bass3}
 for all the null components of the correction to the  Riemann tensor $\de R=R-R^{(Kerr)}$. 
\begin{thm}\label{PeelingTheorem}
Assume that in $V_*$ the estimates \ref{Oboosestwithdec} and \ref{bass3} hold, then, in $V_*$, we have:
\bea
\de{\cal R}\leq \frac{\ep_0}{2}\ .\eql{Mit-Leff result}
\eea
\end{thm}
\NI{\bf Proof:} The proof goes  basically as in \cite{Ni;Peel}. We repeat here the main steps to show in detail how that the various norms are bounded. Let us present the various lemmas which prove the theorem.

\NI{\bf Remark:} {\em In the following lemmas \ref{Lemma1},...,\ref{Lemma5} we prove the estimates for the $\sup$ norms of the null components of $\de R$. To complete the bootstrap we need analogous estimates for their derivatives up to $l=q-1$. Again, apart from notational complications, the more delicate part is the control of the non derived components. The remaining estimates are just a repetition and follow the pattern explicitely written in \cite{Kl-Ni:book} .}
\begin{Le}\label{Lemma1}
Assuming the estimates \ref{Oboosestwithdec} 
and the condition
\bea
\frac{M}{R_0}<<1\eql{extregcond3}
\eea
 it follows that the various null components of $\tilde R=\lie_{T_0}R$ satisfy the following inequalities with an ``independent" constant $c_2>c_1$
and  $\ep=\ga$,
\begin{eqnarray}
&&\sup_{\mathcal{K}}r^{\frac{7}{2}}|u|^{\frac{5}{2}+\frac{\e}{2}}|\a(\lie_{T_0}R)| \leq c_2\big(\varepsilon+\frac{\ep_0}{N_0}\big)\nn\\
&&\sup_{\mathcal{K}}r^{\frac{7}{2}}|u|^{\frac{5}{2}+\frac{\e}{2}} |\b(\lie_{T_0}R)|\leq c_2\big(\varepsilon+\frac{\ep_0}{N_0}\big)\nn\\
&&\sup_{\mathcal{K}}r^3|u|^{3+\frac{\e}{2}}|\ro(\lie_{T_0}R)-\overline{\ro(\lie_{T_0}R)}| \leq c_2\big(\varepsilon+\frac{\ep_0}{N_0}\big)\eql{LTRdec1}\\
&&\sup_{\mathcal{K}} r^3|u|^{3+\frac{\e}{2}}|\si(\lie_{T_0}R)-\overline{\si(\lie_{T_0}R)}|\leq c_2\big(\varepsilon+\frac{\ep_0}{N_0}\big)\nn\\
&&\sup_{\mathcal{K}}r^2|u|^{4+\frac{\e}{2}}|\bb(\lie_{T_0}R)|\leq c_2\big(\varepsilon+\frac{\ep_0}{N_0}\big)\nn\\
&&\sup_{\mathcal{K}}r|u|^{5+\frac{\e}{2}}|\aa(\lie_{T_0}R)|\leq c_2\big(\varepsilon+\frac{\ep_0}{N_0}\big)\ .\nn
\end{eqnarray}
\end{Le}

\NI{\bf Proof:} Under the estimates \ref{Oboosestwithdec} and the condition
\beaa
\frac{M}{R_0}<\frac{1}{N_0^2}<<1\eql{extregcond3}
\eeaa
with $N_0$ integer, we proved in Theorem \ref{TH2.2} the inequalities
\bea
&&\ML\tilde{\cal Q}\leq c_1\!\!\left(\varepsilon^2+\frac{M}{R_0}\ep_0^2\right)\leq c_1\!\!\left(\varepsilon^2+\frac{\ep_0^2}{N_0^2}\right) 
\leq c_1\!\!\left(\varepsilon+\frac{\ep_0}{N_0}\right)^{\!\!2}\nn\\
&&\ML\sup_{V_*}|r^{3}|u|^{3+\frac{\ep}{2}}\overline{\ro({\tilde R})}|\leq c_1^{\frac{1}{2}}\bigg(\varepsilon+\left(\frac{M}{R_0}\right)^{\!\!\frac{1}{2}}\!\!\ep_0\!\bigg)\leq c_1^{\frac{1}{2}}\!\!\left(\varepsilon+\frac{\ep_0}{N_0}\right)
\eea
\NI From it proceeding as in Chapter 5 of \cite{Kl-Ni:book} the thesis follows.

\NI{\bf Remark:} {\em Observe that to get from the $\tilde{\cal Q}$ norms the estimates for the null components of the $\lie_{T_0}R$ tensor we have to use the Bianchi equations, this requires the control of the connection coefficients op to fourth order derivatives as requires in the bootstrap assumptions.}
\smallskip

\NI Next lemma shows that integrating along the incoming cones in $V_*$ we can transform the decay in $|u|$ proved in the previous lemma in a better decay in $r$.

\begin{Le}\label{Lemma2}
From the results of Lemma \ref{Lemma1}, using the assumptions  \ref{Oboosestwithdec} and also the condition
\bea
\kappa\frac{M}{R_0}\leq 1\ ,\eql{extregcond3}
\eea
the following estimates hold:
\begin{eqnarray}
&&\ML\ML\sup_{V_*}r^{5}|u|^{1+\frac{\e}{2}}|\a(\lie_{T_0}R)| \leq \tilde{c}_4\!\left(\varepsilon+\frac{\ep_0}{N_0}\right)\nn\\
&&\ML\ML\sup_{V_*}r^{4}|u|^{2+\frac{\e}{2}} |\b(\lie_{T_0}R)|\leq \tilde{c}_3\!\left(\varepsilon+\frac{\ep_0}{N_0}\right)\!\nn\\
&&\ML\ML\sup_{V_*}r^3|u|^{3+\frac{\e}{2}}|\ro(\lie_{T_0}R)-\overline{\ro(\lie_{T_0}R)}| \leq c_3\!\left(\varepsilon+\frac{\ep_0}{N_0}\right)\nn\\
&&\ML\ML\sup_{V_*} r^3|u|^{3+\frac{\e}{2}}|\si(\lie_{T_0}R)-\overline{\si(\lie_{T_0}R)}|\leq c_3\!\left(\varepsilon+\frac{\ep_0}{N_0}\right)\!\nn\\
&&\ML\ML\sup_{V_*}r^2|u|^{4+\frac{\e}{2}}\bb(\lie_{T_0}R)\leq c_3\!\left(\varepsilon+\frac{\ep_0}{N_0}\right)\nn\\
&&\ML\ML\sup_{V_*}r|u|^{5+\frac{\e}{2}}|\aa(\lie_{T_0}R)|\leq c_3\!\left(\varepsilon+\frac{\ep_0}{N_0}\right)\!\ \ .\ \eql{LTWdec11a}
\end{eqnarray} 
with an ``independent"  constant $c_3>c_2>c_1>c_0$, and $\tilde{c}_3, \tilde{c}_4$ satisfying
\bea
\tilde{c}_3\geq c(1+c''+c'''+c'''')
\eea
where, as usual with $c$ we indicate a generic constant $>1$, independent from the remaining parameters which can be different in different inequalities, $c'',c''',c''''$ satisfy
\bea
c'''\geq c\kappa \frac{M^2}{R_0^2}\left(c_2+c_0\frac{M}{R_0}\right)\ \ ,\ \ c''''\geq cc_2\kappa\frac{M^2}{R_0^{2}}
\eea
\bea
{\tilde c}_4\geq c(1+c_4)
\eea
and
\bea
{c}_4\geq {\tilde c}_3\left(1+\kappa\frac{M^2}{R_0^2}\right)+\kappa c_1^{\frac{1}{2}}\frac{M^2}{R_0^2}\ .
\eea
\end{Le}

\NI{\bf Proof:} This Lemma is basically Theorem 2.3 of \cite{Ni;Peel}. As it is one of the central step of the whole result we repeat its proof here.
\smallskip

\NI{\bf Proof of the second line of \ref{LTWdec11a}:}

\NI From the Bianchi equations, see (3.2.8) of \cite{Kl-Ni:book}, it follows that
$\b=\b(\lie_{T_0}R)$ satisfies, along the incoming null hypersurface $\Cb(\nu)$, the evolution equation
\bea
\dddd_3\b+\tr\chib\b=\nabb\ro+\left[2\omb\b+\dual\nabb\si+2\hat{\chi}\c\bb+
3(\eta\ro+\dual\eta\si)\right]\eql{1.20a}
\eea
which can be rewritten as \footnote{Using a Fermi transported frame, see later for its definition and a discussion on it. Observe that $\b_a=\b(\lie_{T_0})(e_a)$
and the same for the remaining Riemann components.}
\bea
\frac{\partial\b_a}{\partial\la}+\oom\tr\chib\b_a=2\oom\omb\b_a+\oom\left[\nabb_a\ro+\dual\nabb_a\si+2(\hat{\chi}\c\bb)_a+
3(\eta\ro+\dual\eta\si)_a\right]\ \ \ \eql{1.8}
\eea
where\! \footnote{All the notations used in
this paper without an explicit definition are those already introduced in \cite{Kl-Ni:book}. The moving frame compatible with equation \ref{1.8} is the Fermi transported one, see the detailed discussion in \cite{Kl-Ni:book}, Chapter 3.} all the null Riemann components refer to ${\tilde R}=\lie_{T_0}R$. From this equation, see Chapter 4 of \cite{Kl-Ni:book}, we obtain the following inequality, 
\bea
\frac{d}{d\la}|r^{(2-\frac{2}{p})}\b|_{p,S}
&\leq&||2\oom\omb-{(1-1/p)}(\oom\tr\chib-\overline{\oom\tr\chib})||_{\infty}|r^{(2-\frac{2}{p})}\b|_{p,S}\eql{4.7}\\
&+&\|\oom\|_{\infty}\!\left(|r^{(2-\frac{2}{p})}\nabb\ro|_{p,S}
+3|r^{(2-\frac{2}{p})}\eta\ro|_{p,S}+|r^{(2-\frac{2}{p})}\tilde F|_{p,S}\right)\ ,\nn
\eea
where ${\tilde F}(\c)=2\hat{\chi}\c\bb+(\dual\nabb\si+3\dual\eta\si)$.\footnote{The term $\dual\nabb\si+3\dual\eta\si$ behaves as
the term $\nabb_a\ro+3\eta_a\ro$ and, therefore, we will not consider it explicitely.} Integrating along $\Cb(\nu)$, with $\la_1\!=\!u|_{\Cb(\nu)\cap\Si_0}$, we obtain
\bea
&&|r^{(2-\frac{2}{p})}\b|_{p,S}(\la,\nu)\leq|r^{(2-\frac{2}{p})}\b|_{p,S}(\la_1)\nn\\
&&\ \  +\int_{\la_1}^{\la}||2\oom\omb-{(1-1/p)}(\oom\tr\chib-\overline{\oom\tr\chib})||_{\infty}
|r^{(2-\frac{2}{p})}\b|_{p,S}(\la',\nu)\nn\\
&&\ \ +\  \|\oom\|_{\infty}\!\left(\int_{\la_1}^{\la}|r^{(2-\frac{2}{p})}\nabb\ro|_{p,S}\!
+\!3\!\int_{\la_1}^{\la}|r^{(2-\frac{2}{p})}\eta\ro|_{p,S}
\!+\!\frac{1}{2}\!\int_{\la_1}^{\la}|r^{(2-\frac{2}{p})}\tilde F|_{p,S}\right)\  .\nn
\eea 
In $V_*$ the previous assumptions imply the following estimates:
\[\|r|\la|\oom\omb\|_{\infty}\leq \kappa M\ \  \mbox{and}\ \ 
\|r|\la|\oom(\tr\chib-\overline{\oom\tr\chib})\|_{\infty}\leq \kappa M\ .\]
Therefore we can apply the Gronwall's Lemma obtaining:
\bea
|r^{2-\frac{2}{p}}\b|_{p,S}(\la,\nu)&\leq&
c\left[|r^{2-\frac{2}{p}}\b|_{p,S}(\la_1)+\|\oom\|_{\infty}\!\left(\!\int_{\la_1}^{\la}|r^{2-\frac{2}{p}}\nabb\ro|_{p,S}
\right.\right.\nn\\
&+&\left.\left.3\!\int_{\la_1}^{\la}|r^{2-\frac{2}{p}}\eta\ro|_{p,S}
+\frac{1}{2}\!\int_{\la_1}^{\la}|r^{2-\frac{2}{p}}\tilde{F}|_{p,S}\right)\right]
\eea
The constant $c$ can be chosen as an ``independent" constant for the following reason: in this application of the Gronwall Lemma the constant $c$ has to bound the following exponent\footnote{This estimate could be significantly improved.}
\[exp\left\{\int_{\la_1}^{\infty}\frac{\kappa M}{\la^2}\right\}\leq \exp{\frac{kM}{R_0}}\]
therefore under the assumption of the lemma we can choose $c\geq e$.

\NI Recalling that from inequality \ref{O0est} and the explicit expression of $\oom^{(Kerr)}$, $\|\oom\|_{\infty}\leq c$.  Multiplying both sides by
$r^2|\la|^{2+\frac{\e''}{2}}$, with $\ep>\ep''>0$, remembering that $r(\la,\nu)<r(\la_1,\nu)$ and $|\la|<|\la_1|$, we obtain
\bea
&&\ML|r^{4-\frac{2}{p}}|\la|^{2+\frac{\e''}{2}}\b|_{p,S}(\la,\nu)
\leq c\left(|r^{4-\frac{2}{p}}|\la|^{2+\frac{\e''}{2}}\b|_{p,S}(\la_1)\right.\eql{6.17a}\\
&&\ML+\!\int_{\la_1}^{\la}\!|r^{4-\frac{2}{p}}|\la'|^{2+\frac{\e''}{2}}\nabb\ro|_{p,S}
\!+\!\left.3\!\int_{\la_1}^{\la}|r^{4-\frac{2}{p}}|\la'|^{2+\frac{\e''}{2}}\eta\ro|_{p,S}
\!+\!\frac{1}{2}\!\int_{\la_1}^{\la}|r^{4-\frac{2}{p}}|\la'|^{2+\frac{\e''}{2}}\tilde{F}|_{p,S}\!\right)\ .\nn
\eea
We examine the integrals in \ref{6.17a} .
\medskip

\NI a) $\int_{\la_1}^{\la}|r^{4-\frac{2}{p}}|\la'|^{2+\frac{\e''}{2}}\nabb\ro({\tilde R})|_{p,S}$:
\medskip

\NI This first integral has the following estimate we prove in the appendix:
\bea
\sup_{{\cal K}}\big|r^{4-\frac{2}{p}}|\la|^{3+\frac{\ep}{2}}\nabb\ro({\tilde R})\big|_{p,S}\leq c'\!\left(\varepsilon+\frac{\ep_0}{N_0}\right)\ .\eql{estq}
\eea
Therefore
\bea
\int_{\la_1}^{\la}|r^{4-\frac{2}{p}}|\la'|^{2+\frac{\e''}{2}}\nabb\ro({\tilde R})|_{p,S}
\!&\leq&\! c'\!\left(\varepsilon+\frac{\ep_0}{N_0}\right)\int_{\la_1}^{\la}\frac{1}{|\la'|^{1+\frac{\ep-\ep''}{2}}}\nn\\
\!&\leq&\!c''\!\left(\varepsilon+\frac{\ep_0}{N_0}\right)\!\frac{1}{|\la|^{\frac{\ep-\ep''}{2}}}\ .\eql{}
\eea
\medskip

\NI b) $\int_{\la_1}^{\la}|r^{4-\frac{2}{p}}|\la'|^{2+\frac{\e''}{2}}\eta\ro|_{p,S}$\ :
Recalling assumption  \ref{Oboosestwithdec},  $\eta$ satisfies 
\bea
|r^{2-2/p}|\la|\eta|_{p,S}(\la,\nu)\leq \kappa{M^2}\ \ ,\ p\in [2,\infty]\ .\eql{6.19q}
\eea
Using the previous estimate for $\overline{\ro(\tilde R)}$ in $V_*$ and the results of the previous lemma
we can conclude that
\beaa
\sup_{V_*}\left|r^3|\la|^{3+\frac{\ep}{2}}\ro({\tilde R})\right| \leq \left(c_2+c_0\frac{M}{R_0}\right)\!\!\left(\varepsilon+\frac{\ep_0}{N_0}\right)\!\ ;
\eeaa
it follows immediately
\bea
&&\ML\int_{\la_1}^{\la}|r^{4-\frac{2}{p}}|\la'|^{2+\frac{\e''}{2}}\eta\ro|_{p,S}
\leq \kappa M^2\left(c_2+c_0\frac{M}{R_0}\right)\!\!\left(\varepsilon+\frac{\ep_0}{N_0}\right)\!\c
\int_{\la_1}^{\la}\frac{1}{r|\la'|^{2+\frac{\ep-\ep''}{2}}}\nn\\
&&\ML\leq c\kappa \frac{M^2}{R_0^2|\la|^{\frac{\ep-\ep''}{2}}}\left(c_2+c_0\frac{M}{R_0}\right)\!\!\left(\varepsilon+\frac{\ep_0}{N_0}\right)
\leq c\kappa \frac{M^2}{R_0^{2}}\left(c_2+c_0\frac{M}{R_0}\right)\!\!\left(\varepsilon+\frac{\ep_0}{N_0}\right)\!\frac{1}{|\la|^{\frac{\ep-\ep''}{2}}}\nn\\
&&\ML\leq c'''\!\!\left(\varepsilon+\frac{\ep_0}{N_0}\right)\!\frac{1}{|\la|^{\frac{\ep-\ep''}{2}}}\nn
\eea
with
\bea
c'''\geq c\kappa \frac{M^2}{R_0^2}\left(c_2+c_0\frac{M}{R_0}\right)\ .\eql{extregcond4}
\eea
\smallskip

\NI c) $\int_{\la_1}^{\la}|r^{4-\frac{2}{p}}|\la'|^{2+\frac{\e''}{2}}\tilde{F}|_{S,p}$\ :
\medskip

\NI From the expression ${\tilde F}(\c)=\dual\nabb\si+3\dual\eta\si+2\hat{\chi}\c\bb$ and the previous remark concerning
$\dual\nabb\si+3\dual\eta\si$, we are left to prove that{Oboosestwithdec}
\bea
\int_{\la_1}^{\la}|r^{4-\frac{2}{p}}|\la'|^{2+\frac{\e''}{2}}\chih\bb|_{p,S}\leq c
\eea
This is easy, as, from the estimates \ref{LTRdec1} and \ref{Oboosestwithdec}, we have
\bea
\sup_{V_*}|r^{2}|\la|^{4+\frac{\ep}{2}}\bb|\leq c_2\!\left(\varepsilon+\frac{\ep_0}{N_0}\right)\!\ \ \ \ ,
\ \  \sup_{V_*}||\la|r^{2-2/p}\chih|_{p,S}\leq \kappa{M^2}\ \ \ \ \ p\in [2,\infty]\ .\ \ \ \ \ 
\eea
Therefore
\bea
&&\ML\int_{\la_1}^{\la}|r^{4-\frac{2}{p}}|\la'|^{2+\frac{\e''}{2}}\tilde{F}|_{p,S}\leq c_2\!\left(\varepsilon+\frac{\ep_0}{N_0}\right)\!\kappa{M^2}\!\!
\int_{\la_1}^{\la}\frac{1}{|\la'|^{3+\frac{\ep-\ep''}{2}}}\nn\\
&&\ML\leq cc_2\!\left(\varepsilon+\frac{\ep_0}{N_0}\right)\!\kappa\frac{M^2}{R_0^2}\frac{1}{|\la|^{\frac{\ep-\ep''}{2}}}
\leq c''''\!\left(\varepsilon+\frac{\ep_0}{N_0}\right)\!\frac{1}{|\la|^{\frac{\ep-\ep''}{2}}}\ \ \ \ \ \ 
\eea
with
\bea
c''''\geq cc_2\kappa\frac{M^2}{R_0^{2}}\geq cc_2\kappa\frac{M^2}{R_0^2}\ .\eql{extregcond5}
\eea
Collecting all these estimates for the integrals in \ref{6.17a} we infer that
\bea
&&|r^{4-\frac{2}{p}}|\la|^{2+\frac{\e''}{2}}\b|_{p,S}(\la,\nu)\\
&&\leq c\left(|r^{4-\frac{2}{p}}|\la|^{2+\frac{\e''}{2}}\b|_{p,S}(\la_1)
+(c''+c'''+c'''')\!\left(\varepsilon+\frac{\ep_0}{N_0}\right)\!\frac{1}{|\la|^{\frac{\ep-\ep''}{2}}}\right)\nn\\
&&\leq c(1+c''+c'''+c'''')\!\left(\varepsilon+\frac{\ep_0}{N_0}\right)\!\frac{1}{|\la|^{\frac{\ep-\ep''}{2}}}
\leq {\tilde c}_3\!\left(\varepsilon+\frac{\ep_0}{N_0}\right)\!\frac{1}{|\la|^{\frac{\ep-\ep''}{2}}}\nn
\eea
and finally
\bea
|r^{4-\frac{2}{p}}|\la|^{2+\frac{\e}{2}}\b|_{p,S}(\la,\nu)\leq {\tilde c}_3\!\left(\varepsilon+\frac{\ep_0}{N_0}\right)\!\eql{finestbetarightep}
\eea
where we used the initial data assumptions ${\cal Q}_{\Si_0}\leq \varepsilon$ which implies, assuming $\ep\leq\ga$, 
\beaa
|r^{4-\frac{2}{p}}|\la|^{2+\frac{\e}{2}}\b|_{p,S}(\la_1,\ub=|\la_1|)\leq c\varepsilon\ .
\eeaa
To prove the $\sup$ estimate in \ref{LTWdec11a} we have to repeat for $\nabb\b$ the previous estimate for $\b$. This requires the transport equation for $\nabb\b$ along the $\Cb$ ``cones" which at its turn requires the control of an extra derivative for $\ro$ and $\si$. This is the reason why we need a greater regularity in the initial data which translates in the introduction of $\cal Q$ norms with more Lie derivatives than in \cite{Kl-Ni:book}. We do not write the proof here as it goes, with the obvious changes, exactly as for the $\b$ estimate. Therefore we have proved the following inequality,
\bea
\sup_{\mathcal{K}}r^{4}|u|^{2+\frac{\e}{2}} |\b(\lie_TR)|\leq {\tilde c}_3 \!\left(\varepsilon+\frac{\ep_0}{N_0}\right) \eql{bLTest}
\eea
with 
\bea
{\tilde c}_3\geq c(1+c''+c'''+c'''')
\eea
where, as usual with $c$ we indicate a generic constant independent from the remaining parameters which can be different in different inequalities, $c''',c''''$ satisfy
\bea
c'''\geq c\kappa \frac{M^2}{R_0^2}\left(c_2+c_0\frac{M}{R_0}\right)\ \ ,\ \ c''''\geq cc_2\kappa\frac{M^2}{R_0^{2}}\ .
\eea

\NI{\bf Proof of the $\a$ estimate in \ref{LTWdec11a}:}
We look at the transport equation for $|r^{(1-\frac{2}{p})}\a(\lie_{T_0}R)|_{p,S}$\ .
From the evolution equation satified by $\a$, see \cite{Kl-Ni:book}, Chapter 3, equation (3.2.8),
\beaa
\frac{\partial\a}{\partial\la}+\frac{1}{2}\oom\tr\chib\a=4\oom\omb\a+\oom\left[\nabb\hot\b+\left(-3(\hat{\chi}\ro+
\dual\hat{\chi}\si)+(\zeta+4\eta)\hot\b\right)\right]\ ,\ \ \ \ \ 
\eeaa
it follows that we obtain the following inequality, see Chapter 4 of \cite{Kl-Ni:book},
\bea
\ML\ML\frac{d}{d\la}|r^{(1-\frac{2}{p})}\a|_{p,S}
\!&\leq&\!\|\oom\|_{\infty}\left(4p|\omb|_{\infty}+\big(\frac{1}{2}-\frac{1}{p}\big)\!|\tr\chib-\overline{\tr\chib}|_{\infty}\right)|r^{(1-\frac{2}{p})}\a|_{p,S}
\eql{atrasa}\nn\\
&+&\|\oom\|_{\infty}\frac{1}{r^{4}|\la|^{2+\frac{\ep}{2}}}\|r^{5}|\la|^{2+\frac{\ep}{2}}\nabb\b\|_{\infty}+p\|rF\|_{\infty}\ .
\eea
where,  recalling assumptions \ref{Oboosestwithdec},  $rF=\oom r\!\left(-3(\hat{\chi}\ro+\dual\hat{\chi}\si)+(\zeta+4\eta)\hot\b\right)$
satisfies the following bound,
\bea
&&\|rF\|_{\infty}\leq cr\|\oom|_{\infty}\|\hat{\chi}\|_{\infty}\left(\|\ro\|_{\infty}+\|\si\|_{\infty}\right)+cr(\|\zeta\|_{\infty}+\|\eta\|_{\infty})\|\b\|_{\infty}\nn\\
&&\leq cr\!\left[\frac{1}{r^5|\la|^{4+\frac{\ep}{2}}}\|r^2|\la|\hat{\chi}\|_{\infty}\|r^3|\la|^{3+\frac{\ep}{2}}(\ro,\si)\|_{\infty}\right.\nn\\
&&\left.\ \ \ \ \ \ \ +\frac{1}{r^6|\la|^{3+\frac{\ep}{2}}}\bigg(\|r^2|\la|\zeta\|_{\infty}
+\|r^2|\la|\eta\|_{\infty}\bigg)\|r^4|\la|^{2+\frac{\ep}{2}}\!\b\|_{\infty}\right]\\
&&\leq c\!\left[\left(\kappa{M^2}c_1^{\frac{1}{2}}\!\left(\varepsilon+\frac{\ep_0}{N_0}\right)\right)\frac{1}{r^4|\la|^{4+\frac{\ep}{2}}}
+\kappa{M^2}{\tilde c}_3\!\left(\varepsilon+\frac{\ep_0}{N_0}\right)\!\frac{1}{r^5|\la|^{3+\frac{\ep}{2}}}\right)\ .\nn
\eea
Therefore 
\bea
&&\|rF\|_{\infty}\leq c\!\frac{1}{r^4|\la|^{1+\frac{\ep}{2}}}\!\left[\left(\kappa{M^2}c_1^{\frac{1}{2}}\right)\frac{1}{|\la|^{3}}
+\kappa{M^2}{\tilde c}_3\frac{1}{r|\la|^{2}}\right]\!\left(\varepsilon+\frac{\ep_0}{N_0}\right)\nn
\eea
and the previous inequality becomes
\bea
&&\ML\frac{d}{d\la}|r^{(1-\frac{2}{p})}\a|_{p,S}\leq\|\oom\|_{\infty}\left(4p|\omb|_{\infty}+\big(\frac{1}{2}-\frac{1}{p}\big)\!|\tr\chib-\overline{\tr\chib}|_{\infty}\right)
|r^{(1-\frac{2}{p})}\a|_{p,S}\nn\\
&&\ML\ \ \ \ \ \ \ \ +\|\oom\|_{\infty}\frac{1}{r^{4}|\la|^{2+\frac{\ep}{2}}}\|r^{5}|\la|^{2+\frac{\ep}{2}}\nabb\b\|_{\infty}\nn\\
&&\ML\ \ \ \ \ \ \ \ +c\!\frac{1}{r^4|\la|^{2+\frac{\ep}{2}}}\!\left[\left(\kappa{M^2}c_1^{\frac{1}{2}}\right)\frac{1}{|\la|^2}
+\kappa{M^2}{\tilde c}_3\frac{1}{r|\la|}\right]\!\!\left(\varepsilon+\frac{\ep_0}{N_0}\right)\eql{atras2}\nn\\
&&\leq\frac{cM}{r|\la|}|r^{(1-\frac{2}{p})}\a|_{p,S}\nn\\
&&+\frac{1}{r^{4}|\la|^{2+\frac{\ep}{2}}}\left\{{\tilde c}_3\!\left(\varepsilon+\frac{\ep_0}{N_0}\right)
+\!\left[\left(\kappa{M^2}c_1^{\frac{1}{2}}\right)\frac{1}{|\la|^2}
+\kappa{M^2}{\tilde c}_3\frac{1}{r|\la|}\right]\!\right\}\!\left(\varepsilon+\frac{\ep_0}{N_0}\right)\nn\\
&&\leq\frac{cM}{|\la|^2}|r^{(1-\frac{2}{p})}\a|_{p,S}+\frac{1}{r^{4}|\la|^{2+\frac{\ep}{2}}}\left\{{\tilde c}_3+\kappa c_1^{\frac{1}{2}}\frac{M^2}{R_0^2}
+{\tilde c}_3\kappa\frac{M^2}{R_0^2}\right\}\!\left(\varepsilon+\frac{\ep_0}{N_0}\right)\nn\\
&&\leq \frac{cM}{|\la|^2}|r^{(1-\frac{2}{p})}\a|_{p,S}+\frac{{c}_4}{r^{4}|\la|^{2+\frac{\ep}{2}}}\!\left(\varepsilon+\frac{\ep_0}{N_0}\right)\nn
\eea
with
\bea
{c}_4\geq {\tilde c}_3\left(1+\kappa\frac{M^2}{R_0^2}\right)+\kappa c_1^{\frac{1}{2}}\frac{M^2}{R_0^2}\ .
\eea
Integrating along $\Cb(\nu)$ we obtain
\bea
|r^{(1-\frac{2}{p})}\a|_{p,S}(\la,\nu)\leq c\left(|r^{(1-\frac{2}{p})}\a|_{p,S}(\la_1,\nu)+\frac{{c}_4}{r^{4}|\la|^{1+\frac{\ep''}{2}}}\!\left(\varepsilon+\frac{\ep_0}{N_0}\right)\right)
\eea
where $|\la_1|=|u_{\Si_0\cap\Cb(\nu)}|>R_0$\ , and multiplying the inequality by $r^4|\la|^{1+\frac{\ep''}{2}}$ with $\ep''<\ep$,
we obtain
\bea
\ML|r^{(5-\frac{2}{p})}|\la|^{1+\frac{\ep''}{2}}\a(\lie_TR)|_{p,S}(\la,\nu)\leq c\!\left(|r^{(5-\frac{2}{p})}|\la|^{1+\frac{\ep''}{2}}\a(\lie_TR)|_{p,S}(\la_1,\nu)
+c_4\!\left(\varepsilon+\frac{\ep_0}{N_0}\right)\frac{1}{|\la|^{\frac{\ep-\ep''}{2}}}\right)\ \ \ \ \ \ \ 
\eea
with $c\geq e$ and as 
\bea
|r^{(1-\frac{2}{p})}\a|_{p,S}(\la_1,\nu)\leq \frac{c\varepsilon}{r^{4}|\la|^{1+\frac{\ep}{2}}}\ \ ,\ \ 
\eea
we obtain
\bea
|r^{(5-\frac{2}{p})}|\la|^{1+\frac{\ep}{2}}\a(\lie_{T_0}R)|_{p,S}(\la,\nu)\leq c(1+c_4)\!\left(\varepsilon+\frac{\ep_0}{N_0}\right)
\leq {\tilde c}_4\!\left(\varepsilon+\frac{\ep_0}{N_0}\right)\ .
\eea
Next Lemma allows us to go from the estimates of $\a(\lie_{T_0}{\tilde R}),.....$ to the estimates for $\a(\Lie_{T_0}{\tilde R}),.....$; it is (a simplified version of  \footnote{The reason is that in $V_*$ the bootstrap assumptions are stronger than those which can be done ``ab initio" on the whole spacetime in \cite{Ni;Peel}, see the  initial discussion in the introduction there.}) Theorem 2.4 of \cite{Ni;Peel} which we repeat here to control the size of the norms involved.
To prove the next lemma we have to use some norm estimates on the various components of the ${^{(T_0)}\!\pi}$ deformation tensor based on the $\de{\cal O}$ bootstrap assumptions. Moreover in its proof we have to estimate the norms of the Riemann tensor $R=R^{(Kerr)}+\de R$. To control these norms we use the bootstrap assumptions for $\de R$ and we need also the control of the various components of the Kerr part of the Riemann tensor, which we discuss in the sequel.

\subsubsection{The ${^{(T_0)}\!\pi}$ deformation tensor estimates}
As we said we need to control the norms of the null components 
of the ${^{(T_0)}\!\pi}$ deformation tensor, components which are identically zero in Kerr spacetime.
The control of these  norms follows from  the previous bootstrap assumptions in $V_*$, \ref{Oboosestwithdec}.
Under these assumptions we prove that these norms are bounded by $c\ep_0$, with $\ep_0>\varepsilon$.
More precisely we prove the following estimates 
\bea
&&|r^2|u|^{2+\de}{^{(T_0)}\!}\pi(e_3,e_4)|\leq c\ep_0\nn\\
&&|r^2|u|^{2+\de}{^{(T_0)}\!}\pi(e_3,e_a)|\leq c\ep_0\nn\\
&&|r^2|u|^{2+\de}{^{(T_0)}\!}\pi(e_4,e_a)|\leq c\ep_0\eql{Pnormsbootass}\\
&&|r^2|u|^{2+\de}{^{(T_0)}\!}\pi(e_a,e_b)|\leq c\ep_0\nn
\eea
where the explicit expressions for the various null components are the following ones
\bea
&&{^{(T_0)}\!}\pi(e_3,e_3)={^{(T_0)}\!}\pi(e_4,e_4)=0\nn\\
&&{^{(T_0)}\!}\pi(e_3,e_4)=4(\om+\omb)-\left({\bf g}(\dd_3X,e_4)+{\bf g}(\dd_4X,e_3)\right)\nn\\
&&{^{(T_0)}\!}\pi(e_3,e_a)=2\oom\ze(e_a)-{\bf g}(\dd_3X,e_a)-{\bf g}(\dd_aX,e_3)\eql{pito}\\
&&{^{(T_0)}\!}\pi(e_4,e_a)=-2\oom\ze(e_a)-{\bf g}(\dd_4X,e_a)-{\bf g}(\dd_aX,e_4)\nn\\
&&{^{(T_0)}\!}\pi(e_a,e_b)={\oom}(\chi+\chib)(e_a,e_b)-\big({\bf g}(\dd_aX,e_b)+{\bf g}(\dd_bX,e_a)\big)\nn
\eea
Observe that in the Kerr spacetime $X^{(Kerr)}=\om_B\frac{\pr}{\pr\phi}$ and 
\bea
&&\left({\bf g}(\dd_3X,e_4)+{\bf g}(\dd_4X,e_3)\right)^{(Kerr)}=0\nn\\
&&\big({\bf g}(\dd_\la X,e_\la)+{\bf g}(\dd_\la X,e_\la)\big)^{(Kerr)}=0\nn\\
&&\big({\bf g}(\dd_\phi X,e_\phi)+{\bf g}(\dd_\phi X,e_\phi)\big)^{(Kerr)}=0\eql{i(T)J(T)}
\eea
Therefore we expect that it is possible to prove inequalities \ref{Pnormsbootass} assuming the bootstrap assumptions \ref{Oboosestwithdec}. This will be shown in detail later one. Observe, moreover, that  this implies that the following combination of the connection coefficients are identically zero in Kerr and, therefore, in the perturbed Kerr we can assume they satisfy the following inequalities:
\bea
&& |r^2|u|^{2+\de}(\om+\omb)|\leq c\ep_0\ \ ,\ \  |r^2|u|^{2+\de}(\tr\chi+\tr\chib)|\leq c\ep_0 \eql{bass2a}\\
&&|r^2|u|^{2+\de}(\chi+\chib)_{\la\la}|\leq c\ep_0\ \ , \ \ |r^2|u|^{2+\de}(\chi+\chib)_{\phi\phi}|\leq c\ep_0\  .\nn
\eea
\NI{\bf Remark:}\  \ {\em The estimates of the ${^{(T_0)}\!}\pi$ components are required to go from the estimates of the $\lie_{T_0}$ components to those of the $\Lie_{T_0}$ components and later on to the the estimates of the $\pr_{T_0}$ components and finally of the $\de R$ ones. To prove that they have the right smallness and the appropriate decay it is required that we control the corrections of the connection coefficients $\de O$. It is here that to close the bootstrap mechanism we need the trasport equations for the $\de O$ parts and this requires the cumbersome subtraction of the Kerr part from the transport equations and the Hodge elliptic systems, we discuss in subsection \ref{SS2.5}.}

\subsubsection{The Riemann null components in the Kerr spacetime.}\label{SSrieminKerr}
The ``principal null directions" frame $\{l,n,{\tilde e}_{\th},{\tilde e}_\phi\}$ defined, for instance, in Chandrasekar book, \cite{Ch}, 
is made by the following vector fields, where in the whole subsection $r$ denotes the Boyer-Lindquist radial coordinate $r_b$,
\bea
&&l=\frac{r^2+a^2}{\Delta}\frac{\partial}{\partial t}+\frac{a}{\Delta}\frac{\partial}{\partial\phi}+\frac{\partial}{\partial r}\nn\\
&&n=\frac{\De}{2\Si}\left(\frac{r^2+a^2}{\Delta}\frac{\partial}{\partial t}+\frac{a}{\Delta}\frac{\partial}{\partial\phi}-\frac{\partial}{\partial r}\right)\nn\\
&&{\tilde e}_{\theta}=\frac{1}{\sqrt{\Si}}\frac{\partial}{\partial\theta}\\
&&{\tilde e}_{\phi}=\frac{1}{\sqrt{\Si}}\left(a\sin\theta\frac{\partial}{\partial t}+\frac{1}{\sin\theta}\frac{\partial}{\partial\phi}\right)\nn
\eea
where $l,n$ are the principal null directions and
\bea
&&\De=r^2+a^2-2Mr\nn\\
&&\Si=r^2+a^2\cos^2\theta\nn\\
&&\Si R^2={(r^2+a^2)^2-\Delta a^2\sin^2\theta}\ .
\eea
As the Kerr spacetime is of type D, following the Petrov classification, in this null frame where the null vector fields are the principal null directions, the only null Riemann component are $\ro$ and $\si$ or, in the Newman-Penrose notations, $\Psi_2$ and their explicit expression is 
\bea
\Psi_2=\ro(R)+i\si(R)= \frac{1}{(r-i\cos\theta)^3}=\frac{1}{r^3}+i\frac{3a\cos\theta}{r^4}+O\left(\frac{1}{r^5}\right)
\eea
so that
\bea
&&\ro(R)=\frac{1}{r^3}+O\left(\frac{1}{r^5}\right)\nn\\
&&\si(R)=\frac{3a\cos\theta}{r^4}+O\left(\frac{1}{r^6}\right)\ .
\eea
Beside the fact that our initial data are not exactly those of the Kerr spacetime due to corrections $\de{^{(3)}\!}{\bf g}$ and $\de{\bf k}$, one has to observe that the null orthonormal frame we use is not the one associated to the principal null directions. Therefore, in the frame proposed by Israel and Pretorius, see \cite{Is-Pr},
all the Riemann components of the Kerr spacetime are different from zero. Recall that the frame adapted to the double null foliation of $V_*$  is, see \ref{2.37}, 
\bea
&&e^{(Kerr)}_4=2\oom L=\sqrt{\frac{R^2}{\De}}\left(\frac{\partial}{\partial\ub}+\om_B\frac{\partial}{\partial\phi}\right)\nn\\
&&e^{(Kerr)}_3=2\oom\Lb=\sqrt{\frac{R^2}{\De}}\left(-\frac{\partial}{\partial u}+\om_B\frac{\partial}{\partial\phi}\right)\nn\\
&&e^{(Kerr)}_{\la}=\frac{R}{{\mathcal L}}\frac{\partial}{\partial\la} \ \ ,\ \ e_{\phi}=\frac{1}{R\sin\theta}\frac{\partial}{\partial\phi}\ .\nn
\eea
Its relation with the previous one is
\bea
&&e_4
=\frac{\sqrt\De}{2\Si R}\left[\left(r^2+a^2+\frac{R\Si r}{Q}\right)l+\left(r^2+a^2-\frac{R\Si r}{Q}\right)\frac{2\Si}{\De}n-2{\sqrt\Si}a\sin\theta{\tilde e}_{\phi}\right]\nn\\
&&e_3=\frac{\sqrt\De}{2\Si R}\left[\left(r^2+a^2-\frac{R\Si r}{Q}\right)l+\left(r^2+a^2+\frac{R\Si r}{Q}\right)\frac{2\Si}{\De}n-2{\sqrt\Si}a\sin\theta{\tilde e}_{\phi}\right]\nn\\
&&e_\la=\frac{Q}{{\sqrt\Si}R}{\tilde e}_\theta-\frac{\De P}{2\Si R}\left(l-\frac{2\Si}{\De}n\right)\nn\\
&&e_{\phi}=\frac{r^2+a^2}{{\sqrt\Si}R}{\tilde e}_{\phi}-\frac{\De}{2R\Si}a\sin\theta\left(l+\frac{2\Si}{\De}n\right)\ .
\eea
where
\bea
Q^2=(r^2+a^2)^2-a^2\la\lap\ \ ,\ \ P^2=a^2(\la-\sin\th^2)\ \ ,\ \ {\cal L}=\mu PQ\ \ ,\ \ \la=\sin^2\th_*\ \ \ ,\ \ \ \ \eql{PQLdef}
\eea
$\mu$ is an integrating factor defined in \cite{Is-Pr} equation (25) and $\th_*$ is in the $M\rightarrow 0$ limit the spherical $\th$ coordinate of the Minkowski spacetime.

\NI Assuming $M/r\leq M/R_0$ small the previous relations become approximately
\bea
&&e_4=\left[l+O\left(\frac{M^2}{r^2}\right)n+O\left(\frac{M}{r}\right){\tilde e}_{\phi}\right]\nn\\
&&e_3=\left[n+O\left(\frac{M^2}{r^2}\right)l+O\left(\frac{M}{r}\right){\tilde e}_{\phi}\right]\nn\\
&&e_\la={\tilde e}_\theta-O\left(\frac{M}{r}\right)\left(l-2n\right)\eql{PIPN}\\
&&e_{\phi}={\tilde e}_{\phi}-O\left(\frac{M}{r}\right)\left(l+2n\right)\ .\nn
\eea
Let us denote with the upperscript ${^{(PN)}}$ the Riemann components in the ``Principal null directions frame", we have easily:
\bea
&&\ML\ML\a^{(Kerr)}(e_a,e_b)=R(e_a,e_4,e_b,e_4)
=O\left(\frac{M^2}{r^2}\right)\ro^{(PN)}+O\left(\frac{M^2}{r^2}\right)\ep_{ab}\si^{(PN)}\nn\\
&&\ML\ML\b^{(Kerr)}(e_a)=2^{-1}R(e_a,e_4,e_3,e_4)=O\left(\frac{M}{r}\right)\ro^{(PN)}+O\left(\frac{M}{r}\right)\ep_{ab}\si^{(PN)}\ \ \ \ \ \ \eql{Kerrparts}
\eea
which implies the following estimates  
\bea
\|r^5\a(R^{(Kerr)})\|_{\infty}\leq cM^3\ \ ,\ \ \|r^4\b(R^{(Kerr)})\|_{\infty}\leq cM^2\ .\eql{Kerrest}
\eea
These estimates plus the ``Bootstrap assumptions" for the Riemann components in $V_*$ imply \footnote{It has to be pointed out that the estimates \ref{Kerrest} 
refer to the various components of the (Kerr) Riemann tensor in the orthonormal frame associated to the Kerr spacetime, what in fact we have to consider 
here are the null Riemann components relative to the null orthonormal frame associated to the perturbed Kerr spacetime, see eqs. \ref{ortnullpertframe}. 
It is easy to see, using the bootstrap assumptions for the metric components, that estimates \ref{Kerrest} still holds possibly with a different ${\tilde c}$ constant.}
\bea
\|r^5\a(R)\|_{\infty}\leq cM^3+\tilde{c}\ep_0\ \ ,\ \ \|r^4\b(R)\|_{\infty}\leq cM^2+\tilde{c}\frac{\ep_0}{R_0}\eql{Rest0}
\eea

\begin{Le}\label{Lemma3}
Under the same assumptions as in Lemma \ref{Lemma2}, using the results proved there we have in the region $V_*$ the following inequalities:
\beaa
&&\ML\sup_{\mathcal{K}}r^{5}|u|^{1+\frac{\e}{2}}|\a(\Lie_{{T_0}}R)| \leq {\tilde c}_6\!\left(\varepsilon+\frac{\ep_0}{N_0}\right)+c\frac{M^2}{R_0^2}\ep_0\nn\\
&&\ML\sup_{\mathcal{K}}r^{4}|u|^{2+\frac{\e}{2}} |\b(\Lie_{T_0}R)|\leq {\tilde c}_5\!\left(\varepsilon+\frac{\ep_0}{N_0}\right) + c\frac{M}{R_0}\ep_0\nn\\
&&\ML\sup_{\mathcal{K}}r^3|u|^{3+\frac{\e}{2}}|\ro(\Lie_{T_0}R)-{\overline{\ro(\Lie_{T_0}R)}}| \leq c_5\!\left(\varepsilon+\frac{\ep_0}{N_0}\right)\nn\\
&&\ML\sup_{\mathcal{K}} r^3|u|^{3+\frac{\e}{2}}|\si(\Lie_{T_0}R)-{\overline{\si(\Lie_{T_0}R)}}|\leq c_5\!\left(\varepsilon+\frac{\ep_0}{N_0}\right)\ \ \ \ \eql{T51}\\
&&\ML\sup_{\mathcal{K}}r^2|u|^{4+\frac{\e}{2}}|\bb(\Lie_{T_0}R)|\leq c_5\!\left(\varepsilon+\frac{\ep_0}{N_0}\right)\nn\\
&&\ML\sup_{\mathcal{K}}r|u|^{5+\frac{\e}{2}}|\aa(\Lie_{T_0}R)|\leq c_5\!\left(\varepsilon+\frac{\ep_0}{N_0}\right)\nn\ \ .
\eeaa
where, we assumed $\de\geq \frac{\ep}{2}$,
\bea
{\tilde c}_5\geq {\tilde c}_3+\tilde{c}\ \ ;\ \ {\tilde c}_6\geq \left({\tilde c}_4+{c\tilde{c}}\right)\ .
\eea
\end{Le}
\NI{\bf Proof:}
We start recalling the following expressions:
\bea
\Lie_{T_0}R= \lie_{T_0}R+\frac 1 2 {^{(T_0)}}[R]-\frac 3 8(\tr{^{(T_0)}}\pi)R\ ,\eql{piT1}
\eea
where
\bea
{^{(T_0)}}[R]_{\a\b\gamma\delta}={^{(T_0)}}\pi_\a^\mu R_{\mu\b\gamma\delta}
+{^{(T_0)}}\pi_\b^\mu R_{\a\mu\gamma\delta}+{^{(T_0)}}\pi_\gamma^\mu R_{\a\b\mu\delta}
+{^{(T_0)}}\pi_\delta^\mu R_{\a\b\gamma\mu}\ .\ \ \ \ \ \ \ \eql{piT2a}
\eea
From these equations it follows:
\bea
&&\ML\ML\a(\Lie_{T_0}R)_{ab}=\a(\lie_{T_0}R)_{ab}+\frac 1 2 {^{(T_0)}}[R]_{a4b4}-\frac 3 8(\tr{^{(T_0)}}\pi)\a(R)_{ab}\nn\\
&&\ML\ML\b(\Lie_{T_0}R)_{a}=\b(\lie_{T_0}R)_{a}+\frac 1 2 {^{(T_0)}}[R]_{a434}-\frac 3 8(\tr{^{(T_0)}}\pi)\b(R)_a
\eea
and, observing that ${^{(T_0)}}\pi_{44}=0$ we easily obtain
\bea
&&{^{(T_0)}}[R]_{a4b4}=-\frac{1}{2}{^{(T_0)}}{\hat \pi}_{a4}R_{34b4}+\big({^{(T_0)}}{\hat \pi}_{ac}R_{c4b4}
+{^{(T_0)}}{\hat\pi}_{bc}R_{a4c4}+\frac{3}{4}({\tr{^{(T_0)}}\pi})R_{a4b4}\big)\nn\\
&&\ \ \ -\frac{1}{2}{^{(T_0)}}{\hat\pi}_{43}R_{a4b4}+{^{(T_0)}}{\hat \pi}_{4c}(R_{acb4}+ R_{a4bc})-\frac{1}{2}{^{(T_0)}}{\hat \pi}_{b4}R_{a434}\ .\nn
\eea
Therefore estimating ${^{(T_0)}}[R]_{a4b4}$ we obtain
\bea
&&\ML\|{^{(T_0)}}[R]_{a4b4}\|_{\infty}\leq c\left((\||\li{T_0}|\|_{\infty}+\|\lj{T_0}\|_{\infty})\|\a(R)\|_{\infty}
+\|\lm{T_0}\|_{\infty}\|\b(R)\|_{\infty}\right)\nn\\
&&\ML\leq c\left(\frac{1}{r^{2+5}|\la|^{2+\de}}(\|r^2|\la|^{2+\de}\li{T}\|_{\infty}
+\|r^2|\la|^{2+\de}\lj{T}\|_{\infty})\|r^{5}\a(R)\|_{\infty}\right.\nn\\
&&\left.+\frac{1}{r^{2+4}|\la|^{2+\de}}\|r^2|\la|^{2+\de}\lm{T}\|_{\infty}\|r^4\b(R)\|_{\infty}\right)\nn\\
&&\ML\leq c\ep_0\left(\frac{1}{r^{7}|\la|^{2+\de}}\|r^{5}\a(R)\|_{\infty}
+\frac{1}{r^{6}|\la|^{2+\de}}\|r^4\b(R)\|_{\infty}\right)\nn\\
&&\ML\leq c\ep_0\left(\frac{1}{r^{7}|\la|^{2+\de}}(cM^3+\tilde{c}\ep_0)
+\frac{1}{r^{6}|\la|^{2+\de}}(cM^2+\tilde{c}\frac{\ep_0}{R_0})\right)\nn\\
&&\ML\leq \frac{c\ep_0}{r^{5}|\la|^{1+\frac{\ep}{2}}}\left(\frac{(cM^3+\tilde{c}\ep_0)}{r^2|\la|^{1+\de-\frac{\ep}{2}}}
+\frac{(cM^2+\tilde{c}\frac{\ep_0}{R_0})}{r|\la|^{1+\de-\frac{\ep}{2}}}\right)
\eea
where in the last two lines we used estimates \ref{Rest0}
so that, finally,
\bea
\|r^{5}|\la|^{1+\frac{\e}{2}}{^{(T_0)}}[R]_{a4b4}\|_{\infty}\leq 
c\ep_0\!\left(\frac{(cM^3+\tilde{c}\ep_0)}{r^2|\la|^{1+\de-\frac{\ep}{2}}}
+\frac{(cM^2+\tilde{c}\frac{\ep_0}{R_0})}{r|\la|^{1+\de-\frac{\ep}{2}}}\right)\ .\eql{aest1}
\eea
Moreover
\bea
\|(\tr{^{(T_0)}}\pi)\a(R)\|_{\infty}\!&\leq&\!\frac{c}{r^{7}|\la|^{2+\de}} |r^{2}|\la|^{2+\de}\tr{^{(T_0)}\!}\pi|_{\infty}\|r^5\a(R)\|_{\infty}\nn\\
\!&\leq&\!\frac{c\ep_0}{r^{5}|\la|^{1+\frac{\ep}{2}}}\frac{(cM^3+\tilde{c}\ep_0)}{r^2|\la|^{1+\de-\frac{\ep}{2}}}
\nn
\eea
and finally
\bea
\|r^{5}|\la|^{1+\frac{\e}{2}}(\tr{^{(T_0)}}\pi)\a(R)\|_{\infty}\leq {c\ep_0}\frac{(cM^3+\tilde{c}\ep_0)}{r^2|\la|^{1+\de-\frac{\ep}{2}}}\ \eql{aest2}
\eea
using estimates \ref{aest1} and \ref{aest2} it follows immediately, using condition \ref{assneeded1}, that
\bea
&&\ML\sup_{\mathcal{K}}|r^{5}|u|^{1+\frac{\e}{2}}\a(\Lie_{T}R)|\leq\left({\tilde c}_4\!\left(\varepsilon+\frac{\ep_0}{N_0}\right)+c\ep_0\!\left(\frac{(cM^3+\tilde{c}\ep_0)}{r^2|\la|^{1+\de-\frac{\ep}{2}}}
+\frac{(cM^2+\tilde{c}\frac{\ep_0}{|\la|})}{r|\la|^{1+\de-\frac{\ep}{2}}}\right)\right)\nn\\
&&\ \ \ \ \ \ \ \ \ \ \ \ \ \ \ \ \ \ \ \ \ \ \ \ \ \leq{\tilde c}_6\!\left(\varepsilon+\frac{\ep_0}{N_0}\right)+c\frac{M^2}{R_0^{2+\de-\frac{\ep}{2}}}\ep_0
\leq {\tilde c}_6\!\left(\varepsilon+\frac{\ep_0}{N_0}\right)+c\frac{M^2}{R_0^2}\ep_0 \ .\ \ \ \ \ \ \ \ \ \ \ 
\eea

\NI Let us repeat the previous estimate for the $\b$ term.
\bea
&&{^{(T_0)}}[R]_{a434}={^{(T_0)}}\pi_a^\mu R_{\mu434}+{^{(T_0)}}\pi_4^\mu R_{a\mu 34}+{^{(T_0)}}\pi_3^\mu R_{a4\mu4}
+{^{(T_0)}}\pi_4^\mu R_{a43\mu}\nn\\
&&=-\frac{1}{2}{^{(T_0)}}\pi_{a4}R_{3434}+{^{(T_0)}}\pi_{ac}R_{c434}-\frac{1}{2}{^{(T_0)}}\pi_{43}R_{a434}+{^{(T_0)}}\pi_{4c} R_{ac34}\nn\\
&&\ \ \ -\frac{1}{2}{^{(T_0)}}\pi_{34}R_{a434}+{^{(T_0)}}\pi_{3c}R_{a4c4}
-\frac{1}{2}{^{(T_0)}}\pi_{43}R_{a434}+{^{(T_0)}}\pi_{4c}R_{a43c}\eql{piT3aa}\\
&&=-\frac{1}{2}{^{(T_0)}}{\hat \pi}_{a4}R_{3434}+\big({^{(T_0)}}{\hat \pi}_{ac}R_{c434}+\frac{1}{4}({\tr{^{(T_0)}}\pi})R_{a434}\big)
-\frac{1}{2}\big({^{(T_0)}}{\hat\pi}_{43}R_{a434}-\frac{1}{2}({\tr{^{(T_0)}}\pi})R_{a434}\big)
\nn\\
&&\ \ \ +{^{(T_0)}}{\hat \pi}_{4c} R_{ac34}-\frac{1}{2}{^{(T_0)}}{\hat \pi}_{34}R_{a434}
+\big({^{(T_0)}}{\hat\pi}_{3c}R_{a4c4}+\frac{1}{4}({\tr{^{(T_0)}}\pi})R_{a434}\big)\nn\\
&&\ \ \ -\frac{1}{2}\big({^{(T_0)}}{\hat \pi}_{43}R_{a434}-\frac{1}{2}({\tr{^{(T_0)}}\pi})R_{a434}\big)\nn\\
&&=-\frac{1}{2}{^{(T_0)}}{\hat \pi}_{a4}R_{3434}+\big({^{(T_0)}}{\hat \pi}_{ac}R_{c434}
+{^{(T_0)}}{\hat\pi}_{3c}R_{a4c4}+\frac{3}{4}({\tr{^{(T_0)}}\pi})R_{a434}\big)\nn\\
&&\ \ \ -\frac{1}{2}{^{(T_0)}}{\hat\pi}_{43}R_{a434}+{^{(T_0)}}{\hat \pi}_{4c} R_{ac34}-{^{(T_0)}}{\hat \pi}_{34}R_{a434}
+{^{(T_0)}}{\hat \pi}_{4c}R_{a43c}\nn\\
&&=\left(-\frac{1}{2}{^{(T_0)}}{\hat \pi}_{a4}R_{3434}+{^{(T_0)}}{\hat\pi}_{3c}R_{a4c4}+{^{(T_0)}}{\hat \pi}_{4c} R_{ac34}
+{^{(T_0)}}{\hat \pi}_{4c}R_{a43c}\right)\nn\\
&&+\left({^{(T_0)}}{\hat \pi}_{ac}R_{c434}+\frac{3}{4}({\tr{^{(T_0)}}\pi})R_{a434}-\frac{1}{2}{^{(T_0)}}{\hat\pi}_{43}R_{a434}
-{^{(T_0)}}{\hat \pi}_{34}R_{a434}\right)\nn
\eea
Therefore estimating ${^{(T_0)}}[R]_{a434}$ we obtain 
\bea
&&\ML\|{^{(T_0)}}[R]_{a434}\|_{\infty}\leq c\left((\|\lm{T_0}\|_{\infty}+\|\lmm{T_0}\|_{\infty})\|\ro(R)\|_{\infty}
+\big(\|\li{T_0}\|_{\infty}+\|\lj{T_0}\|_{\infty}+(\tr{^{(T_0)}\!}\pi)\big)\|\b\|_{\infty}\right)\nn\\
&&\ML\leq c\left(\frac{\ep_0}{r^{5}|\la|^{2+\de}}\left(\|r^2|\la|^{2+\de}\lm{T_0}\|_{\infty}+\|r^2|\la|^{1+\de}\lmm{T_0}\|_{\infty}\right)\|r^{3}\ro(R)\|_{\infty}\right.\nn\\
&&\left.+\frac{1}{r^{6}|\la|^{2+\de}}\big(\|r^2|\la|^{2+\de}\li{T_0}\|_{\infty}+\|r^2|\la|^{2+\de}\lj{T_0}\|_{\infty}
+|r^2|\la|^{2+\de}\tr{^{(T_0)}\!}\pi|\big)\|r^{4}\b(R)\|_{\infty}\right)\nn\\
&&\ML\leq c\!\left(\frac{\ep_0}{r^{5}|\la|^{2+\de}}\|r^{3}\ro(R)\|_{\infty}
+\frac{\ep_0}{r^{6}|\la|^{2+\de}}\|r^4\b(R)\|_{\infty}\right)\nn\\
&&\ML\leq c\!\left(\frac{\ep_0}{r^{5}|\la|^{2+\de}}\|r^{3}\ro(R)\|_{\infty}
+\frac{\ep_0}{r^{6}|\la|^{2+\de}}\|r^4\b(R)\|_{\infty}\right)
\leq \frac{cM\ep_0}{r^{5}|\la|^{2+\de}}+\frac{c\ep_0}{r^{6}|\la|^{2+\de}}(cM^2+\tilde{c}\frac{\ep_0}{R_0})\ \ \ \ \ 
\eea
where in the last two lines we used estimates \ref{Rest0}
and finally
\bea
\|{r^4|\la|^{(2+\de)}}{^{(T_0)}}[R]_{a434}\|_{\infty}\leq c\left(\frac{M\ep_0}{r}+\frac{\ep_0^2}{R_0^3}\right)\leq c\ep_0\frac{M}{R_0}+c\varepsilon\ ,
\eea
which we can rewrite
\bea
\|{r^4|\la|^{(2+\de)}}{^{(T_0)}}[R]_{a434}\|_{\infty}\leq c\varepsilon+\frac{1}{8}\ep_0\eql{aest11}
\eea
provided we require
\bea
c\frac{M}{R_0}\leq \frac{1}{8}\ .
\eea
Moreover
\bea
\|(\tr{^{(T_0)}}\pi)\b(R)\|_{\infty}\!&\leq&\!\frac{1}{r^{6}|\la|^{2+\de}} |r^{2}|\la|^{2+\de}\tr{^{(T_0)}\!}\pi|_{\infty}\|r^4\b(R)\|_{\infty}\nn\\
\!&\leq&\!\frac{\ep_0}{r^{6}|\la|^{2+\de}}\left(cM^2+\tilde{c}\frac{\ep_0}{R_0}\right)\ \ \ \ \ 
\eea
and finally
\bea
&&\ML\ML\ML\ML\|r^{4}|\la|^{2+\de}(\tr{^{(T_0)}}\pi)\b(R)\|_{\infty}\leq \frac{\ep_0}{R_0^2}\left(cM^2+\tilde{c}\frac{\ep_0}{R_0}\right)\nn\\
&&\ \ \ \ \ \ \ \ \ \ \ \ \ \ \ \ \leq c\ep_0\frac{M^2}{R_0^2}+\tilde{c}\frac{\ep^2_0}{R^3_0}\leq \frac{1}{8}\ep_0+\tilde{c}\varepsilon \ .\eql{aest22}
\eea
using estimates \ref{aest11} and \ref{aest22} we obtain choosing $\de$ such that
\bea
\de\geq \frac{\ep}{2} \ ,
\eea
\bea
\ML\sup_{V_*}|r^4|u|^{2+\frac{\e}{2}}\b(\Lie_{T}R)| \leq ({\tilde c}_3+\tilde{c})\!\left(\varepsilon+\frac{\ep_0}{N_0}\right)
+\frac{1}{4}\ep_0 \leq {\tilde c}_5\!\left(\varepsilon+\frac{\ep_0}{N_0}\right)+\frac{1}{4}\ep_0
\eea
with
\bea
{\tilde c}_5\geq {\tilde c}_3+\tilde{c}\ \  .
\eea
For the other terms there is no need to repeat this computation. In fact the $R$ null components already satisfy the peeling and going from $\lie_{T_0} $ to $\Lie_{T_0}$  the decay factor does not become worst. 
\smallskip

\NI Next step is to obtain from the estimates for $\a(\Lie_{T_0}R),\b(\Lie_{T_0}R) $ the estimates for $\pr_{T_0}\a(R), \pr_{T_0}\b(R)$.
This requires first the control of  $[{T_0},e_a]$ , $[{T_0},e_4]$ and $[{T_0},e_3]$.

\begin{Le}\label{Comm1}
The following expressions hold
\bea
[T_0,e_a]\!\!&=&\frac{\oom}{2}\de\Pi^{\si}_{\la}(\dd_{e_3}+\dd_{e_4})e^{\la}_a\frac{\pr}{\pr\om^{\si}}
+\frac{\hat{\oom}}{2}{\hat\Pi}^{\si}_{\la}(\hat{e}_3^{\mu}+\hat{e}_4^{\mu})e^{\tau}_a(\Ga^{\la}_{\mu\tau}-\hat{\Ga}^{\la}_{\mu\tau})\frac{\pr}{\pr\om^{\si}}
+\frac{\hat{\oom}}{2}\left(\widehat{\ddb_{e_3}}+\widehat{\ddb_{e_4}}\right)\de e_a\nn\\
\!&-&\!\left[\frac{\de\oom}{2}(\chi+\chib)_{ac}e_c+\frac{\hat{\oom}}{2}(\de\chi+\de\chib)_{ac}e_c
+\frac{\hat{\oom}}{2}(\widehat{(\chi)}+\widehat{(\chib)})_{ac}\de e_c\right]\nn\\
\!&+&\!\frac{\hat{\oom}}{2}\left[(\widehat{(\chi)}-\chi^{(Kerr)})_{ac}+(\widehat{(\chib)}-\chi^{(Kerr)})_{ac}\right]{\hat e}_c-[\de X,e_a]-[X,\de e_a]\eql{Teacom}
\eea
\bea
\ML[T_0,e_4]\!&=&\!\oom(\omb+\om)e_4+2\de\oom\ze(e_a)e_a+2\hat{\oom}\de[\ze(e_a)e_a]\nn\\
\ML\!&-&\!\de X^c\left(\frac{\pr_c\oom}{\oom}\right)e_4+\frac{\de X^c}{\oom}\left(\pr_cX^d\right)\frac{\pr}{\pr\om^d}
-e_4(\de X^c)\frac{\pr}{\pr\om^c}\nn\\
\ML\!&+&\!\om_B\frac{\pr}{\pr\phi}\left(\frac{\de\oom}{\oom\hat{\oom}}(\de^{\mu}_{\ub}+\de^{\mu}_cX^c)-\frac{\de X^d}{\oom^{(Kerr)}}\de^{\mu}_d\right)\frac{\pr}{\pr x^{\mu}}+\de e_4(\hat{X}^d)\frac{\pr}{\pr\om^d}\ \ \ \ \ \ \ \eql{Te4com}
\eea
\bea
\ML\ML\ML\ML[T_0,e_3]\!&=&\!\oom(\omb+\om)e_3-2\de\oom\ze(e_a)e_a-2\hat{\oom}\de[\ze(e_a)e_a]\nn\\
\ML\ML\ML\ML\!&+&\!\de X^c\left(\frac{\pr_c\oom}{\oom}\right)e_3-\frac{\de X^c}{\oom}(\pr_cX_{(Kerr)}^d)\frac{\pr}{\pr\om^d}
+e_3(\de X^c)\frac{\pr}{\pr\om^c}\nn\\
\ML\ML\ML\ML\!&+&\!\frac{\om_B\oom}{\oom^{(Kerr)}}\frac{\pr}{\pr\phi}\!\left(\frac{\de\oom}{\oom}\right)\!e_3+\de e_3(X_{(Kerr)})\frac{\pr}{\pr\phi}\ .\eql{Te3com}
\eea
\end{Le}
\NI{\bf Proof:} See the appendix.

\begin{Le}\label{Comm2}
The following expressions hold
\bea
\ML{\bf g}([{T_0},e_a],e_d)\!\!&=&\!\!\frac{\oom}{2}\de\Pi^{\si}_{\la}(\dd_{e_3}+\dd_{e_4})e^{\la}_a{\bf g}(\frac{\pr}{\pr\om^{\si}},e_d)
+\frac{\hat{\oom}}{2}{\hat\Pi}^{\si}_{\la}(\hat{e}_3^{\mu}
+\hat{e}_4^{\mu})e^{\ro}_a(\Ga^{\la}_{\mu\ro}-\hat{\Ga}^{\la}_{\mu\ro}){\bf g}(\frac{\pr}{\pr\om^{\si}},e_d)\nn\\
\!&+&\!\frac{\hat{\oom}}{2}\left({\bf g}(\widehat{\dd_{e_3}}\de e_a,e_d)+{\bf g}(\widehat{\ddb_{e_4}}\de e_a,e_d)\right)\nn\\
\!\!&-&\!\!\frac{\de\oom}{2}(\chi+\chib)_{ad}-\frac{\hat{\oom}}{2}(\de\chi+\de\chib)_{ad}
-\frac{\hat{\oom}}{2}(\chi^{(Kerr)}+\chib^{(Kerr)})_{ac}{\bf g}(\de e_c,e_d)\nn\\
\!\!&-&\!\!{\bf g}([\de X,e_a],e_d)-{\bf g}([X,\de e_a],e_d)\eql{commT0eab}\\
&&\nn\\
\ML\ML{\bf g}([{T_0},e_a],e_4)\!\!&=&\!\!{\bf g}([{T_0},e_a],e_3)=0\eql{commt0e4b}
\eea
\bea
\ML{\bf g}([T_0,e_4],e_d)\!\!&=&\!\!2\de\oom\ze(e_d)+2\hat{\oom}{\bf g}(\de[\ze(e_a)e_a],e_d)\nn\\
\!\!&+&\!\!\left[\frac{\de X^c}{\oom}\left(\pr_cX^e\right)-e_4(\de X^e)-\frac{\om_B}{\hat\oom}\pr_{\phi}{\de X^e}+\de e_4(\hat{X}^e)\right]{\bf g}(\frac{\pr}{\pr\om^e},e_d))
\ \ \ \ \ \ \ \ \ \ \ \ \ \ \ \eql{2.184}
\eea
\bea
\ML{\bf g}([T_0,e_4],e_4)\!\!&=&\!\!-{\bf g}([\de X,e_4],e_4)-{\bf g}([\hat{X},\de e_4],e_4)=0\nn\\
&&\nn\\
\ML{\bf g}([T_0,e_4],e_3)\!\!&=&\!\!-2\oom(\om+\omb)+2\de X^c\!\left(\frac{\pr_c\oom}{\oom}\right)
+2\frac{\om_B\de\oom}{{\oom^{(Kerr)}}\oom^2}\pr_{\phi}\oom-2\frac{\om_B}{{\oom^{(Kerr)}}\oom}\pr_{\phi}\de\oom\ .\ \ \ \ \ \ \ \ \ \ \ \ \ \eql{Te4com2}
\eea
\bea
\ML\ML\ML{\bf g}([T_0,e_3],e_d)\!\!&=&\!-2\de\oom\ze(e_d)-2\hat{\oom}{\bf g}(\de[\ze(e_a)e_a],e_d)
+e_3(\de X^c){\bf g}(\frac{\pr}{\pr\om^c},e_d)\nn\\
\ML\ML\ML\!&-&\!\frac{\de X^c}{\oom}(\pr_c\om_B){\bf g}(\frac{\pr}{\pr\phi},e_d)
+\de e_3(\hat{X}){\bf g}(\frac{\pr}{\pr\phi},e_d)\eql{2.184b}
\eea
\bea
\ML\ML\ML\ML{\bf g}([T_0,e_3],e_4)\!&=&\!-2\oom(\omb+\om)-2\de X^c\left(\frac{\pr_c\oom}{\oom}\right)-2\frac{\om_B\oom}{\oom^{(Kerr)}}\frac{\pr}{\pr\phi}\!\left(\frac{\de\oom}{\oom}\right)\eql{2.185}\\
&&\nn\\
\ML\ML\ML\ML{\bf g}([T_0,e_3],e_3)\!&=&\! 0\ .\eql{2.186}
\eea
\end{Le}
\NI{\bf Proof:} See the appendix.

\medskip

\begin{Le}\label{T0comest}
Under the bootstrap assumptions in the region $V_*$ the following estimates hold
\bea
&&|{\bf g}([T_0,e_a],e_d)| \leq c\!\left(1+\frac{M^2}{R_0^2}\right)\frac{\ep_0}{r^2|u|^{2+\de}}\nn\\
&&|{\bf g}([T_0,e_4],e_3)| \leq c\!\left(1+\frac{M^2}{R_0^2}\right)\frac{\ep_0}{r^2|u|^{2+\de}}\nn\\ 
&&|{\bf g}([T_0,e_4],e_d)| \leq c\!\left[1+\frac{M^2}{R_0^2}\left(1+\frac{\ep_0}{R_0^3}\right)\right]\frac{\ep_0}{r^2|u|^{2+\de}}\nn\\ 
&&|{\bf g}([T_0,e_3],e_d)| \leq c \!\left(1+\frac{M^2}{R_0^2}\right)\frac{\ep_0}{r^2|u|^{2+\de}}\nn\\ 
&&|{\bf g}([T_0,e_3],e_4)| \leq\!c\left(1+\frac{M^2}{R_0^2}\right)\frac{\ep_0}{r^2|u|^{2+\de}}\nn
\eea
\end{Le}
\NI{\bf Proof :} See the appendix.
\medskip

\NI In the next Lemma \ref{Lemma4} we obtain  the estimates we are looking for relative to $\pr_{T_0}(\a(R)(e_a,e_b))$ and $\pr_{T_0}(\b(R)(e_a))$.

\begin{Le}\label{Lemma4}
Under the same assumptions as in Lemma \ref{Lemma2}, using the results proved there and in Lemmas \ref{Lemma3}, \ref{Comm1}, \ref{Comm2}, \ref{T0comest},
we have in the region $V_*$ the following inequalities:
\bea
&&\sup_{\mathcal{K}}r^{5}|u|^{1+\frac{\ep}{2}}|\pr_{T_0}(\a(R)(e_a,e_b))| \leq
 {\tilde c}_7\!\left(\varepsilon+\frac{\ep_0}{N_0}\right)+c\frac{M^2}{R_0^2}\ep_0\ \ .\nn\\
&&\sup_{\mathcal{K}}r^{4}|u|^{2+\frac{\ep}{2}}|\pr_{T_0}(\b(R)(e_a))|\leq {\tilde c}_8\!\left(\varepsilon+\frac{\ep_0}{N_0}\right) + c\frac{M}{R_0}\ep_0\ .
\eea
with
\bea
{\tilde c}_7\geq \left({\tilde c}_6+c\right)\ \ ;\ \ {\tilde c}_8\geq \left({\tilde c}_5+c\right)
\eea
\end{Le}
\NI{\bf Proof:} From the relation
\bea
&&\ML\a(\Lie_{T_0}R)(e_a,e_b)=(\Lie_{T_0}R)(e_a,e_4,e_b,e_4)\eql{Tcomm}\\
&&\ML=\pr_{T_0}(\a(R)(e_a,e_b))+R([{T_0},e_a],e_4,e_b,e_4)+ R(e_a,[{T_0},e_4],e_b,e_4)\nn\\
&&\ML\b(\Lie_{T_0}R)(e_a)=(\Lie_{T_0}R)(e_a,e_4,e_3,e_4)\nn\\
&&\ML=\pr_{T_0}(\b(R)(e_a))+R([{T_0},e_a],e_4,e_3,e_4)+R(e_a,[{T_0},e_4],e_3,e_4)+R(e_a,e_4,[{T_0},e_3],e_4)\nn
\eea
it follows
\bea
&&\ML\a(\Lie_{T_0}R)(e_a,e_b)
=\pr_{T_0}(\a(R)(e_a,e_b))\nn\\
&&\ML+\sum_d{\bf g}([{T_0},e_a],e_d)R(e_d,e_4,e_b,e_4)-\frac{1}{2}{\bf g}([{T_0},e_a],e_4)R(e_3,e_4,e_b,e_4)\nn\\
&&\ML+\sum_d{\bf g}([{T_0},e_b],e_d)R(e_a,e_4,e_d,e_4)-\frac{1}{2}{\bf g}([{T_0},e_b],e_4)R(e_a,e_4,e_3,e_4)\nn\\
&&\ML+\sum_d{\bf g}([{T_0},e_4],e_d)R(e_a,e_d,e_b,e_4)+{\bf g}([{T_0},e_4],e_3)R(e_a,e_4,e_b,e_4)\nn\\
&&\ML+\sum_d{\bf g}([{T_0},e_4],e_d)R(e_a,e_4,e_b,e_d)+{\bf g}([{T_0},e_4],e_3)R(e_a,e_4,e_b,e_4)\nn\\
&&\nn\\
&&\ML=\pr_{T_0}(\a(R)(e_a,e_b))\nn\\
&&\ML+\sum_d{\bf g}([{T_0},e_a],e_d)\a(R)(e_d,e_b)-{\bf g}([{T_0},e_a],e_4)\b(e_b)\nn\\
&&\ML+\sum_d{\bf g}([{T_0},e_b],e_d)\a(R)(e_a,e_d)-{\bf g}([{T_0},e_b],e_4)\b(e_a)\nn\\
&&\ML-2\de_{ab}\sum_d{\bf g}([{T_0},e_4],e_d)\b(e_d)+2{\bf g}([{T_0},e_4],e_3)\a(e_a,e_b)\ .\nn
\eea
Therefore, recalling inequalities \ref{Rest0} and Lemma \ref{T0comest}
\bea
&&\ML|\pr_{T_0}(\a(R)(e_a,e_b))|\leq |\a(\Lie_{T_0}R)(e_a,e_b)|\\
&&\ML+4\left(\sup_d|{\bf g}([{T_0},e_a],e_d)|+|{\bf g}([{T_0},e_4],e_3)|\right)|\a(e_a,e_b)|\nn\\
&&\ML+\left(\sup_d|{\bf g}([{T_0},e_4],e_d)|+\sup_d|{\bf g}([{T_0},e_d],e_3)|\right)(|\b(e_a)|+|\b(e_b)|)\nn\\
&&\ML\leq |\a(\Lie_{T_0}R)(e_a,e_b)|
+c\frac{\ep_0}{r^2|u|^{2+\de}}\frac{(cM^3+\tilde{c}\ep_0)}{r^5}+c\frac{\ep_0}{r^2|u|^{2+\de}}\frac{(cM^2+\tilde{c}\frac{\ep_0}{R_0})}{r^4}\nn\\
&&\ML\leq |\a(\Lie_{T_0}R)(e_a,e_b)|
+c\frac{\ep_0}{r^5|u|^{1+\de}}\left(\frac{(cM^3+\tilde{c}\ep_0)}{r^2|u|}+\frac{(cM^2+\tilde{c}\frac{\ep_0}{R_0})}{r|u|}\right)\nn\\
&&\ML\leq |\a(\Lie_{T_0}R)(e_a,e_b)|
+c\frac{\ep_0}{r^5|u|^{1+\de}}\left(\frac{(M^3+\tilde{c}\ep_0)}{R_0^3}+\frac{(cM^2+\tilde{c}\frac{\ep_0}{R_0})}{R_0^2}\right)\nn\\
&&\ML\leq |\a(\Lie_{T_0}R)(e_a,e_b)|+c\frac{\ep_0}{r^5|u|^{1+\de}}\left(\frac{\ep_0}{R_0^3}+\frac{M^2}{R_0^2}+\frac{M^3}{R_0^3}\right)\nn\\
&&\ML\leq |\a(\Lie_{T_0}R)(e_a,e_b)|+\frac{1}{r^5|u|^{1+\de}}\left(c\varepsilon+c\ep_0\frac{M^2}{R_0^2}\right)\ .
\eea
where we used \ref{assneeded1}. Therefore using the result of Lemma \ref{Lemma3} we have, assuming again
\[\de\geq\frac{\ep}{2}\ ,\]
\bea
&&\ML\sup_{V_*}r^{5}|u|^{1+\frac{\e}{2}}|\pr_{T_0}(\a(R)(e_a,e_b))|\leq {\tilde c}_6\!\left(\varepsilon+\frac{\ep_0}{N_0}\right)+c\frac{M^2}{R_0^{2+\de-\frac{\ep}{2}}}\ep_0+
\frac{1}{R_0^{{\de-\frac{\ep}{2}}}}\left(c\varepsilon+c\ep_0\frac{M^2}{R_0^2}\right)\nn\\
&&\ML\leq \left({\tilde c}_6+\frac{c}{R_0^{{\de-\frac{\ep}{2}}}}\right)\!\left(\varepsilon+\frac{\ep_0}{N_0}\right)+c\frac{M^2}{R_0^{2+\de-\frac{\ep}{2}}}\ep_0\leq
 {\tilde c}_7\!\left(\varepsilon+\frac{\ep_0}{N_0}\right)+c\frac{M^2}{R_0^2}\ep_0\ \ .
\eea
From the relation
\bea
&&\ML\b(\Lie_{T_0}R)(e_a)=(\Lie_{T_0}R)(e_a,e_4,e_3,e_4)\nn\\
&&\ML=\pr_{T_0}(\b(R)(e_a))+R([{T_0},e_a],e_4,e_3,e_4)+R(e_a,[{T_0},e_4],e_3,e_4)+R(e_a,e_4,[{T_0},e_3],e_4)\nn
\eea
it follows
\bea
&&\ML\b(\Lie_{T_0}R)(e_a)=\pr_{T_0}(\b(R)(e_a))\nn\\
&&\ML+\sum_d{\bf g}([{T_0},e_a],e_d)R(e_d,e_4,e_3,e_4)-\frac{1}{2}{\bf g}([{T_0},e_a],e_4)R(e_3,e_4,e_3,e_4)\nn\\
&&\ML+\sum_d{\bf g}([{T_0},e_4],e_d)R(e_a,e_d,e_3,e_4)+{\bf g}([{T_0},e_4],e_3)R(e_a,e_4,e_3,e_4)\nn\\
&&\ML+\sum_d{\bf g}([{T_0},e_3],e_d)R(e_a,e_4,e_d,e_4)-\frac{1}{2}{\bf g}([{T_0},e_3],e_4)R(e_a,e_4,e_3,e_4)\nn\\
&&\ML+\sum_d{\bf g}([{T_0},e_4],e_d)R(e_a,e_4,e_3,e_d)+{\bf g}([{T_0},e_4],e_3)R(e_a,e_4,e_3,e_4)\nn\\
&&\ML=\pr_{T_0}(\b(R)(e_a,e_b))\nn\\
&&\ML+2\sum_d{\bf g}([{T_0},e_a],e_d)\b(R)(e_d)-2{\bf g}([{T_0},e_a],e_4)\ro(R)\nn\\
&&\ML+2\sum_d{\bf g}([{T_0},e_4],e_d)\ep_{ab}\si(R)+2{\bf g}([{T_0},e_4],e_3)\b(R)(e_a)\\
&&\ML+\sum_d{\bf g}([{T_0},e_3],e_d)\a(R)(e_a,e_d)-{\bf g}([{T_0},e_3],e_4)\b(R)(e_a)\nn\\
&&\ML+\sum_d{\bf g}([{T_0},e_4],e_d)(\de_{ad}\ro(R)+\ep_{ad}\si(R))+2{\bf g}([{T_0},e_4],e_3)\b(R)(e_a)\nn
\eea
Therefore, recalling inequalities \ref{Rest0} and Lemma \ref{T0comest}
\bea
&&\ML\b(\Lie_{T_0}R)(e_a)=\pr_{T_0}(\b(R)(e_a))\nn\\
&&\ML+2\sum_d{\bf g}([{T_0},e_a],e_d)\b(R)(e_d)-2{\bf g}([{T_0},e_a],e_4)\ro(R)\nn\\
&&\ML+2\sum_d{\bf g}([{T_0},e_4],e_d)\ep_{ab}\si(R)+2{\bf g}([{T_0},e_4],e_3)\b(R)(e_a)\\
&&\ML+\sum_d{\bf g}([{T_0},e_3],e_d)\a(R)(e_a,e_d)-{\bf g}([{T_0},e_3],e_4)\b(R)(e_a)\nn\\
&&\ML+\sum_d{\bf g}([{T_0},e_4],e_d)(\de_{ad}\ro(R)+\ep_{ad}\si(R))+2{\bf g}([{T_0},e_4],e_3)\b(R)(e_a)\nn
\eea
\bea
&&\ML|\pr_{T_0}(\b(R)(e_a))|\leq |\b(\Lie_{T_0}R)(e_a)|+2\sup_d|{\bf g}([{T_0},e_3],e_d)||\a(R)(e_a,e_d)|\nn\\
&&\ML+4\left(\sup_d|{\bf g}([{T_0},e_a],e_d)|+|{\bf g}([{T_0},e_4],e_3)|+|{\bf g}([{T_0},e_3],e_4)|\right)|\b(R)(e_a)|\nn\\
&&\ML+4\left(\sup_d|{\bf g}([{T_0},e_4],e_d)|+\sup_d|{\bf g}([{T_0},e_d],e_4)|\right)(|\ro(R)|+|\si(R))|)\nn\\
&&\ML\leq |\b(\Lie_{T_0}R)(e_a)|
+c\frac{\ep_0}{r^2|u|^{2+\de}}\frac{(cM^3+\tilde{c}\ep_0)}{r^5}+c\frac{\ep_0}{r^2|u|^{2+\de}}\frac{(cM^2+\tilde{c}\frac{\ep_0}{R_0})}{r^4}
+c\frac{\ep_0}{r^2|u|^{2+\de}}\frac{(cM+\tilde{c}\frac{\ep_0}{R^2_0})}{r^3}\nn\\
&&\ML\leq |\b(\Lie_{T_0}R)(e_a)|
+c\frac{\ep_0}{r^4|u|^{2+\de}}\left(\frac{(cM^3+\tilde{c}\ep_0)}{r^3}+\frac{(cM^2+\tilde{c}\frac{\ep_0}{R_0})}{r^2}+\frac{(cM+\tilde{c}\frac{\ep_0}{R^2_0})}{r}\right)\nn\\
&&\ML\leq |\b(\Lie_{T_0}R)(e_a)|
+c\frac{\ep_0}{r^4|u|^{2+\de}}\left(\frac{(M^3+\tilde{c}\ep_0)}{R_0^3}+\frac{(cM^2+\tilde{c}\frac{\ep_0}{R_0})}{R_0^2}+\frac{(cM+\tilde{c}\frac{\ep_0}{R^2_0})}{R_0}\right)\nn\\
&&\ML\leq |\b(\Lie_{T_0}R)(e_a)|+c\frac{\ep_0}{r^4|u|^{2+\de}}\left(\frac{\ep_0}{R_0^3}+\frac{M}{R_0}+\frac{M^2}{R_0^2}+\frac{M^3}{R_0^3}\right)\nn\\
&&\ML\leq |\b(\Lie_{T_0}R)(e_a)|+\frac{1}{r^4|u|^{2+\de}}\left(c\varepsilon+c\ep_0\frac{M}{R_0}\right)\ .
\eea
\subsubsection{The estimate of $\de R$} 
\NI The final step is to integrate along the integral curves of $T_0$. It is clear that the $|u|$ weight factors will allow to bound uniformily these integrals. 
We have, denoting by $\ga(s)$ the integral curve of $T_0$ starting from $\Si_0$ at a distance $r^*_0$ from the origin and $\a(t)=\a(t)(e_a,e_b)$,
\bea
\a({\overline t},{\overline r_*})=\a^{(Kerr)}(0,{\overline r_*}_0)+\de\a(0,{\overline r_*}_0)+\int_0^{\overline s}(\pr_{T_0}\a)(\ga(s))
\eea
where
\bea
&&{\overline r_*}=r_*(u,\ub)={\ub-u}=\ga^{r_*}(\overline s)\ \ ,\ \ {\overline t}=t(u,\ub)={\ub+u}=\ga^{0}(\overline s)\nn\\
&&u=u({\overline t},{\overline r}_*)=u(0,{r_*}_1)=-\frac{{r_*}_1}{2}\ \ ,\ \ \ub=\ub({\overline t},{\overline {r_*}})=\ub(0,{r_*}_2)=\frac{{r_*}_2}{2}\nn\\
&&{\overline t}=t(u,\ub)=\frac{{r_*}_2-{r_*}_1}{2}=\frac{{\overline t}+{\overline {r_*}}-{r_*}_1}{2}\ .
\eea
As
\bea
T_0=\frac{\pr}{\pr\ub}+\frac{\pr}{\pr u}
\eea
it follows that
\bea
\frac{d\ga^{\mu}}{ds}=T_0^{\mu}=\de^{\mu}_u+\de^{\mu}_{\ub}
\eea
therefore, with these definitions, in the coordinates $\{x^{\mu}\}=\{u,\ub,\om^1,\om^2\}$,
\bea
&&\frac{d}{ds}u(\ga(s))=\frac{d\ga^{\mu}}{ds}\frac{\pr u}{\pr x^{\mu}}=\frac{d\ga^{u}}{ds}=1\nn\\
&&\frac{d}{ds}r_*(\ga(s))=\frac{d\ga^{\mu}}{ds}\left(\frac{\pr u}{\pr x^{\mu}}-\frac{\pr\ub}{\pr x^{\mu}}\right)=\frac{d\ga^{u}}{ds}-\frac{d\ga^{\ub}}{ds}=0
\eea
and, as $\om^1,\om^2$ do not change along $\ga(s)$
\bea
\frac{d}{ds}r(\ga(s))=\frac{dr}{dr_*}\frac{d}{ds}r_*(\ga(s))=0
\eea
and
\bea
&&u(\ga(s;{\overline r}_*))=u(\ga(0;{\overline r}_*))+s\ \ ,\ \ \ub(\ga(s;{\overline r}_*))=\ub(\ga(0;{\overline r}_*))+s\nn\\
&& t=u+\ub=\left(u(\ga(0;{\overline r}_*))+\ub(\ga(0;{\overline r}_*))\right)+2s=2s\ .
\eea
It follows
\bea
&&\de\a(u,\ub)=\a(u,\ub)-\a^{(Kerr)}(u,\ub)=\de\a(0,{\overline r_*})+\int_0^{\overline s}(\pr_{T_0}\a)(\ga(s))\nn
\eea
and
\bea
&&\ML|\de\a(u,\ub)|\leq|\de\a(0,{\overline r_*})|+\int_0^{\overline s}|(\pr_{T_0}\a)(\ga(s))|\leq {\hat c}\frac{\varepsilon}{{r_*}^5}\\
&&\ \ \ \ \ \ \ \ \ 
+\left({\tilde c}_7\varepsilon+c\frac{M^2}{R_0^2}\ep_0\right)\!\int_0^{\overline s}\frac{1}{r(\ga(s))^5|u(\ga(s))|^{1+\frac{\ep}{2}}}ds\ .\nn
\eea
Therefore
\bea
&&|r^5\de\a(u,\ub)|\leq {\hat c}\frac{r^5(u,\ub)}{r^5_*(u,\ub)}\varepsilon+\left({\tilde c}_7\varepsilon
+c\frac{M^2}{R_0^2}\ep_0\right)\!\int_0^{\overline s}\frac{1}{|u(\ga(s))|^{1+\frac{\ep}{2}}}ds\nn\\
&&\leq{\hat c}\frac{r^5(u,\ub)}{r^5_*(u,\ub)}\varepsilon
+\left({\tilde c}_7\varepsilon+c\frac{M^2}{R_0^2}\ep_0\right)\!\int_0^{\overline s}\frac{1}{|u(\ga(0;{\overline r}_*))+s|^{1+\frac{\ep}{2}}}ds\nn\\
&&\leq  {\hat c}_1\varepsilon+c\!\left({\tilde c}_7\varepsilon+c\frac{M^2}{R_0^2}\ep_0\right)\leq {\tilde c}_9\!\left(\varepsilon+\frac{\ep_0}{N_0}\right)\ ,
\eea
where we have chosen ${\hat c}_1$ such that
\bea
 {\hat c}\frac{r^5(u,\ub)}{r^5_*(u,\ub)}\leq {\hat c}_1\ ,
\eea
which is possible as we have proved that $r_*$ and $r(u,\ub)$ stay near, and  chosen ${\tilde c}_9$ such that
\bea
{\tilde c}_9\varepsilon\geq \left({\hat c}_1+{\tilde c}_7\right)\varepsilon\ .
\eea
The proof for $\de\b$ goes exactly in the same way and we do not repeat it.
\medskip

\NI Therefore we have proved the following lemma
\begin{Le}\label{Lemma5}
Under the same assumptions as in Lemma \ref{Lemma2}, using the results proved there and in Lemmas \ref{Lemma3}, \ref{Comm1}, \ref{Comm2}, \ref{T0comest}, \ref{Lemma4} 
we have in the region $V_*$ the following inequalities:
\bea
&&|r^5\de\a(R)|\leq {\tilde c}_9\!\left(\varepsilon+\frac{\ep_0}{N_0}\right)\nn\\
&&|r^4\de\b(R)|\leq {\tilde c}_{10}\!\left(\varepsilon+\frac{\ep_0}{N_0}\right)
\eea
where
\bea
{\tilde c}_9\geq {\hat c}_1+{\tilde c}_7\  \ ;\ \ {\tilde c}_{10}\geq {\hat c}_2+{\tilde c}_8\ \ \ .
\eea
\end{Le}

\subsection{IV step: The control of the $\de\cal O$ norms. }\label{SS2.5}

\NI To prove better estimate for these norms in $V_*$ allowing to show, by a boostrap argument, that the region $V_*$ is unbounded, we have to use the transport equations along the incoming and outgoing cones. The use of the outgoing cones is made, as in \cite{Kl-Ni:book}, obtaininging the estimates starting from ``scri", here the upper boundary of the region $V_*$ which is a portion of an incoming cone.

\NI Therefore we have first to control the not underlined connection coefficients\footnote{basically we denote as not underlined connection coefficients those coefficients whose transport equations we use are those along the outgoing cones, the opposite for the underlined ones. Remember, as discusssed in detail in \cite{Kl-Ni:book} that the choice of the transport equations to use is not arbitrary and is uniquely fixed by the request of avoiding any loss of derivatives which will make the bootstrap mechanism to fail.} on this last slice and to avoid a fatal loss of derivatives we have to prove the existence on it of an appropriate foliation we called the ``last slice canonical foliation", see \cite{Niclast}, \cite{Kl-Ni:book} and \cite{Ch-Kl:book} where the original idea was first stated. 

\NI Therefore the weight factors we can assume in $V_*$ for the various (not underlined) $\de\cal O$  norms  have to be consistent with the weight factors of the norms we can control on the ``last slice". The estimate for the underlined connection coefficients is made, viceversa, starting from the initial data hypersurface $\Si_0$, in this case also an appropriate (canonical) foliation has to be introduced on $\Si_0$.\footnote{This could be in principle avoided requiring more regularity for the initial data.}

\NI As anticipated in subsection \ref{S.S.stepsfortheproof} to prove that in $V_*$ the $\de\cal O$  norms satisfy better estimates than those assumed in the ``Bootstrap assumptions" we need estimates for the corrections to the Kerr metric. These estimates follow once we have the bootstrap assumptions for the connection coefficients and are the content of the following lemma:
\begin{Le}\label{firstOOest}
Assume that in $V_*$ the norms $\de{\cal O}$ satisfy the bootstrap assumptions
\[\de{\cal O}\leq {\ep_0}\ ,\]
then, assuming for the $\de{\cal O}^{(0)}$ norms appropriate initial data conditions, see subsection \ref{SS3.6}, we prove in $V_*$ the following estimates:
\bea
|r|u|^{2+\de}\de\oom|_{\infty}\leq c{\ep_0}\  \ ;\ \ |r^2|u|^{2+\de}\de X|_{\infty}\leq  c{\ep_0}\ \ ;\ \ ||u|^{1+\de}\de \ga|_{\infty}\leq  c{\ep_0}\ .
\eea
\end{Le}
\NI The proof of this lemma is given later on,\footnote{In fact we prove the equivalent Lemma \ref{finOOest}.} see subsection \ref{OOest}. It is important to remark the order of the various proofs: 
\medskip

\NI{\bf a)}  The bootstrap assumptions $\de{\cal O}\leq \ep_0$ imply, Lemma  \ref{firstOOest}, the metric correction estimates $\de{\cal O}^{(0)}\leq c\ep_0$\ .
\smallskip

\NI{\bf b)} The metric correction estimates $\de{\cal O}^{(0)}\leq c\ep_0$, the initial data assumptions and the Riemann bootstrap assumptions imply better estimates for the $\de O$ connection coefficients $\de{\cal O}\leq c\frac{\ep_0}{N_0}\leq \frac{\ep_0}{2}$ with $N_0$ sufficently large.
\smallskip

\NI{\bf c)} The improved estimates for the connection coefficients imply better estimates for the metric correction, see Lemma \ref{finOOest}, $\de{\cal O}^{(0)}\leq \frac{\ep_0}{2}$\ . 

\subsubsection{The control of the $\de{\cal O}$ norms}\label{ss341}
\smallskip

\NI The $\de{\cal O}$ norms involve also the tangential derivatives up to fifth order of (corrections to) the connection coefficients to prove the bootstrap and the peeling, see \cite{Kl-Ni:book} and \cite{Kl-Ni:peeling}. Nevertheless the proof we sketch here is restricted to the zero and first derivatives of the connection coefficients as the control of the higher derivatives is simpler and is just a repetition of what has been done in \cite{Kl-Ni:book}.

\NI To control the $\de{\cal O}$ norms we have to subtract to the connection coefficients their Kerr part so to obtain some transport equations and some Hodge equations relative to the $\de O$ correction parts. This operation which we call ``Kerr decoupling" is a central step of the whole procedure and we first discuss it  in some generality. Then we will examine a typical case.

\NI The connection coefficients in \cite{Kl-Ni:book} have been divided in two sets, the underlined ones which use transport equation along incoming cones and the not underlined  estimated using transport equations along outgoing cones. Let us look to the transport equation for one of the not underlined connection coefficient. The connection coefficients are covariant tensors fields ``belonging" to the tangent spaces $TS$. Therefore let us consider one of them, namely the second null fundamental form $\chi$, as a $TS$ tensor we can write it as
\bea
{\bf g}(\dd_{e_a}e_4,e_b)=\chi(e_a,e_b)=\chi_{\mu\nu}e_a^{\mu}e_b^{\nu}\ .
\eea
Denoting $\{\th^b(\c)\}$ the one forms dual to the $TS$ orthonormal frame $\{e_a\}$,
\bea
&&\chi_{\mu\nu}=\sum_{a,b}\chi(e_a,e_b)\th^a_{\mu}\th^b_{\nu}=\sum_{a,b}{\bf g}(\dd_{e_a}e_4,e_b)\th^a_{\mu}\th^b_{\nu}\nn\\
&&=\sum_{a,b}g_{\ro\si}e_a^{\tau}(\dd_{\tau}e_4)^{\ro} e^{\si}_b\th^a_{\mu}\th^b_{\nu}=\Pi^{\tau}_{\mu}\Pi^{\si}_{\nu}(\dd_{\tau}e_4)^{\ro}g_{\ro\si}\nn
=\Pi^{\tau}_{\mu}(\dd_{\tau}e_4)^{\ro}g_{\ro\si}\Pi^{\si}_{\nu}\nn\\
&&=\Pi^{\tau}_{\mu}(\dd_{\tau}e_4)^{\ro}\Pi_{\ro\nu}=\Pi^{\tau}_{\mu}(\dd_{\tau}e_4)_{\si}\Pi^{\si}_{\nu}\eql{first}
\eea
and
\bea
\chi=\chi_{\mu\nu}dx^{\mu}\!\otimes\!dx^{\nu}=\left((\dd_{\tau}e_4)_{\si}\Pi^{\tau}_{\mu}\Pi^{\si}_{\nu}\right)dx^{\mu}\!\otimes\!dx^{\nu}
\eea
where $\{x^{\mu}\}=\{u,\ub,\th,\phi\}$.\footnote{This $\th$ is approximately, the $\th_*$ of P-I.} More in general we write for a two covariant tensor connection coefficient
\bea
O= O_{\mu\nu}dx^{\mu}\!\otimes\!dx^{\nu}\ .
\eea
and as $O$ is tangent to $S$, it follows that $O_{\mu\nu}=O_{\ro\si}\Pi^{\ro}_{\mu}\Pi^{\si}_{\nu}$. Nevertheless it is important to remark that it is not true that $O=O_{ab}d\om^a\otimes d\om^b$. This can be easily recognized looking at \ref{first}. In fact 
\[\th^a_{\mu}={\bf g}(e_a,\c)=g_{\mu c}e_a^c=\!\left(\de_{\mu}^a\ga_{ac}+\de_{\mu}^u(\ga_{bc}X_{(Kerr)}^b)+\de_{\mu}^{\ub}(\ga_{bc}{X}^b)\right)\!e_a^c\ .\]
The generic structure equations for the connection coefficients $O$ are of two types: equations which are ``transport equations" along the outgoing or incoming cones and   ``elliptic Hodge type" equations on the $S$ surfaces. How many and how these equations, together with first order equations for the metric components, are equivalent to the Einstein equations is discussed elsewhere. 
As a typical example we examine in the following  the transport equation for $\nabb\tr\chi$ and the Hodge equation for $\chih$. The transport equation has the the following general structure, indicating with $O$ a connection coefficient or a tangential derivative of it,
\bea
\ddb_4 O+k\tr\chi O= F\eql{trasp1}
\eea
where the integer $k$ depends on the connection coefficient $O$ we are considering, $\ddb_4$ is the projection on $TS$ of the differential operator $\dd_4=\dd_{e_4}$; therefore
\bea
&&\ddb_4 O=(\ddb_4 O)_{\mu\nu}dx^{\mu}\!\otimes\!dx^{\nu}\nn\\
&&(\ddb_4 O)_{\mu\nu}=\Pi^{\ro}_{\mu}\Pi^{\si}_{\nu}(\dd_4O)_{\ro\si}\ .
\eea
Finally $F$ is a covariant ``$S$-tangent" tensor whose components are quadratic or cubic functions  of the (components of the) connection coefficients and possibly, if $O$ is a tangential derivative of a connection coefficient, of the Riemann tensor,
\[F=F_{\mu\nu}dx^{\mu}\!\otimes\!dx^{\nu}\ \ ,\ \ F_{\mu\nu}=F_{\mu\nu}(\{O\}, \{R\})\ .\]
The Hodge equations for the connection coefficients or their tangential derivatives applied for instance to a two $S$-tangent covariant tensor field have the form
\bea
\divv O= G+R
\eea
Where $\divv$ is the the divergence associated to $\nabb$, the covariant derivative associated to the induced metric, $\ga^{(S)}$, on $S$,
\bea
\divv O=(\divv O)_{\si}dx^{\si}\ \ ,\ \ (\divv O)_{\si}={\ga^{(S)}}^{\mu\nu}(\nabb_{\mu}O)_{\nu\si}\ \ ,
\eea
$G$ is a covariant tensor (of order 1 if $O$ is of order 2) and with $R$ we denote a null component of the Riemann tensor tangent to $S$ of the same degree. To have a specific example of structure equations with this structure we will consider in the following the transport equation for $\nabb\tr\chi$ and the Hodge equation for $\chih$, see also equations (4.3.6) and (4.3.13) of \cite{Kl-Ni:book}.

\NI Observe now that the coordinates $\{x^{\mu}\}$ are the ``same" in the Kerr spacetime and in the perturbed Kerr spacetime, the only difference being the metric we are considering, the Kerr metric \ref{Kerrmet} or the ``perturbed" one, \ref{pertKerr1}. Therefore depending the metric we are considering the region $V_*$ can be thought as a region in the Kerr spacetime or a region of the perturbed Kerr spacetime whose existence we are proving. This simple, but important observation allows us to subtract the Kerr part in the structure equations.

\NI Let us go back to the transport equation \ref{trasp1} and introduce in $V_*$ the tensor field  $O^{(Kerr)}$ which is the connection coefficient analogous to $O$, but associated to the Kerr spacetime (which can therefore expressed in terms of first derivatives of the Kerr metric \ref{Kerrmet}). Obviously we have
\[O^{(Kerr)}=O^{(Kerr)}_{\mu\nu}dx^{\mu}\!\otimes\!dx^{\nu}\]
and we would like to subtract the Kerr part and define
\bea
\de O\equiv O-O^{(Kerr)}\ .\eql{corrO}
\eea
To obtain a transport equation for $\de O$ we apply to it $\ddb_4$ and we write
\bea
\ddb_4\de O\!&=&\!\ddb_4 O-(\ddb_4-\ddb_4^{(Kerr)})O^{(Kerr)}-\ddb_4^{(Kerr)}O^{(Kerr)}\nn\\
\!&=&\!-k\tr\chi O+k\tr\chi^{(Kerr)}O^{(Kerr)}+F-F^{(Kerr)}
\eea
as
\[-\ddb_4^{(Kerr)}O^{(Kerr)}-k\tr\chi^{(Kerr)}O^{(Kerr)}+F^{(Kerr)}=0\ .\]
The transport equation for $\de O$ has, therefore, the following structure
\bea
\ddb_4\de O+k\tr\chi\de O=H(\de O^{(0)},O^{(Kerr)})+\de F(\de O, O^{(Kerr)},\de R)\ \ \ \ \ \eql{deOtrasp0}
\eea
where
\bea
&&H(\de O^{(0)},O^{(Kerr)})=-k\big(\tr\chi-\tr\chi^{(Kerr)}\big)O^{(Kerr)}-(\ddb_4-\ddb_4^{(Kerr)})O^{(Kerr)}\nn\\
&&\de F(\de O, O^{(Kerr)},\de R)=F-F^{(Kerr)}\ .
\eea
From the transport equation \ref{deOtrasp0} we can get an estimate for the $\de O$ norm integrating along the outgoing cones. 
There is, nevertheless, a technical modification to do, in fact when we introduce $\de O$, see \ref{corrO}, and we want to use the modified transport equations for $\de O$ instead that for $O$ we have to remember that $O$ is a vector field tangent to the $S(u,\ub)$ two dimensional surfaces intersections of the outgoing and incoming cones of the double null cone foliation assumed in $V_*$, thought as a region of the perturbed Kerr spacetime; in other words $O_{\mu\nu}$ can be written
\bea
O_{\mu\nu}=\Pi^{\ro}_{\mu}\Pi^{\si}_{\nu}H_{\ro\si}\eql{Otg1}
\eea
where $H$ is a $(0,2)$ tensor in $(V_*,{\bf g})$ a priori not $S$-tangent and $\Pi^{\mu}_{\nu}$ projects from $TV_*$ to $TS$. On the other side the $O^{(Kerr)}$ tensor field is tangent to the $S$ two dimensional surfaces with respect to the ${\bf g}^{(Kerr)}$ metric instead that with respect to the ${\bf g}$ metric. This means that, as in \ref{Otg1}, we have
\bea
O^{(Kerr)}_{\mu\nu}={\Pi^{(Kerr)}}^{\ro}_{\mu}{\Pi^{(Kerr)}}^{\si}_{\nu}H^{(Kerr)}_{\ro\si}\eql{Otg2}\ 
\eea
which is not $S$-tangent in $(V_*,{\bf g})$. On the other side in the transport equations the $\de O$ correction terms have to be, as  the $O$ connection coefficients, $S$-tangent with respect to the ${\bf g}$ metric. Therefore the definition in \ref{corrO} has to be modified in the following way:
\bea
\de O\equiv O-{\hat O}\ .\eql{corrOnew}
\eea
where
\bea
{\hat O}_{\mu\nu}=\Pi^{\ro}_{\mu}\Pi^{\si}_{\nu}H^{(Kerr)}_{\ro\si}\ .\eql{Otg11}
\eea
To be more precise let us look at the explicit expression of some of the connection coefficients, see \cite{Kl-Ni:book}, Chapter 3, the torsion coefficient,
\bea
\ze(e_a)=\frac{1}{2}{\bf g}(\dd_{e_a}e_4,e_3)
\eea
which in the Kerr spacetime can be rewritten as
\bea
&&\ze^{(Kerr)}(e_a^{(Kerr)})=\frac{1}{2}{\bf g}^{(Kerr)}(\dd^{(Kerr)}_{e^{(Kerr)}_a}e^{(Kerr)}_4,e^{(Kerr)}_3)\nn\\
&&=\frac{1}{2}g^{(Kerr)}_{\mu\nu}{e^{(Kerr)}}_a^{\ro}D^{(Kerr)}_{\ro}{e^{(Kerr)}_4}^{\mu}{e^{(Kerr)}}_3^{\nu}\nn\\
&&={e^{(Kerr)}}_a^{\ro}\left({\Pi^{(Kerr)}}_{\ro}^{\si}g^{(Kerr)}_{\mu\nu}D^{(Kerr)}_{\si}{e^{(Kerr)}_4}^{\mu}{e^{(Kerr)}}_3^{\nu}\right)
\eea
Therefore
\bea
\ze^{(Kerr)}_{\ro}={\Pi^{(Kerr)}}_{\ro}^{\si}\left(g^{(Kerr)}_{\mu\nu}D^{(Kerr)}_{\si}{e^{(Kerr)}_4}^{\mu}{e^{(Kerr)}}_3^{\nu}\right)
\equiv {\Pi^{(Kerr)}}_{\ro}^{\si}H^{(Kerr)}_{\si}\ \ \ \ \
\eea
and we define
\bea
{\hat \ze}_{\ro}={\Pi}_{\ro}^{\si}H^{(Kerr)}_{\si}\ .
\eea
Analogously we have
\bea
&&{\chi^{(Kerr)}}({e^{(Kerr)}}_a,{e^{(Kerr)}}_b)={\bf g}^{(Kerr)}(\dd_{{e^{(Kerr)}}_a}e^{(Kerr)}_4,{e^{(Kerr)}}_b)\nn\\
&&={g^{(Kerr)}}_{\mu\nu}{e^{(Kerr)}}_a^{\ro}D^{(Kerr)}_{\ro}{e^{(Kerr)}}^{\mu}_4{e^{(Kerr)}}^{\nu}_b\nn\\
&&={e^{(Kerr)}}_a^{\ro}{e^{(Kerr)}}^{\nu}_b\left({\Pi^{(Kerr)}}_{\ro}^{\si}{\Pi^{(Kerr)}}_{\nu}^{\tau}g^{(Kerr)}_{\mu\tau}D^{(Kerr)}_{\si}{e^{(Kerr)}}^{\mu}_4\right)\nn\\
&&={e^{(Kerr)}}_a^{\mu}{e^{(Kerr)}}^{\nu}_b\left({\Pi^{(Kerr)}}_{\mu}^{\si}{\Pi^{(Kerr)}}_{\nu}^{\tau}g^{(Kerr)}_{\la\tau}D^{(Kerr)}_{\si}{e^{(Kerr)}}^{\la}_4\right)\nn\\
&&={e^{(Kerr)}}_a^{\mu}{e^{(Kerr)}}^{\nu}_b\left[{\Pi^{(Kerr)}}_{\mu}^{\si}{\Pi^{(Kerr)}}_{\nu}^{\tau}\left(g^{(Kerr)}_{\la\tau}D^{(Kerr)}_{\si}{e^{(Kerr)}}^{\la}_4\right)\right]\ .\nn
\eea
Therefore
\bea
{\chi^{(Kerr)}}_{\mu\nu}={\Pi^{(Kerr)}}_{\mu}^{\si}{\Pi^{(Kerr)}}_{\nu}^{\tau}\left(g^{(Kerr)}_{\la\tau}D^{(Kerr)}_{\si}{e^{(Kerr)}}^{\la}_4\right)
={\Pi^{(Kerr)}}_{\mu}^{\si}{\Pi^{(Kerr)}}_{\nu}^{\tau}H^{(Kerr)}_{\si\tau}\ \ \ \ \ \ 
\eea
and
\bea
{\widehat{(\chi)}}_{\mu\nu}={\Pi}_{\mu}^{\si}{\Pi}_{\nu}^{\tau}\left(g^{(Kerr)}_{\la\tau}D^{(Kerr)}_{\si}{e^{(Kerr)}}^{\la}_4\right)
={\Pi}_{\mu}^{\si}{\Pi}_{\nu}^{\tau}H^{(Kerr)}_{\si\tau}\ \ \ \ \ \ .
\eea
$\om$ and $\omb$ are scalar functions and therefore $\om^{(Kerr)}={\hat\om}\ \ ,\ \ \omb^{(Kerr)}={\hat\omb}$.

\NI It will be needed in the following to use the fact that, under the bootstrap assumption, the norm estimates of the ``hat" quantities are as those for the Kerr connection coefficients, possibly with a different constant.  This is proved in the following lemma
\begin{Le}\label{hatestim}
In $V_*$ under the bootstrap assumptions it follows that all the norm for the ``hat" quantities satisfy the following estimates, denoting with $|\c|$ either the pointwise norm or the $|\c|_{p,S}$ norm,
\bea
|{\hat O}|\leq |O^{(Kerr)}|\left(1+c\frac{\ep_0}{r^2|u|^{(1+\de)}}\right)\eql{hates}
\eea
and
\bea
|{\hat O}(e_a)-O^{(Kerr)}(e_a^{(Kerr)})|\leq c\left(\frac{\ep_0}{r^3|u|^{1+\de}}\right)|O^{(Kerr)}|\ .
\eea
\end{Le}
\NI{\bf Proof:}
Let assume for simplicity that $O$ be a covariant tensor then
\bea
&&O^{(Kerr)}={\Pi^{(Kerr)}}^{\ro}_{\mu}H^{(Kerr)}_{\ro}\ \ ,\ \ {\hat O}_{\mu}={\Pi}^{\ro}_{\mu}H^{(Kerr)}_{\ro}\nn\\
&&O^{(Kerr)}(e_a^{(Kerr)})={e_a^{(Kerr)}}^{\ro}H^{(Kerr)}_{\ro}\ \ {\hat O}(e_a)=e_a^{\ro}H^{(Kerr)}_{\ro}
\eea
As $e_a$ is $S$-tangent we have 
\bea
&&e_a^{(Kerr)}=\frac{1}{\sqrt{\ga^{(Kerr)}_{aa}}}\frac{\pr}{\pr\om^a}\nn\\
&&e_a=c_a^b\frac{\pr}{\pr\om^b}=c_a^b\sqrt{\ga^{(Kerr)}_{bb}}e_b^{(Kerr)}
\eea
It is easy to prove, see the appendix, that
\bea
&&\ML c_1^1\sqrt{\ga^{(Kerr)}_{11}}=1+O\left(\frac{\ep_0}{r^2|u|^{1+\de}}\right)\ \ ,\ \ c_1^2\sqrt{\ga^{(Kerr)}_{22}}=O\left(\frac{\ep_0}{r^2|u|^{1+\de}}\right)\\
&&\ML c_2^2\sqrt{\ga^{(Kerr)}_{22}}=1+O\left(\frac{\ep_0}{r^2|u|^{1+\de}}\right)\ \ ,\ \ c_2^1\sqrt{\ga^{(Kerr)}_{11}}=O\left(\frac{\ep_0}{r^2|u|^{1+\de}}\right)\ .\ \ \ \ \nn
\eea
Therefore
\bea
{\hat O}(e_a)=e_a^{\ro}H^{(Kerr)}_{\ro}=c_a^b\sqrt{\ga^{(Kerr)}_{bb}}O^{(Kerr)}(e_b^{(Kerr)})
\eea
and
\bea
|{\hat O}(e_a)-O^{(Kerr)}(e_a^{(Kerr)})|\!&\leq&\! |c_a^a\sqrt{\ga^{(Kerr)}_{aa}}-1||O^{(Kerr)}(e_a^{(Kerr)})|\nn\\
\!&+&\!|c_a^b\sqrt{\ga^{(Kerr)}_{bb}}||O^{(Kerr)}(e_b^{(Kerr)})|\nn\\
\!&\leq&\! c\left(\frac{\ep_0}{r^3|u|^{1+\de}}\right)\ ,
\eea
\bea
\sup_a|{\hat O}(e_a)|\leq \left(1+O\left(\frac{\ep_0}{r^2|u|^{1+\de}}\right)\right)\sup_b|O^{(Kerr)}(e_b^{(Kerr)})|
\eea
Moreover this result can be immediately extended to any norm with the appropriate weights.
As before  to obtain a transport equation for $\de O$ we apply to it $\ddb_4$ and write
\bea
&&\ddb_4\de O=\ddb_4 O-(\ddb_4-\ddb_4^{(Kerr)}){\hat O}-\ddb_4^{(Kerr)}{\hat O}\nn\\
&&=-k\tr\chi O+F-(\ddb_4-\ddb_4^{(Kerr)}){\hat O}-\ddb_4^{(Kerr)}({\hat O}-O^{(Kerr)})-\ddb_4^{(Kerr)}O^{(Kerr)}\nn\\
&&=-k\tr\chi O+F-(\ddb_4-\ddb_4^{(Kerr)}){\hat O}-\ddb_4^{(Kerr)}({\hat O}-O^{(Kerr)})+k\tr\chi^{(Kerr)}O^{(Kerr)}-F^{(Kerr)}\nn\\
&&=-k\tr\chi\de O+\de F\nn\\
&&+\left[-(\ddb_4-\ddb_4^{(Kerr)}){\hat O}-\ddb_4^{(Kerr)}({\hat O}-O^{(Kerr)})-k\de\tr\chi{\hat O}-k\tr\chi^{(Kerr)}({\hat O}-O^{(Kerr)})\right]\nn
\eea
which we rewrite as
\bea
\ddb_4\de O+k\tr\chi\de O=H(\de O^{(0)},{\hat O},O^{(Kerr)})+\de F(\de O,{\hat O},O^{(Kerr)},\de R)\ \ \ \ \ \eql{deOtrasp}
\eea
where
\bea
&&\ML H(\de O^{(0)},{\hat O},O^{(Kerr)})\nn\\
&&\ML=\left[-(\ddb_4-\ddb_4^{(Kerr)}){\hat O}-\ddb_4^{(Kerr)}({\hat O}-O^{(Kerr)})-k\de\tr\chi{\hat O}-k\tr\chi^{(Kerr)}({\hat O}-O^{(Kerr)})\right]\nn\\
&&\ML\de F(\de O, O^{(Kerr)},\de R)=F-F^{(Kerr)}\ .
\eea
\NI{\bf Remark:} {\em Observe that in $H(\de O^{(0)},{\hat O},O^{(Kerr)})$ the parts not explicitely known are in $(\ddb_4-\ddb_4^{(Kerr)})$, $\de\tr\chi$ and in the $\de\Pi$ present in $({\hat O}-O^{(Kerr)})$; therefore there is no loss of derivatives as expected.}
\smallskip

\NI Here we proceed exactly as in \cite{Kl-Ni:book}, Chapter 4, and we just sketch the argument. Let $|\de O|$ be a $|\c|_{p,S}$ norm, applying Gronwall inequality and  Lemma 4.1.5 of \cite{Kl-Ni:book} we obtain the following estimate, with $\si>0$,
\bea
&&\ML\ML\||u|^{2+\de}r^{(2-\si)-\frac{2}{p}}\de O|_{p,S}(u,\ub)\leq c_0\!\left(||u|^{2+\de}r^{(2-\si)-\frac{2}{p}}\de O|_{p,S}(u,\ub_*)\right.\nn\\
&&\ML\ML+\left.\int_{\ub}^{\ub_*}\!\left[||u|^{2+\de}r^{(2-\si)-\frac{2}{p}}H|_{p,S}
+||u|^{2+\de}r^{(2-\si)-\frac{2}{p}}\de F|_{p,S}\right]\!\!(u,\ub')\!\right)\ .\eql{3.198}
\eea
To control the right hand side of \ref{3.198} we have to estimate the norm \[||u|^{2+\de}r^{3-\frac{2}{p}}\de O|_{p,S}(u,\ub_*)\]  on the last slice and the norms
of $H$ and $\de F$. The norm on the last slice will be discussed later on when we prove the existence of the ``last slice canonical foliation". 
The bounds for the norms
$||u|^{2+\de}r^{3-\frac{2}{p}}H|_{p,S}$ and $||u|^{2+\de}r^{3-\frac{2}{p}}\de F|_{p,S}$ are proved in the following lemma:
\begin{Le}\label{L3.9}
Under the bootstrap assumptions, the following estimates hold in $V_*$
\bea
&&||u|^{2+\de}r^{3-\frac{2}{p}}H|_{p,S}\leq c\ep_0\!\left(\frac{M}{R_0}\right)^{\!\!2}\\
&&||u|^{2+\de}r^{3-\frac{2}{p}}\de F|_{p,S}\leq c\ep_0\!\left(\frac{M}{R_0}\right)^{\!\!2}\ .\nn
\eea
\end{Le}
\NI Once Lemma \ref{L3.9} is proved, assuming that on the last slice we have an analogous estimate 
\bea
||u|^{2+\de}r^{2-\frac{2}{p}}\de O|_{p,S}(u,\ub_*)\leq c\ep_0\!\left(\frac{M}{R_0}\right)^{\!\!2}\eql{lastsliceest}
\eea
we integrate obtaining
\bea
\||u|^{2+\de}r^{(2-\si)-\frac{2}{p}}\de O|_{p,S}(u,\ub)\!&\leq&\!c_0\!\left(||u|^{2+\de}r^{(2-\si)-\frac{2}{p}}\de O|_{p,S}(u,\ub_*)\right.\nn\\
\!&+&\!\left.c\ep_0\!\left(\frac{M}{R_0}\right)^{\!\!2}\!\!\int_{\ub}^{\ub_*}\!\!\frac{1}{r^{1+\si}}\!\right)\ \eql{3.198b}
\eea
and the result \footnote{We recall that with $c$ we denote different adimensional constants.}
\bea
||u|^{2+\de}r^{2-\frac{2}{p}}\de O|_{p,S}(u,\ub)\leq c\ep_0\!\left(\frac{M}{R_0}\right)^{\!\!2}\ .
\eea
Choosing $R_0$ such that
\[c\!\left(\frac{M}{R_0}\right)^{\!\!2}\leq \frac{1}{N_0}<\frac{1}{2}\ ,\]
we have proved that the norms $\de{\cal O}$ satisfy in $V_*$ a better estimate than those in the ``Bootstrap assumptions".
\smallskip

\NI In the previous discussion we  have shown how to obtain appropriate bounds for some $|\c|_{p,S}$ norms of the correction to some connection coefficients.
Although complicated by the need of subtracting the Kerr part, the strategy follows the one described in \cite{Kl-Ni:book}. There to complete the bootstrap mechanism beside the transport equations also those structure equations which are elliptic Hodge systems on $S$ are used. Therefore also for these equations the ``Kerr decoupling" has to be performed, namely the Kerr part has to be subtracted. In performing this subtraction there are no new ideas different from those already described, therefore instead of discussing how to treat these elliptic Hodge systems in the whole generality we show it in detail when we estimate (the correction) $\de(\nabb\tr\chi)$.
\medskip

\NI{\bf Remark:} {\em Recall that the existence proof requires better estimate not only for the (correction to the connection coefficients, but also for their tangential derivatives up to fifth order. We do not prove here the estimates for the higher derivatives as the proofs proceed exactly in the same way. Details are given in \cite{Kl-Ni:book}, Chapter 4 in the case of the whole connection coefficients.}
\medskip

\NI{\bf Proof of Lemma \ref{L3.9}:}
Let us examine first the terms in $H$; 
For the first term in $H$ the explicit expression of  the term $(\ddb_4-\ddb_4^{(Kerr)}){\hat O}$ is given in the following lemma:
\begin{Le}\label{deDexpr1}
With the previous definitions the following expression holds:
\bea
&&\ML\big((\ddb_4-\ddb^{(Kerr)}_4){\hat O}\big)_{\mu}=\eql{deD4a}\\
&&\ML(\de\Pi^{\ro}_{\mu})\dd_4{\hat O}_{\ro}+{\Pi^{(Kerr)}}^{\ro}_{\mu}\left(-\frac{{\de\oom}}{\oom}\dd_4^{(Kerr)}+\frac{\de X^c}{\oom}\dd_c\right)\!{\hat O}_{\ro}
+{\Pi^{(Kerr)}}^{\ro}_{\mu}{e^{(Kerr)}}_4^{\la}(\de\Ga^{\tau}_{\la\ro}){\hat O}_{\tau}\nn
\eea
where
\bea
&&\de\Pi^{\ro}_{\mu}=\Pi^{\ro}_{\mu}-{\Pi^{(Kerr)}}^{\ro}_{\mu}\nn\\
&&\de\Ga^{\tau}_{\la\ro}=\Ga^{\tau}_{\la\ro}-{\Ga^{(Kerr)}}^{\tau}_{\la\ro}\ .
\eea
\end{Le}
\NI{\bf Proof:} See the appendix.
\medskip

\NI To bound the norm of  $(\ddb_4-\ddb^{(Kerr)}_4){\hat O}$ we have to use the bootstrap assumptions in $V_*$ for the connection coefficients, their tangential derivatives and the estimates which follow for the metric components. In particular to control $\de\Pi^{\ro}_{\mu}=(\Pi^{\ro}_{\mu}-{\Pi^{(Kerr)}}^{\ro}_{\mu})$  we need its explicit expression. 
\begin{Le}\label{dePiexpr}
With the previous definitions the following expression holds:
\bea
(\Pi^{\si}_{\nu}-{\Pi^{\si}_{\nu}}^{(Kerr)})=\de\oom A^{\si}_{\nu}+(r\de X^c)B^{\si}_{c\nu}
\eea
where
\bea
&&A^{\si}_{\nu}=\frac{1}{{\oom^{(Kerr)}}}\left[\left({\th^3}^{(Kerr)}_{\nu}{e_3^{\si}}^{(Kerr)}\!+{\th^3}_{\nu}{e_3^{\si}}\right)+\left({\th^4}^{(Kerr)}_{\nu}{e_4^{\si}}^{(Kerr)}\!+{\th^4}_{\nu}{e_4^{\si}}\right)\right]\nn\\
&&B^{\si}_{c\nu}=\frac{1}{{\oom^{(Kerr)}}}\th^4_{\nu}\de^{\si}_c
\eea
\end{Le}
\NI{\bf Proof:} See the appendix.
\smallskip

\NI Therefore from the expression \ref{deD4a} we have
\bea
&&\ML|r^{3-\frac{2}{p}}|u|^{2+\de}\big((\ddb_4-\ddb^{(Kerr)}_4){\hat O}\big)|_{p,S}\leq c\left(|r|u|^{2+\de}\de\Pi|_{\infty}|r^{2-\frac{2}{p}}\dd_4{\hat O}|_{p,S}\right.\eql{deD4ab}\\
&&\ML\left.+|r|u|^{2+\de}(r\de X)|_{\infty}|r^{2-\frac{2}{p}}\dd^{(Kerr)}_c{\hat O}|_{p,S}
+|r^3|u|^{1+\de}\de\Ga|_{\infty}|r^{1-\frac{2}{p}}{\hat O}|_{p,S}\right)\leq c\ep_0\frac{M^2}{R_0^2}\nn
\eea
In fact from the bootstrap assumptions and recalling that $\pr_c=\pr_{x^c}$ we have immediately
\bea
&&|r|u|^{2+\de}\de\Pi|_{\infty}\leq c\ep_0\ \ ,\ \ |r|u|^{2+\de}(r\de X)|_{\infty}\leq c\ep_0\nn\\
&&|r^3|u|^{1+\de}\de\Ga|_{\infty}\leq c\ep_0\ ,
\eea
the last line following as 
\bea
\de\Ga=\ga^{-1}\pr_c\de\ga=\ga^{-1}\pr_{x^c}\de\ga=O(r^{-2})O(r^{-1})O(\ep_0|u|^{-(1+\de)})=O\left(\frac{\ep_0}{r^3|u|^{1+\de}}\right)\ .\ \ \ \ \eql{3266}
\eea
{\bf Remark:} {\em Let us make a more detailed estimate of the term $|{\Pi^{(Kerr)}}^{\ro}_{\mu}{e^{(Kerr)}}_4^{\la}(\de\Ga^{\tau}_{\la\ro}){\hat O}_{\tau}|$. Observe first that
$[{\Pi^{(Kerr)}}^{\ro}_{\mu}]=L^0$ due to the fact that the tensor $\Pi$ has dimension zero, $[\Pi]=L^0$, and 
\[\Pi=\Pi^{\mu}_{\nu}\frac{\pr}{\pr x^{\mu}}\otimes dx^{\nu}\ .\]
Analogously as $[e_4]=L^{-1}$ and 
\[e_4=\frac{\pr}{\pr \ub}+rX^c\frac{1}{r}\frac{\pr}{\pr\om^c}\]
it follows that  $[e^{\mu}_4]=L^{0}$. Finally each covariant derivative has dimension $L^{-1}$, always in ``dimensional coordinates". Therefore
\[[|{\Pi^{(Kerr)}}^{\ro}_{\mu}{e^{(Kerr)}}_4^{\la}(\de\Ga^{\tau}_{\la\ro}){\hat O}_{\tau}]=L^{-2}\] On the other side as this term is a correction and therefore proportional to $\ep_0$ it must behave as 
\[|{\Pi^{(Kerr)}}^{\ro}_{\mu}{e^{(Kerr)}}_4^{\la}(\de\Ga^{\tau}_{\la\ro}){\hat O}_{\tau}|=\frac{\ep_0}{r^3|u|^{2+\de}}\ .\]
Let us now look at it in a more explicit way, the term ${e^{(Kerr)}}_4^{\la}(\de\Ga^{\tau}_{\la\ro})$ has a term of the following kind (the other terms can be treated in a similar way)
\bea
{e^{(Kerr)}}_4^{\la}(\de\Ga^{\tau}_{\la\ro})=\ga^{\tau\si}\pr_{\si}\de\ga_{u,\ub}{\hat O}_{\tau}+\c\c\c\c
\eea
where the indices $u,\ub$ arise from the saturation with $\Pi$ and $e_4$ (not the only possible component anyway). $\de\ga_{u,\ub}$ has dimension $L^0$ and is propoertional to $\ep_0$ therefore $|\de\ga_{u,\ub}|=O(\ep_0r^{-2}|u|^{-(1+\de)})$. If the derivative $\pr_\si$ is $\pr_{\om^c}$ there is not extra decay factor, but in this case the term we are considering is $\ga^{\tau c}\pr_{\om^c}\de\ga_{u,\ub}{\hat O}_{\tau}$. Let us consider the index $\tau$ we can imagine $\tau$ an index associated to a cartesian coordinate, therefore here $u$ or $\ub$, in this case $\ga^{u c}$ has dimension $[\ga^{u c}]=L^{-1}$ which means that $|\ga^{u c}|=O(r^{-1})$; analogously $[{\hat O}_{u}]=L^{-1}$ which implies $|{\hat O}_{u}|=O(r^{-1})$ and collecting together
\bea
\ML|\ga^{u c}\pr_{\om^c}\de\ga_{u,\ub}{\hat O}_{u}|\leq |\ga^{u c}||\pr_{\om^c}\de\ga_{u,\ub}||{\hat O}_{u}|
= O(r^{-1})O((\ep_0r^{-2}|u|^{-(1+\de)})O(r^{-1})=O\frac{\ep_0}{r^4|u|^{1+\de}}\ .\nn
\eea
If viceversa the index $\tau$ were associated to $\om_d$ then it follows that  $|{\hat O}_{d}|=O(r^{0})$, but in this case $\ga^{d c}$ has dimension $[\ga^{d c}]=L^{-2}$ and therefore $|\ga^{d c}|=O(r^{-2})$ which compensates the loss in the decay of $|{\hat O}_{d}|$. }

\NI We are left to examine the term $\de F$; to describe its general  structure we proceed in the following way,  any connection coefficient has dimension $L^{-1}$ any metric component (respect to the ``cartesian" coordinates) has dimension $L^0$ any derivative decreases the dimension by 1.
Therefore the dimension of the right hand side of the equation \ref{deOtrasp} has to be the same of the left hand side, but we know that in the right hand side  the terms cannot be more than quadratic if we are considering a  connection coefficient (with dimension $L^{-1}$), cubic if we are considering a tangential derivative of a connection coefficient (with dimension $L^{-1}$) and so on and so for. Moreover, as already said, the transport equations chosen to prove the existence theorem do not have any loss of derivatives, therefore if we consider the transport equation for the tangential derivative of (the correction of ) a connection coefficient 
in the right hand side there cannot be $\de O$ terms involving second tangential derivatives, therefore terms with dimension $L^{-3}$.

\NI The previous considerations imply the following general structure for the transport equations for $\de O$: Let us denote $O=O^{(1)}$, a connection coefficient, $O^{(2)}$ a tangential  derivative, $O^{(k)}$ the $k$th tangential derivatives, then
\bea
&&\de F=f(O^{(0)},O^{(1)})-f({O^{(Kerr)}}^{(0)},{O^{(Kerr)}}^{(1)})\nn\\
&&=\left[c_{11}\de O^{(1)}({\hat O}^{(1)})+c_{02}\de O^{(0)}({\hat O}^{(2)})\right]\eql{exdef}
\eea
If viceversa $O=O^{(2)}$ is the tangential derivative of a connection coefficient the general structure is more complicated, but the way to control it is the same which we will discuss  in a relevant specific example. From the expression \ref{exdef} using the bootstrap assumptions and the Kerr part estimates, we obtain
\footnote{The bootstrap assumptions allow to control the norm $|r^{1-\frac{2}{p}}|u|^{2+\de}\de O^{(0)}|_{p,S}$ provided that with $\de O^{(0)}$ we denote $\de\oom$, $r\de X$ and $r^{-2}\de\ga_{ab}$.}
\bea
|r^{3-\frac{2}{p}}|u|^{2+\de}\de F|_{p,S}\!&\leq&\! c\left(|r^{2-\frac{2}{p}}|u|^{2+\de}\de O^{(1)}|_{p,S}|r{O^{(Kerr)}}^{(1)}|_{\infty}\right.\\
\!&+&\! \left.|r^{1-\frac{2}{p}}|u|^{2+\de}\de O^{(0)}|_{p,S}|r^2{O^{(Kerr)}}^{(2)}|_{\infty}\right)\leq c\ep_0\frac{M^2}{R_0^2}\ .\nn
\eea
\subsubsection{Detailed estimate of the norm $|r^{3-\frac{2}{p}}|u|^{2+\de}\de(\nabb\tr\chi)|$\ .}\label{ss342}

\NI Let us look specifically at one of the $\de{\cal O}$ norms, the one associated to \[\Us=\oom^{-1}\left(\nabb\tr\chi+\tr\chi\ze\right)\ ,\] and show in detail for its transport equation the structure  we have sketched in general. The transport equation $\Us$ satisfies is the following one:
\bea
\oom\ddb_4\Us+\frac{3}{2}{\oom}\tr\chi\Us=\Fs\ .
\eql{Usev}
\eea
Proceeding as discussed in general, see subsection \ref{ss341}, we write $\Us$ in the following way 
\bea
\Us={\hat {\Us}}+\de\Us
\eea
where
\bea
{\hat {\Us}}_{\mu}=\Pi^{\nu}_{\mu}\left({\oom^{(Kerr)}}^{-1}D_{\nu}\tr\chi^{(Kerr)}+\tr\chi^{(Kerr)}H^{(Kerr)}_{\nu}\right)
\eea
where
\bea
H^{(Kerr)}_{\nu}=\left(g^{(Kerr)}_{\mu\ro}(D^{(Kerr)}_{\nu}{e^{(Kerr)}_4})^{\mu}{e^{(Kerr)}}_3^{\ro}\right)\ .
\eea
Therefore
\bea
{\hat {\Us}}=\frac{1}{{\oom^{(Kerr)}}}\left(\nabb\tr\chi^{(Kerr)}+\tr\chi^{(Kerr)}{\hat \ze}\right)\eql{Usdef}
\eea
and
\bea
\de\Us\!&=&\!-\frac{\de\oom}{\oom{\oom^{(Kerr)}}}\left(\nabb\tr\chi+\tr\chi\ze\right)\\
\!&+&\!\frac{1}{{\oom^{(Kerr)}}}\left(\nabb\de\tr\chi+\de\tr\chi\ze+\tr\chi^{(Kerr)}\de\ze\right)\nn
\eea
where
\bea
\de\ze=\ze-{\hat \ze}\ .
\eea
Analogously we write
\bea
\Fs={\hat \Fs}+\de\Fs\ .
\eea
Therefore
equation \ref{Usev} can be written as
\bea
&&\ML\oom\ddb_4\de\Us+\frac{3}{2}{\oom}\tr\chi\de\Us=\de\Fs+({\hat \Fs}-\Fs^{(Kerr)})+\nn\\
&&\ML-\left[\oom\ddb_4({\hat\Us}-\Us^{(Kerr)})+\oom(\de\ddb_4)\Us^{(Kerr)}
+\de\oom\ddb_4^{(Kerr)}\Us^{(Kerr)}\right]\nn\\
&&\ML-\left[\frac{3}{2}{\oom}\tr\chi({\hat\Us}-\Us^{(Kerr)})
+\frac{3}{2}{\oom}\de\tr\chi\Us^{(Kerr)}+\frac{3}{2}{\de\oom}\tr\chi^{(Kerr)}\Us^{(Kerr)}+({\hat \Fs}-\Fs^{(Kerr)})\right]\nn\\
&&\ML+\left[-\oom^{(Kerr)}\ddb_4^{(Kerr)}\Us^{(Kerr)}-\frac{3}{2}{\oom^{(Kerr)}}\tr\chi^{(Kerr)}\Us^{(Kerr)}+\Fs ^{(Kerr)}\right]\nn\\
&&\nn\\
\eea
where the crucial fact is that the last line is identically zero being the structure equation satisfied in Kerr spacetime; therefore we can rewrite the equation as
\bea
&&\oom\ddb_4\de\Us+\frac{3}{2}{\oom}\tr\chi\de\Us=\de\Fs+[(\de\Us),{\hat\Us},\Us^{(Kerr)}]\eql{treqnabbdechi}
\eea
where
\bea
&&\ML[(\de\Us),{\hat\Us},\Us^{(Kerr)}]=-\left[\oom(\de\ddb_4){\hat\Us}+\oom\ddb^{(Kerr)}_4({\hat\Us}-\Us^{(Kerr)})+
+\de\oom\ddb_4^{(Kerr)}\Us^{(Kerr)}\right]\nn\\
&&\ML-\left[\frac{3}{2}{\oom}\tr\chi({\hat\Us}-\Us^{(Kerr)})
+\frac{3}{2}{\oom}\de\tr\chi\Us^{(Kerr)}+\frac{3}{2}{\de\oom}\tr\chi^{(Kerr)}\Us^{(Kerr)}\right]\nn
\eea
We have 
\begin{Le}\label{deUest}
Under all the previous assumptions the following inequality holds in $V_*$
\bea
&&\big|r^{4-\frac{2}{p}}|u|^{2+\de}[(\de\Us),{\hat\Us},\Us^{(Kerr)}]\big|_{p,S}\leq c\left(|r^{4-\frac{2}{p}}|u|^{2+\de}(\de\ddb_4){\hat\Us}|_{p,S}\right.\nn\\
&&\leq +\left.|r^{4-\frac{2}{p}}|u|^{2+\de}\oom\ddb^{(Kerr)}_4({\hat\Us}-\Us^{(Kerr)})|_{p,S}
+|r|u|^{2+\de}\de\oom|_{\infty}|r^{4-\frac{2}{p}}\ddb_4^{(Kerr)}\Us^{(Kerr)}|_{p,S}\right.\nn\\
&&+\frac{3}{2}|r{\oom}\tr\chi|_{\infty}|r^{3-\frac{2}{p}}|u|^{2+\de}({\hat\Us}-\Us^{(Kerr)})|_{p,S}+|r^2|u|^{2+\de}\de\tr\chi|_{\infty}|r^{3-\frac{2}{p}}\Us^{(Kerr)}|_{p,S}\nn\\
&&\left.
+|r|u|^{2+\de}\de\oom|_{\infty}|r\tr\chi^{(Kerr)}|_{\infty}|r^{3-\frac{2}{p}}\Us^{(Kerr)}|_{p,S}+|r^{4-\frac{2}{p}}|u|^{2+\de}(({\hat \Fs}-\Fs^{(Kerr)}))|_{p,S}\right)\nn\\
&&\leq \frac{c\ep_0}{r^4|u|^{2+\de}}\frac{M^2}{R_0^2}\ .
\eea
\end{Le}
\NI{\bf Proof:}
The easy part of the estimates, we do not report here, simply follows by the bootstrap assumptions and the explicit expression of $\Us^{(Kerr)}$,
The bound of the norm $|r^{4-\frac{2}{p}}|u|^{2+\de}(\de\ddb_4){\hat \Us}|_{p,S}$ is done exactly as the previous estimate of
$|r^{3-\frac{2}{p}}|u|^{2+\de}\big((\ddb_4-\ddb^{(Kerr)}_4)O^{(Kerr)}\big)|_{p,S}$, equation \ref{deD4a}, with the only difference that ${\hat \Us}$ has dimension $L^{-2}$ and therefore an extra $r$ decay, therefore we obtain
\bea
|r^{4-\frac{2}{p}}|u|^{2+\de}(\de\ddb_4)\Us^{(Kerr)}|_{p,S}\leq c\ep_0 \frac{M^2}{r^2}\ .\eql{toprove1}
\eea
The term $|r^{4-\frac{2}{p}}|u|^{2+\de}\oom\ddb^{(Kerr)}_4({\hat\Us}-\Us^{(Kerr)})|_{p,S}$ requires the control of $|r^{3-\frac{2}{p}}|u|^{2+\de}({\hat\Us}-\Us^{(Kerr)})|_{p,S}$, the effect of  $\ddb^{(Kerr)}_4$ is only that of adding a power of $r$ in the decay. On the other side 
using Lemma \ref{hatestim}, we have immediately
\bea
|r^{3-\frac{2}{p}}|u|^{2+\de}({\hat\Us}-\Us^{(Kerr)})|_{p,S}\leq c\ep_0\frac{M^2}{R^2_0}
\eea
so that finally we have
\bea
\big|r^{4-\frac{2}{p}}|u|^{2+\de}[(\de\Us),{\hat\Us},\Us^{(Kerr)}]\big|_{p,S}\leq c\ep_0\frac{ M^2}{R_0^2}\ .
\eea
Let us now examine $\de\Fs+({\hat \Fs}-\Fs^{(Kerr)})=\Fs-\Fs^{(Kerr)}$ whose explicit expression is, 
\bea
\de\Fs\!&=&\!-\oom\chih\c\de\Us-\oom\de\chih\Us^{(Kerr)}-\de\oom\chih^{(Kerr)}\Us^{(Kerr)}\nn\\
\!&-&\!2\chih\nabb\de\chih-2\chih(\de\nabb)\chih^{(Kerr)}-2\de\chih\ \!\nabb^{(Kerr)}\chih^{(Kerr)}\nn\\
\!&-&\!\eta\chih\c\de\chih-\eta\de\chih\c\chih^{(Kerr)}-\de\eta|\chih^{(Kerr)}|^2\nn\\
\!&+&\!\tr\chi\ \chih\c\de\etab+\tr\chi\ \de\chih\c\etab^{(Kerr)}+\de\tr\chi\ \chih^{(Kerr)}\c\etab^{(Kerr)}\nn\\
\!&-&\!\tr\chi\de\beta-\de\tr\chi\b^{(Kerr)}\ .\eql{deF}
\eea
The first term of $\de\Fs$ has the following estimate
\bea
|r^{4-\frac{2}{p}}|u|^{2+\de}\oom\chih\c\de\Us|_{p,S}\leq c\frac{M^2}{r^3}|r^{4-\frac{2}{p}}|u|^{2+\de}\de\Us|\eql{esta1}
\eea
The norms of all the other terms can be estimated as follows, the proof is in the appendix:
\bea
|r^{4-\frac{2}{p}}|u|^{2+\de}(\de\Fs-\oom\chih\c\de\Us) |_{p,S}\leq {c\ep_0}\!\left(\frac{M^2}{R_0^2}+\frac{M^4}{R_0^4}\right)\ .\eql{esta2}
\eea

\NI Again proceeding as we did in general for $\de O$, applying Gronwall inequality and  Lemma 4.1.5 of \cite{Kl-Ni:book} we obtain the following estimate, with $\si>0$,
\bea
&&\ML\ML\||u|^{2+\de}r^{(3-\si)-\frac{2}{p}}\de\Us|_{p,S}(u,\ub)\leq c_0\!\left(||u|^{2+\de}r^{(3-\si)-\frac{2}{p}}\de\Us|_{p,S}(u,\ub_*)\right.\eql{deUseq}\\
&&\ML\ML+\left.\int_{\ub}^{\ub_*}\!\left[||u|^{2+\de}r^{(3-\si)-\frac{2}{p}}[(\de U)U^{(Kerr)}]|_{p,S}
+||u|^{2+\de}r^{(3-\si)-\frac{2}{p}}\de\Fs|_{p,S}\right]\!\!(u,\ub')\!\right)\ .\nn
\eea 
Integrating, provided $M/R_0<<1$ and provided the last slice canonical problem is under control we obtain that 
\bea
|\de\Us|\leq \frac{c\varepsilon}{r^3|u|^{2+\de}}+c\frac{M}{R_0}\frac{c\varepsilon}{r^3|u|^{2+\de}}\leq \frac{\ep_0}{2r^3|u|^{2+\de}}\ .
\eea
if
\bea
c(\varepsilon+\frac{M}{R_0}\ep_0)\leq \frac{\ep_0}{2}\ .
\eea
{\bf Remark:} {\em
Observe that when starting from the transport equation for $\de\Us$, equation \ref{treqnabbdechi}, we write a transport equation for the $|\c|_{p,S}$ norms we use a Fermi trasported frame. This Fermi transported frame is used only at this stage, while the ``Kerr decoupling" is performed at the level of the tensorial equations.}
\smallskip

\NI Next step is to obtain from the knowledge of the bounds for the norms of $\Us$ the bounds for the norms of $\nabb\tr\chi$ and for $\chih$.\footnote{Observe that the way we prove this result here is slightly different from what has been done in \cite{Kl-Ni:book}, see in particular remark 1 at page 132.}
$\chih$ satisfies the following equation, see \cite{Kl-Ni:book} eq. (4.3.13),
\bea
\divv\chih+\ze\c\chih-\frac{\oom}{2}\Us+\b=0\ .
\eea
Therefore:
\bea
\divv\de\chih=-(\de\divv){\widehat{(\chih)}}-\ze\c\de\chih-\de\ze\c{\widehat{(\chih)}} +\frac{\oom}{2}\de\Us+\!\frac{\de\oom}{2}{\hat \Us}\!-\de\b\ \ \ \ \ \eql{Hodge1}
\eea
where
\bea
\de\b=\b-{\hat \b}
\eea
and $\hat \b$ is defined as the corresponding connection coefficients.
As we have
\bea
(\de\divv){\widehat{(\chih)}}\!&=&\!\de\ga^{\si\mu}\nabb_{\si}{\widehat{(\chih)}}_{\mu\nu}+{\ga^{(Kerr)}}^{\si\mu}\de\nabb_{\si}{\widehat{(\chih)}}_{\mu\nu}
\eea
it follows that to use \ref{Hodge1} to estimate $\nabb\de\chih$ we need to be able to estimate the right hand side and for that we need to have only a norm estimate for  the first derivatives of  $\de\ga$ which are present in $\de\nabb_{\si}$. 
From \ref{Hodge1} we obtain immediately the following estimate
\bea
&&\ML||u|^{2+\de}r^{3-\frac{2}{p}}\nabb\de\chih|_{p,S}\leq c\left(||u|^{2+\de}r^{3-\frac{2}{p}}(\de\divv){\widehat{(\chih)}}|_{p,S}+||u|^{2+\de}r^{3-\frac{2}{p}}\ze\c\de\chih|_{p,S}\right.\nn\\
&&\ML\left.+||u|^{2+\de}r^{3-\frac{2}{p}}\de\ze\c{\widehat{(\chih)}}|_{p,S}+||u|^{2+\de}r^{3-\frac{2}{p}}\frac{\oom}{2}\de\Us|_{p,S}+||u|^{2+\de}r^{3-\frac{2}{p}}\frac{\de\oom}{2}{\hat \Us}|_{p,S}\right.\nn\\
&&\ML\left.+||u|^{2+\de}r^{3-\frac{2}{p}}\de\b|_{p,S}\right)\nn\\
&&\ML\leq c\left(||u|^{2+\de}r^{3-\frac{2}{p}}(\de\divv){\widehat{(\chih)}}|_{p,S}+||u|^{2+\de}r^{2-\frac{2}{p}}\de\chih|_{p,S}|r\ze|_{\infty}
+||u|^{2+\de}r^{2-\frac{2}{p}}\de\ze|_{p,S}|r{\widehat{(\chih)}}|_{\infty}\nn\right.\\
&&\ML\left.+||u|^{2+\de}r^{3-\frac{2}{p}}\de\Us|_{p,S}|\oom|_{\infty}+||u|^{2+\de}r^{1-\frac{2}{p}}{\de\oom}|_{p,S}|r^2{\hat \Us}|_{\infty}\right)\nn\\
&&\ML\leq c\left(\ep_0\frac{M^2}{R_0^2}+{\tilde c}_8\varepsilon+c\ep_0\frac{M}{R_0}\right)\ \leq \frac{\ep_0}{2}\eql{dechihest}
\eea
provided
\bea
c\ep_0\left(\frac{M}{R_0}+\frac{M^2}{R_0^2}\right)+{\tilde c}_8\varepsilon\leq \frac{\ep_0}{2}\ .
\eea

\NI{\bf Remarks:} {\em 

\NI a): The estimates of the quantities with the hat  is the same as the one for the Kerr terms. In fact their difference is a small correction as it has been proved in 
Lemma \ref{hatestim}.
\smallskip

\NI b): It is here when we want to close the bootstrap that we have to require $\de=\frac{\ep}{2}$, in fact looking at the inequality \ref{dechihest} it follows that the term depending on $\de\b$ is bounded only if $\de\leq \frac{\ep}{2}$. As on the other side we required before, when we control the boundedness of the $\tilde{\cal Q}$ norms, that $\de\geq \frac{\ep}{2}$, see \ref{gaderel}, therefore the conclusion is that in the bootstrap assumptions we must choose
\bea
\de=\frac{\ep}{2} \ .\eql{fincond}
\eea }
\subsubsection{The various radial coordinates}\label{rcoord}

\NI Before examining the transport equations which the $\de O^{(0)}$ coefficients and the estimates for their norms we look at the various quantities that we use as ``radial coordinates" and compared them. The Kerr metric written in the Pretorius Israel coordinates is
\beaa
&&\ML{\bf g}_{(Kerr)}\nn\\
&&\ML=-4\oom_{(Kerr)}^2dud\ub+\ga^{(Kerr)}_{ab}\!\left(d\om^a-X_{(Kerr)}^a(du+d\ub)\right)\!\!\left(d\om^b-X_{(Kerr)}^b(du+d\ub)\right)\nn
\eeaa
there we define
\bea
r_*^{(Kerr)}=\ub-u \eql{defr*kerr}
\eea
Moreover the $\{r_*^{(Kerr)},\th_*\}$ Pretorius Israel coordinates (for Kerr spacetime) can be expressed in terms of th Boyer Lindquist coordinates
\bea
r_*^{(Kerr)}=r_*^{(Kerr)}(\th,\rb)\ \ ,\ \ \th_*=\th^*(\th,\rb)\eql{rbdef}
\eea
where $\rb$ is the Boyer Lindquist radial coordinate, see \cite{Is-Pr}. In the perturbed Kerr metric, see \ref{newmeta},
\beaa
&&\ML{\bf g}_{(pert.Kerr)}\nn\\
&&\ML=-4\oom^2dud\ub+\ga_{ab}\!\left(d\om^a-({X^a_{(Kerr)}}du+X^ad\ub)\right)\!\!\left(d\om^b-({X^b_{(Kerr)}}du+X^bd\ub)\right)\ .\ \ \ \ \ \ \ \ \
\eeaa
and again
\bea
r_*=\ub-u \ .\eql{defr*pertkerr}
\eea
Then we define a radial function $r(u,\ub)$ as proportional to the square root of the corresponding surface $S(u,\ub)$, both for the Kerr and for the perturbed Kerr case,
more precisely
\bea
&&r_{(Kerr)}=r_{(Kerr)}(u,\ub)=(\sqrt{4\pi})^{-\frac{1}{2}}|S_{(Kerr)}(u,\ub)|^{\frac{1}{2}}\nn\\
&&r=r(u,\ub)=(\sqrt{4\pi})^{-\frac{1}{2}}|S(u,\ub)|^{\frac{1}{2}}\ \ .\eql{rdef}
\eea
We use all these radial functions in a interchangeable way, this is possible as we can control their norm differences.

\NI First of all we want to compare $r^{(Kerr)}_*$ with $\rb$. It follows from the explicit expression for $r^{(Kerr)}_*$ in \cite{Is-Pr} that we have
\bea
r^{(Kerr)}_*\leq \rb\left(1+c\frac{M}{R_0}\right)\ ,\eql{rest1}
\eea
We want now to compare $r_{(Kerr)}(u,\ub)$ with $r^{(Kerr)}_*$. We have
\bea
&&4\pi r_{(Kerr)}^2(u,\ub)\equiv|S_{(Kerr)}(u,\ub)|=\int_{S(u,\ub)}d\mu_{\ga}=\int_0^{2\pi}d\phi\int\sqrt{\frac{{\cal L}^2}{R^2}R^2\sin\th^2}d\la d\phi\nn\\
&&=2\pi\int_0^{\pi}{\cal L}\sin\th\frac{\pr\la}{\pr\th}\bigg|_{r_*}d\th
\eea
As from \cite{Is-Pr} it follows easily
\bea
\frac{\pr\la}{\pr\th}\bigg|_{r_*}=\frac{Q}{\cal L}
\eea
we have
\bea
&&4\pi r_{(Kerr)}^2(u,\ub)\equiv|S_{(Kerr)}(u,\ub)|=2\pi\int_0^{\pi}Q\sin\th d\th\leq 2\pi\int_0^{\pi}(\rb^2+a^2)\sqrt{1-\frac{a^2\la\lap}{\rb^2+a^2}}\sin\th d\th\nn\\
&&\leq 2\pi\int_0^{\pi}\rb^2(1+\frac{a^2}{\rb^2})\sqrt{1-\frac{a^2\la\lap}{\rb^2+a^2}}\sin\th d\th
\leq 2\pi\int_0^{\pi}r_*^2(1+c\frac{M}{R_0}+c\frac{\ep_0}{r^2|u|^{1+\de}})\sin\th d\th\nn\\
&&\leq 4\pi {r_*^{(Kerr)}}^2\left(1+c\frac{M}{R_0}\right)\ .
\eea
Finally we can easily show that, under the bootstrap assumptions for the metric components, $r(u,\ub)$ is ``near" to $r_{(Kerr)}(u,\ub)$
\bea
|r(u,\ub)-r_{(Kerr)}(u,\ub)|\leq c\frac{\ep_0}{r|u|^{1+\de}}\ .
\eea
Therefore, as expected, if $\frac{M}{R_0}<<1$, we can identify $r(u,\ub),r_{(Kerr)}(u,\ub),r_*,r_*^{(Kerr)}, \rb$.

\subsubsection{The estimates of the $\de{\cal O}^{(0)}$ norms}\label{OOest}
To obtain this result is slightly more delicate than the control of the connection coefficients (the norms of $\de{\cal O}^{(1)}$). The reason being that we have first to find the analogous of the transport equations for the corrections to the Kerr components of the metric $\de\oom$, $\de X^a$ and $\de\ga_{ab}$.
We prove the following lemma:
\begin{Le}
The corrections to the Kerr components of the metric $\de\oom$, $\de X^a$ and $\de\ga_{ab}$, satisfy the following equations:
\bea
&&\ML\pr_{\Nb}(\de\oom)=-2({{\oom^{(Kerr)}}}+\de\oom)^2\de\omb+\left(\frac{\de\oom(\oom+{{\oom^{(Kerr)}}})}{{{\oom^{(Kerr)}}}^2}\right)\pr_{\Nb}{{\oom^{(Kerr)}}}\eql{traspdeOo}\\
&&\nn\\
&&\ML\pr_{\Nb}\de X^a-\oom(\pr_cX_{(Kerr)}^a)\de X^c=-\oom\left(\frac{2}{\oom^{(Kerr)}}\frac{Q\De}{\Si R^2}\frac{\pr}{\pr r_b}{X_{(Kerr)}^a}\right)\de\oom+4\oom^2{\de\ze}^a\ \ \ \ \ \ \ \ \ \  \eql{traspdeX}\\
&&\nn\\
&&\ML\pr_{\Nb}(\de\ga_{ab})-\oom\tr\chib(\de\ga_{ab})=-\left[\frac{\pr X_{(Kerr)}^c}{\pr\om^a}(\de\ga_{cb})+\frac{\pr X_{(Kerr)}^c}{\pr\om^b}(\de\ga_{ac})\right]\eql{traspdega}\\
&&\ML\ \ \ \ \ \ \ \ \ \ \ \ \ \ \ 
+\left[\oom\ga^{(Kerr)}_{ab}\de\tr\chib+\de\oom\ga^{(Kerr)}_{ab}\tr\chib^{(Kerr)}+2{\oom^{(Kerr)}}\de\chibh_{ab}+2\de\oom\chibh^{(Kerr)}_{ab}\right]
\ .\nn
\eea
\end{Le}
\smallskip

\NI{\bf Proof:}
{\bf The equation for $\de X^a$:}
The first thing to remark is that here we are writing equations satisfied by the components of the metric written in the $\{u,\ub,\om^1,\om^2\}$ coordinates.
\NI Observe that with the definition of the $\bf g$ metric, \ref{pertKerr1}, the commutation relation \ref{comm}
has the following aspect
\bea
\oom\left(\pr_{e_4}X_{(Kerr)}^a-\pr_{e_3}X^a\right)=-4\oom^2\ze(e_c)e_c^a
\eea
which we can write as
\bea
&&\ML\oom\pr_{e_3}\de X^a=\oom\left(\pr_{e_4}X_{(Kerr)}^a-\pr_{e_3}X_{(Kerr)}^a\right)+4\oom^2\ze(e_c)e_c^a\nn\\
&&\ML=\oom(\de e_4^{\mu}-\de e_3^{\mu})\pr_{\mu}X_{(Kerr)}^a+4\oom^2{\de\ze}^a+\left[\oom\left(\pr_{{\hat e}_4}X_{(Kerr)}^a-\pr_{{\hat e}_3}X_{(Kerr)}^a\right)
+4\oom^2{\ze^{(Kerr)}}^a\right]\nn\\
&&\ML=\oom(\de e_4^{\mu}-\de e_3^{\mu})\pr_{\mu}X_{(Kerr)}^a+4\oom^2{\de\ze}^a
\eea
This is the equation to use to estimate $\de X$:
\bea
\pr_{e_3}\de X^a=\oom(\de e_4^{\mu}-\de e_3^{\mu})\pr_{\mu}X_{(Kerr)}^a+4\oom^2{\de\ze}^a\ ,\eql{findeX}
\eea
where
\bea
\de e_4=-\frac{\de\oom}{\oom^{(Kerr)}}e_4+\frac{\de X}{\oom}\ \ ;\ \ \de e_3=-\frac{\de\oom}{\oom^{(Kerr)}}e_3\ \ .
\eea
therefore we can rewrite it as
\bea
\pr_{e_3}\de X^a=(\pr_cX_{(Kerr)}^a)\de X^c+\frac{\oom}{\oom^{(Kerr)}}\left(\pr_{e_3}X_{(Kerr)}^a-\pr_{e_4}X_{(Kerr)}^a\right)\de\oom+4\oom^2{\de\ze}^a\ .\ \ \ \ \ \ 
\eea
Observe that
\bea
&&\oom\left(\pr_{e_3}X_{(Kerr)}^a-\pr_{e_4}X_{(Kerr)}^a\right)=\left(\pr_{u}X_{(Kerr)}^a-\pr_{\ub}X_{(Kerr)}^a\right)=-2\frac{\pr}{\pr r_*}X_{(Kerr)}^a\nn\\
&&=-2\frac{\pr\rb}{\pr r_*}\frac{\pr}{\pr\rb}X_{(Kerr)}^a=-2\frac{Q\De}{\Si R^2}\frac{\pr}{\pr\rb}X_{(Kerr)}^a
\eea
and the final expression is
\bea
\pr_{e_3}\de X^a-(\pr_cX_{(Kerr)}^a)\de X^c=-\left(\frac{2}{\oom^{(Kerr)}}\frac{Q\De}{\Si R^2}\frac{\pr}{\pr\rb}X_{(Kerr)}^a\right)\de\oom+4\oom^2{\de\ze}^a\ .\ \ \ \ \ \ \eql{findeX2}
\eea
\smallskip

\NI{\bf Remark:} {\em Observe that these equations refer to the metric components in the $\{u,\ub,\om^1,\om^2\}$ coordinates. We do not have here the previous problem of considering tensor fields tangent to $S$, therefore we do not have to introduce the ``auxiliary" $\hat O$ quantities. }
\medskip

\NI{\bf The equation for $\de\oom$:} We start from the equation 
\[-\frac{1}{2}\dd_3\log\oom=\omb\]
From it
\bea
&&-\frac{1}{2}\dd_3\log\oom=-\frac{1}{2\oom^2}\pr_{\Nb}({\oom^{(Kerr)}}+\de\oom)\nn\\
&&=-\frac{1}{2{\oom^{(Kerr)}}^2}\pr_{\Nb}{\oom^{(Kerr)}}-\left(\frac{1}{2\oom^2}-\frac{1}{2{\oom^{(Kerr)}}^2}\right)\pr_{\Nb}{\oom^{(Kerr)}}-\frac{1}{2\oom^2}\pr_{\Nb}\de\oom\nn
\eea
Therefore
\bea
&&=-\left(\frac{1}{2\oom^2}-\frac{1}{2{\oom^{(Kerr)}}^2}\right)\pr_{\Nb}{\oom^{(Kerr)}}-\frac{1}{2\oom^2}\pr_{\Nb}\de\oom=\left(\frac{1}{2{\oom^{(Kerr)}}^2}\pr_{\Nb}{\oom^{(Kerr)}}+{\hat\omb}\right)+\de\omb\nn
\eea
and the final equation we have for $\de\oom$ is
\bea
\pr_{\Nb}\de\oom=-2({\oom^{(Kerr)}}\!+\de\oom)^2\de\omb+\left(\frac{\de\oom(\oom+{\oom^{(Kerr)}})}{{\oom^{(Kerr)}}^2}\right)\pr_{\Nb}{\oom^{(Kerr)}}\ .\eql{findeOom}
\eea
\medskip

\NI{\bf The equation for $\de\ga_{ab}$:}
\smallskip

\NI After the transport equations for $\de X^a$ and $\de\oom$ we look for the equation for $\ga_{ab}$.\footnote{We use here small latin letters $a,b,...$ to indicate the $\th, \phi$ coordinates, the frame vector fields tangential to $S$ will be indicated with $e_A,e_B...$.}
We recall first the definition of the induced metric on the generic $S(u,\ub)$, whose components we denote $\{\ga^{(S)}_{\ \ \ro\si}\}$,
\bea
\Pi^{\ro}_{\mu}\Pi^{\si}_{\nu}g_{\ro\si}=\ga^{(S)}_{\ \ \mu\nu}
\eea
where
\bea
\Pi^{\ro}_{\mu}=\de^{\ro}_{\mu}-(\th^3_{\mu}e_3^{\ro}+\th^4_{\mu}e_4^{\ro})\ .\eql{Pieq1}
\eea
{\bf Remark:} {\em The ``metric" whose tensor is $\ga^{(S)}_{\mu\nu}$ is the induced metric on $S$; it is easy to see that, nevertheless, $\ga^{(S)}$ is not a $2\times2$ matrix (with $\ga_{ab}$ the only components different from zero); this is not due to a wrong choice of coordinates, but to the fact that the the spacetime is not static; in fact this is not possible even in the Kerr spacetime. Nevertheless the following holds
\bea
\ga^{(S)}_{\ \ ab}=\ga_{ab}\ \ \mbox{and}\ \ \ga^{(S)}(e_A,e_B)=\ga^{(S)}_{\ \ ab}e_A^ae_B^b=\ga_{ab}e_A^ae_B^b\ .
\eea} 
\NI To prove it let us compute the components of $\th^3$ and $\th^4$
\bea
&&\th^3(\c)=-\frac{1}{2}{\bf g}(e_4,\c)\ \ \mbox{therefore}\ \ \th^3_{\mu}=-\frac{1}{2}g_{\mu\nu}e_4^{\nu}\nn\\
&&\th^3_{u}=-\frac{1}{2}g_{u\ub}e_4^{\ub}-\frac{1}{2}g_{uc}e_4^{c}=-\frac{1}{2\oom}(-2\oom^2+X\c X_{(Kerr)})-\frac{1}{2\oom}(-\ga_{cd}X_{(Kerr)}^dX^{c})=\oom\nn\\
&&\th^3_{\ub}=-\frac{1}{2}g_{\ub\ub}e_4^{\ub}-\frac{1}{2}g_{\ub c}e_4^{c}=-\frac{1}{2\oom}X\c X_{(Kerr)}-\frac{1}{2}(-\ga_{cd}{\hat X}^d\frac{X^{c}}{\oom})=0\nn\\
&&\th^3_{a}=-\frac{1}{2}g_{a\ub}e_4^{\ub}-\frac{1}{2}g_{ac}e_4^{c}=-\frac{1}{2\oom}(-\ga_{ac}X^c)-\frac{1}{2\oom}\ga_{ac}X^{c}=0\eql{Pieq2}
\eea
Analogously
\bea
\th^4_{\ub}=\oom\ \ ,\ \ \th^4_{u}=\th^4_{c}=0\ .
\eea
Therefore we have
\bea
&&\Pi^a_{\ro}=\de^a_{\ro}-(\th^3_{\ro}e_3^{a}+\th^4_{\ro}e_4^{a})=\de^a_{\ro}-(\de^u_{\ro}X_{(Kerr)}^{a}+\de^{\ub}_{\ro}X^{a})\nn\\
&&\Pi^u_{\ro}=\de^u_{\ro}-(\th^3_{\ro}e_3^{u}+\th^4_{\ro}e_4^{u})=\de^u_{\ro}-\th^3_{\ro}e_3^{u}=\de^u_{\ro}-\de^u_{\ro}=0\nn\\
&&\Pi^{\ub}_{\ro}=\de^{\ub}_{\ro}-(\th^3_{\ro}e_3^{\ub}+\th^4_{\ro}e_4^{\ub})=\de^{\ub}_{\ro}-\th^4_{\ro}e_4^{\ub}=\de^{\ub}_{\ro}-\de^{\ub}_{\ro}=0\eql{Pieq3}
\eea
and
\bea
\Pi^{\mu}_{\ro}=\de^{\mu}_c\de^c_{\ro}-\de^{\mu}_c(\de^u_{\ro}X_{(Kerr)}^{c}+\de^{\ub}_{\ro}X^{c})\ .
\eea
Therefore
\bea
&&\ML\ga^{(S)}_{\ \ro\si}=\Pi^{\mu}_{\ro}\Pi^{\nu}_{\si}g_{\mu\nu}=\left(\de^{\mu}_c\de^c_{\ro}-\de^{\mu}_c(\de^u_{\ro}X_{(Kerr)}^{c}+\de^{\ub}_{\ro}X^{c})\right)
\left(\de^{\nu}_d\de^d_{\si}-\de^{\nu}_d(\de^u_{\si}X_{(Kerr)}^{d}+\de^{\ub}_{\si}X^{d})\right)g_{\mu\nu}\\
&&\ML=\ga_{cd}\de^c_{\ro}\de^d_{\si}-\ga_{cd}\left((\de^u_{\ro}X_{(Kerr)}^{c}+\de^{\ub}_{\ro}X^{c})\de^d_{\si}+\de^c_{\ro}(\de^u_{\si}X_{(Kerr)}^{d}+\de^{\ub}_{\si}X^{d})\right)\nn\\
&&\ML
+\ga_{cd}(\de^u_{\ro}X_{(Kerr)}^{c}+\de^{\ub}_{\ro}X^{c})(\de^u_{\si}X_{(Kerr)}^{d}+\de^{\ub}_{\si}X^{d})\nn\\
&&\ML=\ga_{cd}\de^c_{\ro}\de^d_{\si}-\ga_{cd}\left((\de^u_{\ro}X_{(Kerr)}^{c}+\de^{\ub}_{\ro}X^{c})\de^d_{\si}+\de^c_{\ro}(\de^u_{\si}X_{(Kerr)}^{d}+\de^{\ub}_{\si}X^{d})\right)+\de^u_{\ro}\de^u_{\si}|X_{(Kerr)}|^2+\de^{\ub}_{\ro}\de^{\ub}_{\si}|X|^2\nn\\
&&\ML+(\de^u_{\ro}\de^{\ub}_{\si}+\de^{\ub}_{\ro}\de^u_{\si})X_{(Kerr)}\c X\ .\nn
\eea

\NI{\bf The equation for $\de\ga_{ab}$\ \!:}
Let us start observing that 
\[\ga_{ab}\equiv {\bf g}\left(\frac{\pr}{\pr{\om^a}},\frac{\pr}{\pr{\om^b}}\right)\ .\]
Then we have
\bea
&&\ML\pr_{\Nb}\ga_{ab}=\dd_{\Nb}\!\left({\bf g}\left(\frac{\pr}{\pr{\om^a}},\frac{\pr}{\pr{\om^b}}\right)\right)=
{\bf g}\left(\dd_{\Nb}\frac{\pr}{\pr{\om^a}},\frac{\pr}{\pr{\om^b}}\right)+{\bf g}\left(\frac{\pr}{\pr{\om^a}},\dd_{\Nb}\frac{\pr}{\pr{\om^b}}\right)\nn\\
&&\ML{\bf g}\left([\Nb,\frac{\pr}{\pr{\om^a}}],\frac{\pr}{\pr{\om^b}}\right)+{\bf g}\left(\frac{\pr}{\pr{\om^a}},[\Nb,\frac{\pr}{\pr{\om^b}}]\right)
+{\bf g}\left(\dd_{\frac{\pr}{\pr{\om^a}}}\Nb,\frac{\pr}{\pr{\om^b}}\right)+{\bf g}\left(\frac{\pr}{\pr{\om^a}},\dd_{\frac{\pr}{\pr{\om^b}}}\Nb\right)\nn\\
&&\ML=2\oom\chi_{ab}+{\bf g}\left([\Nb,\frac{\pr}{\pr{\om^a}}],\frac{\pr}{\pr{\om^b}}\right)+{\bf g}\left(\frac{\pr}{\pr{\om^a}},[\Nb,\frac{\pr}{\pr{\om^b}}]\right)\ .
\eea
We have
\bea
\left[\Nb,\frac{\pr}{\pr{\om^a}}\right]=-\frac{\pr X_{(Kerr)}^c}{\pr\om^a}\frac{\pr}{\pr{\om^c}}\ ,
\eea
therefore
\bea
\pr_{\Nb}\ga_{ab}=2\oom\chib_{ab}-\frac{\pr X_{(Kerr)}^c}{\pr\om^a}\ga_{cb}-\frac{\pr X_{(Kerr)}^c}{\pr\om^b}\ga_{ca}\ .
\eea
Analogously
\bea
\pr_{\Nb}\ga^{(Kerr)}_{ab}=2{\oom^{(Kerr)}}\chib^{(Kerr)}_{ab}-\frac{\pr X_{(Kerr)}^c}{\pr\om^a}\ga^{(Kerr)}_{cb}-\frac{\pr X_{(Kerr)}^c}{\pr\om^b}\ga^{(Kerr)}_{ca}\ .
\eea
and, subtracting,
\bea
&&\pr_{\Nb}(\de\ga_{ab})-\oom\tr\chib(\de\ga_{ab})=-\left[\frac{\pr X_{(Kerr)}^c}{\pr\om^a}(\de\ga_{cb})+\frac{\pr X_{(Kerr)}^c}{\pr\om^b}(\de\ga_{ac})\right]\\
&&+\left[\oom\ga^{(Kerr)}_{ab}\de\tr\chib+\de\oom\ga^{(Kerr)}_{ab}\tr\chib^{(Kerr)}+2{\oom^{(Kerr)}}\de\chibh_{ab}+2\de\oom\chibh^{(Kerr)}_{ab}\right]\ .\nn
\eea

\subsubsection{A discussion about the  loss of derivatives}\label{SSloss}

The previous equations seem to imply a loss of derivatives potentially dangerous. In fact on the right hand side of all the three equations there are terms of the order of first derivatives of the metric components, namely connection coefficients like $\om,\ze,\tr\chi$. Nevertheless this loss of derivatives is not harmful as it does not propagate when we estimate the connection coefficients and their tangential derivatives up to the order we need. To show it let us look first to the transport equation for $\de\ga$, equation \ref{traspdega}. This equation tells us that to control the norm of $\de\ga$ we have to control the norm of $\de\chih$ so that to control the norm of 
$\prb\de\ga$ we have to control $\nabb\de\chih$. What is therefore needed is that to control the norm of $\nabb\chi$ we do not have to use the second derivative 
$\prb^2\de\ga$ as this will start a never ending recursive mechanism. To control $\nabb\de\chih$ we have first to control $\de\Us$ which has been already done and looking at these estimates it is immediately to recognize that the control of second tangential derivatives of $\de\ga$ is never required. Then one has to look to how we control $\nabb\de\chih$ once we control $\de\Us$. This has also been done previously see equation \ref{Hodge1} and after and it follows that for these estimates no second derivatives of $\de\ga$ are required.
\smallskip

\NI Let us examine now the transport equation for $\de\oom$. From equation \ref{traspdeOo} it follows that to control the norm of $\de\oom$ we have to control the norm of $\de\omb$. On the other side to control the norm of 
\bea
\nabb\de\oom=\nabb\oom-(\nabb-\nabb^{(Kerr)}){\oom^{(Kerr)}}-\nabb^{(Kerr)}{\oom^{(Kerr)}}=\nabb\oom-\nabb^{(Kerr)}{\oom^{(Kerr)}}\nn
\eea
we have just to recall that 
\bea
\nabb\de\oom=2^{-1}(\de\eta+\de\etab)
\eea
which implies we do not derive this equation, but we use the estimates for $\de\eta$ and $\de\etab$ implying that to control the first derivative of $\de\oom$ we need to control $\de\eta$ and $\de\etab$. therefore we are only left to show that to control these (corrections) of the connection coefficients we do not need second derivatives of the (corrections of) the metric components. This follows easily as looking at the Hodge systems for $\eta$ and $\etab$, see equations (4.3.34), (4.3.35) in \cite{Kl-Ni:book} it follows that as only first derivatives are there only first derivatives of the metric corrections can be present, moreover on the right hand side of these equations we have to control two scalar functions called $\mu$ and $\mub$ respectively. Then looking at their transport equations it is easy to recognize that again there only first derivatives $\nabb\eta$ and $\nabb\etab$ appear and when we look for the equations for the corrections $\de\mu$ and $\de\mub$ again only first derivatives of the metric corrections can be present. Therefore we are only left to investigate the transport equation for $\de X$., \ref{traspdeX}. Again to estimate the first derivative of $\de X$ we need to control the first derivative of $\de\ze$ or equivalently the first derivative of $\de\eta$ and $\de\etab$. This quantities are controlled again by the Hodge systems for $\eta$ and $\etab$, equations (4.3.34), (4.3.35) in \cite{Kl-Ni:book}, and we have already seen that the right hand side needs only the control of the first derivatives of the (corrections of the) metric components and in the left hand side the only term depending on the metric is $\de\nabb$ which depends again only on first derivatives. Therefore the argument is complete and there is not any loss of derivatives, as expected.
\smallskip

\NI Next lemma proves the first part of the ``main Theorem", Theorem \ref{Theorem to prove} . His statement is consistent with the discussion made in subsection \ref{S.S.stepsfortheproof}
\begin{Le}\label{finOOest}
Assume that in $V_*$ we have already proved that the norms $\de{\cal O}$ satisfy better estimates than those of the bootstrap assumptions, namely
\[\de{\cal O}\leq \frac{\ep_0}{N_0}\ ,\]
with $N_0$ a large integer number, then, assuming for the ${\cal O}^{(0)}$ norms appropriate initial data conditions, we prove that in $V_*$ better estimates hold:
\bea
|r|u|^{2+\de}\de\oom|\leq  \frac{\ep_0}{2}\  \ ;\ \ |r^2|u|^{2+\de}\de X|\leq  \frac{\ep_0}{2}\ \ ;\ \ ||u|^{1+\de}\de \ga|\leq  \frac{\ep_0}{2}\ .
\eea
\end{Le}
\NI{\bf Proof:}
\smallskip

\NI{\bf $\{\de\oom\}$:} $\de\oom$ satisfies the equation
\bea
\pr_{\Nb}(\de\oom)=-2(\oom)^2\de\omb+(\de\oom)\!\left(\frac{(\oom+{\oom^{(Kerr)}})}{{\oom^{(Kerr)}}^2}\right)\pr_{\Nb}{\oom^{(Kerr)}}\ .\eql{findeOom}
\eea
It follows
\bea
\pr_{\Nb}|\de\oom|\leq F|\de\oom|+2\oom^2|\de\omb|
\eea
where
\bea
F=\big|{(\oom+{\oom^{(Kerr)}})}{{\oom^{(Kerr)}}^{-2}}\pr_{\Nb}{\oom^{(Kerr)}}\big|=O\!\left(\frac{M}{r^2}\right)\ .
\eea
and integrating on $\Cb(\ub;[u_0,u])$ we obtain
\bea
|\de\oom|(\ub,u)\leq c\!\left(|\de\oom|_{\Cb(\ub)\cap\Si_0}+\int_{u_0}^udu'|2\oom^2\de\omb|(\ub,u')\right)\  .
\eea
As we have already proved that the estimates for $\de{\cal O}^{(1)}$ are better than the bootstrap assumptions, then  for $\de\omb$ the following estimate holds
\[|r^2|u|^{2+\de}\de\omb|\leq \frac{\ep_0}{N_0}\]
and we can obtain from the previous inequality the following one
\bea
&&\ML|r^2|u|\de\oom|(\ub,u)\leq |r^{3+\de}\de\oom|_{\Cb(\ub)\cap\Si_0}+\int_{u_0}^udu'|2\oom^2\de\omb|(\ub,u')r(u',\ub)^2|u'|\\
&&\ML\leq c\!\left(|r^3\de\oom|_{\Cb(\ub)\cap\Si_0}+\frac{\ep_0}{N_0}\int_{u_0}^udu'\frac{1}{|u'|^{1+\de}}\right)
\leq c\!\left(|r^3\de\oom|_{\Cb(\ub)\cap\Si_0}+\frac{\ep_0}{N_0|u|^{\de}}\right)\ .\ \ \ \ \ \ \ \ \ \ \nn
\eea
From it
\bea
|r^2|u|^{1+\de}\de\oom|(\ub,u)\leq c\!\left(|r^{3+\de}\de\oom|_{\Cb(\ub)\cap\Si_0}+\frac{\ep_0}{N_0}\right)\leq \frac{\ep_0}{N_1}\ \ .
\eea
with $2^{-1}>N^{-1}_1>N^{-1}_0$, choosing the initial data sufficiently small and $N_0$ sufficiently large.
\smallskip

\NI{\bf $\{\de X\}$:} $\de X$ satisifies the following equation
\bea
\pr_{e_3}\de X^a-(\pr_cX_{(Kerr)}^a)\de X^c=-\left(\frac{2}{\oom^{(Kerr)}}\frac{Q\De}{\Si R^2}\frac{\pr}{\pr\rb}X_{(Kerr)}^a\right)\de\oom+4\oom^2{\de\ze}^a\ \ \ \ \ \eql{findeX2a}
\eea
where we have immediately, with $\si>0$,
\bea
|\pr_cX_{(Kerr)}^a|=|(\pr_c\om_B)\de_{\phi}^a|=\left|(\pr_c\frac{2Ma\rb}{\Si R^2})\right|\leq \frac{M^4}{R_0^{4-\si}}\frac{1}{r^{1+\si}}
\eea
and
\bea
\left|\left(\frac{2}{\oom^{(Kerr)}}\frac{Q\De}{\Si R^2}\frac{\pr}{\pr\rb}X_{(Kerr)}^a\right)\right|\leq \frac{M^2}{R_0^2}\frac{1}{r^2}\ .
\eea
Finally using the improved estimates for the $\de{\cal O}^{(1)}$ norms, see also Lemma \ref{Comm3}, we have
\bea
\left|\de\ze^a\right|\leq c\frac{\ep_0}{N_0}\frac{1}{r^3|u|^{2+\de}}\ .
\eea
Using these estimates from the definition
\[|\de X|=\sqrt{\sum_a|\de X^a|^2}\]
we write, defining,
\bea
&&G=-\left(\frac{2}{\oom^{(Kerr)}}\frac{Q\De}{\Si R^2}\frac{\pr}{\pr\rb}X_{(Kerr)}^a\right)\ ,
\eea
\bea
&&\ML\ML\pr_{\Nb}|\de X|^2=2|\de X|\pr_{\Nb}|\de X|=2\oom\sum_a\de X^a\pr_{e_3}\de X^a
=2\oom\sum_a\de X^a\left[(\pr_cX_{(Kerr)}^a)\de X^c+G\de\oom+4\oom^2{\de\ze}^a\right]\nn\\
&&\ML\ML\leq 4\oom(\sum_a|(\pr_cX_{(Kerr)}^a)|)|\de X|^2+|G||\de\oom||\de X|+4\oom^2(\sum_a|{\de\ze}^a|)|\de X|
\eea
and immediately
\bea
\pr_{\Nb}|\de X|\!&\leq&\! 4\oom(\sum_a|(\pr_cX_{(Kerr)}^a)|)|\de X|+|G||\de\oom|+4\oom^2(\sum_a|{\de\ze}^a|)\nn\\
\!&\leq&\! c\frac{M^4}{R_0^{4-\si}}\frac{1}{r^{1+\si}}|\de X|+\left[\frac{M^2}{R_0^2}\frac{\ep_0}{r^4|u|^{1+\de}}+\frac{\ep_0/N_0}{r^3|u|^{2+\de}}\right]\nn\\
\!&\leq&\! c\frac{M^4}{R_0^{4-\si}}\frac{1}{r^{1+\si}}|\de X|+\frac{\ep_0}{r^3|u|^{2+\de}}\left(\frac{1}{N_0}+c\frac{M^2}{R_0^2}\right)\ .
\eea
Applying the Gronwall Lemma we obtain
\bea
|\de X|\!&\leq&\! \exp\left\{{c\frac{M^4}{R_0^{4-\si}}\int_{u_0}^u\frac{1}{r^{1+\si}}}\right\}\left[c\ep_0\left(\frac{1}{N_0}+\frac{M^2}{R_0^2}\right)\int_{u_0}^u\frac{1}{r^3|u'|^{2+\de}}du'\right]
\nn\\
\!&\leq&\!  \exp\left\{c\frac{M^4}{R_0^{4}}\right\}c\ep_0\left(\frac{1}{N_0}+\frac{M^2}{R_0^2}\right)\frac{1}{r^3|u|^{1+\de}}\leq \frac{\ep_0}{2}\frac{1}{r^3|u|^{1+\de}}
\eea
which proves the result.
\medskip

\NI{\bf $\{\de\ga\}$:} 
The equation \ref{traspdega} satisfied by $\de\ga_{ab}$, 
\bea
&&\ML\pr_{\Nb}(\de\ga_{ab})-\oom\tr\chib(\de\ga_{ab})=-\left[\frac{\pr X_{(Kerr)}^c}{\pr\om^a}(\de\ga_{cb})+\frac{\pr X_{(Kerr)}^c}{\pr\om^b}(\de\ga_{ac})\right]\nn\\
&&\ML\ \ \ \ \ \ \ \ \ \ \ \ \ \ \ 
+\left[\oom\ga^{(Kerr)}_{ab}\de\tr\chib+\de\oom\ga^{(Kerr)}_{ab}\tr\chib^{(Kerr)}+2{\oom^{(Kerr)}}\de\chibh_{ab}+2\de\oom\chibh^{(Kerr)}_{ab}\right]\nn
\eea
can be rewritten as
\bea
\ML\ML\pr_{\Nb}(\de\ga_{ab})-\overline{\oom\tr\chib}(\de\ga_{ab})=(\oom\tr\chib-\overline{\oom\tr\chib})(\de\ga_{ab})+\left(G^c_a(\de\ga_{cb})+G^c_b(\de\ga_{ca})\right)
+F(O^{(Kerr)},\de\oom,\de\tr\chib,\de\chib)\ \ \ \ \ \ \ \ \ 
\eea
where
\bea
\ML F(O^{(Kerr)},\de\oom,\de\tr\chib,\de\chib)=\left[\oom\ga^{(Kerr)}_{ab}\de\tr\chib+\de\oom\ga^{(Kerr)}_{ab}\tr\chib^{(Kerr)}+2{\oom^{(Kerr)}}\de\chibh_{ab}+2\de\oom\chibh^{(Kerr)}_{ab}\right]\ .\ \ \ \ \ \ \ 
\eea
As 
\bea
\pr_{\Nb}r(u,\ub)=\frac{\pr}{\pr u}r(u,\ub)=\frac{r(u,\ub)}{2}\overline{\oom\tr\chib}
\eea
it follows
\bea
&&\pr_{\Nb}\frac{(\de\ga_{ab})}{r^2}=\frac{1}{r^2}\pr_{\Nb}(\de\ga_{ab})-\frac{2}{r^3}\frac{r}{2}\overline{\oom\tr\chib}(\de\ga_{ab})=\frac{1}{r^2}\left[\pr_{\Nb}(\de\ga_{ab})-\overline{\oom\tr\chib}(\de\ga_{ab})\right]\nn\\
&&=\frac{1}{r^2}\left[(\oom\tr\chib-\overline{\oom\tr\chib})(\de\ga_{ab})+\left(G^c_a(\de\ga_{cb})+G^c_b(\de\ga_{ca})\right)
+F(O^{(Kerr)},\de\oom,\de\tr\chib,\de\chib)\right]\nn
\eea
Therefore
\bea
\pr_{\Nb}(r^{-2}\de\ga_{ab})=(\oom\tr\chib-\overline{\oom\tr\chib})(r^{-2}\de\ga_{ab})+\left(G^c_a(r^{-2}\de\ga_{cb})+G^c_b(r^{-2}\de\ga_{ca})\right)+\frac{F}{r^2}\ \ \ \ \ \ 
\eea
Defining
\[|\de\ga|\equiv\sum_{ab}|\de\ga_{ab}|\]
we have the following inequality:
\bea
&&\ML\pr_{\Nb}\frac{|\de\ga|^2}{r^4}=\frac{2|\de\ga|}{r^2}\pr_{\Nb}\frac{|\de\ga|}{r^2}
=\frac{1}{r^4}\left[2|\de\ga|\pr_{\Nb}|\de\ga|-\frac{4}{r}\frac{r\overline{\oom\tr\chib}}{2}|\de\ga|\right]\nn\\
&&\ML=\sum_{ab}(\de\ga_{ab})\left[\frac{1}{r^4}\left(-\overline{\oom\tr\chib}(\de\ga_{ab})+\pr_{\Nb}(\de\ga_{ab})\right)\right]\nn\\
&&\ML=\sum_{ab}\frac{(\de\ga_{ab})}{r^2}\left[(\oom\tr\chib-\overline{\oom\tr\chib})\frac{(\de\ga_{ab})}{r^2}+\left(G^c_a\frac{(\de\ga_{cb})}{r^2}+G^c_b\frac{(\de\ga_{ca})}{r^2}\right)+\frac{F(O^{(Kerr)},\de\oom,\de\tr\chib,\de\chib)}{r^2}\right]\nn\\
&&\leq \frac{|\de\ga|}{r^2}\left[|(\oom\tr\chib-\overline{\oom\tr\chib})|\frac{|\de\ga|}{r^2}+2|G|\frac{|\de\ga|}{r^2}+\frac{|F|}{r^2}\right]\nn
\eea
which implies
\bea
\pr_{\Nb}\left(\frac{|\de\ga|}{r^2}\right)\leq \left[|(\oom\tr\chib-\overline{\oom\tr\chib})+2|G|\right]\left(\frac{|\de\ga|}{r^2}\right)+\frac{|F|}{r^2}\ .
\eea
As before we have for $|G|$ and $|(\oom\tr\chib-\overline{\oom\tr\chib})|$ the following estimates, with $\si>0$
\bea
|G|=\sum_{ac}|G^a_c|\leq c\frac{M^4}{R_0^{4-\si}}\frac{1}{r^{1+\si}}\ \ ,\ \ |(\oom\tr\chib-\overline{\oom\tr\chib})|\leq \frac{c}{r^2}\ ,
\eea
therefore, applying Gronwall Lemma,
\bea
\left(\frac{|\de\ga|}{r^2}\right)(u,\ub)\!&\leq&\!\left(\frac{|\de\ga|}{r^2}\right)(u_0,\ub)+\exp\left\{{c\frac{M^4}{R_0^{4-\si}}\int_{u_0}^u\frac{1}{r^{1+\si}}}\right\}
\int_{u_0}^u\frac{|F|}{r^2}(u',\ub)du'\nn\\
\!&\leq& c\int_{u_0}^u\frac{|F|}{r^2}(u',\ub)du'
\eea
Observe now that the following inequality holds
\bea
\frac{|F|}{r^2}\!&\leq&\!\frac{1}{r^2}\left|\oom\ga^{(Kerr)}_{ab}\de\tr\chib+\de\oom\ga^{(Kerr)}_{ab}\tr\chib^{(Kerr)}+2{\oom^{(Kerr)}}\de\chibh_{ab}+2\de\oom\chibh^{(Kerr)}_{ab}\right|\nn\\
\!&\leq&\!\frac{1}{r^2}\left[|\oom|r^2|\de\tr\chib|+|\de\oom|r^2|\tr\chib^{(Kerr)}|+2|{\oom^{(Kerr)}}||\de\chibh|+2|\de\oom||\chibh^{(Kerr)}|\right]\nn\\
\!&\leq&\!c\frac{\ep_0}{r^2|u|^{2+\de}}\ .
\eea
Therefore integrating the final result is
\bea
||u|^{1+\de}\de\ga|\leq c{\ep_0}\left(\frac{1}{N_0}+\frac{1}{N_1}\right)\leq \frac{\ep_0}{2}\ ,\eql{estga}
\eea
proving the result.

\subsection{The last slice canonical foliation}\label{lastlslice}
To control $\nabb\de\tr\chi$ we use trasport equation \ref{treqnabbdechi} and this requires an estimate of $\nabb\de\tr\chi$ on the last slice of $V_*$ we denote $\Cb_*$, this as discussed in \cite{Kl-Ni:book} and in \cite{Niclast} requires a delicate choice of its foliation. Let us recall the main problem we have to cure; the equation for $\tr\chi$ along an incoming cone is
\bea
&&\dd_3\tr\chi+\frac{1}{2}\tr\chib{\tr\chi}+(\dd_3\log\oom)\tr\chi+\chibh\c\chih-2|\zeta|^2-4\zeta\c\nabb\log\oom-2|\nabb\log\oom|^2\nn\\
&&\ \ \ \ \ \ \ \ \ \ \ \ \ \ \ \ \ \ \ \ \ \ \ =2(\lapp\log\oom+\divv\zeta+\ro)\  \ .\eql{lastslice1}
\eea
Looking at this expression it is clear that there is a loss of derivatives due to the term in the right hand side. To cure this problem in \cite{Kl-Ni:book} we require that $\log\oom$ satisfies the equation
\bea
\lapp\log\oom=-\divv\ze-\ro+\overline{\ro}\ .
\eea
To satisfy it we introduce a background foliation whose leaves are the level surfaces of the affine parameter $v$,
associated to the null geodesics generators of  $\Cb_{*}$. Then we look for a new foliation $u_{*}=u_{*}(v)$, expressed 
relatively to the background one, such that, relatively to it, $\oom$ satisfies the equations
\bea
&&\lapp\log\oom=-\divv\ze-\ro+\overline{\ro}\ \ ;\ \  \overline{\log 2\oom}=0\nn\\
&&\frac{du_{*}}{dv}=(2\oom^{2})^{-1};\ \ u_{*}|_{S_{*}(0)}=\la_1
\eql{lastsl}
\eea
where $S_{*}(0)=\Cb_{*}\cap\Si_0$.
Once these conditions are satisfied, the evolution equation for $\tr\chi$ becomes
\beaa
\dd_3\tr\chi+\frac{1}{2}\tr\chib{\tr\chi}+(\dd_3\log\oom)\tr\chi+\chibh\c\chih-2|\zeta|^2-4\zeta\c\de\nabb\log\oom-2|\nabb\log\oom|^2=2\overline{\ro}\  \ 
\eeaa
and the loss of derivatives is disappeared when we derive tangentially. The proof of the existence of this foliation is in \cite{Niclast}. It is then clear that the last slice transport equation for $\nabb\tr\chi$ has the following expression:
\beaa
&&\ML\ML\dd_3\nabb\tr\chi+\frac{1}{2}\tr\chib\nabb{\tr\chi}+(\dd_3\log\oom)\nabb\tr\chi+[\nabb,\dd_3]\tr\chi+\frac{1}{2}(\nabb\tr\chib){\tr\chi}+(\dd_3\nabb\log\oom)\tr\chi\nn\\
&&\ML\ML+[\nabb,\dd_3]\log\oom\tr\chi+\nabb\chibh\c\chih+\chibh\c\nabb\chih-4\ze\c\nabb\ze-4\nabb\zeta\c\nabb\log\oom-4\eta\c(-\divv\ze-\ro+\overline{\ro})=0\  .
\eeaa
In the present case the problem has to be worked in a  slight different way; Let us consider again the transport equation \ref{lastslice1},
\beaa
&&\dd_3\tr\chi+\frac{1}{2}\tr\chib{\tr\chi}+(\dd_3\log\oom)\tr\chi+\chibh\c\chih-2|\zeta|^2-4\zeta\c\nabb\log\oom-2|\nabb\log\oom|^2\nn\\
&&\ \ \ \ \ \ \ \ \ \ \ \ \ \ \ \ \ \ \ \ \ \ \ =2(\lapp\log\oom+\divv\zeta+\ro)\  \ .
\eeaa
and subtract to it the ``Kerr part" obtaining
\beaa
&&\ML\dd_3\de\tr\chi+\frac{1}{2}\tr\chib{\de\tr\chi}+\frac{1}{2}\de\tr\chib\tr\chi^{(Kerr)}-2\omb\de\tr\chi-2\de\omb\tr\chi^{(Kerr)}+\chibh\c\de\chih+\de\chibh\c\chih^{(Kerr)}\nn\\
&&\ML-2\ze\c\de\ze-2\de\ze\c\ze^{(Kerr)}-4\ze\c\de\nabb\log\oom-4\de\ze\c(\nabb\log\oom)^{(Kerr)}-4\nabb\log\oom\c\de\nabb\log\oom\nn\\
&&\ML-4\de\nabb\log\oom\c(\de\nabb\log\oom)^{(Kerr)}\nn\\
&&\ML=2(\lapp\de\log\oom+(\lapp-\lapp^{(Kerr)})(\log\oom)^{(Kerr)}+\divv\de\zeta+(\divv-\divv^{(Kerr)})\ze^{(Kerr)}+\de\ro)\  \ .
\eeaa
To avoid the loss of derivatives $\de\log\oom$ has to satisfy the following equation
\bea
\lapp\de\log\oom+\divv\de\zeta+\de\ro-\overline{\de\ro}=0\ ,
\eea
and the important point is that the term 
\[(\lapp-\lapp^{(Kerr)})(\log\oom)^{(Kerr)}+(\divv-\divv^{(Kerr)})\ze^{(Kerr)}\]
does not produce loss of derivatives as it contains second derivatives only of $(\log\oom)^{(Kerr)}$ which is a given function.

\NI Using the transport equations for the not underlined connection coefficients on the last slice, $\Cb_*$, allows to control their norms on the last slice in terms of the corresponding norms in the intersection $\Cb_*\cap\Si_0$ and prove they are bounded again by \footnote{The details of the ``last slice problem", first discussed and solved for a different foliation in \cite{Ch-Kl:book}, are discussed in detail in \cite{Kl-Ni:book}, Chapter 7 and also in \cite{Niclast}.} 
\[c\!\left(\varepsilon+\ep_0\frac{M^2}{R_0^2}\right)\ .\]

\subsection{V step: The initial data}\label{SS3.6}
The global existence and the peeling is proved here assuming a strong regularity for the initial data. The reason for it is that our initial data have to guarantee that the norms of the connection coefficients and   tangential derivatives up to third order are small and bounded on $\Si_0$ and moreover, the $\tilde{\cal Q}$ norms written in terms of initial data have to be finite and small.  
We collect in the next subsections all the initial data conditions we have used in the various proofs. Finally all these conditions can be expressed in terms of quantities relative only to the initial data hypersurface, namely the three dimensional metric ${^{(3)}}{\bf g}$ and the  the second fundamental form ${^{(3)}\!}{\bf k}$, together with their covariant derivatives.\footnote{Covariant with respect to the three dimensional metric ${^{(3)}}{\bf g}$.} This will allows to express the initial data smallness conditions requiring that a $L^2$ integral on $\Si_0/B_{R_0}$, whose integrand depends only on $\de{^{(3)}\!}{\bf k}={^{(3)}\!}{\bf k}-{^{(3)}\!}{\bf k}^{(Kerr)}$ and on $\de {^{(3)}\!\!\!}\mbox{ Ricci}={^{(3)}\!\!\!}\mbox{ Ricci}-{^{(3)}\!\!\!}\mbox{ Ricci}^{(Kerr)}$.
\subsubsection{The asymptotic conditions on the initial data metric}
From equation \ref{pertKerr1} it follows
\bea
{\bf g}|_{\Si_0}(\c,\c)\!&=&\!\oom^2dr_*^2+\ga_{ab}\big(d\om^a+\frac{1}{2}({X_{(Kerr)}}^a-X^a)dr_*\big)\!\big(d\om^b+\frac{1}{2}({X_{(Kerr)}}^b-X^b)dr_*\big)\nn\\
\!&=&\!(\oom^2+\frac{1}{4}\de X^2)dr_*^2-\frac{1}{2}\ga_{ab}\de X^adr_*d\om^b+\ga_{ab}d\om^ad\om^b \eql{pertKerr10}\\
\!&=&\!\left[\oom_{(Kerr)}^2dr_*^2+\ga^{(Kerr)}_{ab}d\om^ad\om^b\right]\nn\\
\!&+&\!\left((2\oom_{(Kerr)}\de\oom+\de\oom^2+\frac{1}{4}\de X^2)dr_*^2-\frac{1}{2}\ga_{ab}\de X^adr_*d\om^b+\de\ga_{ab}d\om^ad\om^b \right)\nn
\eea
Therefore
\bea
({\bf g}|_{\Si_0}-{\bf g}_{(Kerr)}|_{\Si_0})(\c,\c)=(2\oom_{(Kerr)}\de\oom+\de\oom^2+\frac{1}{4}\de X^2)dr_*^2-\frac{1}{2}\ga_{ab}\de X^adr_*d\om^b+\de\ga_{ab}d\om^ad\om^b\nn
\eea
and the asymptotic conditions on the metric components are
\bea
\de\oom=o_5\left(\frac{1}{r^{3+\ga}}\right)\ \ ;\ \ \de X^a=o_5\left(\frac{1}{r^{4+\ga}}\right)\ \ ;\ \ \de\ga_{ab}=o_5\left(\frac{1}{r^{1+\ga}}\right)\ ,
\eea
where $r$ is defined as in equation \ref{rdefa} and the $S$'s are the two dimensional surfaces associated to the canonical foliation of $\Si_0$ we are going to define.
We call these conditions: {\bf{``Kerr asymptotic flatness"}}.

\subsubsection{The smallness conditions on the initial data metric}
The smallness conditions for the metric components are 
\bea
\sup_{\Si_0/{B_{R_0}}}|r^{3+\de}\de\oom|\leq  \varepsilon\  \ ;\ \ \sup_{\Si_0/{B_{R_0}}} |r^{4+\de}\de X|\leq   \varepsilon\ \ ;\ \ \sup_{\Si_0/{B_{R_0}}}|r^{1+\de}\de \ga|\leq   \varepsilon\ .\ \ \ \ 
\eea

\subsubsection{The  smallness conditions on the initial data connection coefficients}
As the estimates of the connection coefficients in $V_*$ are obtained in terms of the initial data ones,\footnote{Remark, see the discussion about the last canonical slice, that the way in which the connection coefficients norms depend on the corresponding ones on the initial data, is different from the underlined and the not underlined ones.} the initial data have to be such that the following estimates hold, with $l\leq 4$:
\bea
&&|r^{4+l+\de}\nabb^l\de\tr\chi|\leq  \varepsilon\ \ ;\ \ |r^{4+l+\de}\nabb^l\de\tr\chib|\leq  \varepsilon\nn\\
&&|r^{4+l+\de}\nabb^l\de\chih|\leq  \varepsilon\ \ ;\ \ |r^{4+l+\de}\nabb^l\de\chibh|\leq  \varepsilon\nn\\
&&|r^{4+l+\de}\nabb^l\de\ze|\leq  \varepsilon\eql{deOcond1}\\
&&|r^{4+l+\de}\nabb^l\de\om|\leq \varepsilon\ \ ,\ \ |r^{4+l+\de}\nabb^l\de\omb| \leq \varepsilon\ .\nn
\eea

\subsubsection{The  smallness conditions on the initial data Riemann components}
Our initial data have to guarantee that the norms of the connection coeffcients and their tangential derivatives are small and bounded on $\Si_0$, but, more than that, the $\tilde{\cal Q}$ norms defined in terms of initial data have to be finite; this implies that they have to be such that
\bea
{\tilde{\cal Q}}_{\Si_0/B_{R_0}}\leq \varepsilon^2\eql{indata1}
\eea
and  also, on $\Si_0/B_{R_0}$ the following condition must hold,
\bea
\sup_{\Si_0/B_{R_0}}|r^{6+\frac{{\ga}}{2}}\overline{\ro({\tilde R})}|^2\leq \varepsilon^2 \ .\eql{indata2}
\eea

\subsubsection{The canonical foliation on $\Si_0$}
A foliation of $\Si_0/B_{R_0}$ in terms of two dimensional surfaces $\{S_0\}$ is specified through a function $\ub_0(p)$, the generic $S_0$ is defined as
\bea
S_0(\ub_0=\nu)=\{p\in \Si_0/B_{R_0}| \ub_0(p)=\nu\}\ .
\eea
The solution of the eikonal equation $\ub(p)$ with initial data $\ub_0(p)$ on $\Si_0/B_{R_0}$ is such that the level hypersurfaces $\ub(p)=\nu$ defines the incoming cones of the double null foliation. As the norms of the ``underlined" connection coefficients, see \cite{Kl-Ni:book} Chapter 3 for all the detailed definitions, are estimated, using the transport equations along the incoming cones in terms of the same norms on the initial hypersurface, while the opposite happens for the not underlined coefficients estimated in terms of the norms on ``Scri", an analogous altough mild problem appears in this case. To avoid a loss of derivatives we have to choose an appropriate (canonical) foliation of $\Si_0/B_{R_0}$. We do not go here through the details as the way to obtain it is analogous to what was done in \cite{Kl-Ni:book}, Chapter 7. Moreover differently from the last slice case, here in principle we could even avoid the choice of the canonical foliation admitting a loss of regularity going from the initial data to the solution, while this could not be allowed when we prove that $V_*$ can be extended.

\NI The previous discussion is not completely precise. In fact the intersections of the outgoing cones which are defined starting from ``Scri" with the hypersurface $\Si_0$
are not the $S_0$ surfaces. This minor problem requires to define a spacelike hypersurface $\Si'_0$ near to $\Si_0$ and control the various norms in the strip between these two hypersurfaces. This has been done carefully in \cite{Kl-Ni:book}, subsection 4.1.3, and we do not repeat it here.

\subsubsection{The initial data condition in terms of $\de{^{(3)}}{\bf g}$ and  $\de{^{(3)}}{\bf k}$ }
As we said all the initial data conditions can be reexpressed requiring that the metric stays near to the Kerr metric in a definite way, a condition similar to the ``Strong asymptotic flat condition" defined in \cite{Ch-Kl:book} and that, moreover,  an $L^2$ integral over $\Si_0/{B_{R_0}}$, $\bf{\cal J}$, whose integrand is made by (the square of) $\de{^{(3)}\!}{\bf k}$, $\de {^{(3)}\!\!\!}\mbox{ Ricci}$ and their derivatives is bounded by $\varepsilon$. Its explicit expression is
\bea
&&\ML{\bf{\cal J}(\de{^{(3)}}{\bf g},\de{^{(3)}}{\bf k})}=\sup_{\Si_0/{B_{R_0}}}\left[|r^{3+\de}\de\oom|+|r^{4+\de}\de X|+|r^{1+\de}\de \ga|\right] \eql{3.412a}\\
&&\ML+\left[\int_{\Si_0/{B_{R_0}}}\sum_{l=0}^5(1+d^2)^{(1+l)+\frac{5}{2}+\de}|\nabb^l\de k|^2
+\int_{\Si_0/{B_{R_0}}}\sum_{l=0}^3(1+d^2)^{(3+l)+\frac{5}{2}+\de}|\nab^l\de B|^2\right]^{\frac{1}{2}}\ .\nn
\eea
To express the tensor quantities of the four dimensional spacetime restricted to $\Si_0$ in terms of the three dimensional quantities and their norm bounds in term of the corresponding three dimensional ones, requires some work. We just sketch here the main steps as this is a repetition, with some obvious modifications, of what has been done in \cite{Kl-Ni:book}, (mainly in Chapter 7) and in the original work \cite{Ch-Kl:book}, (mainly in Chapter 5). 
\bigskip

\NI {\bf Step 1) The choice of a foliation and initial data on the Kerr spacetime:}

\NI Let us denote simply $\{\Si_0,{^{(3)}\!}{\bf g}\}$ the manifold with boundary $\Si_0/B_{R_0}$. A foliation is defined assigning on $\Si_0$ a function  $w(p)$, with some appropriate properties, so that the leaves of the foliation are defined by the level surfaces of this function
\bea
S_0(\nu)=\{p\in\Si_0|w(p)=\nu\}\ .
\eea
The unit length normal to a generic point of $S_0(\nu)$ is 
\bea
{\tilde N}^i=\frac{1}{|\pr w|}g^{ij}\pr_jw\ .
\eea
We can use $w(p)$ as the ``normal coordinate, in this case the metric can be written as
\bea
{^{(3)}\!}{\bf g}(\c,\c)=a^2dw^2+{\tilde\ga}_{ab}d\om^ad\om^b\ ,
\eea
and we have still the freedom of choosing the angular variables. Specifically if we consider the Kerr spacetime whose metric restricted to $\Si_0$ is
\bea
{^{(3)}\!}{\bf g}_{(Kerr)}(\c,\c)=\oom_{(Kerr)}^2dr_*^2+\ga^{(Kerr)}_{ab}d\om^ad\om^b\ .\eql{Kerr1ap}
\eea
we are suggested to choose $r_*(p)$ as the function $w(p)$ defining the foliation, so that
\bea
S_0(\nu)=\{p\in\Si_0|r_*(p)=\nu\}\ \ ,\ \ \tilde{N}=\frac{1}{\oom_{(Kerr)}}\frac{\pr}{\pr r_*}\ .
\eea
On the other  side in the Kerr spacetime the null vector of the null frame adapted to the double null foliation is
\bea
&&e_3=\frac{1}{\oom_{(Kerr)}}\left(\frac{\pr}{\pr u}+\om_B\frac{\pr}{\pr\phi}\right)\ \ ,\ \ 
e_4=\frac{1}{\oom_{(Kerr)}}\left(\frac{\pr}{\pr\ub}+\om_B\frac{\pr}{\pr\phi}\right)\nn
\eea
where
\bea
\frac{\pr}{\pr\ub}=\frac{\pr}{\pr t}+\frac{\pr}{\pr r_*}\ \ \ \ ,\ \ \ \ \frac{\pr}{\pr u}=\frac{\pr}{\pr t}-\frac{\pr}{\pr r_*}\ .
\eea
Therefore
\bea
&&\ML e_4=\frac{1}{\oom_{(Kerr)}}\frac{\pr}{\pr r_*}+\frac{1}{\oom_{(Kerr)}}\left(\frac{\pr}{\pr t}+\om_B\frac{\pr}{\pr\phi}\right)
=\tilde{N}+\frac{1}{2}(e_3+e_4)=\tilde{N}+T\nn\\
&&\ML e_3=-\frac{1}{\oom_{(Kerr)}}\frac{\pr}{\pr r_*}+\frac{1}{\oom_{(Kerr)}}\left(\frac{\pr}{\pr t}+\om_B\frac{\pr}{\pr\phi}\right)
=-\tilde{N}+\frac{1}{2}(e_3+e_4)=-\tilde{N}+T\nn
\eea
where 
\[T=\frac{1}{2}(e_3+e_4)=\frac{1}{\oom_{(Kerr)}}\left(\frac{\pr}{\pr t}+\om_B\frac{\pr}{\pr\phi}\right)\ ,\]
is a vector field orthogonal to the hypersurface $\Si_0$ of uniti length, \footnote{In fact
$$
-g^{\mu 0}\frac{\pr}{\pr x^{\mu}}=-g^{00}\frac{\pr}{\pr t}-g^{\phi 0}\frac{\pr}{\pr \phi}
=\frac{1}{\lap\sin^2\th}\left(g_{\phi\phi}\frac{\pr}{\pr t}-g_{\phi 0}\frac{\pr}{\pr \phi}\right)
=\frac{1}{\oom_{(Kerr)}^2}\left(\frac{\pr}{\pr t}+\om_B\frac{\pr}{\pr \phi}\right)
$$}
in fact
\bea
{\bf g}(T,T)=-1\ .
\eea
Therefore we define the second fundamental form of the hypersurface $\Si_0$ and the second fundamental form of the leaves $S_0(\nu)$ as
\bea
&&k(e_a,e_b)=-{\bf g}(\dd_{e_a}T,e_b)\\
&&\th(e_a,e_b)={^{(3)}\!}{\bf g}(\nab_{e_a}{\tilde N},e_b)\ .\nn
\eea
Some details have to be given here; first of all on the hypersurface $\Si_0$ the metric induced by the Kerr metric ${\bf g}$ has the following components
\bea
&&{^{(3)}\!}{g}_{\mu\nu}=\Pi^{\ro}_\mu\Pi^{\si}_\nu{g}_{\ro\si}=(\de^{\ro}_\mu+T^{\ro}T_\mu)(\de^{\si}_\nu+T^{\si}T_\nu){g}_{\ro\si}\nn\\
&&=g_{\mu\nu}+g_{\mu\si}T^{\si}T_{\nu}+T^{\ro}T_{\mu}g_{\ro\nu}-T_{\mu}T_\nu\ .
\eea
On the other side, given $p\in\Si_0$ the metric tensor ${^{(3)}\!}{g}(p)$ can be applied only on the  vectors of the subspace $T\Si_0|_p$ of $T{\cal M}_p$, orthogonal to $T|p$. These vector fields con be expressed in terms of the orthonormal basis 
\[\{e_1,e_2,e_3\}=\{e_{\th_*}=\frac{R}{\cal L}\frac{\pr}{\pr\th_*},\ \ e_{\phi}=\frac{1}{R\sin\th}\frac{\pr}{\pr\phi},\ \ {\tilde N}=\frac{1}{\oom_{(Kerr)}}\frac{\pr}{\pr r_*}\}\ \]
and we recall that
\bea
&&\Pi\frac{\pr}{\pr t}=-\om_B\frac{\pr}{\pr\phi}\\
&&{^{(3)}\!}{g}_{ij}=\Pi^{\ro}_i\Pi^{\si}_j{g}_{\ro\si}=(\de^{\ro}_i-T^{\ro}T_i)(\de^{\si}_j-T^{\si}T_j){g}_{\ro\si}=\de^{\ro}_i\de^{\si}_j{g}_{\ro\si}={g}_{ij}\ .\nn
\eea
where the last line follows from
\bea
T_i=g_{i\si}T^{\si}=\de_i^{\phi}\left(g_{\phi0}T^0+g_{\phi\phi}T^{\phi}\right)=\de_i^{\phi}\left(-\om_Bg_{\phi\phi}+g_{\phi\phi}\om_B\right)=0\ .
\eea
Therefore the metric induced on $\Si_0$ is exactly ${^{(3)}\!}{g}$.
Moreover
\bea
&&\nab_{\ro}{\tilde N}^{\si}=\Pi_{\tau}^{\si}\Pi_{\ro}^{\la}D_{\la}{\tilde N}^{\tau}\\
&&e_A^a\nab_{a}{\tilde N}^{\si}=\Pi_{\tau}^{\si}e_A^a(D_{a}{\tilde N})^{\tau}\nn
\eea
so that
\bea
{^{(3)}\!}{\bf g}(\nab_{e_a}{\tilde N},e_b)={\bf g}(\dd_{e_a}{\tilde N},e_b)\ .
\eea
It follows immediately that
\bea
&&\chi(e_a,e_b)={\bf g}(\dd_{e_a}e_4,e_b)={\bf g}(\dd_{e_a}{\tilde N},e_b)+{\bf g}(\dd_{e_a}T,e_b)\nn\\
&&={^{(3)}\!}{\bf g}(\nab_{e_a}{\tilde N},e_b)-k(e_a,e_b)=\th(e_a,e_b)-k(e_a,e_b)\ .
\eea
Therefore
\bea
&&\chi(e_a,e_b)=\th(e_a,e_b)-k(e_a,e_b)\nn\\
&&\chib(e_a,e_b)=-\th(e_a,e_b)-k(e_a,e_b)\nn\\
&&\ze(e_a)=k(e_a,{\tilde N})\\
&&\om=-\frac{1}{2\oom_{(Kerr)}}\pr_{r_*}\log{\oom_{(Kerr)}}\ \ ,\ \ \omb=\frac{1}{2\oom_{(Kerr)}}\pr_{r_*}\log{\oom_{(Kerr)}}\ .\nn
\eea
It turns out, therefore, that to specify the initial data for the connection coefficients we have to assign $\oom_{(Kerr)}$ and the second fundamental form $k$ which in the Kerr spacetime are known, but in general as we will see can be assigned in a free way; moreover we have to assign $\th(e_a,e_b)$ which cannot be assigned freely on $\Si_0$ as it satisfies on it a transport equation, \ref{7.1.10b}. Therefore $\th(e_a,e_b)$ depends on the choice of the foliation and satisfies a transport equation; to know it on $\Si_0$ we have to satisfy it giving ``initial data" on $S_0(0)=\pr B_{R_0}$. On the other side looking at the trasport equation satisfied by $\tr\th$ and $\hat\th$ one sees that on the right hand side the components of the Ricci tensor of $\Si_0$, ${^{(3)}\!}Rij$ and the scalar curvature are present. As on $\Si_0$ the constraint equations hold the scalar curvature can be expressed in terms of $k$, for the components of ${^{(3)}\!}R_{ij}$ we use the Bianchi equations and together with the trasport equations for $\th$ we have a system of equations which can be solved once we control $B=\curl\hat{{^{(3)}\!}Ricci}$, (this will be discussed better later on when we discuss the real situation of initial data not exactly Kerr). Therefore in principle we need ``initial conditions" on $B$ and $k$, together with their derivatives as we require conditions also on the derivatives of the connection coefficients. We need also initial conditions on the Riemann components, writing the Gauss and Codazzi-Mainardi equations expressing the four dimensional Riemann tensor in terms of the three dimensional Ricci tensor and $k$ we are able to write the Riemann components in terms again of the Ricci three dimensional tensor, the second fundamental form $k$ and the various connection coefficients which, at the end, implies again in terms of $k$, $B$  and their derivatives up to a fixed order. As we will see in a moment an analogous procedure can be done when our initial data are only ``near" to Kerr.
\medskip

\NI {\bf Step 2) The choice of a foliation and initial data on the perturbed Kerr spacetime:}

\NI The metric in the perturbed spacetime is
\bea
&&\ML{\bf g}(\c,\c)\nn\\
&&\ML=-\oom^2(dt^2-dr_*^2)+\ga_{ab}\!\left(d\om^a-({X^a_{(Kerr)}}du+X^ad\ub)\right)\!\!\left(d\om^b-({X^b_{(Kerr)}}du+X^bd\ub)\right)\ .\nn
\eea
Again the normal to the hypersurface $t=0$ is $T$ where
\bea
T^{\mu}=g^{\mu\nu}\de^0_{\nu}=g^{\mu0}
\eea
\bea
g^{\mu\nu}=-\frac{1}{2\oom^2}\left[\left(\de^{\mu}_{\ub}\de^{\nu}_{u}+\de^{\mu}_{u}\de^{\nu}_{\ub}\right)
+X^c\left(\de^{\mu}_{c}\de^{\nu}_{u}+\de^{\nu}_{c}\de^{\mu}_{u}\right)+{X_{(Kerr)}}^d\!\left(\de^{\mu}_{d}\de^{\nu}_{\ub}+\de^{\nu}_{d}\de^{\mu}_{\ub}\right)\right]+\ga^{ab}\de^{\mu}_a\de^{\nu}_b\ .\nn
\eea
\bea
g^{\mu0}=-\frac{1}{4\oom^2}\left[\left(\de^{\mu}_{\ub}+\de^{\mu}_{u}\right)
+X^c\de^{\mu}_{c}+{X_{(Kerr)}}^d\de^{\mu}_{d}\right]\ ,\nn
\eea
and normalized to minus one,
\bea
&&T=\frac{1}{2\oom}\left[\left(\frac{\pr}{\pr\ub}+\frac{\pr}{\pr u}\right)+(2\om_B\de^c_{\phi}+\de X^c)\frac{\pr}{\pr\om^c}\right]
=\frac{1}{2\oom}\left[2\frac{\pr}{\pr t}+(2\om_B\de^c_{\phi}+\de X^c)\frac{\pr}{\pr\om^c}\right]\nn\\
&&=\frac{1}{\oom}\left[\frac{\pr}{\pr t}+(\om_B\de^c_{\phi}+\frac{\de X^c}{2})\frac{\pr}{\pr\om^c}\right]
=\frac{1}{\oom}\left[\frac{\pr}{\pr t}+\om_B\frac{\pr}{\pr\phi}+\frac{\de X^c}{2}\frac{\pr}{\pr\om^c}\right]\eql{Tdef1}\\
&&=\frac{\oom_{(Kerr)}}{\oom}T^{(Kerr)}+\frac{\de X^c}{2\oom}\frac{\pr}{\pr\om^c}=T^{(Kerr)}-\frac{\de\oom}{\oom}T^{(Kerr)}+\frac{\de X^c}{2\oom}\frac{\pr}{\pr\om^c}
=\frac{1}{2}(e_3+e_4)\ .\nn
\eea
Proceeding as before on $\Si_0/{B_{R_0}}$, hereafter denoted  simply $\Si_0$, the assigned metric is
\bea
{^{(3)}\!}{\bf g}(\c,\c)=(\oom^2+\frac{1}{4}|\de X|^2)dr_*^2-\frac{1}{2}\ga_{ab}(\de X^adr_*d\om^b+\de X^bd\om^adr_*)+\ga_{ab}d\om^ad\om^b\ \ \ \ \ \  \ \eql{pertKerr1ap}
\eea
near to the Kerr $\Si_0$ metric
\bea
{^{(3)}\!}{\bf g}_{(Kerr)}(\c,\c)=\oom_{(Kerr)}^2dr_*^2+\ga^{(Kerr)}_{ab}d\om^ad\om^b\ .\eql{Kerr1ap}
\eea
Moreover
\bea
&&{^{(3)}\!}g^{11}=\frac{1}{|D|}(\ga_{11}\ga_{22}-\ga^2_{12})\ \ ,\ \ {^{(3)}\!}g^{22}=\frac{1}{|D|}((\oom^2+|\de X|^2)\ga_{22}-\de X^2_2)\\
&&{^{(3)}\!}g^{12}=\frac{1}{|D|}(-\de X_1\ga_{22}+\de X_2\ga_{12})\ \ ,\ \ {^{(3)}\!}g^{33}=\frac{1}{|D|}((\oom^2+|\de X|^2)\ga_{11}-\de X_1^2)\nn\\
&&{^{(3)}\!}g^{13}=\frac{1}{|D|}(-\de X_1\ga_{12}+\de X_2\ga_{11})\ \ ,\ \ {^{(3)}\!}g^{23}=-\frac{1}{|D|}((\oom^2+|\de X|^2)\ga_{12}-\de X_1\de X_2)\nn
\eea
\bea
|D|\!&=&\!(\oom^2+|\de X|^2)\det\ga-\de X_1^2\ga_{22}-\de X_2^2\ga_{11}+2\de X_1\de X_2\ga_{12}\nn\\
\!&=&\!\det\ga\left[\oom^2+|\de X|^2-\de X_1^2\frac{\ga_{22}}{\det\ga}-\de X_2^2\frac{\ga_{11}}{\det\ga}+2\de X_1\de X_2\frac{\ga_{12}}{\det\ga}\right]\nn\\
\!&=&\!\det\ga\oom^2
\eea
From the initial data assumptions we can write
\bea
&&{^{(3)}\!}g^{11}=\frac{1}{\oom^2}\ \ ,\ \ {^{(3)}\!}g^{22}=\frac{\ga_{22}}{\det\ga}+O\left(\frac{\varepsilon}{r^{4+\de}}\right)^2\ \ ,\ \ 
{^{(3)}\!}g^{33}=\frac{\ga_{11}}{\det\ga}+O\left(\frac{\varepsilon}{r^{4+\de}}\right)^2\nn\\
&&{^{(3)}\!}g^{1,\{\th,\phi\}}=O\left(\frac{\varepsilon}{r^{4+\de}}\right)\ \ ,\ \ {^{(3)}\!}g^{\th,\phi}=O\left(\frac{\varepsilon}{r^{5+\de}}\right)\ .
\eea
Therefore
\bea
&&\ML{^{(3)}\!}{\bf g}(\c,\c)-{^{(3)}\!}{\bf g}_{(Kerr)}(\c,\c)\eql{deffol0}\\
&&\ML=\left[\de\oom(\oom+\oom_{(Kerr)})+\frac{1}{4}|\de X|^2\right]dr_*^2-\frac{1}{2}\ga_{ab}(\de X^adr_*d\om^b+\de X^bd\om^adr_*)+\de\ga_{ab}d\om^ad\om^b\nn 
\eea
From the previous definition \ref{Tdef1}
\bea
T\!&=&\!\frac{1}{2\oom}\left[\left(\frac{\pr}{\pr\ub}+\frac{\pr}{\pr u}\right)+(X^c+X_{(Kerr)}^c)\frac{\pr}{\pr\om^c}\right]\nn\\
\!&=&\!\frac{1}{\oom}\left[\frac{\pr}{\pr t}+\frac{1}{2}(X^c+X_{(Kerr)}^c)\frac{\pr}{\pr\om^c}\right]=\frac{1}{2}(e_3+e_4)
\eea
it follows that
\bea
&&e_4
=\frac{1}{\oom}\left(\frac{\pr}{\pr t}+\frac{\pr}{\pr r_*}+X^c\frac{\pr}{\pr\om^c}\right)=T+\frac{1}{\oom}\left(\frac{\pr}{\pr r_*}+\frac{1}{2}\de X^c\frac{\pr}{\pr\om^c}\right)\nn\\
&&e_3
=\frac{1}{\oom}\left(\frac{\pr}{\pr t}-\frac{\pr}{\pr r_*}+X_{(Kerr)}^c\frac{\pr}{\pr\om^c}\right)=T-\frac{1}{\oom}\left(\frac{\pr}{\pr r_*}+\frac{1}{2}\de X^c\frac{\pr}{\pr\om^c}\right)
\nn
\eea
Therefore
\bea
e_4=T+{\tilde N}\ \ ,\ \ e_3=T-{\tilde N}
\eea
where
\bea
{\tilde N}=\frac{1}{\oom}\!\left(\frac{\pr}{\pr r_*}+\frac{1}{2}\de X^c\frac{\pr}{\pr\om^c}\right)\ \ .\eql{defN1}
\eea
It is easy to prove that with the metric \ref{pertKerr1ap} 
\beaa
{^{(3)}\!}{\bf g}(\c,\c)=(\oom^2+\frac{1}{4}|\de X|^2)dr_*^2-\frac{1}{2}\ga_{ab}(\de X^adr_*d\om^b+\de X^bd\om^adr_*)+\ga_{ab}d\om^ad\om^b\ 
\eeaa
we have
\[{^{(3)}\!}{\bf g}({\tilde N},{\tilde N})=1\ \ , \ \ {^{(3)}\!}{\bf g}(T,T)=-1\ \ ,\ \ {^{(3)}\!}{\bf g}(T,{\tilde N})=0\ \ .\]
An easy calculation gives
\bea
{\tilde n}={\tilde n}_idx^i =\oom dr_*\ \ \ ,\ \ \ {\tilde n}_i={^{(3)}\!}g_{ij}{\tilde N}^j\ .\eql{gradformr*}
\eea
Again
\bea
&&{^{(3)}\!}{g}_{\mu\nu}=\Pi^{\ro}_\mu\Pi^{\si}_\nu{g}_{\ro\si}=(\de^{\ro}_\mu+T^{\ro}T_\mu)(\de^{\si}_\nu+T^{\si}T_\nu){g}_{\ro\si}\nn\\
&&=g_{\mu\nu}+g_{\mu\si}T^{\si}T_{\nu}+T^{\ro}T_{\mu}g_{\ro\nu}-T_{\mu}T_\nu\ .
\eea
and also
\bea
T_0=-\oom\ \ ,\ \ T_i=g_{i\si}T^{\si}=0
\eea
so that
\[{^{(3)}\!}{g}_{ij}=g_{ij}\ .\]
Therefore the metric induced on $\Si_0$ is exactly ${^{(3)}\!}{g}$.
Moreover
\bea
&&\nab_{\ro}{\tilde N}^{\si}=\Pi_{\tau}^{\si}\Pi_{\ro}^{\la}D_{\la}{\tilde N}^{\tau}\\
&&e_A^a\nab_{a}{\tilde N}^{\si}=\Pi_{\tau}^{\si}e_A^a(D_{a}{\tilde N})^{\tau}\nn
\eea
so that
\bea
{^{(3)}\!}{\bf g}(\nab_{e_a}{\tilde N},e_b)={\bf g}(\dd_{e_a}{\tilde N},e_b)\ .
\eea
It follows immediately that
\bea
&&\chi(e_a,e_b)={\bf g}(\dd_{e_a}e_4,e_b)={\bf g}(\dd_{e_a}{\tilde N},e_b)+{\bf g}(\dd_{e_a}T,e_b)\nn\\
&&={^{(3)}\!}{\bf g}(\nab_{e_a}{\tilde N},e_b)-k(e_a,e_b)=\th(e_a,e_b)-k(e_a,e_b)\ .
\eea
Therefore
\bea
&&\chi(e_a,e_b)=\th(e_a,e_b)-k(e_a,e_b)\nn\\
&&\chib(e_a,e_b)=-\th(e_a,e_b)-k(e_a,e_b)\nn\\
&&\ze(e_a)=k(e_a,{\tilde N})\\
&&\om=-\frac{1}{2\oom}\pr_{T}\log{\oom-\frac{1}{2\oom}}\pr_{\tilde N}\log{\oom}\nn\\
&&\omb=-\frac{1}{2\oom}\pr_{T}\log{\oom+\frac{1}{2\oom}}\pr_{\tilde N}\log{\oom}\ .\nn
\eea
Next step is to prove the following lemma
\begin{Le}\label{Lchangeofcooord}
It is possible to perform a change of the coordinates 
\[\{r_*,\om^1,\om^2\}\rightarrow\{{\tilde r}_*,{\tilde\om}^1,{\tilde\om}^2\}\]
such that the vector field ${\tilde N}$ is in these coordinates
\bea
{\tilde N}=\frac{1}{\oom}\frac{\pr}{\pr {\tilde r}_*}
\eea
and the metric has the form
\bea
{^{(3)}\!}{\bf g}(\c,\c)=\oom^2d{\tilde r}_*^2+{\tilde\ga}_{ab}d{\tilde\om}^ad{\tilde\om}^b\ .
\eea
Moreover
\bea
{\tilde \ga}_{ab}=\ga_{ab}+O\left(\frac{\varepsilon}{r^{2+\de}}\right)\ \ ,\ \ {\tilde\om}^a=\om^a+O\left(\frac{\varepsilon}{r^{3+\de}}\right)\ .
\eea
\end{Le}
\NI{\bf Proof:} 
Let us consider the following coordinate transformation:
\bea
&&r_*={\tilde r}_*\nn\\
&&\om^b={\tilde\om}^b+f^b(r_*,{\tilde\om})
\eea
where $f^b$ satisfies
\bea
\frac{\pr f^b}{\pr r_*}(r_*,{\tilde\om}(r_*,\om))=\frac{\de X^b}{2}(r_*,\om)\ .
\eea
Due to the ``smallness" of $\de X$ it follows immediately that the previous map is invertible on $\Si_0/{B_{R_0}}$ and therefore is a coordinate transformation. Moreover it follows that in the $\{{\tilde r}_*,{\tilde\om}^b\}$ coordinates the vector field ${\tilde N}$ has the expression
\[{\tilde N}=\frac{1}{\oom}\frac{\pr}{\pr{\tilde r}_*}\ .\]
From the relation
\[{^{(3)}\!}{\tilde g}_{ij}=\frac{\pr x^l}{\pr y^i}\frac{\pr x^s}{\pr y^j}{^{(3)}\!}g_{ls}\ ,\]
denoting
\[\{y^i\}=\{{\tilde r}_*,{\tilde\om}^1,{\tilde\om}^2\}\ \ ;\ \ \{x^l\}=\{r_*,{\om}^1,{\om}^2\}\ ,\]
it follows immediately
\bea
&&{^{(3)}\!}{\tilde g}_{11}=\oom^2\ \ ,\ \ {^{(3)}\!}{\tilde g}_{1b}=0\\
&&{^{(3)}\!}{\tilde g}_{cd}\equiv{\tilde\ga}_{cd}=\ga_{cd}+\frac{\pr f^a}{\pr y^c}\ga_{ad}+\frac{\pr f^b}{\pr y^d}\ga_{cb}+\frac{\pr f^a}{\pr y^c}\frac{\pr f^b}{\pr y^d}\ga_{ab}\ \ .\nn
\eea
\medskip

\NI Therefore in these coordinates the metric has the expression
\bea
{^{(3)}\!}{\bf g}(\c,\c)=\oom^2d{\tilde r}_*^2+{\tilde\ga}_{ab}d{\tilde\om}^ad{\tilde\om}^b\ .
\eea
and
\bea
{\tilde N}=\frac{1}{\oom}\frac{\pr}{\pr {\tilde r}_*}\ \ ,\ \ {\tilde n}=\oom d{\tilde r}_*
\eea
so that in these coordinates the leaves of the foliation are defined by the level surfaces of the  same function ${\tilde r}_*(p)=r_*(p)$,
\bea
S_0(\nu)=\{p\in\Si_0|{\tilde r}_*(p)=\nu\}\ .
\eea
Moreover it is important to observe that although the coordinates ${\tilde\om}^a$ are different from $\om^a$ nevertheless we still have, as expected,
\[{^{(3)}\!}{\bf g}({\tilde N},\frac{\pr}{\pr{\tilde\om}^a})={^{(3)}\!}{\bf g}({\tilde N},\frac{\pr}{\pr{\om}^a})\ \ , \ a\in\{1,2\}\ .\]
We proceed as in the Kerr case, the second fundamental form $k$, apart from satisfying the constraint equations, can be assigned freely and it can be decomposed in the obvious way as
\bea
k(e_i,e_j)=k_{(Kerr)}(e_i,e_j)+\de k(e_i,e_j)
\eea
and the smallness of the $\de k(e_i,e_j)$ part together with its ($\Si_0$-) tangential derivatives will be assigned in a direct way. The situation is more delicate for the connection coefficients which depend on $\th(e_a,e_b)$, the second fundamental form of the leaves $S_0(\nu)$ as they satisfy some transport equations.
As $\th(e_a,e_b)$ depends on the choice of the foliation and satisfies a transport equation, to know it on $\Si_0$ we have to satisfy it giving ``initial data" on $S_0(0)=\pr B_{R_0}$. On the other side looking at the trasport equations satisfied by $\tr\th$ and $\hat\th$ one sees that on the right hand side the components of the Ricci tensor of $\Si_0$, ${^{(3)}\!}R_{ij}$ and the scalar curvature are present. As on $\Si_0$ the constraint equations hold, the scalar curvature can be expressed in terms of $k$, for the components of ${^{(3)}\!}R_{ij}$ we use the Bianchi equations and together with the trasport equations for $\th$ we have a system of equations which can be solved once we control $B=\curl\widehat{{^{(3)}\!}Ricci}$. Therefore in principle we need ``initial conditions" on $B$ and $k$, together with their derivatives as we require conditions also on the derivatives of the connection coefficients. Moreover we need also initial conditions on the Riemann components; to do it we write the Gauss and Codazzi-Mainardi equations expressing the four dimensional Riemann tensor in terms of the three dimensional Ricci tensor and $k$ and we are able to write the Riemann components in terms  of the Ricci three dimensional tensor, the second fundamental form $k$ and the various connection coefficients which, at the end, are estimated again in terms of $k$, $B$  and their derivatives up to a fixed order. This general picture, nevertheless, does not take into account that here all this procedure has to be done subtracting the various ``Kerr parts". Let us sketch it in some more detail. The first point is the transport equation for the second fundamental form $\theta$. It comes from the contraction of the Codazzi-Mainardi equation, see \cite{Kl-Ni:book}, Chapter 7, (7.1.9). 

\NI Nevertheless as done in $V_*$ to subtract the Kerr part we use the same coordinates therefore we have to interpret Lemma \ref{Lchangeofcooord}. It is clear that nothing prevents to rename the coordinates ${\tilde r}_*,{\tilde\om}^b$ as $r_*,\om^b$ and, therefore we have for the initial data the following situation on $\Si_0/{B_{R_0}}$:
\bea
&&{^{(3)}\!}{\bf g}_{(Kerr)}(\c,\c)=\oom^2d{r}_*^2+{\ga}_{ab}d{\om}^ad{\om}^b\nn\\
&&{^{(3)}\!}{\bf g}(\c,\c)=\oom^2d{r}_*^2+{\tilde\ga}_{ab}d{\om}^ad{\om}^b\\
&&{\tilde\ga}_{ab}=\ga_{ab}+\frac{\pr f^c}{\pr y^a}\ga_{cb}+\frac{\pr f^d}{\pr y^b}\ga_{ad}+\frac{\pr f^c}{\pr y^a}\frac{\pr f^d}{\pr y^b}\ga_{cd}\nn\\
&&{\tilde N}={\tilde N}_{(Kerr)}=\frac{1}{\oom}\frac{\pr}{\pr r_*}\ \ ,\ \ e_a=e^{(Kerr)}_a+\de e_a\nn
\eea
where in the last equation the relation between $e_a$ and $e^{(Kerr)}_a$ is exactly as discussed in Lemma \ref{hatestim}.
The equations for $\th_{ab}$ have the following form:
\bea
&&\nabb_{\tilde N}{\hat\theta}_{ac}+\tr\theta{\ \hat\theta}_{ac}=
-\oom^{-1}{\widehat{\nabb_a\nabb_c}}\oom
-\left[{^{(3)}\!R}_{ac}+2^{-1}\ga_{ac}({^{(3)}\!R}_{{\tilde N}{\tilde N}}-{^{(3)}\!R})\right]\nn\\
&&\nab_{\tilde N}\tr\theta+\frac{1}{2}\tr\theta^2=-\oom^{-1}\lapp \oom
-|\hat\theta|^2-{^{(3)}\!R}_{{\tilde N}{\tilde N}}\eql{7.1.10b}
\eea
where ${\widehat{\nabb_a\nabb_c}}=\nabb_a\nabb_c-\frac{1}{2}\ga_{ac}\lapp$\ .

\NI As in the general case we can subtract to ${\theta}_{ab}$ its Kerr part \footnote{Observe that here we do not have to make a distinction between $\hat{\th}$ and $\th ^{(Kerr)}$ as here the leaves of the foliation are the same.} and observing that equations \ref{7.1.10b} are satisfied in Kerr we obtain
\bea
&&\nabb_{\tilde N}{\de\hat\theta}_{ac}+\tr\theta{\ \de\hat\theta}_{ac}=\de F_{ac}(\de\oom,\de\ga)
-\left[\de{^{(3)}\!R}_{ac}+2^{-1}\ga_{ac}(\de{^{(3)}\!R}_{{\tilde N}{\tilde N}}-\de{^{(3)}\!R})\right]\nn\\
&&\nab_{\tilde N}\tr\de\theta+\tr\theta\ \tr\de\theta=\de G(\de\oom,\de\ga,\de{\hat\th})
-|\de\hat\theta|^2-\de{^{(3)}\!R}_{{\tilde N}{\tilde N}}\eql{7.1.10bc}
\eea
where $\de F_{ac}(\de\oom,\de\ga)$ and $\de G(\de\oom,\de\ga,\de{\hat\th})$ depend also on the derivatives of their arguments and on the corresponding Kerr parts. These terms can be estimated exactly as done for the transport equations of the connection coefficients in $V_*$, see for instance subsection \ref{ss342}, and we do not report the analogous estimates. The next steps are analogous to those discussed in Chapter 7 of \cite{Kl-Ni:book} with the obvious modifications, equations \ref{7.1.10bc} have to be supplemented with the analogous of equations for the components of the three dimensional Ricci tensor to which the Kerr parts have been subtracted. They follow from the second Bianchi identities; defining
\bea
S_{AB}\equiv\widehat{{^{(3)}\! R}}_{AB}\ ,\  P_A\equiv\widehat{{^{(3)}\! R}}_{A{\tilde N}}\ ,\ 
Q\equiv\widehat{{^{(3)}\! R}}_{{\tilde N}{\tilde N}}\ ,\eql{7.1.13b}
\eea
\footnote{$(\curl\ \widehat{{^{(3)}\! R}})_{ij}\equiv\ep_j^{\ ls}\nab_l\left({^{(3)}\! R}_{is}-\frac{1}{3}g_{is}{^{(3)}\!
R}\right)$.}
\bea
B_{ij}=(\curl\ \widehat{{^{(3)}\! R}})_{ij}
\eea
and assuming the adapted frame with $\oom=1$ and Fermi transported\index{Fermi transported frame}
\footnote{$\nabb_{\!{\tilde N}}e_A=0$.} they have the form, see \cite{Kl-Ni:book} eqs. (7.1.16),
\bea
&&\divv P=\frac{1}{6}\nab_{\tilde N}R-\nab_{\tilde N}Q-\frac{3}{2}\tr\theta Q+\hat{S}\c\hat{\theta}\nn\\
&&\curll P=B_{{\tilde N}{\tilde N}}+\hat{\theta}\wedge\hat{S}\nn\\
&&\nabb_{\!N}P+\tr\theta P=\frac{1}{12}\nabb_aR+{^*\!\BBb}_{\tilde N}+\nabb Q-2\hat{\theta}\c P\nn\\
&&\divv\hat{S}=\left(\frac{1}{12}\nabb R-{^*\!\BBb}_{\tilde N}\right)-\frac{1}{12}\nabb Q+(\hat{\theta}\c P)
-\frac{1}{2}\tr\theta P\nn\\
&&\nabb_{\tilde N}\hat{S}+\frac{1}{2}\tr\theta \hat{S}={^*\!\BBb}+\frac{1}{2}\nabb\hat{\otimes}P+\frac{3}{2}\hat{\theta}Q 
\eea
where $\nabb_{\tilde N}X$ is the projection of $\nab_{\tilde N}X$ on $TS_0$, $\BBb$ is the $S$-tangent symmetric two tensor
$\BBb_{AB}=B_{AB}$ and ${^*\!\BBb}_{ab}=\ep_a^{\ c}\ep_b^{\ d}\BBb_{cd}$ . 

\NI Again from these equations to which  we subtract the Kerr part and equations \ref{7.1.10bc} we prove in the same way as in \cite{Kl-Ni:book} that the $\de\th$ part of the connection coefficients on $\Si_0$ satisfies the required initial data conditions provided we require some decay and smallness conditions for $\de k$ (and its derivatives) and $\de B$ (and its derivatives).

\NI We are left to discuss a final point to complete the initial data choice, namely the fact that the initial conditions on the Riemann components, equation \ref{indata1} and \ref{indata2} require the control on $\Si_0$ of the null components of $\lie_{T_0}{^{(4)}\!R}$, more precisely of $\a(\lie_{T_0}{^{(4)}\!R}),\b(\lie_{T_0}{^{(4)}\!R}),.....$. Again these conditions can be expressed in terms of conditions on $\de k$, $\de B$ and their $\Si_0$-tangential derivatives. In fact from lemmas \ref{Lemma3}, \ref{Lemma4} it folllows that we control  $\a(\lie_{T_0}{^{(4)}\!R}),\b(\lie_{T_0}{^{(4)}\!R}),.....$ once we control the null components of ${^{(4)}\!R}$ and their $\pr_{T_0}$ derivatives. The null components of ${^{(4)}\!R}$ are controlled in terms of their Kerr part and of the norms for $\de k$ and $\de B$ which allow to control $\de\a({^{(4)}\!R}),\de\b({^{(4)}\!R}),.....$, the control of the $\pr_{T_0}$ derivatives, $\pr_{T_0}\a({^{(4)}\!R}),\pr_{T_0}\b({^{(4)}\!R}),.....$ follows observing that $T_0=\frac{\pr}{\pr t}$ and that
\bea
\pr_t\a({^{(4)}\!R})=\pr_t{^{(4)}\!R}(e_a,e_4,e_b,e_4)\!&=&\!\pr_t\de{^{(4)}\!R}(e_a,T_0,e_b,T_0)+\pr_t\de{^{(4)}\!R}(e_a,T_0,e_b,{\tilde N})\nn\\
\!&+&\!\pr_t\de{^{(4)}\!R}(e_a,{\tilde N},e_b,T_0)+\pr_t\de{^{(4)}\!R}(e_a,{\tilde N},e_b,{\tilde N})\nn
\eea
and we can decompose each $\de{^{(4)}\!R}$ component in a $\de{^{(3)}\!R}$ and the product of two $\de k$ tensor or $k\de k$. We are left, therefore with the need to control $\pr_t\de k$ and $\pr_t\de{^{(3)}\!R}$. To do it we simply observe that in a $\Si_t$ foliation the Einstein equations can be written as equations for ${^{(3)}\!R}g_{ij}$ and $k_{ij}$, see \cite{Kl-Ni:book} equation (1.3.17), which in the present case have the form
\bea
&&\partial_t g_{ij} = -2 \oom k_{ij} +\Liee_{\Shift} g_{ij}\label{evol}\\
&&\partial_t k_{ij} = - \nabla_i \nabla_j \oom + \oom({^{(3)}\!R}_{ij} +\trr kk_{ij} - 2 k_{im}k^m_{\ j})  + \Liee_{\Shift} k_{ij}\nn
\eea
where
\[\Shift=X_{(Kerr)}+\frac{\de X}{2}\ .\]
Again subtracting the Kerr part from both sides of the equations we obtain
\bea
&&\partial_t \de g_{ij} = -2 \de\oom k_{ij} -2\oom_{(Kerr)}\de k_{ij} +\left[ \Liee_{\Shift} g_{ij}- \Liee_{X_{(Kerr)}} g^{(Kerr)}_{ij}\right]\nn\\
&&\partial_t \de k_{ij} = - (\de\nabla_i) \pr_j \oom - (\nabla^{(Kerr)}_i) \pr_j\de\oom+ \de\oom({^{(3)}\!R}_{ij} +\trr kk_{ij} - 2 k_{im}k^m_{\ j})  + \Liee_{\Shift} k_{ij}\nn\\
&&+\oom^{(Kerr)}(\de{^{(3)}\!R}_{ij} +(\de\trr k)k_{ij} +\trr k^{(Kerr)}\de k_{ij}- 2 \de k_{im}k^m_{\ j}- 2k^{(Kerr)}_{im}\de k^m_{\ j})\nn\\
&&+\left[ \Liee_{\Shift} k_{ij}- \Liee_{X_{(Kerr)}} k^{(Kerr)}_{ij}\right]\ .
\eea
This allows to control $\partial_t \de g_{ij} $ and $\partial_t \de k_{ij}$ in terms of the Kerr expressions for $k$ and ${^{(3)}\!R}_{ij}$ their corrections together with their $\Si_0$ tangential derivatives. The final result is , therefore, that the smallness of the initial data can be expressed requiring that an $L^2$ integral over $\Si_0/{B_{R_0}}$, $\bf{\cal J}$, whose integrand is made by (the square of) $\de{^{(3)}\!}{\bf k}$, $\de {^{(3)}\!\!\!}\mbox{ Ricci}$ and their derivatives is bounded by $\varepsilon$. More precisely
\bea
&&{\bf{\cal J}(\de{^{(3)}}{\bf g},\de{^{(3)}}{\bf k})}=\sup_{\Si_0/{B_{R_0}}}\left[|r^{3+\de}\de\oom|+|r^{4+\de}\de X|+|r^{1+\de}\de \ga|\right] \eql{3.412}\\
&&+\left[\int_{\Si_0/{B_{R_0}}}\sum_{l=0}^5(1+d^2)^{(1+l)+\frac{5}{2}+\de}|\nabb^l\de k|^2
+\int_{\Si_0/{B_{R_0}}}\sum_{l=0}^3(1+d^2)^{(3+l)+\frac{5}{2}+\de}|\nab^l\de B|^2\right]^{\frac{1}{2}}\ .\nn
\eea
\medskip

\NI{\bf Step 3: The construction of a canonical foliation}

\NI The strategy is exactly the same as the one discussed in \cite{Kl-Ni:book}, we just sketch the main difference, looking at the equation for $\tr\de\th$ one realizes that on the right hand side there is a term proportional to $\de\ro$, this implies a loss of derivatives on the initial data unless the foliation is such that in this equation $\de\ro$ is substituted by $\overline{\de\ro}$. This implies that $\de\oom$ has to satisfy an elliptic equation, the analogous of equation (7.1.26) of \cite{Kl-Ni:book}. Therefore once we have the original foliation on $\Si_0$ we can  construct starting from it a new foliation we call ``canonical" such that the incoming cones in $V_*$ are defined as the level surfaces $w=const$ of a function $w$ solution of the eikonal equation with initial data given by a function $w_0(p)$ defined on $\Si_0$, such that the surfaces $w_0(p)=\nu$ are the leaves of the canonical foliation.

\subsection{VI step: The extension of the region $V_*$}\label{extension}

As previously said once we have proved that in the region $V_*$ the estimates for the norms of the connection coefficients and for the Riemann tensor are better than those assumed in the ``Bootstrap assumptions" we have done the basic step to conclude that a region larger than $V_*$ does exist where the previous quantity satisfy again the ``Bootstrap assumptions". To obtain this result, nevertheless, something more is needed. In fact the bootstrap argument requires that two conditions be satisfied. First that a region with the assumed properties of the $V_*$ region does exists, possibly a very  small one, second it requires to prove that this region can be extended. 

\NI The first condition  can be implemented by a local existence result in a small strip above $\Si_0/B_{R_0}$ and, in fact, of this solution we are interested only to the dependence region $V$ associated to the annulus in $\Si_0$ with $r\in[R_0, R_0+\de R]$, with $\de R$ arbitrary small and it is clear that starting from appropriate initial data we can satisfy in $V$ all the ``bootstrap assumptions". 

\NI The second condition to be satisfied requires to solve again an existence problem starting from the upper boundary of $V_*$. The strategy is not completely standard, but it has been already discussed in \cite{Kl-Ni:book}, Chapter 3, section 3.7.8, see also \cite{Niclast}. Therefore we sketch here only the more relevant points and the differences fron the situation discussed there. 

\NI The region whose existence we have to prove is a ``strip" above the last slice of the region $V_*$ of arbitrary small width. Nevertheless as there is no bound for the size of the region $V_*$ this existence problem is not a local problem in both directions, it is local in the ``outgoing cones" direction, but not local in the ``incoming cones" directions. Moreover this is a characteristic problem as its initial data are on the intersection of  the portions of two cones, one outgoing starting at the intersection of the last slice of $V_*$ with $\Si_0$, the other one being exactly the last slice. As the last slice can be very extended, to prove this result will require again a bootstrap argument, the difference being now that once we assume (again by a bootstrap assumption) that a portion of this strip does exist then to prove that it can be extended, therefore proving that the whole strip does exists requires in this case a real characteristic local problem which is easy to obtain. We do not give more details on it as this problem has already been treated an solved in two previous papers by the same authors, see \cite{Ca-Ni:char}, \cite{Ca-Ni:exist}. Observe finally that the initial data in this case automatically satisfy the constraints for the characteristic probelm as they are the restriction on these hypersurfaces of the Einstein solutions; the difference with respect to the discussion on \cite{Ca-Ni:exist} is that in this case the ``initial data" are near to Kerr instead than to Minkowski, but this is not a problem.

\NI Finally to complete the result we have to prove that, again, in the extended region, $V_*+\de V$, the estimates $\de{\cal R}\leq \ep_0, \de{\cal O}\leq \ep_0$ hold. This requires some care, in fact from the ``last slice initial data" one would be tempted to infer that the $\tilde{\cal Q}$ norms, to be finite avoiding possible logarithmic divergences, would require a weight $|u|^{\tilde\ga}$ with $\tilde\ga<\ga$, the analogous of the loss of decay from the initial data to the solutions. The problem is cured observing that once we have a solution in $V_*+\de V$ and a new double null canonical foliation we can, exactly as before, to estimate the  $\tilde{\cal Q}$ on the outgoing and incoming cones in terms of the same norms on $\Si_0$ and repeat all the previous lemmas to reobtain the correct decay for the $\de{\cal R}$ norms. Once this is done we repeat exactly the same procedure done in $V_*$ for the connection coefficients so that, finally, we have proved that in $V_*+\de V$ all the bootstrap assumptions are still valid and, by contradiction, that $V_*$ has to coincide with the global (external) spacetime.

\section{The final result}\label{Final result section}
We can state now with all the details our final result:
\begin{thm}\label{final version}
Assume that initial data  are given on $\Si_0$ such that, outside of a ball centered in the origin of radius $R_0$, they are different from the ``Kerr initial data of a Kerr spacetime with mass $M$ satisfying 
\[\frac{M}{R_0}<<1\ \ ,\ \ J\leq M^2\]
for some metric corrections decaying faster than $r^{-3}$ toward spacelike infinity together with its derivatives up to an order $q\geq 4$, namely
\footnote{The components of the metric tensor written in dimensional coordinates.}
\bea
g_{ij}=g^{(Kerr)}_{ij}+o_{q+1}(r^{-(3+\frac{\ga}{2})})\ \ ,\ \ k_{ij}=k^{(Kerr)}_{ij}+o_{q}(r^{-(4+\frac{\ga}{2})})
\eea
where $\ga>0$. Let us assume that the metric correction $\de g_{ij}$, the second fundamental form correction $\de k_{ij}$ are sufficiently small, namely the function ${\cal J}$ equation \ref{3.412} made by $L^2$ norms on $\Si_0$ of these quantities is small, \footnote{This will also imply a slightly stronger condition on the decay of the metric and second fundamental form components, basically that $\int_{R_0}^{\infty}dr r^{5+\ga}|\de g_{ij}|^2<\infty$,  $\int_{R_0}^{\infty}dr r^{7+\ga}|\de k_{ij}|^2<\infty$.}
\bea
{\cal J}(\Si_0,R_0; \de{^{(3)}\!}{\bf g}, \de{^{(3)}\!}{\bf k})\leq \varepsilon\ ,
\eea
then this initial data set has a unique development, ${\widetilde{\M}}$, defined outside the domain of influence of $B_{R_0}$ with the following properties:

\NI
i)\,\,\, ${\widetilde{\M}}={\widetilde{\M}}^{+}\cup{\widetilde{\M}}^{-}$ where ${\widetilde{\M}}^{+}$ consists of the part of
${\widetilde{\M}}$ which is in the future of $\Si/B_{R_0}$, ${\widetilde{\M}}^{-}$ the one to the past. 

\medn
\NI
ii) \,\,\,$({\widetilde{\M}}^{+},g)$ can be foliated by a canonical double null foliation
$\{C(u),\Cb(\ub)\}$  whose outgoing leaves $C(u)$ are complete
\footnote{By this we mean that the null geodesics generating $C(u)$ can be indefinitely
extended toward the future.}
for all $|\la|\geq |u_0|=R_0$. The boundary of $B_{R_0}$ can be chosen to be the intersection of $C(u_0)$ with $\Si_0$.
\medn

\NI
iii) \,\,\, The various null components of the Riemann tensor relative to the null frame associated to the double null canonical foliation, decay along the outgoing ``cones" in agreement with the ``Peeling Theorem".
\end{thm}

\NI {\bf Remark:} {It is clear that, from the way this result has been obtained, the condition $\frac{M}{R_0}<<1$ has to be such that the development we prove is far from the event horizon we assume to exist in a spacetime, near to the Kerr spacetime, which describes the complete outer region. In fact trying to go near to the event horizon or even to the ``photosphere region", see \cite{Blue} and references therein,  we would immediately find serious problems trying to control the $\tilde{\cal Q}$ norms in terms of the initial data ones.} 

\section{Conclusions}
As said in the introduction the global existence proof is separated from the ``peeling result". The global existence near Kerr spacetime required, in a broad sense, to subtract the Kerr part. This is done concerning the Riemann components looking for the estimates for the ``time derivative" of the Riemann tensor, which eliminates the contribution of the Kerr spacetime, or more in general of any stationary spacetime. The subtraction for the connection coefficients is viceversa made in a more general way as the subtracted part has not to be time independent.\footnote{In principle one could use the same strategy used for the Riemann components, namely to defined ``time derivatives" of the connection coefficients, write for them the structure equations, estimate their norms and recover by a time integration the connection coefficients, this method, although less general, should give the same result we have obtained.} This is in some sense the more original part as, in a more external region, the peeling decay has been already proved in \cite{Ni;Peel}.

\section{Appendix}
\subsection{Proof of various inequalities and lemmas}

\NI {\bf Proof of inequality \ref{estq}:}

\NI This result is just one of the standard estimates used in Chapter 5 of \cite{Kl-Ni:book} to show how from the control of the $\cal Q$ norms one can obtain the control of the $\sup$ norms of the null Riemann components. The only difference  is that here we are considering the $\tilde R$ Weyl field instead of the Riemann tensor $R$ and that the $\cal Q$ norms are substituted by the $\tilde{\cal Q}$ norms. We write the proof explicitely to make the paper more consistent.
\smallskip

\NI Observe that from Lemma 5.1.1 in \cite{Kl-Ni:book} we have
\bea
&&\ML\int_{S(\la,\nu)}|\la|^{5+\ep}r^4r^2|\nabb\ro(\tilde R)|^2\leq \int_{S(\la,\nu)}|\la|^{5+\ep}r^4|\Lie_O\ro(\tilde R)|^2\eql{CB1}\\
&&\ML\leq\int_{S(\la,\nu)}|\la|^{5+\ep}r^4|\ro(\lie_O\tilde R)|^2+\int_{S(\la,\nu)}|\la|^{5+\ep}r^4|\tr{^{(O)}\!}\pi|^2|\ro(\tilde R)|^2+{\{good\ corrections\}}\ .\nn
\eea
From the estimate for $\ro({\tilde R})$ in Theorem \ref{TH2.2} and Lemma \ref{Lemma1}
\[\sup_{{\cal K}}|r^3|\la|^{3+\frac{\ep}{2}}\ro({\tilde R})|\leq \hat{c}\varepsilon\]
with \[\hat{c}\geq \left(c_0\frac{M}{R_0}+c_2\right)\]
and from the estimate 
\[\sup_{{\cal K}}|r^2{^{(O)}\!}\pi|\leq c{M^2}\ ,\]
the second integral in \ref{CB1} can be estimated by
\bea
&&\int_{S(\la,\nu)}|\la|^{5+\ep}r^4|\tr{^{(O)}\!}\pi|^2|\ro(\tilde R)|^2\leq
\hat{c}M^4\varepsilon^2\int_{S(\la,\nu)}|\la|^{5+\ep}\frac{1}{r^6|\la|^{6+\ep}}\nn\\
&&\leq \hat{c}M^4\varepsilon^2|\la|^{5+\ep}\frac{1}{r^4|\la|^{6+\ep}}\leq 4\pi\hat{c}M^4\varepsilon^2\frac{1}{r^4|\la|}
\eea
where for notational simplicity we write
\[\varepsilon_{N_0}\equiv \!\left(\varepsilon+\frac{\ep_0}{N_0}\right)\ .\]
This implies 
\bea
&&\int_{\Cb(\nu)}|\la|^{5+\ep}r^4r^2|\nabb\ro({\tilde R})|^2\leq \int_{\Cb(\nu)}|\la|^{5+\ep}r^4|\ro(\lie_O{\tilde R})|^2+4\pi\hat{c}\varepsilon_{N_0}^2\frac{M^4}{R_0^4}\nn\\
&&\leq c{\tilde{\cal Q}}_{\Si_0}+4\pi\hat{c}\varepsilon_{N_0}^2\frac{M^4}{R_0^4}
\leq \left(c+4\pi\hat{c}\frac{M^4}{R_0^4}\right)\varepsilon_{N_0}^2\eql{2.85ff}
\eea
Using Corollary 4.1.1 of \cite{Kl-Ni:book} we have
\bea
\left(\int_{S(\la,\nu)}r^2|\la|^2|F|^4\right)^{\frac{1}{4}}\leq \left(\int_{S(\la_1,\nu)=\Cb(\nu)\cap\Si_0}r^2|\la|^2|F|^4\right)^{\frac{1}{4}}
+\left(\int_{\Cb(\nu)} \left[|F|^2+\c\c\c\c \right]\right)^{\frac{1}{2}}\ \ \ \ \ \ \eql{bbqq}
\eea
Choosing \ $F=|\la|^{\frac{5+\ep}{2}}r^3\nabb\ro({\tilde R})$ we obtain
\bea
&&\ML|r^{4-\frac{2}{4}}|\la|^{3+\frac{\ep}{2}}\nabb\ro({\tilde R})|_{p=4,S}
=\left(\int_{S(\la,\nu)}||\la|^{3+\frac{\ep}{2}}r^{4-\frac{2}{4}}\nabb\ro({\tilde R})|^4\right)^{\!\frac{1}{4}}\eql{newest}\\
&&\ML\leq \left(\int_{S(\la_1,\nu)=\Cb(\nu)\cap\Si_0}||\la|^{3+\frac{\ep}{2}}r^{4-\frac{2}{4}}\nabb\ro({\tilde R})|^4\right)^{\!\frac{1}{4}}
\!+\left(\int_{\Cb(\nu)}\left[||\la|^{\frac{5+\ep}{2}}r^{3}\nabb\ro({\tilde R})|^2+\c\c\c\c\right]\right)^{\!\frac{1}{2}}\nn\\
&&\ML\leq \left(\int_{S(\la_1,\nu)=\Cb(\nu)\cap\Si_0}||\la|^{3+\frac{\ep}{2}}r^{4-\frac{2}{4}}\nabb\ro({\tilde R})|^4\right)^{\!\frac{1}{4}}
\!+\left(c+4\pi\hat{c}\frac{M^4}{R_0^4}\right)^{\!\!\frac{1}{2}}\!\varepsilon_{N_0}\nn\\
&&\ML\leq \hat{c}_0\varepsilon\!+\left(c+4\pi\hat{c}\frac{M^4}{R_0^4}\right)^{\!\!\frac{1}{2}}\!\varepsilon_{N_0}=\hat{c}_0\varepsilon+\hat{c}_1\varepsilon_{N_0}
=\hat{c}_2\varepsilon_{N_0}\ ,\nn
\eea

\NI where in the the last inequality we assume that initial data are such that
\bea
\left(\int_{S(\la_1,\nu)=\Cb(\nu)\cap\Si_0}||\la|^{3+\frac{\ep}{2}}r^{4-\frac{2}{4}}\nabb\ro({\tilde R})|^4\right)^{\!\frac{1}{4}}
\leq \hat{c}_0\varepsilon\ .\ \ \ \eql{Initcond2}
\eea
Observe now that inequality \ref{2.85ff} 
\bea
&&\int_{\Cb(\nu)}|\la|^{5+\ep}r^4r^2|\nabb\ro({\tilde R})|^2=\int_{\la_1}^{\la}||\la|^{\frac{5+\ep}{2}}r^3\nabb\ro({\tilde R})|^2_{p=2,S}
\leq \hat{c}_1^2\varepsilon_{N_0}^2\nn
\eea
imply that, 
\bea
||\la|^{3+\frac{\ep}{2}}r^{4-\frac{2}{p}}\nabb\ro({\tilde R})|_{p=2,S}\leq \hat{c}_1\varepsilon_{N_0}\ .\eql{2.93ff}
\eea
Interpolating between \ref{newest} and \ref{2.93ff} we obtain for $p\in[2,4]$,
\bea
|r^{4-\frac{2}{p}}|\la|^{3+\frac{\ep}{2}}\nabb\ro({\tilde R})|_{p,S}\leq c'\varepsilon_{N_0} \eql{2.94ff}
\eea
proving the lemma. 
\smallskip

\NI In the proofs of various Lemmas, for the metric components we can use directly the difference between a quantity and its Kerr counterpart, recall that going from $O$ to $\hat O$ and from $\hat O$ to $O^{(Kerr)}$ in two steps is required for the connection coefficients and for the null components of the Riemann tensor. Moreover it has been already proved that the difference ${\hat O}-O^{(Kerr)}$ is bounded and ``small", see Lemma \ref{hatestim}. Therefore for the metric components with $\hat{}$ we denote the Kerr part , $\hat{O}_{(0)}=O^{(Kerr)}_{(0)}$, while their meaning is different for the connection coefficients and for the null components of the Riemann tensor. We also write
$\widehat{(\chi)}\ ,\ \widehat{(\chib)}$ to indicate the ``hat" part of $\chi$ and $\chib$ not to be confused with their traceless part $\chih$, $\chibh$.
\smallskip

\NI{\bf Proof of inequalities \ref {Pnormsbootass}}

\NI Looking at the explicit expressions \ref{pito} and \ref{bass2a} it is immediate to realize that inequalities \ref{bass2a} are immediately derived by the bootstrap assumptions \ref{Oboosestwithdec} while for the right hand side terms in \ref{pito} we only need to control $\dd_3 \de X,\dd_4\de X$ and $\nabb \de X$. The control of the  first two terms arises from the explicit expression of $\dd_3 \de X$, see \ref{findeX2}, Lemma \ref{finOOest}. The control of $\nabb \de X$ requires to derive equation \ref{findeX2} and obtain from this equation an estimate for $\nabb\de X$ proceeding as in Lemma \ref{finOOest}. Observe that the  loss of derivatives due to the presence on the right end side of $\nabb\ze$ is only apparent as to estimate $\nabb\ze$ there is not any loss of derivatives, see the discussion in subsection \ref{SSloss}.
\medskip

\NI{\bf Proof of Lemma \ref{Comm1}:}
 We have
\bea
[{T_0},e_a]=\frac{\oom}{2}(\dddd_3e_a+\dddd_4e_a)-\frac{\oom}{2}(\chi+\chib)_{ac}e_c-[X,e_a]\ \ .
\eea
This follows observing that 
\[T_0=\frac{\oom}{2}(e_3+e_4)-X\]
and, see \cite{Kl-Ni:book} eqs. (3.1.45),
\bea
&&[\oom e_3,e_a]=\oom\ddb_{e_3}e_a-\oom\chib_{ac}e_c\nn\\
&&[\oom e_4,e_a]=\oom\ddb_{e_4}e_a-\oom\chi_{ac}e_c\ .
\eea
As $[{T_0},e_a]^{(Kerr)}=0$ we can write the commutator in terms of the $\de O$ correction terms, more precisely:
\bea
[{T_0},e_a]\!\!&=&\!\!\frac{\de\oom}{2}(\ddb_{e_3}+\ddb_{e_4})e_a+\frac{\hat{\oom}}{2}\left((\ddb_{e_3}-\widehat{\ddb_{e_3}})e_a+(\ddb_{e_4}-\widehat{\ddb_{e_4}})e_a\right)\nn\\
\!&+&\!\frac{\hat{\oom}}{2}\left(\widehat{\ddb_{e_3}}+\widehat{\ddb_{e_4}}\right)\de e_a\nn\\
\!&-&\!\left[\frac{\de\oom}{2}(\chi+\chib)_{ac}e_c+\frac{\hat{\oom}}{2}(\de\chi+\de\chib)_{ac}e_c
+\frac{\hat{\oom}}{2}(\widehat{(\chi)}+\widehat{(\chib)})_{ac}\de e_c\right]\\
\!&+&\!\frac{\hat{\oom}}{2}\!\left[(\widehat{(\chi)}-\chi^{(Kerr)})_{ac}+(\widehat{(\chib)}-\chi^{(Kerr)})_{ac}\right]\!{\hat e}_c-[\de X,e_a]-[X,\de e_a]\ .\nn
\eea
The following relation holds:
\bea
&&(\ddb_{e_3}-\widehat{\ddb_{e_3}})e_a^{\ro}=\de\Pi^{\ro}_{\si}(\ddb_{e_3}e_a)^{\si}+{\hat\Pi}^{\ro}_{\si}(\dd_{e_3}-\widehat{\dd_{e_3}})e_a^{\si}
+{\hat\Pi}^{\ro}_{\si}\widehat{\dd_{e_3}}\de e_a^{\si}\ .\nn
\eea
As
\bea
\ML\ML(\dd_{e_3}-\widehat{\dd_{e_3}})e_a\!&=&\!\left[(e_3-\hat{e}_3)^{\mu}\dd_{\mu}+\hat{e}_3^{\mu}(\dd_{\mu}-\hat{\dd}_{\mu})e_a\right]\eql{Rel1}\\
\ML\ML\!&=&\!\left[\left(\frac{1}{\oom}-\frac{1}{{\hat{\oom}}}\right)(\de^{\mu}_u+\om_B\de^{\mu}_{\phi})+\hat{e}_3^{\mu}e^{\ro}_a(\Ga^{\si}_{\mu\ro}-\hat{\Ga}^{\si}_{\mu\ro})\frac{\pr}{\pr\om^{\si}}\right]\nn\\
\!&=&\!-\frac{\de\oom}{{\hat{\oom}}}\dd_{e_3}e_a+\hat{e}_3^{\mu}e^{\ro}_a(\Ga^{\si}_{\mu\ro}-\hat{\Ga}^{\si}_{\mu\ro})\frac{\pr}{\pr\om^{\si}}\ ,\nn
\eea
therefore
\bea
[{T_0},e_a]^{\si}\!\!&=&\!\!\frac{\de\oom}{2}(\ddb_{e_3}+\ddb_{e_4})e_a^{\si}+\frac{\hat{\oom}}{2}\de\Pi^{\si}_{\la}(\dd_{e_3}+\dd_{e_4})e^{\la}_a\nn\\
\!\!&+&\!\!\frac{\hat{\oom}}{2}{\hat\Pi}^{\si}_{\la}\left[-\frac{\de\oom}{{\hat{\oom}}}(\dd_{e_3}+\dd_{e_4})e^{\la}_a+(\hat{e}_3^{\mu}+\hat{e}_4^{\mu})e^{\tau}_a(\Ga^{\la}_{\mu\tau}-\hat{\Ga}^{\la}_{\mu\tau})\right]+\frac{\hat{\oom}}{2}\left(\widehat{\ddb_{e_3}}+\widehat{\ddb_{e_4}}\right)\de e^{\si}_a\nn\\
\!&-&\!\left[\frac{\de\oom}{2}(\chi+\chib)_{ac}e^{\si}_c+\frac{\hat{\oom}}{2}(\de\chi+\de\chib)_{ac}e^{\si}_c
+\frac{\hat{\oom}}{2}(\widehat{(\chi)}+\widehat{(\chib)})_{ac}\de e^{\si}_c\right]\nn\\
\!&+&\!\!\frac{\hat{\oom}}{2}\!\left[(\widehat{(\chi)}-\chi^{(Kerr)})_{ac}+(\widehat{(\chib)}-\chi^{(Kerr)})_{ac}\right]\!{\hat e}^{\si}_c-[\de X,e_a]^{\si}-[X,\de e_a]^{\si}\nn\\
\!\!&=&\!\!\left[\frac{\de\oom}{2}\Pi^{\si}_{\la}+\frac{\hat{\oom}}{2}\de\Pi^{\si}_{\la}-\frac{\de\oom}{2}{\hat\Pi}^{\si}_{\la}\right](\dd_{e_3}+\dd_{e_4})e^{\la}_a\nn\\
\!\!&+&\!\!\frac{\hat{\oom}}{2}{\hat\Pi}^{\si}_{\la}(\hat{e}_3^{\mu}+\hat{e}_4^{\mu})e^{\tau}_a(\Ga^{\la}_{\mu\tau}-\hat{\Ga}^{\la}_{\mu\tau})+\frac{\hat{\oom}}{2}\left(\widehat{\ddb_{e_3}}+\widehat{\ddb_{e_4}}\right)\de e^{\si}_a\nn\\
\!&-&\!\left[\frac{\de\oom}{2}(\chi+\chib)_{ac}e^{\si}_c+\frac{\hat{\oom}}{2}(\de\chi+\de\chib)_{ac}e^{\si}_c
+\frac{\hat{\oom}}{2}(\widehat{(\chi)}+\widehat{(\chib)})_{ac}\de e^{\si}_c\right]\nn\\
\!&+&\!\!\frac{\hat{\oom}}{2}\!\left[(\widehat{(\chi)}-\chi^{(Kerr)})_{ac}+(\widehat{(\chib)}-\chi^{(Kerr)})_{ac}\right]\!{\hat e}^{\si}_c-[\de X,e_a]^{\si}-[X,\de e_a]^{\si}\nn\\
\!\!&=&\frac{\oom}{2}\de\Pi^{\si}_{\la}(\dd_{e_3}+\dd_{e_4})e^{\la}_a
+\frac{\hat{\oom}}{2}{\hat\Pi}^{\si}_{\la}(\hat{e}_3^{\mu}+\hat{e}_4^{\mu})e^{\tau}_a(\Ga^{\la}_{\mu\tau}-\hat{\Ga}^{\la}_{\mu\tau})+\frac{\hat{\oom}}{2}\left(\widehat{\ddb_{e_3}}+\widehat{\ddb_{e_4}}\right)\de e^{\si}_a\nn\\
\!&-&\!\left[\frac{\de\oom}{2}(\chi+\chib)_{ac}e^{\si}_c+\frac{\hat{\oom}}{2}(\de\chi+\de\chib)_{ac}e^{\si}_c
+\frac{\hat{\oom}}{2}(\widehat{(\chi)}+\widehat{(\chib)})_{ac}\de e^{\si}_c\right]\nn\\
\!&+&\!\!\frac{\hat{\oom}}{2}\!\left[(\widehat{(\chi)}-\chi^{(Kerr)})_{ac}+(\widehat{(\chib)}-\chi^{(Kerr)})_{ac}\right]\!{\hat e}^{\si}_c-[\de X,e_a]^{\si}-[X,\de e_a]^{\si}\nn
\eea
and equation \ref{Teacom} follows. The second commutator appearing in \ref{Te4com}can be written,
recalling that in the Kerr spacetime the coomutator is identically zero,
\bea
&&\ML[T_0,e_4]=\frac{1}{2}[\oom(e_3+e_4),e_4]-[X,e_4]=\frac{\oom}{2}[e_3,e_4]-\frac{1}{2}e_4(\oom)\left(e_3+e_4\right)-[X,e_4]\nn\\
&&\ML=\frac{\oom}{2}\big((\dd_4\log\oom)e_3-(\dd_3\log\oom)e_4\big)-\frac{\oom}{2}(\dd_4\log\oom)e_3-\frac{\oom}{2}(\dd_4\log\oom)e_4\nn\\
&&\ML+\de\left[2\oom\ze(e_a)e_a\right]-[\de X,e_4]-[\hat{X},\de e_4]\nn\\
&&\ML=\oom(\omb+\om)e_4+\de\left[2\oom\ze(e_a)e_a\right]-[\de X,e_4]-[\hat{X},\de e_4]\nn\\
&&\ML=\oom(\omb+\om)e_4+2\de\oom\ze(e_a)e_a+2\hat{\oom}\de[\ze(e_a)e_a]-[\de X,e_4]-[\hat{X},\de e_4]\ .\nn
\eea 
We have
\bea
[\de X,e_4]=-\de X^c\left(\frac{\pr_c\oom}{\oom}\right)e_4+\frac{\de X^c}{\oom}\left(\pr_cX^d\right)\frac{\pr}{\pr\om^d}
-e_4(\de X^c)\frac{\pr}{\pr\om^c}\ \ \ \ \ \ \ \ \ \ \ \ \eql{deZ}
\eea
and recalling the expression
\bea
\de e_4=-\frac{\de\oom}{\oom^{(Kerr)}}e_4+\frac{1}{\oom^{(Kerr)}}\de X
=\left(-\frac{\de\oom}{\oom\hat{\oom}}(\de^{\mu}_{\ub}+\de^{\mu}_cX^c)+\frac{\de X^d}{\oom^{(Kerr)}}\de^{\mu}_d\right)\frac{\pr}{\pr x^{\mu}}\nn
\eea
we have also
\bea
[\hat{X},\de e_4]
=\om_B\frac{\pr}{\pr\phi}\left(-\frac{\de\oom}{\oom\hat{\oom}}(\de^{\mu}_{\ub}+\de^{\mu}_cX^c)+\frac{\de X^d}{\oom^{(Kerr)}}\de^{\mu}_d\right)\frac{\pr}{\pr x^{\mu}}-\de e_4(\hat{X})\frac{\pr}{\pr\phi}\ ,\ \ \ 
\eea
therefore 
\bea
\ML[T_0,e_4]\!&=&\! \oom(\omb+\om)e_4+2\de\oom\ze(e_a)e_a+2\hat{\oom}\de[\ze(e_a)e_a]\nn\\
\!&-&\!\de X^c\left(\frac{\pr_c\oom}{\oom}\right)e_4+\frac{\de X^c}{\oom}\left(\pr_cX^d\right)\frac{\pr}{\pr\om^d}
-e_4(\de X^c)\frac{\pr}{\pr\om^c}\nn\\
\ML\!&+&\!\om_B\frac{\pr}{\pr\phi}\left(\frac{\de\oom}{\oom\hat{\oom}}(\de^{\mu}_{\ub}+\de^{\mu}_cX^c)-\frac{\de X^d}{\hat\oom}\de^{\mu}_d\right)\frac{\pr}{\pr x^{\mu}}+\de e_4(\hat{X}^d)\frac{\pr}{\pr\om^d}\ \ \ \ \ \ \ \eql{Te4coma}
\eea
and equation \ref{Te4com} is satisfied.
\bea
&&\ML[T_0,e_3]=\frac{1}{2}[\oom(e_3+e_4),e_3]-[X,e_3]=-\frac{\oom}{2}[e_3,e_4]-\frac{1}{2}e_3(\oom)\left(e_3+e_4\right)-[X,e_3]\nn\\
&&\ML=-\frac{\oom}{2}\big((\dd_4\log\oom)e_3-(\dd_3\log\oom)e_4\big)-\frac{\oom}{2}(\dd_3\log\oom)e_3-\frac{\oom}{2}(\dd_3\log\oom)e_4\nn\\
&&\ML-\de\left[2\oom\ze(e_a)e_a\right]-[\de X,e_3]-[\hat{X},\de e_3]\nn\\
&&\ML=\oom(\omb+\om)e_3-\de\left[2\oom\ze(e_a)e_a\right]-[\de X,e_3]-[\hat{X},\de e_3]\nn\\
&&\ML=\oom(\omb+\om)e_3-2\de\oom\ze(e_a)e_a-2\hat{\oom}\de[\ze(e_a)e_a]-[\de X,e_3]-[\hat{X},\de e_3]\ .\nn
\eea 
We have
\bea
[\de X,e_3]=-\de X^c\left(\frac{\pr_c\oom}{\oom}\right)e_3+\frac{\de X^c}{\oom}(\pr_c X_{(Kerr)}^d)\frac{\pr}{\pr\om^d}
-e_3(\de X^c)\frac{\pr}{\pr\om^c}\ \ \ \ \ \ \ \ \ \ \ \ \eql{deZ}
\eea
and recalling the expression
\bea
\de e_3=-\frac{\de\oom}{\oom^{(Kerr)}}e_3
=-\frac{\de\oom}{\oom\hat{\oom}}(\de^{\mu}_{\ub}+\de^{\mu}_cX_{(Kerr)}^c)\frac{\pr}{\pr x^{\mu}}\nn
\eea
we have also
\bea
[\hat{X},\de e_3]
\!&=&\!\om_B\frac{\pr}{\pr\phi}\left(-\frac{\de\oom}{\oom\hat{\oom}}(\de^{\mu}_{\ub}+\de^{\mu}_c X_{(Kerr)}^c)\right)\!\frac{\pr}{\pr x^{\mu}}-\de e_3(X_{(Kerr)})\frac{\pr}{\pr\phi}\nn\\
\!&=&\!-\frac{\om_B\oom}{\oom^{(Kerr)}}\frac{\pr}{\pr\phi}\!\left(\frac{\de\oom}{\oom}\right)\!e_3-\de e_3(X_{(Kerr)})\frac{\pr}{\pr\phi}\  ,\ \ \ 
\eea
so that finally
\bea
[T_0,e_3]\!&=&\!\oom(\omb+\om)e_3-2\de\oom\ze(e_a)e_a-2\hat{\oom}\de[\ze(e_a)e_a]\nn\\
\!&+&\!\de X^c\left(\frac{\pr_c\oom}{\oom}\right)e_3-\frac{\de X^c}{\oom}(\pr_c X_{(Kerr)}^d)\frac{\pr}{\pr\om^d}
+e_3(\de X^c)\frac{\pr}{\pr\om^c}\nn\\
\!&+&\!\frac{\om_B\oom}{\oom^{(Kerr)}}\frac{\pr}{\pr\phi}\!\left(\frac{\de\oom}{\oom}\right)\!e_3+\de e_3(X_{(Kerr)})\frac{\pr}{\pr\phi}\ .
\eea
proving equation \ref{Te3com}.
\medskip

\NI{\bf Proof of Lemma \ref{Comm2}}
\bea
&&\ML\ML{\bf g}([T_0,e_4],e_d)=2\de\oom\ze(e_d)+2\hat{\oom}{\bf g}(\de[\ze(e_a)e_a],e_d)\\
&&\ \ \ \ \ \ \ \ \ +\left[\frac{\de X^c}{\oom}\left(\pr_cX^e\right)-e_4(\de X^e)-\frac{\om_B}{\hat\oom}\pr_{\phi}{\de X^e}+\de e_4(\hat{X}^e)\right]{\bf g}(\frac{\pr}{\pr\om^e},e_d))\nn
\eea
proving equation \ref{2.184}\ .
\bea
{\bf g}([T_0,e_4],e_3)
\!&=&\!-2\oom(\om+\omb)+2\de X^c\!\left(\frac{\pr_c\oom}{\oom}\right)
+2\frac{\om_B\de\oom}{{\oom^{(Kerr)}}\oom^2}\pr_{\phi}\oom-2\frac{\om_B}{{\oom^{(Kerr)}}\oom}\pr_{\phi}\de\oom\ ,\nn
\eea
proving equation \ref{Te4com2} .
Equations \ref{2.184b}, \ref{2.185}, \ref{2.186} are immediate.

\medskip

\NI{\bf Proof of Lemma \ref{T0comest}:}

\NI We present here some basic dimensional estimates for all these quantities, it should be easy for the careful reader to make all these estimates more precise. In these estimates we do not distinguish between the $r$ and the $|u|$.
\medskip

\NI{\bf $\{|{\bf g}([T_0,e_a],e_d)|$\} :} 
\medskip

\NI Let us estimate all the terms appearing in \ref{commT0eab}.
\medskip

\NI {\bf i)\ \ $\frac{\oom}{2}\de\Pi^{\si}_{\la}(\dd_{e_3}+\dd_{e_4})e^{\la}_a{\bf g}(\frac{\pr}{\pr\om^{\si}},e_d)$:}

\NI $\frac{\oom}{2}\de\Pi^{\si}_{\la}=O\left({\ep_0}r^{-(3+\de)}\right)$\ ,\ $(\dd_{e_3}+\dd_{e_4})e^{\la}_a{\bf g}(\frac{\pr}{\pr\om^{\si}},e_d)=O(r^{-1})$
\medskip

\NI {\bf ii)\ \  $(\hat{e}_3^{\mu}+\hat{e}_4^{\mu})e^{\ro}_a(\Ga^{\si}_{\mu\ro}-\hat{\Ga}^{\si}_{\mu\ro}){\bf g}(\frac{\pr}{\pr\om^{\si}},e_d)$\ :}

\NI $(\hat{e}_3^{\mu}+\hat{e}_4^{\mu})e^{\ro}_a=O(r^{-2})$\ , $(\Ga^{\si}_{\mu\ro}-\hat{\Ga}^{\si}_{\mu\ro})= \ga^{-1}\pr\de\ga+\de\ga^{-1}\pr\ga=O(r^{-2}){\ep_0}{r^{-(1+\de)}}$\ , ${\bf g}(\frac{\pr}{\pr\om^{\si}},e_d)=O(r)$ so that, finally, 
\[\left|(\hat{e}_3^{\mu}+\hat{e}_4^{\mu})e^{\ro}_a(\Ga^{\si}_{\mu\ro}-\hat{\Ga}^{\si}_{\mu\ro}){\bf g}(\frac{\pr}{\pr\om^{\si}},e_d)\right|
=O\!\left(\frac{\ep_0}{r^{4+\de}}\right)\ .\]
\medskip

\NI {\bf iii)\ \  $\frac{\hat{\oom}}{2}\!\!\left({\bf g}(\widehat{\dd_{e_3}}\de e_a,e_d)+{\bf g}(\widehat{\ddb_{e_4}}\de e_a,e_d)\right)$\ :}
\medskip

\NI At each point of any $S$  $\{\hat{e}_a\}$ forms a, not orthogonal, basis. We can write in this basis the $\{e_a\}$ basis and then their difference,  \bea
\de e_a=e_a-\hat{e}_a=c_{ab}{\hat{e}}_b\ .
\eea
$c_{ab}$ is known explicitely and depends on $\de\ga$, more explicitely we have, see Lemma \ref{hatestim},
\bea
e_a={\hat e}_a\!\left(1+c_1\frac{\de\ga}{|\ga|}\right)\!{\hat e}_a+c_2\left(\frac{\de\ga}{|\ga|}\right)\!{\hat e}_b\ \ \;\ \ b\neq a\ , \eql{dea1}
\eea 
which implies
\bea
\de e_a=c_1\frac{\de\ga}{|\ga|}{\hat e}_a+c_2\left(\frac{\de\ga}{|\ga|}\right)\!{\hat e}_b\ \ \;\ \ b\neq a\  \eql{dea2}
\eea
and
\bea
\left|({\bf g}(\widehat{\dd_{e_3}}\de e_a,e_d)\right|\leq O(r^{-1})\left|\frac{\de\ga}{|\ga|}\right||{\bf g}({\hat e}_a,e_d)|\leq c\frac{\ep_0}{r^{4+\de}}\ . \eql{dea3}
\eea

\NI {\bf iv)\ \  $\frac{\de\oom}{2}(\chi+\chib)_{ad}-\frac{\hat{\oom}}{2}(\de\chi+\de\chib)_{ad}$\ :}
\medskip

\NI These terms are easy to control using the bootstrap assumptions for $\de\oom$ and $\de\chi+\de\chib$. In particular the first term is controlled in the following way
\bea
\left|\frac{\de\oom}{2}(\chi+\chib)_{ad}\right|=O\left(\frac{\ep_0}{r^{3+\de}}\right)O\left(\frac{M^2}{r^3}\right)
\leq c\frac{M^2}{R_0^2}\frac{\ep_0}{r^{4+\de}}\ .
\eea
which implies
\bea
\left|\frac{\de\oom}{2}(\chi+\chib)_{ad}-\frac{\hat{\oom}}{2}(\de\chi+\de\chib)_{ad}\right|\leq c\frac{M^2}{R_0^2}\frac{\ep_0}{r^{4+\de}}\ .
\eea
\medskip

\NI {\bf v)\ \  $\frac{\hat{\oom}}{2}(\widehat{(\chi)}+\widehat{(\chib)})_{ac}{\bf g}(\de e_c,e_d)$\ :}
\medskip

\NI To control this term we have to control ${\bf g}(\de e_c,e_d)$. From \ref{dea1},...,\ref{dea3} and Lemma \ref{hatestim} it follows
\bea
&&|{\bf g}(\de e_c,e_d)|=O\left(\frac{\ep_0}{r^{3+\de}}\right)\ \ ,\ \ 
\left|\frac{\hat{\oom}}{2}(\widehat{(\chi)}+\widehat{(\chib)})_{ac}\right|=O\left(\frac{M^2}{r^3}\right)
\eea
\medskip

\NI {\bf v)\ \ ${\bf g}([\de X,e_a],e_d)-{\bf g}([X,\de e_a],e_d)$  \ :}

\NI Observe that 
\bea
&&[\de X,e_a]=\left(\de X^e\frac{\pr e_a^c}{\pr\om^e}-e_a^e\frac{\pr \de X^c}{\pr\om^e}\right)\frac{\pr}{\pr\om^c}\nn\\
&&{\bf g}([\de X,e_a],e_d)=\left(\de X^e\frac{\pr e_a^c}{\pr\om^e}-e_a^e\frac{\pr \de X^c}{\pr\om^e}\right){\bf g}({\pr\om^c},e_d)\nn\\
&&[X,\de e_a]=\left(X^e\frac{\pr \de e_a^c}{\pr\om^e}-\de e_a^e\frac{\pr X^c}{\pr\om^e}\right)\frac{\pr}{\pr\om^c}\\
&&{\bf g}([X,\de e_a],e_d)=\left(X^e\frac{\pr \de e_a^c}{\pr\om^e}-\de e_a^e\frac{\pr X^c}{\pr\om^e}\right){\bf g}({\pr\om^c},e_d)\nn
\eea
and looking at the bootstrap assumptions for $\de X$ and to Lemma \ref{hatestim} for $\de e_a^c$ we obtain easily
\bea
&&\left|\left(\de X^e\frac{\pr e_a^c}{\pr\om^e}-e_a^e\frac{\pr \de X^c}{\pr\om^e}\right)\right|=O\left(\frac{\ep_0}{r^{5+\de}}\right)\nn\\
&&\left|\left(X^e\frac{\pr \de e_a^c}{\pr\om^e}-\de e_a^e\frac{\pr X^c}{\pr\om^e}\right)\right|=O\left(\frac{\ep_0}{r^{5+\de}}\right)\nn\\
&&|{\bf g}({\pr\om^c},e_d)|=O(r)
\eea
and the estimate follows.
\medskip

\NI{\bf $\{|{\bf g}([T_0,e_4],e_3)|$\} :} 
\medskip

\NI Let us estimate all the terms appearing in \ref{Te4com2}.
\medskip

\NI {\bf i)\ \  $-2\oom(\omb+\om)$ \ :}
\medskip

\NI To estimate this term it is enough to use the bootstrap assumptions of $\de{\cal O}_1$ recalling that in Kerr $\om+\omb=0$\ . Therefore
\bea
|2\oom(\omb+\om)|\leq c\frac{\ep_0}{r^{2+\de}|u|^2}
\eea
 \medskip

\NI {\bf ii)\ \  $2\de X^c\!\left(\frac{\pr_c\oom}{\oom}\right)$ \ :}
\medskip

\NI Recall that $\pr_c$ does not give an extra $r$, therefore
\bea
\left|2\de X^c\!\left(\frac{\pr_c\oom}{\oom}\right)\right|\leq c\left|\de X\right|\leq c\frac{\ep_0}{r^{2+\de}|u|^2}
\eea
\medskip

\NI {\bf iii)\ \ $2\frac{\om_B\de\oom}{{\oom^{(Kerr)}}\oom^2}\pr_{\phi}\oom$  \ :}
\medskip

\NI Immediately
\bea
\left|2\frac{\om_B\de\oom}{{\oom^{(Kerr)}}\oom^2}\pr_{\phi}\oom\right|\leq c|\om_B||\de\oom|\leq c\frac{M^2}{r^3}\frac{\ep_0}{r^3}
\leq c\frac{M^2}{R_0^2}\frac{\ep_0}{r^{2+\de}|u|^2}
\eea
\medskip

\NI {\bf iv)\ \  $-2\frac{\om_B}{{\oom^{(Kerr)}}\oom}\pr_{\phi}\de\oom$ \ :}
\medskip

\NI This estimate goes exactly as the previous one, therefore
\bea
\left|2\frac{\om_B}{{\oom^{(Kerr)}}\oom}\pr_{\phi}\de\oom\right|\leq c|\om_B||\de\oom|\leq c\frac{M^2}{r^3}\frac{\ep_0}{r^3}
\leq c\frac{M^2}{R_0^2}\frac{\ep_0}{r^{2+\de}|u|^2}\ \ \ .
\eea
\medskip

\NI{\bf $\{|{\bf g}([T_0,e_4],e_d)|$\} :} 
\medskip

\NI Let us estimate all the terms appearing in \ref{2.184}.
\medskip

\NI {\bf i)\ \  $2\de\oom\ze(e_d)$ \ :}
\medskip

\NI Immediately
\bea
\left|2\de\oom\ze(e_d)\right|\leq c|\de\oom||\ze(e_d)|\leq c\frac{\ep_0}{r^{3+\de}}\frac{M^2}{r^3}\leq c\frac{M^2}{R_0^2}\frac{\ep_0}{r^{2+\de}|u|^2}
\eea
\medskip

\NI {\bf ii)\ \  $2\hat{\oom}{\bf g}(\de[\ze(e_a)e_a],e_d)$ \ :}
\medskip

\NI We have
\bea
2\hat{\oom}{\bf g}(\de[\ze(e_a)e_a],e_d)= 2\hat{\oom}\de\ze(e_d)+2\hat{\oom}{\hat\ze}(e_a){\bf g}(\de e_a,e_d)
\eea
and from the bootstrap assumptions we have
\bea
\left|2\hat{\oom}{\bf g}(\de[\ze(e_a)e_a],e_d)\right|\!&\leq&\! c\left(|\hat{\oom}||\de\ze(e_d)|+|\hat{\oom}||{\hat\ze}(e_a)||{\bf g}(\de e_a,e_d)|\right)\\
\!&\leq&\! c\frac{\ep_0}{r^{2+\de}|u|^2}+c\frac{M^2}{r^3}\frac{\ep_0}{r^{3+\de}}\leq  c\left(1+\frac{M^2}{R_0^2}\right)\frac{\ep_0}{r^{2+\de}|u|^2}\nn
\eea
\medskip

\NI {\bf iii)\ \  $\left[\frac{\de X^c}{\oom}\left(\pr_cX^e\right)-\frac{\om_B}{\hat\oom}\pr_{\phi}{\de X^e}+\de e_4(\hat{X}^e)\right]{\bf g}(\frac{\pr}{\pr\om^e},e_d)$ }

\NI We have
\bea
&&\ML\ML\ML\left|
\frac{\de X^c}{\oom}\left(\pr_cX^e\right)-\frac{\om_B}{\hat{\oom}}\pr_{\phi}\!\left(\de X^e\right)+\de e_4(\hat{X}^e)\right||{\bf g}(\frac{\pr}{\pr\om^e},e_d)|\nn\\
&&\ML\ML\ML\leq c\left[\left(\frac{M^2}{r^3}\right)\frac{\ep_0}{r^{4+\de}}+\frac{\ep_0}{r^{4+\de}}\frac{M^2}{r^3}
+\frac{\ep_0}{r^{4+\de}}\frac{M^2}{r^3}\right]r\leq c\frac{M^2}{R_0^2}\frac{\ep_0}{r^{4+\de}}
\eea
\medskip


\NI {\bf iv)\ \  $e_4(\de X^e){\bf g}(\frac{\pr}{\pr\om^e},e_d)$\ :}
\medskip

\NI We have
\bea
\left|e_4(\de X^e){\bf g}(\frac{\pr}{\pr\om^e},e_d)\right|\leq |e_4(\de X^e)|r \leq c\frac{\ep_0}{r^{4+\de}}
\eea
as the following estimate holds, which we prove later on:
\bea
 |e_4(\de X^e)|\leq c\frac{\ep_0}{r^{3+\de}|u|^2}\ .\eql{toprove}
\eea
\medskip

\NI {\bf v)\ \  $\frac{\de\oom}{\oom^{(Kerr)}}e_4(\om_B){\bf g}(\frac{\pr}{\pr \phi},e_d)$\ :}
\medskip

\NI We have
\bea
\left|\frac{\de\oom}{\oom^{(Kerr)}}e_4(\om_B){\bf g}(\frac{\pr}{\pr \phi},e_d)\right|\leq c\frac{\ep_0}{r^{1+\de}|u|^2}\frac{M^2}{r^4}r
\leq c\frac{M^2}{R_0^2}\frac{\ep_0}{r^{4+\de}}\ .
\eea
We are left to prove the estimate \ref{toprove}, this is the content of the following lemma.
\begin{Le}\label{Comm3-}
Under the bootstrap assumptions the following inequality holds in $V_*$,
\bea
|e_4(\de X^e)|\leq c\frac{\ep_0}{r^3|u|^{2+\de}}\ .
\eea
\end{Le}
\NI{\bf Proof:} Recall that we have an evolution equation along the incoming cones for $\de X$, see the second equation eq. in \ref{traspdeOo}, but we do not have an equation along the outgoing cones. Therefore we have to derive the previous equation along $e_4$ to get a transport equation for $e_4(\de X^e)$. We start from the equation for $\de X$,
\bea
\oom\pr_{e_3}\de X^a-\oom(\pr_c X_{(Kerr)}^a)\de X^c=-F^a\de\oom+4\oom^2{\de\ze}^a\ .\eql{findeX2}
\eea
where
\[F^a=\oom\left(\frac{2}{\oom^{(Kerr)}}\frac{Q\De}{\Si R^2}\frac{\pr}{\pr r}X_{(Kerr)}^a\right)\equiv-\oom f^a\ \]
and $f^a$ is a completely known function which behaves as $O(M^2r^{-4})$ . Denoting 
\[\Psi^c\equiv \oom e_4(\de X^c)\] 
its evolution equation is
\bea
&&\ML\pr_{e_3}\Psi^c=(\pr_dX_{(Kerr)}^c)\Psi^d+\left(\pr_{e_4}(\oom\pr_dX_{(Kerr)}^c)\right)\!\de X^d+\pr_{e_4}\left[\oom f^c\de\oom+ 4\oom^2{\de\ze}^c\right]\nn\\
&&\ML-(\pr_{e_4}\log\oom)\!\left[\oom(\pr_d X_{(Kerr)}^c)\de X^d+\oom f^c\de\oom+ 4\oom^2{\de\ze}^c\right]+(\pr_{e_3}X^e)\pr_e\de X^c\ .\ \ \ \ \ \ \ \ \ \ \ \ \ \ \ \eql{eqPsi}
\eea
\NI{\bf Proof:}
\bea
&&\ML\oom\pr_{e_3}e_4(\de X^c)=\oom\pr_{e_3}\pr_{e_4}\de X^c=\oom\pr_{e_4}\pr_{e_3}\de X^c+\oom(\pr_{e_3}e_4^{\ro})\pr_{\ro}\de X^c\nn\\
&&\ML= \pr_{e_4}(\oom\pr_{e_3}\de X^c)-(\pr_{e_4}\log\oom)(\oom\pr_{e_3}\de X^c)+\oom\left(\left(\pr_{e_3}\frac{1}{\oom}\right)(\de^{\ro}_{\ub}+X^e\de^{\ro}_e)+\frac{1}{\oom}(\pr_{e_3}X^e)\de^{\ro}_e\right)\pr_{\ro}\de X^c\nn\\
&&\ML= \pr_{e_4}(\oom\pr_{e_3}\de X^c)-(\pr_{e_4}\log\oom)(\oom\pr_{e_3}\de X^c)-\oom(\pr_{e_3}\log\oom)(\pr_{e_4}\de X^c)
+(\pr_{e_3}X^e)\pr_e\de X^c\ .
\eea
therefore
\bea
&&\ML\pr_{e_3}\left(\oom e_4(\de X^c)\right)=\pr_{e_4}(\oom\pr_{e_3}\de X^c)-(\pr_{e_4}\log\oom)(\oom\pr_{e_3}\de X^c)+(\pr_{e_3}X^e)\pr_e\de X^c\nn\\
&&\ML=\pr_{e_4}\left[\oom(\pr_d X_{(Kerr)}^c)\de X^d+\oom f^c\de\oom+ 4\oom^2{\de\ze}^c\right]
-(\pr_{e_4}\log\oom)\left[\oom(\pr_d X_{(Kerr)}^c)\de X^d+\oom f^c\de\oom+ 4\oom^2{\de\ze}^c\right]\nn\\
&&+(\pr_{e_3}X^e)\pr_e\de X^c\nn\\
&&\ML=(\pr_dX_{(Kerr)}^c)(\oom{e_4}(\de X^d))+\left(\pr_{e_4}(\oom\pr_dX_{(Kerr)}^c)\right)\!\de X^d+\pr_{e_4}\left[\oom f^c\de\oom+ 4\oom^2{\de\ze}^c\right]\nn\\
&&\ML-(\pr_{e_4}\log\oom)\!\left[\oom(\pr_dX_{(Kerr)}^c)\de X^d+\oom f^c\de\oom+ 4\oom^2{\de\ze}^c\right]+(\pr_{e_3}X^e)\pr_e\de X^c\ .
\eea
Equation \ref{eqPsi} can be rewritten in a more compact way as
\bea
\pr_{e_3}\Psi^c\!&=&\!(\pr_d X_{(Kerr)}^c)\Psi^d
+\left\{\left(\pr_{e_4}(\oom\pr_d X_{(Kerr)}^c)\right)\!\de X^d+\pr_{e_4}(\oom f^c\de\oom)+(\pr_{e_3}X^e)\pr_e\de X^c\right.\nn\\
\!&-&\!\left.(\pr_{e_4}\log\oom)\!\left[\oom(\pr_d X_{(Kerr)}^c)\de X^d+\oom f^c\de\oom\right]\right\}
+(\pr_{e_4}\log\oom)4\oom^2{\de\ze}^c+ \pr_{e_4}(4\oom^2{\de\ze}^c)\ ,\nn
\eea
or
\bea
\pr_{e_3}\Psi^c=(\pr_d X_{(Kerr)}^c)\Psi^d+H^c+(\pr_{e_4}\log\oom)4\oom^2{\de\ze}^c+ \pr_{e_4}(4\oom^2{\de\ze}^c)
\eea
where it is easy to see that
\bea
|H^c|\leq c\left(1+\frac{M^2}{R_0^2}\right)\frac{\ep_0}{r^3|u|^{3+\de}}
\eea
Therefore if we prove that
\bea
|{\de\ze}^c|\leq c\frac{\ep_0}{r^3|u|^{2+\de}}\ \ ;\ \ |\pr_{e_4}(4\oom^2{\de\ze}^c)|\leq c\frac{\ep_0}{r^3|u|^{3+\de}}\eql{Lastproof}
\eea
it follows immediately that integrating along the incoming cones with the assumed initial data $\Psi^c$ satisfies the following bound
\bea
|\oom e_4(\de X^c)|\leq c\frac{\ep_0}{r^3|u|^{2+\de}}
\eea
proving the lemma.

\NI We are still left to prove inequalities \ref{Lastproof}, this is the content of the next lemma.
\begin{Le}\label{Comm3}
Under the bootstrap assumptions in the region $V_*$ the following inequalities hold
\bea
|{\de\ze}^c|\leq c\frac{\ep_0}{r^3|u|^{2+\de}}\ \ ;\ \ |\pr_{e_4}(4\oom^2{\de\ze}^c)|\leq c\frac{\ep_0}{r^3|u|^{3+\de}}\ .\eql{Lastproof1}
\eea
\end{Le}
\NI{\bf Proof:}
The first inequality follows immediately recalling that by the result on the correction of the connection coefficients we have
\[|\de\ze(e_a)|=\frac{\ep_0}{r^3|u|^{1+\de}}\]
and that $|e_a^b|=O(r^{-1})$. The second estimate is slightly more delicate as it involves $\pr_{e_4}\de\ze$, we have therefore to use the expression of $\ddb_4\de\ze$ which, apart not harmful terms, depends on $\de\b$. The estimate for $\de\b$ on the other side has been already obtained and allows to get the expected bound. The only apparent problem seems a loss of derivatives here as $e_4(\de X^a)$ is at the order of connection coefficients while $\b$ of a derivative of them. But, as already discusses, this is not harmful as in the next step when we are interested to the tangential derivatives for the connection coefficients, in this case to $\nabb\de\ze$, the equation we use is a elliptic Hodge equation which allows us to get an estimate for  $\nabb\de\ze$ without loosing derivatives, see \cite{Kl-Ni:book}, Chapter 4, Proposition 4.3.3. Therefore and this is a general phenomenon appearing in this approach, a loss of derivatives can appear in the estimate of the connection coefficients or of the metric components, but does not reproduce itself when we estimate their tangential derivatives.
\medskip

\NI{\bf Proof of Lemma \ref{hatestim}:} 

\NI We diagonalize the matrix $\ga=\{\ga_{ab}\}$ via an orthogonal matrix $U$, $\b=U\ga U^T$\ .

 \bea\mathbf{\ga}=\left(\begin{array}
{ccc}\ga^{(Kerr)}_{11}+\de\ga_{11}\!&\de\ga_{12}\\
\\
\de\ga_{21}&\!\ga^{(Kerr)}_{22}+\de\ga_{22}\\
\end{array}\right)
\eea
\[\mathbf{\b}=\left(\begin{array}{cccc}\cos\a & \sin\a\\
-\sin\a&\cos\a\\
\end{array}\right)\!\c\!\left(\begin{array}
{ccc}\ga^{(Kerr)}_{11}+\de\ga_{11}\!&\de\ga_{12}\\
\\
\de\ga_{21}&\!\ga^{(Kerr)}_{22}+\de\ga_{22}\\
\end{array}\right)\!\c\!\left(\begin{array}{cccc}\cos\a &-\sin\a\\
\sin\a&\cos\a\\
\end{array}\right)\]
The requirement that the matrix $\b$ be diagonal implies the conditions
\bea
&&\ML\ML\cos^2\a(\ga^{(Kerr)}_{11}+\de\ga_{11})+2\cos\a\sin\a\de\ga_{12}+\sin^2\a(\ga^{(Kerr)}_{22}+\de\ga_{22})=\b_{11}=\ga^{(Kerr)}_{11}+\de b_{11}+O(\ep_0^2)\nn\\
&&\ML\ML\sin^2\a(\ga^{(Kerr)}_{11}+\de\ga_{11})-2\cos\a\sin\a\de\ga_{12}+\cos^2\a(\ga^{(Kerr)}_{22}+\de\ga_{22})=\b_{11}=\ga^{(Kerr)}_{22}+\de b_{22}+O(\ep_0^2)\nn\\
&&\ML\ML\sin\a\cos\a(-(\ga^{(Kerr)}_{11}+\de\ga_{11})+(\ga^{(Kerr)}_{22}+\de\ga_{22}))+(\cos^2\a-\sin^2\a)\de\ga_{12}=0
\eea
where $\de\b_{11},\de\b_{22}$ are of order $\ep_0$\ . This implies that $\a=O(\ep_0)$. A standard calculation implies that the orthonormal frame in the region $V_*$ of the  perturbed Kerr spacetime can be written in the following way
\bea
&&\ML\ML e_1=\frac{1}{\sqrt{\ga_{11}\cos^2\a+\ga_{22}\sin^2\a-2\ga_{12}\sin\a\cos\a}}\left(\cos\a\sqrt{\ga^{(Kerr)}_{11}}e^{(Kerr)}_1,-\sin\a\sqrt{\ga^{(Kerr)}_{22}}e^{(Kerr)}_2\right)\nn\\
&&\ML\ML e_2=\frac{1}{\sqrt{\ga_{22}\cos^2\a+\ga_{11}\sin^2\a+2\ga_{12}\sin\a\cos\a}}\left(\sin\a\sqrt{\ga^{(Kerr)}_{11}}e^{(Kerr)}_1,\cos\a\sqrt{\ga^{(Kerr)}_{22}}e^{(Kerr)}_2\right)\nn
\eea
and keeping only the order $\ep_0$ we obtain the thesis.
\medskip

\NI{\bf Proof of Lemma \ref{deDexpr1}:} Observe first that 
\bea
&&\ML(\dd_4-\dd_4^{(Kerr)}){\hat O}=(e_4^{\mu}\dd_{\mu}-{e^{(Kerr)}}_4^{\mu}\dd^{(Kerr)}_{\mu}){\hat O}\\
&&=\left[(e_4^{\mu}-{e^{(Kerr)}}_4^{\mu})\dd_{\mu}+{e^{(Kerr)}}_4^{\mu}(\dd_{\mu}-\dd^{(Kerr)}_{\mu})\right]{\hat O}\nn\\
&&=-\frac{{\de\oom}}{\oom}\dd_4^{(Kerr)}{\hat O}+\frac{r\de X^c}{\oom}\dd_c{\hat O}+{e^{(Kerr)}}_4^{\mu}(\dd_{\mu}-\dd^{(Kerr)}_{\mu}){\hat O}\ \ \ \ .\nn
\eea

\NI Assuming for simplicity that $O$ be a covariant vector tangent to $S$,
\bea
&&\ML\big((\ddb_4-\ddb^{(Kerr)}_4){\hat O}\big)_{\mu}=(\Pi^{\ro}_{\mu}\dd_4-{\Pi^{(Kerr)}}^{\ro}_{\mu}\dd^{(Kerr)}_4){\hat O}_{\ro}\nn\\
&&\ML=(\Pi^{\ro}_{\mu}-{\Pi^{(Kerr)}}^{\ro}_{\mu})\dd_4{\hat O}_{\ro}+{\Pi^{(Kerr)}}^{\ro}_{\mu}(\dd_4-\dd^{(Kerr)}_4){\hat O}_{\ro}\nn\\
&&\ML=(\de\Pi^{\ro}_{\mu})\dd_4{\hat O}_{\ro}+\frac{(r\de X^c)}{\oom}\dd^{(Kerr)}_c{\hat O}_{\ro}
+{\Pi^{(Kerr)}}^{\ro}_{\mu}(\de\Ga^{\tau}_{\la\ro}){e^{(Kerr)}}_4^{\la}{\hat O}_{\tau}\ .\nn
\eea
\smallskip

\NI{\bf Proof of Lemma \ref{dePiexpr}:}
From equation \ref{Pieq1} we have
\bea
\Pi^{\si}_{\nu}=\de^{\si}_{\nu}-(\th^3_{\nu}e_3^{\si}+\th^4_{\nu}e_4^{\si})\ .
\eea
\NI The components of $\th^3$ and $\th^4$ satisfy, see \ref{Pieq2}:
\bea
&&\th^3_{u}=\oom\ \ ,\ \ \th^3_{\ub}=\th^3_{c}=0\nn\\
&&\th^4_{\ub}=\oom\ \ ,\ \ \th^4_{u}=\th^4_{c}=0\ .
\eea
Therefore
\bea
&&(\Pi^{\si}_{\nu}-{\Pi^{\si}_{\nu}}^{(Kerr)})=({\th^3}^{(Kerr)}_{\nu}{e_3^{\si}}^{(Kerr)}+{\th^4}^{(Kerr)}_{\nu}{e_4^{\si}}^{(Kerr)})-(\th^3_{\nu}e_3^{\si}+\th^4_{\nu}e_4^{\si})\nn\\
&&=({\th^3}^{(Kerr)}_{\nu}\!-\th^3_{\nu}){e_3^{\si}}^{(Kerr)}+\th^3_{\nu}({e_3^{\si}}^{(Kerr)}\!-e_3^{\si})+({\th^4}^{(Kerr)}_{\nu}\!-\th^4_{\nu}){e_4^{\si}}^{(Kerr)}+\th^4_{\nu}({e_4^{\si}}^{(Kerr)}\!-e_4^{\si})\nn
\eea
Observe that
\bea
&&({\th^3}^{(Kerr)}_{\nu}\!-\th^3_{\nu})=\de^u_{\nu}(\oom-{\oom^{(Kerr)}})=\de^u_{\nu}\de\oom\nn\\
&&({\th^4}^{(Kerr)}_{\nu}\!-\th^4_{\nu})=\de^{\ub}_{\nu}(\oom-{\oom^{(Kerr)}})=\de^{\ub}_{\nu}\de\oom
\eea
\bea
&&({e_3^{\si}}^{(Kerr)}\!-e_3^{\si})=\frac{\de\oom}{{\oom^{(Kerr)}}}e_3^{\si}\nn\\
&&({e_4^{\si}}^{(Kerr)}\!-e_4^{\si})=\frac{\de\oom}{{\oom^{(Kerr)}}}e_4^{\si}-\frac{1}{\oom}\de^{\si}_c(r\de X^c)\ .\nn
\eea
recalling agin that $[\de\oom]=L^0\ \ ,\ \ [\de X^c]=L^{-1}$. Therefore
\bea
&&\ML\ML(\Pi^{\si}_{\nu}-{\Pi^{\si}_{\nu}}^{(Kerr)})=\de^u_{\nu}\de\oom{e_3^{\si}}^{(Kerr)}+\th^3_{\nu}\frac{\de\oom}{{\oom^{(Kerr)}}}e_3^{\si}
+\de^{\ub}_{\nu}\de\oom{e_4^{\si}}^{(Kerr)}+\th^4_{\nu}\frac{\de\oom}{{\oom^{(Kerr)}}}e_4^{\si}-\th^4_{\nu}\frac{1}{\oom}\de^{\si}_c(r\de X^c)\nn\\
&&\ML\ML=\de^u_{\nu}{\oom^{(Kerr)}}\frac{\de\oom}{{\oom^{(Kerr)}}}{e_3^{\si}}^{(Kerr)}+\th^3_{\nu}\frac{\de\oom}{{\oom^{(Kerr)}}}e_3^{\si}
+\de^{\ub}_{\nu}{\oom^{(Kerr)}}\frac{\de\oom}{{\oom^{(Kerr)}}}{e_4^{\si}}^{(Kerr)}+\th^4_{\nu}\frac{\de\oom}{{\oom^{(Kerr)}}}e_4^{\si}-\th^4_{\nu}\frac{r\de X^c}{{\oom^{(Kerr)}}}\de^{\si}_c\nn\\
&&\ML\ML=\frac{\de\oom}{{\oom^{(Kerr)}}}\left[\left({\th^3}^{(Kerr)}_{\nu}{e_3^{\si}}^{(Kerr)}\!+{\th^3}_{\nu}{e_3^{\si}}\right)+\left({\th^4}^{(Kerr)}_{\nu}{e_4^{\si}}^{(Kerr)}\!+{\th^4}_{\nu}{e_4^{\si}}\right)\right]-\frac{r\de X^c}{{\oom^{(Kerr)}}}\th^4_{\nu}\de^{\si}_c\nn
\eea
Therefore
\bea
(\Pi^{\si}_{\nu}-{\Pi^{\si}_{\nu}}^{(Kerr)})=\de\oom A^{\si}_{\nu}+r\de X^cB^{\si}_{c\nu}
\eea
where
\bea
&&A^{\si}_{\nu}=\frac{1}{{\oom^{(Kerr)}}}\left[\left({\th^3}^{(Kerr)}_{\nu}{e_3^{\si}}^{(Kerr)}\!+{\th^3}_{\nu}{e_3^{\si}}\right)+\left({\th^4}^{(Kerr)}_{\nu}{e_4^{\si}}^{(Kerr)}\!+{\th^4}_{\nu}{e_4^{\si}}\right)\right]\nn\\
&&B^{\si}_{c\nu}=\frac{1}{{\oom^{(Kerr)}}}\th^4_{\nu}\de^{\si}_c
\eea
\medskip

\NI {\bf Proof of the estimates \ref{esta2}:} The following estimates are easily obtained from the bootstrap assumptions \ref{Oboosestwithdec}, using the previous estimates for the Riemann tensor $R$ and for its corrections $\de R$, see Lemma \ref{Lemma5} and inequality \ref{Kerrest}. Observe also that the estimate of
\[|r^{4-\frac{2}{p}}|u|^{2+\de}(\de\nabb)\chih^{(Kerr)}|_{p,S}\]
has been proved following the previous argument, see  the remark after \ref{3266}.
\bea
&&\ML\ML|r^{4-\frac{2}{p}}|u|^{2+\de}\oom\de\chih\Us^{(Kerr)}|_{p,S}\leq c|r^{2-\frac{2}{p}}|u|^{2+\de}\de\chih|_{p,S}|r^2\Us^{(Kerr)}|_{p,S}|_{\infty}\leq c\ep_0\frac{M^2}{R_0^2}\nn\\
&&\ML\ML|r^{4-\frac{2}{p}}|u|^{2+\de}\de\oom\chih^{(Kerr)}\Us^{(Kerr)}|_{p,S}\leq c|r^{1-\frac{2}{p}}|u|^{2+\de}\de\oom|_{p,S}|r\chih^{(Kerr)}|_{\infty}|r^2\Us^{(Kerr)}|_{\infty}\leq c\ep_0\frac{M^4}{R_0^4}\nn\\
&&\ML\ML|r^{4-\frac{2}{p}}|u|^{2+\de}\chih\nabb\de\chih|_{p,S}\leq c|r^{3-\frac{2}{p}}|u|^{2+\de}\nabb\de\chih|_{p,S}|r\chih|_{\infty}\leq c\ep_0\frac{M^2}{R_0^2}\nn\\
&&\ML\ML|r^{4-\frac{2}{p}}|u|^{2+\de}\chih(\de\nabb)\chih^{(Kerr)}|_{p,S}\leq c|r\chih|_{\infty}|r^{3-\frac{2}{p}}|u|^{2+\de}(\de\nabb)\chih^{(Kerr)}|_{p,S}\leq c\ep_0\frac{M^2}{R_0^2}\nn\\
&&\ML\ML|r^{4-\frac{2}{p}}|u|^{2+\de}\eta\chih\c\de\chih|_{p,S}\leq c|r\eta|_{\infty}|r\chih|_{\infty}|r^{2-\frac{2}{p}}|u|^{2+\de}\de\chih|_{p,S}
\leq c\ep_0\frac{M^4}{R_0^4}\nn\\
&&\ML\ML|r^{4-\frac{2}{p}}|u|^{2+\de}\eta\de\chih\c\chih^{(Kerr)}|_{p,S}\leq c|r\eta|_{\infty}|r\chih^{(Kerr)}|_{\infty}|r^{2-\frac{2}{p}}|u|^{2+\de}\de\chih|_{p,S}
\leq c\ep_0\frac{M^4}{R_0^4}\nn\\
&&\ML\ML|r^{4-\frac{2}{p}}|u|^{2+\de}\de\eta|\chih^{(Kerr)}|^2|_{p,S}\leq c|r\chih^{(Kerr)}|^2_{\infty}|r^{2-\frac{2}{p}}|u|^{2+\de}\de\eta|_{p,S}
\leq c\ep_0\frac{M^4}{R_0^4}\nn\\
&&\ML\ML|r^{4-\frac{2}{p}}|u|^{2+\de}\tr\chi\ \chih\c\de\etab|_{p,S}\leq c|r\tr\chi|_{\infty}|r\chih|_{\infty}|r^{2-\frac{2}{p}}|u|^{2+\de}\de\etab|_{p,S}
\leq c\ep_0\frac{M^4}{R_0^4}\nn\\
&&\ML\ML|r^{4-\frac{2}{p}}|u|^{2+\de}\tr\chi\ \de\chih\c\etab^{(Kerr)}|_{p,S}\leq c|r\tr\chi|_{\infty}|r\etab^{(Kerr)}|_{\infty}|r^{2-\frac{2}{p}}|u|^{2+\de}\de\chih|_{p,S}
\leq c\ep_0\frac{M^4}{R_0^4}\nn\\
&&\ML\ML|r^{4-\frac{2}{p}}|u|^{2+\de}\de\tr\chi\ \chih^{(Kerr)}\c\etab^{(Kerr)}|_{p,S}\leq c|r^{2-\frac{2}{p}}|u|^{2+\de}\de\tr\chi|_{p,S}|\chih^{(Kerr)}|_{\infty}|\etab^{(Kerr)}|_{\infty}\leq c\ep_0\frac{M^4}{R_0^4}\nn\\
&&\ML\ML|r^{4-\frac{2}{p}}|u|^{2+\de}\tr\chi\de\beta|_{p,S}\leq c|r\tr\chi|_{\infty}|r^{3-\frac{2}{p}}|u|^{2+\de}\tr\chi\de\beta|_{p,S}\leq c\ep_0\frac{M^2}{R_0^2}\nn\\
&&\ML\ML|r^{4-\frac{2}{p}}|u|^{2+\de}\de\tr\chi\b^{(Kerr)}|_{p,S}\leq c|r^{2-\frac{2}{p}}|u|^{2+\de}\de\tr\chi|_{p,S}|r^2\b^{(Kerr)}|_{\infty}
\leq c\ep_0\frac{M^2}{R_0^2}\nn
\eea
where we used the previous estimates for the Riemann tensor $R$ and for its corrections $\de R$, see Lemma \ref{Lemma5} and inequality \ref{Kerrest}.

\subsection{Various Kerr computations}

\NI Let us define $u,\ub$ as two functions solutions of the eikonal equation
\[g^{\mu\nu}\pr_{\mu}w\pr_{\nu}w=0\ ,\]
\bea
u=\frac{t-r_*(\theta,r)}{2}\ \ ;\ \ \ub=\frac{t+r_*(\theta,r)}{2}\ \ \ .
\eea
The Kerr metric in the Boyer-Lindquist coordinates $\{x^{\mu}\}=\{t,r,\theta,\phi\}$ is 
\bea
{\bf g}_{(Kerr)}(\c,\c)\!&=&\!-\frac{\left(\Delta-a^2\sin^2\theta\right)}{\Si}dt^2
-a\sin^2\theta\frac{(r^2+a^2-\Delta)}{\Si}\left(dt\!\otimes\! d\phi+d\phi\!\otimes\!dt\right)\nn\\ 
&+&\frac{\Si}{\Delta}dr^2+\Si d\theta^2+\frac{(r^2+a^2)^2-{\Delta}a^2\sin^2\theta}{\Si}\sin^2\theta d\phi^2\eql{2.24}
\eea
$a={J}/{M}$ and
\bea
&&\Si=r^2+a^2\cos^2\theta=r^2+a^2-a^2\sin^2\theta\nn\\
&&\Delta=r^2+a^2-2Mr=r^2\bigg(1-\frac{2M}{r}\bigg)+a^2\eql{3.148}\\
&&R^2=\frac{(r^2+a^2)^2-\Delta a^2\sin^2\theta}{\Sigma}=(r^2+a^2)+\frac{2Mra^2\sin^2\theta}{\Si}\nn
\eea
and \footnote{
\bea
&&g^{00}=-\frac{g_{\phi\phi}}{\lap\sin^2\theta}\ ,\ g^{0\phi}=\frac{g_{0\phi}}{\lap\sin^2\theta}\ ,\ g^{\phi\phi}=-\frac{g_{00}}{\lap\sin^2\theta}\nn\\
&&g^{rr}=g_{rr}^{-1}\ ,\ g^{\theta\theta}=g_{\theta\theta}^{-1}\ .\nn
\eea}
in the $\{u,\ub,\th_*,\phi\}$ coordinates
\bea
&&\ML{\bf g}_{(Kerr)}(\c,\c)\eql{Kerrmeta}\\
&&\ML=-4\oom_{(Kerr)}^2dud\ub+\ga^{(Kerr)}_{ab}\left(d\om^a-X_{(Kerr)}^a(du+d\ub)\right)\left(d\om^b-X_{(Kerr)}^b(du+d\ub)\right)\ ,\nn
\eea
where
\bea
X_{(Kerr)}^a=\om_B\de^a_{\phi}\ \ ,\ \ \om_B=\frac{2Mar}{\Si R^2}\ \ ,\ \ \oom=\sqrt{\frac{\Delta}{R^2}}\ .
\eea
The null (radial) geodesic vector field $L,\Lb$
\[L=L^{\ro}\frac{\pr}{\pr x^{\ro}}\ \ ,\ \ \Lb=\Lb^{\ro}\frac{\pr}{\pr x^{\ro}}\]
have the components $\Lb^{\ro}=-g^{\ro\mu}\pr_{\mu}\frac{(t+r_*)}{2}\ \  ,\ \  L^{\ro}=-g^{\ro\mu}\pr_{\mu}\frac{(t-r_*)}{2}$, therefore
\bea
&&\Lb^{0}=-\frac{g^{00}}{2}\ \ \Lb^{r}=-g^{rr}\frac{(\pr_rr_*)}{2} \ \ \Lb^{\theta}=-g^{\theta\theta}\frac{(\pr_{\theta}r_*)}{2}\ \ \Lb^{\phi}=-\frac{g^{\phi0}}{2}\nn\\
&&L^{0}=-\frac{g^{00}}{2}\ \ L^{r}=g^{rr}\frac{(\pr_rr_*)}{2} \ \ L^{\theta}=g^{\theta\theta}\frac{(\pr_{\theta}r_*)}{2}\ \ L^{\phi}=-\frac{g^{\phi0}}{2}
\eea
From these definitions with an easy calculation we obtain 
\bea
&&\Lb=\frac{1}{2\oom^2}\left(\frac{\pr}{\pr u}+\om_B\frac{\pr}{\pr\phi}\right)\nn\\
&&L=\frac{1}{2\oom^2}\left(\frac{\pr}{\pr\ub}+\om_B\frac{\pr}{\pr\phi}\right)
\eea
Let us define
\bea
&&e_3=\frac{1}{\oom}\left(\frac{\pr}{\pr u}+\om_B\frac{\pr}{\pr\phi}\right)\ \ ,\ \ 
e_4=\frac{1}{\oom}\left(\frac{\pr}{\pr\ub}+\om_B\frac{\pr}{\pr\phi}\right)\nn\\
&&\Nb=\left(\frac{\pr}{\pr u}+\om_B\frac{\pr}{\pr\phi}\right)\ \ ,\ \  N=\left(\frac{\pr}{\pr\ub}+\om_B\frac{\pr}{\pr\phi}\right)\ .
\eea
Therefore
\bea
&&T=\frac{\oom}{2}(e_3+e_4)=\frac{1}{2}\!\left(\frac{\pr}{\pr u}+\frac{\pr}{\pr \ub}\right)+\om_B\frac{\pr}{\pr\phi}=\frac{\pr}{\pr t}+\om_B\frac{\pr}{\pr\phi}\nn\\
&&T_0=\frac{\oom}{2}(e_3+e_4)-\om_B\frac{\pr}{\pr\phi}=\frac{\pr}{\pr t}
\eea
Recall that in general the following relation holds:
\bea
[N,\Nb]=-4{\oom}^2\ze(e_a)e_{a}
\eea
and computing explicitely the l.h.s. in the Kerr spacetime we obtain:
\bea
&&-4{\oom}^2\ze(e_a)e_{a}=\left[\left(\frac{\pr}{\pr\ub}+\om_B\frac{\pr}{\pr\phi}\right),\left(\frac{\pr}{\pr u}+\om_B\frac{\pr}{\pr\phi}\right)\right]\nn\\
&&=\left(\frac{\pr}{\pr\ub}-\frac{\pr}{\pr u}\right)\om_B\frac{\pr}{\pr\phi}=2\frac{\pr\om_B}{\pr r_*}\frac{\pr}{\pr\phi}
=2\frac{\pr r}{\pr r_*}\frac{\pr \om_B}{\pr r}\frac{\pr}{\pr\phi}=2\frac{Q\De}{\Si R^2}\frac{\pr \om_B}{\pr r}\frac{\pr}{\pr\phi}\nn\\
&&=2\frac{Q\De}{\Si R^2}R\sin\th\frac{\pr \om_B}{\pr r}e_{\phi}=-4\oom^2\left[-\frac{QR\sin\th}{2\Si}\frac{\pr \om_B}{\pr r}\right]e_{\phi}
\eea
Therefore we have proved that in the Kerr spacetime, with the null frame choice given before we have:
\bea
\ze(e_{\th_*})=0\ \ ;\ \ \ze(e_{\phi})=-\frac{QR\sin\th}{2\Si}\frac{\pr \om_B}{\pr r}\ .\eql{zeest}
\eea
We compute now the null components of the ${^{(T)}\!}\pi$ and the ${^{(T_0)}\!}\pi$ deformation tensors in Kerr, of course the second one is identically zero, but we are interested to see how the various parts compensate.
\smallskip

\NI{\bf The deformation tensors of $T$ and $T_0$.}
\smallskip

\NI {\bf${^{(T)}\!}\pi$:} Recall that
\[T=\frac{\oom}{2}(e_3+e_4)=\frac{\pr}{\pr t}+\om_B\frac{\pr}{\pr\phi}\ .\]
\begin{Le}\label{pi(T)exp}
In the Kerr spacetime ${^{(T)}\!}\pi$ is different from zero, but some components are zero. In fact we have
\bea
&&{^{(T)}\!}\pi(e_3,e_3)={^{(T)}\!}\pi(e_4,e_4)={^{(T)}\!}\pi(e_3,e_4)=0\nn\\
&&{^{(T)}\!}\pi(e_a,e_b)=(\dd_{\la}\om_B)R\sin\th(\de_{c\phi}\de_{d{\la}}+\de_{d\phi}\de_{c{\la}})e_a^ce_b^d\nn\\
&&{^{(T)}\!}\pi(e_a,e_3)=2\oom\ze(e_a);\ \ \ {^{(T)}\!}\pi(e_a,e_4)=-2\oom\ze(e_a)
\eea
\end{Le}
\NI{\bf Proof:} 
\bea
&&\ML\ML{^{(T)}\!}\pi(e_3,e_3)=\frac{1}{2}{\bf g}(\dd_3\oom(e_3+e_4),e_3)=-\oom\dd_3\log\oom+\frac{\oom}{2}{\bf g}(\dd_3e_4,e_3)=2\omb-2\omb=0\nn\\
&&\ML\ML={\bf g}(\dd_3\frac{\pr}{\pr t},e_3)+{\bf g}(\dd_3\om_B\frac{\pr}{\pr\phi},e_3)=\dd_3\om_B{\bf g}(\frac{\pr}{\pr\phi},e_3)=(\dd_3\om_B)R\sin\th{\bf g}(e_{\phi},e_3)=0\ .\ \ \ \ \ \ \ \ \ 
\eea
recalling that as $\frac{\pr}{\pr t}$ and $\frac{\pr}{\pr \phi}$ are Killing and implying
\[{\bf g}(\dd_3\frac{\pr}{\pr t},e_3)={\bf g}(\dd_3\frac{\pr}{\pr \phi},e_3)=0\ .\]
Analogous result for ${^{(T)}\!}\pi(e_4,e_4)$. Moreover
\bea
&&{^{(T)}\!}\pi(e_3,e_4)=\frac{1}{2}{\bf g}(\dd_3\oom(e_3+e_4),e_4)+\frac{1}{2}{\bf g}(\dd_4\oom(e_3+e_4),e_3)\nn\\
&&=-\left((\dd_3\oom)+(\dd_4\oom)\right)+\frac{\oom}{2}{\bf g}(\dd_3\oom e_3,e_4)+\frac{\oom}{2}{\bf g}(\dd_4\oom e_4,e_3)\nn\\
&&=2\oom(\omb+\om)+2\oom(\omb+\om)=4\oom(\om+\omb)\nn\\
&&={\bf g}(\dd_3\frac{\pr}{\pr t},e_4)+{\bf g}(\dd_3\om_B\frac{\pr}{\pr\phi},e_4)+{\bf g}(\dd_4\frac{\pr}{\pr t},e_3)+{\bf g}(\dd_4\om_B\frac{\pr}{\pr\phi},e_3)\nn\\
&&=(\dd_3\om_B){\bf g}(\dd_3\om_B\frac{\pr}{\pr\phi},e_4)+(\dd_4\om_B){\bf g}(\frac{\pr}{\pr\phi},e_3)=0\ .
\eea
\bea
&&{^{(T)}\!}\pi(e_3,e_a)=\frac{1}{2}{\bf g}(\dd_3\oom(e_3+e_4),e_a)+\frac{1}{2}{\bf g}(\dd_a\oom(e_3+e_4),e_3)\nn\\
&&=\frac{\oom}{2}{\bf g}(\dd_3(e_3+e_4),e_a)-\oom\nabb_a\log\oom+\frac{\oom}{2}{\bf g}(\dd_ae_4,e_3)\nn\\
&&=\frac{\oom}{2}{\bf g}(\dd_3e_4,e_a)-\oom\nabb_a\log\oom+\frac{\oom}{2}{\bf g}(\dd_ae_4,e_3)\nn\\
&&=\oom\left(\eta(e_a)-\nabb_a\log\oom+\ze(e_a)\right)=2\oom\ze(e_a)
\eea
\bea
&&{^{(T)}\!}\pi(e_a,e_b)=\frac{1}{2}{\bf g}(\dd_a\oom(e_3+e_4),e_b)+\frac{1}{2}{\bf g}(\dd_b\oom(e_3+e_4),e_a)\nn\\
&&=\frac{\oom}{2}{\bf g}(\dd_a(e_3+e_4),e_b)+\frac{\oom}{2}{\bf g}(\dd_b(e_3+e_4),e_a)={\oom}\left(\chi+\chib\right)(e_a,e_b)\nn\\
&&={\bf g}(\dd_a\frac{\pr}{\pr t},e_b)+{\bf g}(\dd_a\om_B\frac{\pr}{\pr \phi},e_b)
+{\bf g}(\dd_b\frac{\pr}{\pr t},e_a)+{\bf g}(\dd_b\omb\frac{\pr}{\pr \phi},e_a)\nn\\
&&=(\dd_a\om_B){\bf g}(\frac{\pr}{\pr \phi},e_b)+(\dd_b\om_B){\bf g}(\frac{\pr}{\pr \phi},e_a)
=(\dd_{\la}\om_B)R\sin\th(\de_{b\phi}\de_{a{\la}}+\de_{a\phi}\de_{b{\la}})\nn
\eea

\begin{Le}
In the Kerr spacetime the deformation tensor components of $T_0$ are all identically zero.
\end{Le}
\NI{\bf Proof:}
This follows immediately from the stationarity of the Kerr spacetime, we prove it directly by computation as these are useful when we do perturb the spacetime.
\bea
&&{^{(T_0)}\!}\pi(e_3,e_3)=\frac{1}{2}{\bf g}(\dd_3\oom(e_3+e_4),e_3)-{\bf g}(\dd_3X^c\frac{\pr}{\pr\om^c},e_3)\\
&&=-{\oom}\dd_3\log\oom+\frac{\oom}{2}{\bf g}(\dd_3e_4,e_3)+{\bf g}(X^c\frac{\pr}{\pr\om^c},\dd_3e_3)=2\omb-2\omb=0\nn\\
&&{^{(T_0)}\!}\pi(e_4,e_4)=0
\eea
\bea
&&{^{(T_0)}\!}\pi(e_3,e_4)=\frac{1}{2}{\bf g}(\dd_3\oom(e_3+e_4),e_4)+\frac{1}{2}{\bf g}(\dd_4\oom(e_3+e_4),e_3)\nn\\
&&-{\bf g}(\dd_3X^{\phi}\frac{\pr}{\pr\phi},e_4)-{\bf g}(\dd_4X^{\phi}\frac{\pr}{\pr\phi},e_3)\\
&&=4\oom(\om+\omb)-\left({\bf g}(\dd_3X^{\phi}\frac{\pr}{\pr\phi},e_4)+{\bf g}(\dd_4X^{\phi}\frac{\pr}{\pr\phi},e_3)\right)\nn\\
&&-{X^{\phi}}\left({\bf g}(\dd_3\frac{\pr}{\pr\phi},e_4)+\frac{1}{2}{\bf g}(\dd_4\frac{\pr}{\pr\phi},e_3)\right)=-{X^{\phi}}{^{(Z)}\!}\pi(e_3,e_4)=0\ .\nn
\eea
\bea
&&{^{(T_0)}\!}\pi(e_3,e_a)={^{(T)}\!}\pi(e_3,e_a)-{\bf g}(\dd_3X^{\phi}\frac{\pr}{\pr\phi},e_a)-{\bf g}(\dd_aX^{\phi}\frac{\pr}{\pr\phi},e_3)\nn\\
&&={^{(T)}\!}\pi(e_3,e_a)-(\dd_3X^{\phi})R\sin\th{\bf g}(e_{\phi},e_a)-X^{\phi}{\bf g}(\dd_3\frac{\pr}{\pr\phi},e_a)-X^{\phi}{\bf g}(\dd_a\frac{\pr}{\pr\phi},e_3)\nn\\
&&={^{(T)}\!}\pi(e_3,e_a)-(\dd_3X^{\phi})R\sin\th\de_{\phi a}-X^{\phi}{^{(Z)}\!}\pi(e_3,e_a)\\
&&={^{(T)}\!}\pi(e_3,e_a)-(\dd_3X^{\phi})R\sin\th\de_{\phi a}=\left(2\oom\ze(e_{\phi})-(\dd_3X^{\phi})R\sin\th\right)\de_{a\phi}\nn\\
&&=\left(2\oom\ze(e_{\phi})+\frac{1}{\oom}\frac{\pr}{\pr r_*}(\pr_r\om_B)R\sin\th\right)\de_{a\phi}=
\left(2\oom\ze(e_{\phi})+\frac{1}{\oom}\frac{Q\De}{\Si R^2}(\pr_r\om_B)R\sin\th\right)\de_{a\phi}\nn\\
&&=2\oom\left(\ze(e_{\phi})-\left[-\frac{QR\sin\th}{2\Si}\pr_r\om_B\right]\right)\de_{a\phi}=0\ .
\eea
\bea
&&{^{(T_0)}\!}\pi(e_a,e_b)=\frac{1}{2}{\bf g}(\dd_a\oom(e_3+e_4),e_b)+\frac{1}{2}{\bf g}(\dd_b\oom(e_3+e_4),e_a)\nn\\
&&-\left({\bf g}(\dd_aX^{\phi}\frac{\pr}{\pr\phi},e_b)+{\bf g}(\dd_bX^{\phi}\frac{\pr}{\pr\phi},e_a)\right)\nn\\
&&={^{(T)}\!}\pi(e_a,e_b)-\left({\bf g}(\dd_aX^{\phi}\frac{\pr}{\pr\phi},e_b)+{\bf g}(\dd_bX^{\phi}\frac{\pr}{\pr\phi},e_a)\right)\nn\\
&&={^{(T)}\!}\pi(e_a,e_b)-(\pr_{\la}\om_B)R\sin\th(\de_{a\la}\de_{b\phi}+\de_{a\phi}\de_{b\la})-{^{(Z)}\!}\pi(e_a,e_b)\nn\\
&&=\oom(\chi+\chib)(e_a,e_b)-e_{\la}(\om_B)R\sin\th(\de_{a\la}\de_{b\phi}+\de_{a\phi}\de_{b\la})=0\ .
\eea

\begin{cor}\label{Kerrcor123}
From the previous Lemma and \ref{zeest} it follows that in the Kerr spacetime
\bea
&&\oom\left(\chi+\chib\right)(e_a,e_b)=\big(e_{\la}(\om_B)R\sin\th\big)(\de_{b\phi}\de_{a{\la}}+\de_{a\phi}\de_{b{\la}})\nn\\
&&\ze(e_a)=\left(-\frac{QR\sin\th}{2\Si}{\pr_r \om_B}\right)\de_{a\phi}\nn\\
&&\om+\omb=0\ \ ,\ \ \tr\chi+\tr\chib=0
\eea
\end{cor}

\newpage

\end{document}